\pdfoutput=1
\documentclass[10pt, a4paper]{article}
\usepackage{amsmath}
\usepackage{amssymb}
\usepackage{amsfonts}
\usepackage{setspace}
\usepackage[]{graphicx}
\usepackage{tikz}
\usepackage{oldgerm}
\usepackage{fancyhdr} 
\usepackage{vmargin}
\usepackage{longtable}
\usepackage{lscape}
\usepackage{cite}

\usepackage{color}
\definecolor{MyDarkBlue}{rgb}{0.15,0.15,0.45}
\usepackage[linktocpage=true]{hyperref}
\hypersetup{
colorlinks=true,
citecolor=MyDarkBlue,
linkcolor=MyDarkBlue,
urlcolor=MyDarkBlue,
pdfauthor={Paul Richmond},
pdftitle={Multiple M-branes and 3-algebras},
pdfsubject={hep-th}
}

\usepackage[numbers,sort&compress]{natbib}
\usepackage{hypernat}
\usepackage{slashed}
\usepackage{appendix}
\usepackage{tensor}
\numberwithin{equation}{section} 

\newcommand{\be}{\begin{equation}}
\newcommand{\ee}{\end{equation} }
\newcommand{\beqa}{\begin{eqnarray} }
\newcommand{\eeqa}{\end{eqnarray} }
\newcommand{\ba}{\begin{array}}
\newcommand{\ea}{\end{array}}

\newcommand\cL{{\cal L}}

\newcommand\tr{{\rm \, tr}}

\newcommand\sfrac[2]{{\textstyle\frac{#1}{#2}}}
\newcommand{\eref}[1]{Eq.\,(\ref{#1})}

\begin{document}

\allowdisplaybreaks 

\setcounter{secnumdepth}{5}

\setpapersize{A4}
\setmarginsrb{40mm}{20mm}{20mm}{20mm}{12pt}{11mm}{0pt}{11mm}

\pagestyle{fancy}
\rhead{}
\chead{}
\cfoot{\thepage}
\title{Multiple M-branes and 3-algebras}
\author{Paul Richmond \footnote{email: paul.richmond@kcl.ac.uk}\\  \\{\itshape Department of Mathematics, King's College London,\/}\\{\itshape The Strand, London WC2R 2LS, U.K.\/}}

\begin{titlepage}
\begin{flushright}
\end{flushright}
\begin{center}
\vspace{10pt}
\resizebox{5cm}{!}{\includegraphics{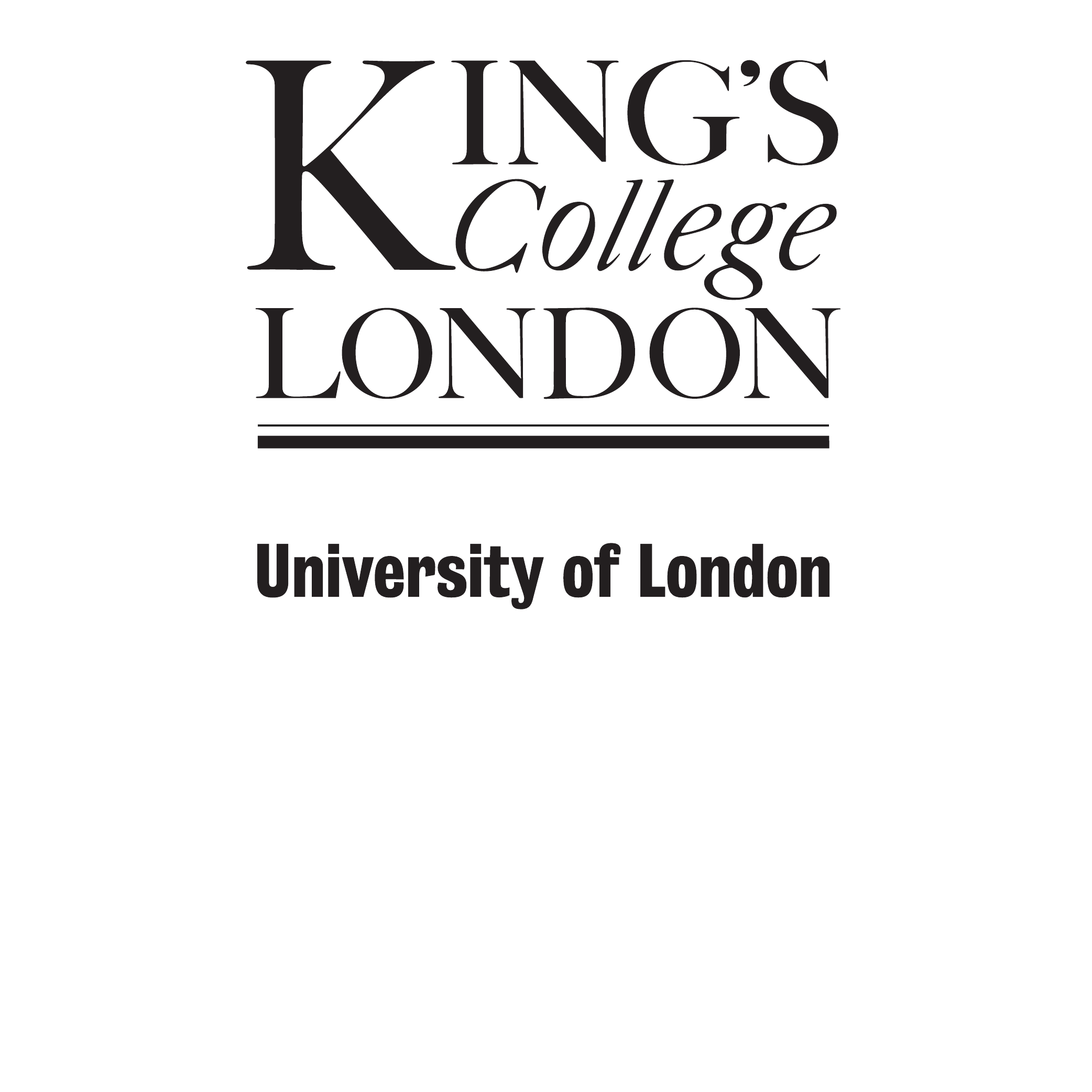}}
\end{center}
\vspace{60pt}
\centering{\LARGE{\bfseries Multiple M-branes and 3-algebras}}\\
\vspace{30pt}
\def\thefootnote{\fnsymbol{footnote}}
Paul Richmond\footnote{\href{mailto:paul.richmond@kcl.ac.uk}{email: paul.richmond@kcl.ac.uk}} \\
\setcounter{footnote}{0}
\vspace{10pt}
{\itshape Department of Mathematics, King's College London,\/}\\
{\itshape Strand, London WC2R 2LS, U.K.\/}\\
\vspace{200pt}
\normalsize{Submitted in partial fulfilment of the requirements for the degree of PhD at}\\
\normalsize{King's College London}\\
\normalsize{Supervisor: Professor Neil Lambert}\\
\normalsize{Submitted in September 2012}\\
\end{titlepage}

\begin{spacing}{1.5}
\clearpage
\newpage

%
%
\begin{center}
\LARGE{Abstract} \\ \ \\
%
%
\end{center}
\normalsize
\begin{slshape}
M-theory is well-known but not well-understood. It arises as an umbrella theory that unifies the various perturbative string theories into a single nonperturbative theory. In its strong coupling phase M-theory does not possess string states but rather M2-branes and M5-branes.
The purpose of this thesis is to explore the properties of multiple coincident M2- and M5-branes. It is based on the author's papers \cite{Lambert:2009qw,Lambert:2011gb} (in collaboration with Neil Lambert), \cite{Jeon:2012fn} (in collaboration with Imtak Jeon and Neil Lambert) and \cite{Richmond:2012by}. 

We begin with a review of the construction of three-dimensional $\mathcal{N}=8$ and $\mathcal{N}=6$ supersymmetric Chern-Simons-matter theories. These include the BLG and ABJM models of multiple M2-branes and our focus will be on their formulation in terms of 3-algebras. 

%
%
We then examine the coupling of multiple M2-branes to the background 3-form and 6-form gauge fields of eleven-dimensional supergravity. In particular we show in detail how a natural generalisation of the Myers flux-terms, along with the resulting curvature of the background metric, leads to mass terms in the effective field theory.

%
%
Working to lowest nontrivial order in fermions, we demonstrate the supersymmetric invariance of the four-derivative order corrected Lagrangian of the Euclidean BLG theory and determine the theory's higher derivative corrected supersymmetry transformations. The supersymmetry algebra is also shown to close on the scalar and gauge fields.

%
%
We also consider periodic arrays of M2-branes in the ABJM model in the spirit of a circle compactification to D2-branes in type IIA string theory. The result is a curious formulation of three-dimensional maximally supersymmetric Yang-Mills theory. Upon further T-duality on a transverse torus we obtain a non-manifest-Lorentz-invariant description of five-dimensional maximally supersymmetric Yang-Mills which can be viewed as an M-theory description of M5-branes on $\mathbb{T}^3$.

%
%
After reviewing work to describe multiple M5-branes using 3-algebras we show how the resulting novel system of equations reduces to one-dimensional motion on instanton moduli space. Quantisation leads to the previous light-cone proposal of the (2,0) theory, generalised to include a potential that arises on the Coulomb branch as well as couplings to background gauge and self-dual 2-form fields.
\end{slshape}

\newpage
\begin{center}
\LARGE{Acknowledgements} \\ \ \\
\end{center}
%
I would like to thank Neil Lambert for his guidance, supervision and support throughout both my MSc and PhD. It has been a pleasure to be his student. I am also grateful to Costis Papageorgakis for the careful explanation of his work and his patience in answering my numerous questions. Thank you also to the other PhD students whose company and insight I have shared over the last four years. In particular, I'm grateful to Finn and Mehmet for their ability to appear interested as I ranted about higher derivative corrections. I reserve my greatest thanks for Bec as this PhD would not have been possible without her support of every conceivable kind. 

Finally, I dedicate this thesis to my late father whose engineering textbooks triggered my fascination with mathematics at an early age. It is an enormous regret to me that he is not here to witness the completion of this work.
\newpage

\tableofcontents
\listoffigures

\newpage
\section{Introduction} \label{c_Introduction}

The search for a unified theory of quantum gravity has lead to the development of string theory. Here the point particles of the familiar four-dimensional world are replaced by one-dimensional vibrating strings. For a consistent, anomaly free, theory of supersymmetric strings one must make the conceptual leap to a world which is ten-dimensional. Once this leap has been made it is a simple matter to accept the possibility of even higher dimensional spacetime and the extended string-like objects known as branes. We now know of the existence of five string theories each with its own idiosyncrasies. It is astonishing that these seemingly disparate string theories are actually all interconnected via dualities.

Branes have become an essential part of string theory. The Dirichlet boundary conditions imposed on open strings imply that the string end points are fixed to a D-brane embedded in the ten-dimensional spacetime. The open strings ending on D-branes carry $U(1)$ charges. When $N$ parallel D-branes become coincident the open strings stretching between them become light and the $U(1)^N$ symmetry is enhanced to $U(N)$. At low energy the dynamics of $N$ D$p$-branes is described by a non-Abelian gauge theory which is simply the dimensional reduction of ten-dimensional supersymmetric $U(N)$ Yang-Mills to $p+1$ dimensions \cite{Witten:1995im}.

String interactions are described by an infinite perturbative expansion in the string self-coupling parameter $g_s$. That this expansion is UV finite is a great success and one of the main reasons for the adoption of string theory. A priori, there is no need for $g_s$ to be small and therefore because $g_s$ could be large, the perturbative expansion cannot tell us everything about string theory. The masses of D-branes are proportional to $1/g_s$ so that in the perturbative regime where the string coupling is small, the D-branes are energetically unfavourable and they can be neglected. Alternatively, as $g_s$ is increased the perturbative string expansion becomes less well behaved but the D-branes are now light and their effects dominate. Therefore the properties of D-branes give us an understanding of string theory beyond its perturbative expansion. The discovery of D-branes has revolutionised string theory, allowing for a refined understanding of how non-Abelian gauge theories may be incorporated into it and other important advances such as the AdS/CFT conjecture \cite{Maldacena:1997re}. 

Another way to explore the string theories is to examine their low energy limits which are the ten-dimensional supergravity theories. Here, only the massless fields in the string spectrum are considered and the theories are not UV finite. In this way we can think of the string theories as being the UV completion of the appropriate ten-dimensional supergravity.
There also exists a supergravity theory in eleven dimensions \cite{Cremmer:1978km}. Under a small set of assumptions, this theory is unique and in keeping with other non-topological gravity theories it has UV divergences. Just as the various ten-dimensional supergravity theories are the low energy limits of the UV complete string theories, eleven-dimensional supergravity is the low energy limit of an eleven-dimensional UV complete theory dubbed M-theory. Beyond eleven dimensions there are no interacting supersymmetric theories without including unwanted massless fields of spin greater than two \cite{Nahm:1977tg}. We now consider M-theory to be the unique umbrella theory that unifies the various perturbative string theories \cite{Witten:1995ex} but its specific formulation is not well understood. 
Of particular relevance to this thesis is the connection between type IIA string theory and M-theory. Quite simply, the value of the string coupling constant in type IIA string theory is given by the radius of an additional circular spatial dimension
\begin{equation}
R_{11} = l_s g_s \, ,
\end{equation}
where $l_s$ is the string length.
In the infinite coupling limit this M-theory circle decompactifies into a genuine non-compact dimension which is indistinguishable from all the others. As $g_s$ was the only coupling parameter in type IIA string theory it implies that M-theory does not have a continuous coupling parameter. Hence it is necessarily strongly coupled and nonperturbative.

Despite being strongly coupled, we may still learn some of the features of M-theory from its low energy limit because of supersymmetry. The superalgebra associated with eleven-dimensional supergravity is
\begin{equation}
\{ Q_\alpha , Q_\beta \} = ( \Gamma^M C^{-1} )_{\alpha \beta} P_M + \frac{1}{2!} ( \Gamma^{MN} C^{-1} )_{\alpha \beta} Z_{MN} + \frac{1}{5!} ( \Gamma^{MNOPQ} C^{-1} )_{\alpha \beta} Z_{MNOPQ} \, , \label{SUGRA_susy}
\end{equation}
where we have chosen the Majorana representation for which $C=\Gamma_0$. The right hand side of \eref{SUGRA_susy} exhausts the possible central charges that my be added. There are two ways of seeing this. Firstly, the left hand side is symmetric in the $\alpha, \beta$ spinor indices and on the right hand side, $\Gamma^M C^{-1}$, $\Gamma^{MN} C^{-1}$ and $\Gamma^{MNOPQ} C^{-1}$ are the only matrices with the same property.\footnote{The matrix $\Gamma^{MNOPQR} C^{-1}$ is also symmetric but in eleven dimensions it is dual to $\Gamma^{MNOPQ} C^{-1}$.} Secondly, the number of independent charge components on each side matches: 528 from the anti-commutator of the $32 \times 32$ Majorana supercharges and $528=11+55+462$ from the eleven-dimensional vector, 2-form and 5-form. From the superalgebra in \eref{SUGRA_susy} we can identify the objects which exist in eleven-dimensional supergravity. For example, the spatial components of the 2-form and 5-form central charges are associated with three-dimensional and six-dimensional hypersurfaces respectively. The existence of these branes can also be deduced from the presence of the background 3-form in the field content of eleven-dimensional supergravity. These branes are examples of an important class of states called BPS states. To see how these states are important we can take $P_M = ( E , 0 , \ldots, 0 )$, where the energy $E$ is the product of the brane's tension $T_q$ and its spatial volume. The central charge $Z_{M_1 \cdots M_q}$ is the product of the brane's charge $Q_q$ and its spatial volume. The left hand side of the supersymmetry algebra is positive definite and consequently from \eref{SUGRA_susy} can be deduced a relationship of the form 
\begin{equation}
T_q \geq | Q_q | \, .
\end{equation}
BPS states saturate this bound i.e.
\begin{equation}
T_q = | Q_q | \, . \label{BPS_saturated}
\end{equation}
The maximally supersymmetric 2- and 5-brane of eleven-dimensional supergravity are BPS states and so are D-branes in string theory. The critical feature of BPS states is that generically the relation in \eref{BPS_saturated} is not altered by quantum effects. Hence we may make claims about BPS states in a low energy effective theory and these claims almost always hold true at strong coupling. This argument shows us that the 2- and 5-brane in eleven-dimensional supergravity also exist in M-theory. These are the M2-brane and the M5-brane, collectively known as M-branes, and it is hoped that a thorough understanding of their properties and dynamics will illuminate M-theory beyond its eleven-dimensional supergravity approximation.\footnote{Other components of the central charges give the Kaluza-Klein monopole and M9-brane, whereas the momentum leads to the M-wave. In this thesis we will focus exclusively on the M2- and M5-brane.} The eleven-dimensional superalgebra in \eref{SUGRA_susy} also allows for M2- and M5-brane intersections and shows, perhaps surprisingly, that M-theory does not contain strings.

As both M-branes and type IIA D-branes are BPS objects we can make precise statements about the relation between them \cite{Townsend:1995af}. If the M-theory circle is transverse to the M2-brane worldvolume then in the low energy limit, $R_{11} \rightarrow 0$, it becomes a D2-brane. An M5-brane whose worldvolume wraps around the M-theory circle becomes a D4-brane in type IIA string theory. We can therefore use D2- and D4-branes as a guide to M-brane physics by uplifting to eleven dimensions.

A single instance of either type of M-brane is fairly well understood, both from the worldvolume action perspective and their behaviour as charged blackhole solutions of supergravity. We can then imagine $N$ parallel M2- or M5-branes becoming coincident and interacting. The supergravity description of these systems is again straightforward because they behave as single M-brane blackhole solutions but with $N$ units of charge. The worldvolume perspective is easy to state: the worldvolume field theory on $N$ M2- or M5-branes is the strong coupling limit of three- or five-dimensional supersymmetric $U(N)$ Yang-Mills respectively. The lack of a continuous coupling parameter in M-theory implies these strong coupling limits are conformal fixed points of the three- and five-dimensional theories. Whilst it is a simple matter to claim what the worldvolume M-brane theories should be, finding an explicit mathematical description of them is not straightforward and in the case of multiple M5-branes remains elusive.

The blackhole entropy of $N$ M2- or M5-branes may be calculated in supergravity and the results show that it scales like $N^{3/2}$ and $N^3$ respectively. The analogous calculation for D-branes yields $N^2$ which are the $U(N)$ matrix degrees of freedom arising from  the open string end points.  The M-brane scalings are unusual although in the case of the M2-branes it could in principle be understood as arising from an $N \times N$ matrix with constraints removing some degrees of freedom. The M5-brane scaling is particularly perplexing as additional degrees of freedom are gained going to strong coupling. Recently there has been work which claims to see the $N^3$ scaling from five-dimensional supersymmetric Yang-Mills. Uplifting the open string ending on a D-brane system to eleven dimensions suggests that the degrees of freedom of M-theory may be explained by looking at the M2-brane ending on an M5-brane.

By considering a configuration in which multiple coincident D1-branes end on a D3-brane in type IIB string theory and lifting to eleven dimensions, Basu and Harvey \cite{Basu:2004ed} were able to propose a BPS equation for multiple M2-branes ending on an M5-brane. In a paper motivated by this work, Bagger and Lambert \cite{Bagger:2006sk} constructed non-gauged supersymmetry transformations from which the Basu-Harvey BPS equation can be derived but which also indicated the presence of a novel gauge symmetry. In a follow-up paper \cite{Bagger:2007jr} they successfully incorporated gauge fields and demonstrated that the supersymmetry transformations closed provided the fields satisfied certain equations of motion.\footnote{Independently, Gustavsson \cite{Gustavsson:2007vu} also suggested a set of multiple M2-brane supersymmetry transformations using an algebraic structure seemingly different to Bagger and Lambert's. However, the two proposals were shown to be equivalent in \cite{Bagger:2007vi}.} These field equations were then used to deduce a candidate Lagrangian for multiple M2-branes,  the striking feature of which was the appearance of an algebraic structure christened a 3-algebra. The discovery of this three-dimensional, interacting, non-Yang-Mills type Lagrangian prompted a cascade of research. Bagger and Lambert's pioneering work has now largely been superseded by the ABJM/ABJ \cite{Aharony:2008ug,Aharony:2008gk} model of multiple M2-branes. The usual description of ABJM/ABJ is as a bifundamental Chern-Simons-matter theory but there is also a formulation in terms of 3-algebras \cite{Bagger:2008se}.

The remainder of this thesis is as follows: in chapter \ref{c_Multiple_M2s_and_3_algebras} we review the construction of theories of multiple M2-branes and some of their properties.
In chapter \ref{c_M2s_and_background_fields} we determine the coupling of multiple M2-branes to the background 3-form field, as reported in \cite{Lambert:2009qw}.
In chapter \ref{c_Higher_Derivative_BLG} we determine the four-derivative order corrections to the Bagger-Lambert Lagrangian and supersymmetry transformations, as reported in \cite{Richmond:2012by}.
In chapter \ref{c_Periodic_arrays_of_M2s} we consider a different way to compactify multiple M2-branes on a circle to multiple D2-branes, as reported in \cite{Jeon:2012fn}.
In chapter \ref{c_Multiple_M5s_and_3_algebras} we review the properties of M5-branes and an attempt to describe their dynamics with a non-Abelian representation of the $(2,0)$ tensor multiplet constructed using 3-algebras.
In chapter \ref{c_Light_cone_M5s} we determine the solutions to the 3-algebra (2,0) equations of motion for the case of a null vacuum expectation value assigned to the auxiliary field and find they give one-dimensional quantum mechanics on the instanton moduli space as reported in \cite{Lambert:2011gb}.
Finally, in chapter \ref{c_Conclusions} we offer some concluding remarks and an outlook for further work.
Also included are appendices which provide further details of the higher derivative calculations featured in the main body of this work.

\newpage
\section{Multiple M2-branes and 3-algebras} \label{c_Multiple_M2s_and_3_algebras}

In this chapter we will review the construction of models of coincident multiple M2-branes. The subject has enjoyed remarkable attention in the last five years and because of this we will only be able to provide a small glimpse of the progress that has been made. A thorough review of multiple M2-branes can be found in \cite{Bagger:2012jb}. The rest of this chapter is as follows. In section \ref{s_BLG_and_real_3_algebras} we will look at the 3-algebra construction of the Bagger-Lambert-Gustavsson (BLG) model closely following the presentation in \cite{Bagger:2006sk,Bagger:2007jr,Bagger:2007vi} (the supersymmetry algebra was independently shown to close in \cite{Gustavsson:2007vu}). We will also mention the interpretation of the BLG model given by the moduli space of the theory and the `novel Higgs mechanism'. We also briefly discuss a wider class of non-Euclidean 3-algebra BLG theories and their drawbacks. In section \ref{s_ABJM_and_complex_3_algebras} we will discuss the ABJM model \cite{Aharony:2008ug} and its formulation in terms in complex 3-algebras \cite{Bagger:2008se}. 

\subsection{BLG and Real 3-algebras} \label{s_BLG_and_real_3_algebras}

The field content of a theory describing multiple M2-branes should possess eight scalar fields, parametrising directions transverse to the worldvolume, as well as their fermionic superpartners which correspond to broken supersymmetries. The presence of one or more branes breaks the Poincar\'e symmetry from $SO(1,10)$ to $SO(1,2) \times SO(8)$ with $SO(8)$ being the R-symmetry which acts on the fields. The M2-branes also preserve/break one-half of the background supersymmetry. This manifests itself as a projection condition on the supersymmetry parameter $\epsilon$ and fermion $\psi$:
\begin{align}
\Gamma_{012} \epsilon \ &= \ + \epsilon \, , \\[10pt]
\Gamma_{012} \psi \ &= \ - \psi \, , 
\end{align}
where the first condition corresponds to the preserved supersymmetries and the second to the broken supersymmetries. The fermion $\psi$ is a Majorana spinor in eleven dimensions and as such has $\sfrac{1}{2} \cdot 2^{[11/2]}=32$ real components. The second projection condition reduces the number of spinor components to 16 and this is further reduced to 8 real components once we are on-shell. Supersymmetry dictates that the on-shell degrees of freedom contributed by the bosonic fields must equal those contributed by the fermionic fields. We see that this is the case for the field content of multiple M2-branes and therefore precludes the addition of degrees of freedom from other fields for example, from gauge fields. Nevertheless, we can proceed by focusing solely on the scalar-spinor sector. 

The supersymmetry transformations for a free M2-brane are \cite{Bergshoeff:1987cm}
\begin{align}
\delta X^I \ &= \ i\bar \epsilon \Gamma^I \psi \label{abelian_M2_s} \, , \\[10pt]
\delta \psi \ &= \ \partial_\mu X^I \Gamma^\mu \Gamma^I \epsilon \, , \label{abelian_M2_f}
\end{align}
where $\mu , \nu , \ldots= 0,1,2$ are the M2-brane worldvolume coordinates and $I, J , \ldots = 3,4, \ldots, 10$ label the eight directions transverse to the worldvolume.
In order to construct an interacting M2-brane theory involving the scalar and fermion fields, we assume that they take values in some real vector space $\mathcal{A}$. This is analogous to the multiple D-brane situation where the fields are valued in a non-Abelian Lie-algebra. However, in the multiple M2-brane case we make no assumptions on the form of the real vector space $\mathcal{A}$. A set of multiple M2-brane supersymmetry transformations must respect the symmetries of the theory and this places restrictions on the possible additions to the free-field terms in Eqs.\,\eqref{abelian_M2_s} and \eqref{abelian_M2_f}. Assuming canonical kinetic terms for the fields we know that in three dimensions the mass dimensions of the fields are
\begin{equation}
[ X ] = + \frac{1}{2} \, , \qquad [ \psi ] = + 1 \, , \qquad [ \epsilon ] = - \frac{1}{2} \, .
\end{equation}
%
In fact, the scalar variation must have the same form as the free theory as this is the only transformation consistent with mass dimensions. Furthermore, we know that $\psi$ and $\epsilon$ have opposite chirality with respect to $\Gamma_{012}$ and the fermion supersymmetry transformation should respect this. This constrains the $\Gamma$-matrix structure in the $\delta \psi$ term to contain an odd number of transverse indices. Dimensional analysis dictates that $\delta \psi$ can contain only a derivative of a scalar field, which is simply the free-field term \eqref{abelian_M2_f}, and a cubic scalar term. With these considerations the supersymmetry transformations must be of the form
\begin{align}
\delta X^I &= i\bar \epsilon \Gamma^I \psi \, , \label{M2_starting_point_s} \\[10pt]
\delta \psi &= \partial_\mu X^I \Gamma^\mu \Gamma^I \epsilon +  \kappa [ X^I, X^J , X^K ] \Gamma^{IJK} \epsilon \, , \label{M2_starting_point_f}
\end{align}
where the triple product $[X^I , X^J , X^K]$ is antisymmetric and linear in each of the fields and $\kappa$ is a dimensionless constant.  We note that there could be other cubic terms that are not totally antisymmetric in $I,J,K$, and will return to this point later in this section. There is another reason for the presence of the cubic scalar term. Setting $\delta \psi = 0$ gives rise to a BPS equation akin to that proposed by Basu and Harvey \cite{Basu:2004ed} as we now show.
If we take an M2-M5 brane configuration in which multiple M2-branes lie in the $x^0,x^1,x^2$ plane and an M5-brane in the $x^0,x^1,x^3,x^4,x^5,x^6$ directions then the preserved supersymmetries satisfy $\Gamma_{012} \epsilon = \epsilon$ and $\Gamma_{013456} \epsilon = \epsilon$. It follows that the common preserved supersymmetries in the M2-M5 system satisfy $\Gamma_2 \epsilon = \Gamma_{3456} \epsilon$. The fluctuations of the M2-branes that lie along the M5-brane are $X^{i}$ with $i \in \{3,4,5,6\}$. We look for solutions in which only $X^{i}$ are nonzero and moreover they depend solely on $x^2$ which is the M2-brane worldvolume direction orthogonal to the M5-brane. The condition $\Gamma_2 \epsilon = \Gamma_{3456} \epsilon$ is equivalent to $\Gamma^{ijk} \epsilon = \varepsilon^{ijkl} \Gamma^2 \Gamma^{l} \epsilon$. 
The BPS equation for this configuration is \cite{Bagger:2007jr,Bagger:2007vi}
\begin{equation}
\frac{d X^{i}}{d (x^2)} = \kappa \, \varepsilon^{ijkl}  [ X^j , X^k , X^l ] \, , \label{Basu-Harvey}
\end{equation}
and is essentially the Basu-Harvey equation \cite{Basu:2004ed}.

%
%
%

Commuting the proposed supersymmetry transformations in \eqref{M2_starting_point_s} and \eqref{M2_starting_point_f} on the scalar fields gives
\begin{equation}
[ \delta_1 , \delta_2 ] X^I = 2i \bar{\epsilon}_2 \Gamma^\mu \epsilon_1 \partial_\mu X^I + 6 \kappa i \bar \epsilon_2 \Gamma_{JK} \epsilon_1 [X^J , X^K , X^I] \, . \label{scalar_closure}
\end{equation}
The first term in \eref{scalar_closure} is a translation. The second term
\begin{equation}
\delta X^I \propto i \bar{\epsilon}_2 \Gamma_{JK} \epsilon_1 [X^J , X^K , X^I] \label{gauget} \, ,
\end{equation} 
can be viewed as a gauged version of the global symmetry transformation
\begin{equation}
\delta X = [\alpha, \beta , X] \, , \label{general_gauget}
\end{equation}
where $\alpha, \beta \in \mathcal{A}$. It is useful to introduce a basis for the algebra $\mathcal{A}$ involving some generators $T^a$ where $a = 1, \ldots , N$ and $N$ is the dimension of $\mathcal{A}$. The structure constants associated with this algebra are defined by
\begin{equation}
[T^a, T^b , T^c] = f^{abc}{}_{d} T^d \label{quibit}
\end{equation}
and they inherit the total antisymmetry of the triple product which immediately implies 
\begin{equation}
f^{abc}{}_{d} = f^{[abc]}{}_{d} \, .
\end{equation}
In this case, the symmetry transformation \eqref{general_gauget} can be expressed generally as
\begin{equation}
\delta X^I_a = f^{cdb}{}_{a} \Lambda_{cd} X^I_b \equiv \tilde{\Lambda}^b{}_{a} X^I_b \, , \label{final_gauget}
\end{equation}
with \eref{gauget} corresponding to the choice $\tilde{\Lambda}^b{}_{a} \propto i \bar{\epsilon}_2 \Gamma_{JK} \epsilon_1 X^J_c X^K_d f^{cdb}{}_a$. In order to promote this global symmetry to a local symmetry a covariant derivative is defined such that 
\begin{equation}
\delta (D_\mu X) = \delta (D_\mu) X + D_\mu (\delta X) \, .
\end{equation}
Due to the form of the local transformation in \eref{final_gauget}, a natural choice for the covariant derivative is
\begin{equation}
(D_\mu X)_a = \partial_\mu X_a - \tilde{A}_\mu{}^b{}_{a} X_b \, , \label{cov4}
\end{equation}
with $\tilde{A}_\mu{}^b{}_{a} = f^{cdb}{}_{a} A_{\mu cd}$. 
The gauge field strength is defined as 
\begin{equation}
([D_\mu , D_\nu] X)_a = \tilde{F}_{\mu \nu}{}^b{}_{a} X_b \, ,
\end{equation}
from which it follows
\begin{equation}
\tilde F_{\mu\nu}{}^b{}_a = \partial_\nu \tilde{A}_\mu{}^b{}_a - \partial_\mu \tilde{A}_\nu{}^b{}_a + \tilde{A}_\nu{}^b{}_c \tilde{A}_\mu{}^c{}_a - \tilde{A}_\mu{}^b{}_c \tilde{A}_\nu{}^c{}_a \, .
\end{equation}
The associated Bianchi identity is $D_{[\mu} \tilde{F}_{\nu \lambda]}{}^b{}_a =0$. Having introduced the gauge field $\tilde{A}_\mu{}^b{}_{a}$ we can write down the following set of supersymmetry transformations
\begin{align}
\delta X^I_a &= i \bar \epsilon \Gamma^I \psi_a \label{BLG_susy_s} \, , \\[10pt]
\delta \psi_a &= D_\mu X^I_a \Gamma^\mu \Gamma^I \epsilon - \frac{1}{6} X^I_b X^J_c X^K_d f^{bcd}{}_a \Gamma^{IJK} \epsilon  \label{BLG_susy_f} \, , \\[10pt]
\delta \tilde{A}_\mu{}^b{}_a &= i \bar \epsilon \Gamma_\mu \Gamma^I X^I_c \psi_d f^{cdb}{}_{a} \label{BLG_susy_g} \, .
\end{align}
The form of the gauge field transformation is fixed by dimensional analysis. Let us make some comments on the fermion transformation. There are two additional types of term that could be included. The first type which is cubic in scalars but not totally antisymmetric in the gauge indices leads to either mixed internal/R-symmetries or gauged R-symmetries both of which are not allowed in rigid supersymmetry. The second type is linear in the scalar field and leads to mass deformations of the theory as we will see in chapter \ref{c_M2s_and_background_fields}. Closing on the scalar and fermion fields leads to
\begin{align}
[\delta_1,\delta_2] X^I_a =& - 2 i ( \bar{\epsilon}_2 \Gamma^\mu \epsilon_1 ) D_\mu X^I_a - i ( \bar{\epsilon}_2 \Gamma_{JK} \epsilon_1 ) X^J_c X^K_d X^I_b f^{cdb}{}_a \, , \\[10pt]
\nonumber
[\delta_1,\delta_2] \psi_a  =& - 2 i ( \bar{\epsilon}_2 \Gamma^\mu \epsilon_1 ) D_\mu \psi_a - i ( \bar{\epsilon}_2 \Gamma_{IJ} \epsilon_1 ) X^I_c X^J_d \psi_b f^{cdb}{}_a \\
\nonumber
&+ i(\bar\epsilon_2\Gamma_\nu\epsilon_1)\Gamma^\nu\left(\Gamma^\mu D_\mu\psi_a +\frac{1}{2}\Gamma_{IJ} X^I_c X^J_d \psi_b f^{cdb}{}_a \right) \\
&- \frac{i}{4}(\bar\epsilon_2\Gamma_{KL}\epsilon_1)\Gamma^{KL}\left(\Gamma^\mu D_\mu\psi_a +\frac{1}{2}\Gamma_{IJ} X^I_c X^J_d \psi_b f^{cdb}{}_a \right)  \, ,
\end{align}
where two terms involving $\bar{\epsilon}_2 \Gamma_\mu \Gamma_{IJKL} \epsilon_1$ in the fermion closure cancel only if the 3-bracket coefficient in $\delta \psi$ is $-1/6$. The gauge field closure is
\begin{align}
\nonumber
[\delta_1,\delta_2] \tilde A_\mu{}^b{}_a =& + 2 i ( \bar{\epsilon}_2 \Gamma^\nu \epsilon_1 ) \varepsilon_{\mu\nu\lambda} \left(X^I_c D^\lambda X^I_d +\frac{i}{2}\bar\psi_c\Gamma^\lambda\psi_d \right) f^{cdb}{}_{a} \\
\nonumber
&- 2i ( \bar{\epsilon}_2 \Gamma_{IJ} \epsilon_1 ) X^I_c D_\mu X^J_d f^{cdb}{}_a \\
&- \frac{i}{3} ( \bar{\epsilon}_2 \Gamma_\mu \Gamma_{IJKL} \epsilon_1 ) X^I_c X^J_e X^K_f X^L_g f^{efg}{}_d f^{cdb}{}_a \, . 
\end{align}
The final term must be zero for the superalgebra to close and this happens if the structure constants satisfy the `fundamental identity'
\begin{equation}
f^{[efg}{}_d f^{c]db}{}_a=0 \, , \quad \text{ i.e. } \quad f^{efg}{}_d f^{abc}{}_g = f^{efa}{}_g f^{bcg}{}_d + f^{efb}{}_g f^{cag}{}_d + f^{efc}{}_g f^{abg}{}_d \, . \label{real_FI}
\end{equation}
Hence the supersymmetries close on to translations and gauge transformations after imposing the following equations of motion
\begin{align}
E_{A_\lambda{}^a{}_b} &= \frac{1}{2} \varepsilon^{\mu\nu\lambda} \tilde F_{\mu\nu}{}^b{}_a - \left(X^I_c D^\lambda X^I_d +\frac{i}{2}\bar\psi_c\Gamma^\lambda\psi_d \right)f^{cdb}{}_{a} = 0 \, , \\[10pt]
E_{\bar{\psi}^a} &= \Gamma^\mu D_\mu \psi_a +\frac{1}{2}\Gamma_{IJ}X^I_cX^J_d \psi_bf^{cdb}{}_{a} =0 \, , 
\end{align}
The scalar equation of motion:
\begin{equation}
E_{X^{Ia}} = D^2X^I_a-\frac{i}{2}\bar\psi_c\Gamma^{IJ} X^J_d\psi_b f^{cdb}{}_a +\frac{1}{2}f^{bcd}{}_{a}f^{efg}{}_{d} X^J_bX^K_cX^I_eX^J_fX^K_g = 0 \, ,
\end{equation}
can be identified by taking the supervariation of the fermion equation of motion.
The fundamental identity ensures that the gauge symmetry acts as a derivation
\begin{equation}
\delta ([X, Y, Z]) = [\delta X , Y , Z] + [X , \delta Y , Z] + [X, Y , \delta Z].
\end{equation}
This is analogous to the Jacobi identity for Lie-algebras where the Jacobi identity arises from demanding that the transformation $\delta X = [\alpha, X]$ acts as a derivation.\footnote{Note that if the fields $X^I$ took values in the Lie-algebra $u(N)$ (as with the D2-brane theory) then $[X^I , X^J , X^K]$ would be given by a nested commutator $[X^I , X^J , X^K] = \frac{1}{3!} [[X^I , X^J], X^K] \pm \textrm{cyclic}$ and would vanish by the Jacobi identity.} It is possible to construct a gauge invariant Lagrangian by defining an inner product on the algebra $\mathcal{A}$. This acts as a bilinear map ${\rm Tr} : \mathcal{A} \times \mathcal{A} \rightarrow \mathbb{C}$ which is symmetric and invariant
\begin{equation}
{\rm Tr} (X Y) = {\rm Tr}(Y X) \, ,
\end{equation} 
\begin{equation}
{\rm Tr}([V, X, Y]  Z) = - {\rm Tr}(V  [X, Y, Z]) \label{real_tr_invariance} \, .
\end{equation}
The inner product provides a notion of metric
\begin{equation}
h^{ab} = {\rm Tr} (T^a  T^b) \, , \label{met}
\end{equation}
which can be used to raise and lower the gauge indices. The invariance relation \eqref{real_tr_invariance} on the inner product together with antisymmetry of the triple-bracket implies
\begin{equation}
f^{abcd} = f^{[abcd]} \, .
\end{equation}
With the notion of a gauge invariant metric we see that the equations of motion can be obtained from the following Lagrangian 
\begin{align}
\mathcal{L} =& -\frac{1}{2} D_\mu X^{aI} D^\mu X^I_a + \frac{i}{2} \bar \psi^a \Gamma^\mu D_\mu \psi_a + \frac{i}{4} \bar \psi_b \Gamma^{IJ} X^I_c X^J_d \psi_a f^{abcd} - V + \mathcal{L}_{CS} \, , \label{BLG_act}
\end{align}
where the bosonic potential is
\begin{equation}
V = \frac{1}{12} f^{abcd} f^{efg}{}_{d} X^I_a X^J_b X^K_c X^I_e X^J_f X^K_g 
\end{equation}
and
\begin{equation}
\mathcal{L}_{CS} = + \frac{1}{2} \varepsilon^{\mu \nu \lambda} \left(f^{abcd} A_{\mu ab} \partial_\nu A_{\lambda cd} + \frac{2}{3} f^{cda}{}_{g} f^{efgb} A_{\mu ab} A_{\nu cd} A_{\lambda ef} \right) \, .
\end{equation}
Alternatively written in terms of the Tr, \eqref{BLG_act} is
\begin{align}
\mathcal{L} = {\rm Tr} \bigg( &-\frac{1}{2} D_\mu X^{I} D^\mu X^I + \frac{i}{2} \bar \psi \Gamma^\mu D_\mu \psi + \frac{i}{4} \bar \psi \Gamma^{IJ} [ X^I , X^J , \psi ] - \frac{1}{12} [X^I,X^J,X^K][X^I,X^J,X^K] \bigg) \nonumber \\
&+ \frac{1}{2} \varepsilon^{\mu \nu \lambda} \left(f^{abcd} A_{\mu ab} \partial_\nu A_{\lambda cd} + \frac{2}{3} f^{cda}{}_{g} f^{efgb} A_{\mu ab} A_{\nu cd} A_{\lambda ef} \right) \, . \label{BLG_act_Tr}
\end{align}

The gauge potential has no canonical kinetic term, but only a Chern-Simons term as suggested in \cite{Schwarz:2004yj}, and hence it has no propagating degrees of freedom. The BLG Lagrangian is the first example of an interacting gauge theory with maximal supersymmetry in three dimensions that is not of Yang-Mills type.

\subsubsection{Interpreting the BLG Theory}

Eleven-dimensional supergravity is parity conserving and M2-branes are expected to inherit this property. In \cite{Bagger:2007jr} the BLG theory was shown to be parity conserving despite the presence of Chern-Simons terms which are usually parity violating. Further, in \cite{Bandres:2008vf} it was verified that the theory possesses $OSp(8|4)$ superconformal symmetry. Thus it would seem that the BLG theory has all the expected properties of a theory describing an arbitrary number of coincident M2-branes. Unfortunately, this turns out not to be the case. 

As constructed above, the BLG theory is classical. Ultimately one is interested in unitary QFTs built from classical Lagrangians. With this in mind, the 3-algebra inner product is taken to have Euclidean signature so that the quantum theory has observables with positive probabilities etc. It turns out that for finite-dimensional representations with Euclidean metric the fundamental identity is a very strong condition and  there is a unique 3-algebra (up to direct sums) \cite{nagy-2007,Papadopoulos:2008sk,Gauntlett:2008uf} for which
\begin{equation}
f^{abcd} = \frac{2 \pi}{k} \, \varepsilon^{abcd} \, , \label{A_4_structure_constants}
\end{equation}
with $a,b,c,d \in \{ 1,2,3,4 \}$ and $k \in \mathbb{Z}$. The factor $2 \pi/k$ in \eref{A_4_structure_constants} is required because the coefficient of the Chern-Simons action is subject to a quantisation condition which ensures that the path integral is well-defined. The integer $k$ is known as the Chern-Simons level. This unique 3-algebra is the so-called ${\mathcal{A}}_4$ 3-algebra which is simply $so(4) \cong su(2) \oplus su(2)$. Whilst there is a single 3-algebra and Lagrangian associated with the Euclidean BLG theory there are two inequivalent gauge groups given by either $SO(4) \cong (SU(2) \times SU(2)) / \mathbb{Z}_2$ or $Spin(4) \cong SU(2) \times SU(2)$ \cite{Lambert:2010ji}. The restricted nature of the gauge algebra is something of a disappointment. One might have hoped that the rank of the gauge algebra could be freely chosen and was related to the number of M2-branes in analogy with D-branes. This rather begs the question what is the Euclidean BLG theory describing? 

To answer this we must look to the vacuum moduli space of the theory as in \cite{Bagger:2007vi,Lambert:2008et,Distler:2008mk,Lambert:2010ji}. This is the space of gauge inequivalent configurations which minimise the potential. With Euclidean signature for the 3-algebra inner product, the potential is positive definite and minimised when $V(X)=0$ i.e.\ $[X^I,X^J,X^K]=0$. This occurs when the scalar field takes the form
\begin{equation}
X^I = \left( \begin{array}{c} v^I_1 \\ v^I_2 \\ 0 \\ 0 \end{array} \right) \, , \label{X_moduli}
\end{equation}
for any two vectors $v^I_1$, $v^I_2 \in \mathbb{R}^8$. Since the two eight-dimensional vectors are arbitrary the starting point for the moduli space is $\mathcal{M} = \mathbb{R}^8 \times \mathbb{R}^8$. We must now identify the gauge transformations which leave the form of $X^I$ in \eqref{X_moduli} unaffected but have a nontrivial action on $v^I_1$, $v^I_2$. There is a discrete symmetry whose action is
\begin{align}
g \left(\begin{array}{c}v^I_1 \\v^I_2\end{array}\right) =& \left(\begin{array}{c}v^I_2 \\v^I_1\end{array}\right) \, .
\end{align}
This is simply a $\mathbb{Z}_2$ identification of $v^I_1$ and $v^I_2$. There is also a continuous $SO(2)$ symmetry which rotates $v^I_1$ and $v^I_2$
\begin{equation}
g_\theta \left(\begin{array}{c}v^I_1 \\v^I_2\end{array}\right) = \left(\begin{array}{c} v^I_1 \cos \theta - v^I_2 \sin \theta \\ v^I_1 \sin \theta + v^I_2 \cos \theta \end{array}\right) \, .
\end{equation} 
By introducing the complex vector $z^I = v^I_1 + i v^I_2$, we can see the continuous symmetry in its $U(1)$ form: $g_\theta(z^I) = e^{i\theta} z^I$.
Determining the effect on the moduli space due to this continuous symmetry is subtle. A careful treatment \cite{Lambert:2008et,Distler:2008mk,Lambert:2010ji} shows that $g_\theta$ can be gauged by the Chern-Simons terms that survive on the moduli space. This leads to the identification
\begin{equation}
z^I \rightarrow e^{i \sigma/k} z^I \, .
\end{equation}
The $U(1)$ gauge field $\sigma$ is periodic due to flux quantisation and the period is dependent upon which of the two BLG gauge groups is chosen. For the choice $SO(4) \cong ( SU(2) \times SU(2) )/\mathbb{Z}_2$ the period was found to be $2\pi$ whereas for $Spin(4) \cong SU(2) \times SU(2)$ it is $\pi$ \cite{Lambert:2010ji}. The periodicity of $\sigma$ together with the gauge fixed value $\sigma=0$, leads to
\begin{align}
z^I \rightarrow
\begin{cases} 
e^{2\pi i/k} z^I &\text{for } SO(4) \, , \\[10pt]
e^{\pi i/k} z^I &\text{for } Spin(4) \, .
\end{cases} 
\end{align}
These are respectively a $\mathbb{Z}_k$ and $\mathbb{Z}_{2k}$ identification of the moduli fields. We must quotient by these discrete and continuous symmetries, which generally do not commute, thereby introducing into the moduli space a dependence on the Chern-Simons level.
%
%
Consequently the moduli space of the $\mathcal{A}_4$ BLG theory is \cite{Lambert:2010ji}
\begin{align}
\mathcal{M}_k =
\begin{cases} 
\dfrac{\mathbb{R}^8 \times \mathbb{R}^8}{ D_{2k} } \ &\text{ for } SO(4) \, , \\[16pt]
\dfrac{\mathbb{R}^8 \times \mathbb{R}^8}{ D_{4k} } \ &\text{ for } Spin(4) \, .
\end{cases} 
\end{align}
Where $D_{2k} = \mathbb{Z}_2 \ltimes \mathbb{Z}_{k}$ is the dihedral group, $k$ is the usual Chern-Simons level and $D_{4k} = D_{2n}$ for $n=2k$. 

Let us examine the moduli space for specific values of $k$. For $k=1$ we have
\begin{align} \label{moduli_k_1}
\mathcal{M}_1 =
\begin{cases} 
\dfrac{\mathbb{R}^8 \times \mathbb{R}^8}{ D_{2} } \cong \dfrac{\mathbb{R}^8 \times \mathbb{R}^8}{ S_{2} } \ &\text{ for } SO(4) \, , \\[16pt]
\dfrac{\mathbb{R}^8 \times \mathbb{R}^8}{ D_{4} } \ &\text{ for } Spin(4) \, .
\end{cases} 
\end{align}
Likewise, for $k=2$ we have
\begin{align}
\mathcal{M}_2 =
\begin{cases} \label{moduli_k_2}
\dfrac{\mathbb{R}^8 \times \mathbb{R}^8}{ D_{4} } \ &\text{ for } SO(4) \, , \\[16pt]
\dfrac{\mathbb{R}^8 \times \mathbb{R}^8}{ D_{8} } \cong \dfrac{\mathbb{R}^8/\mathbb{Z}_2 \times \mathbb{R}^8/\mathbb{Z}_2}{ S_{2} } \ &\text{ for } Spin(4) \, .
\end{cases} 
\end{align}
Here $S_2 \cong \mathbb{Z}_2$ is the symmetric group with two elements. The moduli space of $N$ M2-branes in flat transverse spacetime is $ (\mathbb{R}^8 )^N / S_N$, where the $S_N$ permutes the $N$ indistinguishable branes. It follows that at level $k=1$ the $SO(4)$ theory describes two M2-branes in flat transverse space whereas the $Spin(4)$ theory does not have an M2-brane interpretation. For $k=2$ the $Spin(4)$ theory describes two M2-branes propagating in an $\mathbb{R}^8/\mathbb{Z}_2$ orbifold but now the $SO(4)$ theory does not have an M2-brane interpretation. Although we have not explicitly written it, the moduli space of the $k=4$, $SO(4)$ theory \cite{Bashkirov:2011pt} is the same as $\mathcal{M}_2$ for $Spin(4)$ and also has the interpretation of two M2-branes propagating in an $\mathbb{R}^8/\mathbb{Z}_2$ orbifold. Beyond the cases we have just outlined, the $\mathcal{A}_4$ BLG theory has no spacetime interpretation in terms of M2-branes. In this thesis it is implicit that we are referring to the interpretation given above when we state the BLG theory is a theory of multiple M2-branes. 

If Euclidean BLG describes two M2-branes then via the M-theory/IIA duality it should be related to two D2-branes at strong coupling. This presents a puzzle: how is the nondynamical Chern-Simons gauge field living on M2-branes related to the dynamical Yang-Mills gauge field on D2-branes? The answer to this puzzle is given by the `novel Higgs mechanism' \cite{Mukhi:2008ux} as follows. A large vacuum expectation value (vev), $v$, is given to one of the scalar fields and zero vevs to all other fields. The symmetries of the theory ensure that we can always arrange for the vev to be assigned to $X^8_\phi$ where $\phi$ labels the 4 direction in gauge space. For the gauge sector the nondynamical field $A_{\mu ab}$ can be split into
\begin{align}
A_{\mu a'} = A_{\mu a' \phi} \, , \qquad B_{\mu a'} = \frac{1}{2} \varepsilon_{a' b' c'} A_{\mu}{}^{b' c'} = \frac{k}{4\pi} \tilde{A}_{\mu a' \phi} \, ,
\end{align}
where $a', \ldots = 1,2,3$ and $\varepsilon_{a' b' c'} :=\varepsilon_{a' b' c' \phi}$. The form of the Chern-Simons term in the BLG Lagrangian is such that the derivative of $B_{\mu a'}$ does not appear, nor can it appear from the covariant derivatives in the kinetic terms. Consequently, $B_{\mu a'}$ acts only as an auxiliary field and can be removed by using its equation of motion. The $B_\mu$ field equation has to be found recursively and can be shown to be equal to the field strength of $A_{\mu a'}$ plus infinite corrections \cite{Mukhi:2011jp}. Remarkably on replacing $B_{\mu a'}$, its quadratic mass term is converted into a Yang-Mills kinetic term for $A_{\mu a'}$. The nondynamical gauge field has absorbed the degree of freedom from the veved scalar field and has become dynamical as a result. This should be contrasted with the usual Higgs mechanism where a massless, but dynamical, gauge field absorbs a scalar degree of freedom and becomes massive. Ultimately after some redefinitions the BLG Lagrangian with a single scalar vev $\langle X^8_\phi \rangle = v$ yields
\begin{equation}
\mathcal{L} = \mathcal{L}_{SYM} + \mathcal{L}_{decoupled} + \mathcal{L}_{higher} \, , \label{Higgsed_BLG}
\end{equation}
where
\begin{align}
\nonumber
\mathcal{L}_{SYM} =&\frac{k}{2\pi v^2} \bigg( - \frac{1}{4} F_{\mu \nu a'} F^{\mu \nu a'} - \frac{1}{2} D_\mu X^m_{a'} D^\mu X^{m a'}  - \frac{1}{4} \varepsilon^{b' c' a'} \varepsilon^{d' e'}{}_{a'} X^m_{b'} X^n_{c'} X^m_{d'} X^n_{e'} \\
&\phantom{ \frac{k}{2\pi v^2} \bigg(} + \frac{i}{2} \bar{\psi}_{a'} \Gamma^\mu D_\mu \psi^{a'} + \frac{i}{2} \varepsilon^{a' b' c'} \bar{\psi}_{a'} \Gamma^m \Gamma^8 X^m_{b'} , \psi_{c'} \bigg) \, , \\[10pt]
\mathcal{L}_{decoupled} =& - \frac{1}{2} \partial_\mu X^8_\phi \partial^\mu X^{8 \phi} - \frac{1}{2} \partial_\mu X^m_\phi \partial^\mu X^{m \phi} + \frac{i}{2} \bar{\psi}_\phi \Gamma^\mu \partial_\mu \psi^\phi \, , \\[10pt]
\mathcal{L}_{higher} =& k \, \mathcal{O} \left( \frac{1}{v^3} \right) + \ldots \, . 
\end{align}
Here
\begin{equation}
F_{\mu \nu a'} = \partial_\mu A_{\nu a'} - \partial_\nu A_{\mu a'} - \varepsilon_{a' b' c'} A_\mu^{b'} A_\nu^{c'} \, , \qquad
D_\mu X^m_{a'} = \partial_\mu X^m_{a'} - \varepsilon_{a' b' c'} A_\mu^{b'} X^{m c'} \, ,
\end{equation}
where $\varepsilon_{a' b' c'}$ are the structure constants of $su(2)$ and the transverse indices are now $m,n=1, \ldots ,7$. 
In three dimensions a scalar is dual to an Abelian 2-form. Consequently the first term in $\mathcal{L}_{decoupled}$ may be dualised to give 
\begin{equation}
\mathcal{L}_{decoupled} = - \frac{1}{4} F_{\mu\nu \phi} F^{\mu\nu \phi} - \frac{1}{2} \partial_\mu X^m_\phi \partial^\mu X^{m \phi} + \frac{i}{2} \bar{\psi}_\phi \Gamma^\mu \partial_\mu \psi^\phi \, , \label{Abelian_multiplet}
\end{equation}
so that it describes an Abelian multiplet. Uniquely in three dimensions we have $[X]=1/2$ (before any field redefinitions) and $[g_{YM}]=1/2$. This allows for the identification 
\begin{equation}
v = g_{YM} \sqrt{ \frac{k}{2\pi} } \, .
\end{equation}
The leading term in the Lagrangian \eqref{Higgsed_BLG} is then simply three-dimensional maximally supersymmetric Yang-Mills with gauge algebra $su(2)$. Together with the Abelian multiplet \eqref{Abelian_multiplet}, the full gauge symmetry is $su(2) \oplus u(1) \cong u(2)$ and the Lagrangian is invariant under $U(2)$ i.e.\ it is the theory describing the dynamics of a pair of D2-branes in type IIA string theory. However, the theory is more than just three-dimensional supersymmetric $U(2)$ Yang-Mills  because of the additional presence of higher order corrections in inverse powers of $v$. For finite $k$, sending $v \rightarrow \infty$ results in $\mathcal{L}_{higher}$ tending towards zero but also means $g_{YM} \rightarrow \infty$ and therefore strongly coupled Yang-Mills. However, if we send $v \rightarrow \infty$ and $k \rightarrow \infty$ with $g_{YM} \propto k/v^2$ fixed and finite then the $\mathcal{L}_{higher}$ corrections in \eqref{Higgsed_BLG} are suppressed as $k \, \mathcal{O} \left( 1/v^3 \right) \rightarrow 0$ and we are left precisely with finitely coupled supersymmetric $U(2)$ Yang-Mills.

\subsubsection{Non-Euclidean Real 3-algebras}

%
As we have mentioned the 3-algebra which underpins the Euclidean BLG model is severely restricted so that the theory describes at most a pair of M2-branes. 
%
%
To avoid having such a restricted theory we can consider infinite-dimensional 3-algebras or those with non-Euclidean metrics. Infinite-dimensional representations exist and such algebras may have some relevance to infinite arrays of M2-branes as considered in chapter \ref{c_Periodic_arrays_of_M2s} but are otherwise not needed for this thesis. Relaxing the assumption that the metric on the 3-algebra is positive definite leads to an infinite set of 3-algebras \cite{Gomis:2008uv,Benvenuti:2008bt,Ho:2008ei,deMedeiros:2008bf,Ho:2009nk,deMedeiros:2009hf}. 
%
%
Following \cite{Benvenuti:2008bt} we start by taking any ordinary Lie-algebra $\mathcal{G}$ with basis $T^\alpha$, $\alpha =1, \ldots , {\rm dim} ( \mathcal{G} )$ and structure constants $f^{\alpha \beta}{}_\gamma$. To this vector space we add a pair of time-like generators $T^\pm$ so that the basis is now given by $T^a = \{ T^+ , T^- , T^\alpha \}$ and has dimension ${\rm dim} ( \mathcal{G} ) +2$. One can then use the Lie-algebra to build structure constants with four indices:
\begin{equation}\label{lorentzian_f}
f^{+ \alpha \beta}{}_{\gamma} = - f^{\alpha + \beta}{}_{\gamma} = f^{\alpha \beta +}{}_{\gamma} = f^{\alpha \beta}{}_{\gamma} \, , \quad  f^{\alpha \beta \gamma}{}_{-} = f^{\alpha \beta \gamma} \, , \quad \text{all other components of } f^{a b c}{}_d = 0 \, .
\end{equation}
It is clear from \eref{lorentzian_f} that $f^{a b c}{}_d$ is totally antisymmetric in the $a,b,c$ indices. Moreover, it can be verified that the choice \eref{lorentzian_f} satisfies the fundamental identity.  Hence we have constructed a 3-algebra from an ordinary Lie-algebra.  An invariant metric for this 3-algebra can be given in terms of the standard metric $h^{\alpha \beta}$ on $\mathcal{G}$ 
\begin{align}
{\rm Tr} ( T^+ T^- ) =& -1 \, , \\[10pt]
{\rm Tr} ( T^\alpha T^\beta ) =& h^{\alpha \beta} \, ,
\end{align}
with all other components of $h^{ab}$ vanishing. This metric is clearly not positive definite, having signature $({\rm dim}(\mathcal{G})+1, 1)$ if $\mathcal{G}$ is semi-simple. We will refer to this class of 3-algebras as Lorentzian 3-algebras. By continuing to add a further $t-1$ pairs of time-like generators, one can construct a class of 3-algebras whose metric has $({\rm dim}(\mathcal{G})+t, t)$ signature.

It is straightforward to form BLG Lagrangians based on these non-Euclidean 3-algebras, the hope being that for $\mathcal{G}=su(N)$ they are capable of describing $N$ M2-branes. However, for the Lorentzian theories the fields have the following basis expansion 
\begin{equation}
\phi = \phi_a T^a = \phi_+ T^+ + \phi_- T^- + \phi_\alpha T^\alpha \, .
\end{equation}
After expanding the terms in the BLG Lagrangian the following ghost terms can be identified
\begin{equation}
\mathcal{L}_{ghost} = + \partial_\mu X^I_+ \partial^\mu X^I_- - \frac{i}{2} \bar{\psi}_+ \Gamma^\mu \partial_\mu \psi_- - \frac{i}{2} \bar{\psi}_- \Gamma^\mu \partial_\mu \psi_+ \, .
\end{equation}
Consequently it is not obvious that these Lorentzian 3-algebra theories are unitary. Of course this potential problem carries over to the multiple time-like case as well. In \cite{Bandres:2008kj,Gomis:2008be} it was demonstrated that for the Lorentzian theories these ghost terms can be removed resulting in well-defined theories. The key observation of \cite{Bandres:2008kj,Gomis:2008be} is that there is a global shift symmetry associated with the fields in the `$-$' direction which can be gauged. The new gauge symmetry allows for the choice $X^I_- = 0 =\psi_-$ which eliminates $\mathcal{L}_{ghost}$. Furthermore, the full analysis shows that $X^I_+$ is constant and $\psi_+=0$. Choosing $X^I_+=0$ preserves the $SO(8)$ R-symmetry but results in a free theory. On the other hand, choosing $X^I_+ \neq 0$ breaks the $SO(8)$ R-symmetry to $SO(7)$ as well as breaking the conformal symmetry. In particular setting $X^I_+= v \delta^{I 8}$ reproduces three-dimensional maximally supersymmetric Yang-Mills with fields in the adjoint of $\mathcal{G}$ and coupling parameter $v$. This is an exact result, there are no $\mathcal{O}\left( 1/v^3 \right)$ corrections present which is in contrast to what occurred in using the `novel Higgs mechanism'. As shown in \cite{Ezhuthachan:2008ch}, it is also possible to start from multiple D2-branes and rewrite the theory in terms of Lorentzian 3-algebras. Therefore it seems that the Lorentzian 3-algebras theories are a reformulation of the worldvolume theory of multiple D2-branes rather than bona fide M2-branes. For the multiple time-like case the story is somewhat similar. Once again there is a global shift symmetry associated with the $T^-$ analogues that upon gauging allows the ghost terms to be removed. Fields in the $T^+$-like directions are constant and can be identified with Fourier modes of multiple D$(t+1)$-branes wrapping $\mathbb{T}^{t-1}$ \cite{Ho:2009nk,Kobo:2009gz}.

\subsection{ABJM and Complex 3-algebras} \label{s_ABJM_and_complex_3_algebras}

We now give an alternative, but equivalent, formulation of Euclidean BLG due to van Raamsdonk \cite{VanRaamsdonk:2008ft} who used the relation $so(4) \cong su(2) \oplus su(2)$ to show that the theory can be cast as an ordinary gauge theory with matter in the bifundamental representation of the gauge group i.e.\ a Chern-Simons-matter theory. Under the  $su(2) \oplus su(2)$ decomposition, a vector of $so(4)$ i.e.\ $V_a$, $a=1,2,3,4$ becomes a $2 \times 2$ matrix in the bifundamental of $su(2) \oplus su(2)$ i.e.\ $V_{\alpha \dot{\beta}}$, $\alpha =1,2$, $\dot{\beta}=1,2$ and obeys the reality condition
\begin{equation}
V_{\alpha \dot{\beta}} = \varepsilon_{\alpha \beta} \varepsilon_{\dot{\beta} \dot{\alpha}} (V^\dagger)^{\dot{\alpha} \beta} \, .
\end{equation}
Explicitly, we can write
\begin{equation}
X^{I} = \left(\begin{array}{c} X^I_1 \\ X^I_2 \\ X^I_3 \\ X^I_4 \end{array}\right) \quad \rightarrow \quad X^{I} = \frac{1}{2} ( X^I_4 + i X^I_m \sigma^m ) = \frac{1}{2} \left( \begin{array}{cc} X^I_4 + i X^I_3 & X^I_2 + i X^I_1 \cr - X^I_2 + i X^I_1 & X^I_4 - i X^I_3 \end{array} \right) \, ,
\end{equation}
\begin{equation}
\psi = \left(\begin{array}{c} \psi_1 \\ \psi_2 \\ \psi_3 \\ \psi_4 \end{array}\right) \quad \rightarrow \quad \psi = \frac{1}{2} ( \psi_4 + i \psi_m \sigma^m ) =\frac{1}{2} \left( \begin{array}{cc} \psi_4 + i \psi_3 & \psi_2 + i \psi_1 \cr -\psi_2 + i \psi_1 & \psi_4 - i \psi_3 \end{array} \right) \, ,
\end{equation}
with $m=1,2,3$ and the Pauli matrices $\sigma^m$ are normalised so that ${\rm tr}(\sigma^m \sigma^n) = 2 \delta^{mn}$ and ${\rm tr}(\sigma^1 \sigma^2 \sigma^3) = 2i$, where ${\rm tr}$ is now the usual matrix trace.
The gauge field $A_{\mu ab}$ can be separated into self-dual and anti-self-dual parts
\begin{equation}
A_{\mu a b} = - \frac{k}{4\pi} (A^+_{\mu a b} + A^-_{\mu a b}) \, , \qquad A_{\mu a b}^\pm = \pm \frac{1}{2} \varepsilon_{abcd} A_\mu^{\pm cd} \, ,
\end{equation}
so that
\begin{equation}	
\tilde{A}_\mu{}^{cd} = - A_\mu^{+ cd} + A_\mu^{- cd} \, .
\end{equation}
The duality conditions reduce the number of independent components of $A_{\mu a b}^\pm$ from six to three, which we can take to be $A_{\mu 14}^\pm$, $A_{\mu 24}^\pm$, $A_{\mu 34}^\pm$. From these components we can define
\begin{equation}
A^L_\mu = A^+_{\mu 4 m} \sigma^m \, , \qquad A^R_\mu = A^-_{\mu 4 m} \sigma^m \, .
\end{equation}
The gauge covariant derivatives are now
\begin{equation}
D_\mu X^I = \partial_\mu X^I - i A^L_\mu X^I + i X^I A^R_\mu \, , \qquad D_\mu \psi = \partial_\mu \psi - i A^L_\mu \psi + i \psi A^R_\mu \, .
\end{equation}
After substituting all these replacements the 3-algebra based Lagrangian given in \eref{BLG_act_Tr} becomes
\begin{align}
\nonumber
\mathcal{L} = {\rm tr} \Bigg( &- (D^\mu X^I)^\dagger D_\mu X^I + i \bar{\psi}^\dagger \Gamma^\mu D_\mu \psi \\
\nonumber
&- \frac{2i}{3} \left( \frac{2\pi}{k} \right) \bar{\psi}^\dagger \Gamma_{IJ} (X^I X^{J \dagger} \psi + X^J \psi^\dagger X^I + \psi X^{I \dagger} X^J) + \frac{8}{3} \left( \frac{2\pi}{k} \right)^2 X^{[I} X^{J \dagger} X^{K]} X^{I \dagger} X^J X^{K \dagger} \\
&+ \frac{k}{4\pi} \epsilon^{\mu \nu \lambda} \left( A^L_\mu \partial_\nu A^L_\lambda - \frac{2i}{3} A^L_\mu A^L_\nu A^L_\lambda \right)  - \frac{k}{4\pi} \epsilon^{\mu \nu \lambda} \left( A^R_\mu \partial_\nu A^R_\lambda - \frac{2i}{3} A^R_\mu A^R_\nu A^R_\lambda \right) \Bigg) \, . \label{BLG_bifundamental}
\end{align}
The $\mathcal{N}=8$ supersymmetry transformations may also be decomposed in this way.

In \cite{Aharony:2008ug} Aharony, Bergman, Jafferis and Maldacena (ABJM) constructed an infinite class of brane configurations whose low energy effective Lagrangian is a Chern-Simons-matter theory with an SO(6) R-symmetry, manifest $\mathcal{N}= 6$ supersymmetry and conformal invariance. The gauge group is $U(N)\times U(N)$ for arbitrary $N$ and matter is in the bifundamental representation, $(\mathbf{N} , \bar{\mathbf{N}})$.\footnote{The original ABJM paper examines gauge groups of the form $U(N)\times U(N)$ and $SU(N)\times SU(N)$ but subsequent work in \cite{Lambert:2010ji} shows that they are related. There are also the $\mathcal{N}=6$ ABJ models \cite{Aharony:2008gk} with gauge groups of the form $U(M) \times U(N)$.} The moduli space of the $U(N)\times U(N)$ ABJM theory is $\mathcal{M}_k = (\mathbb{C}^4/\mathbb{Z}_k)^N/S_N$ and consequently the theory has a clear spacetime interpretation - it describes $N$ M2-branes propagating in a $\mathbb{C}^4 / \mathbb{Z}_k$ orbifold background where once again $k$ is the integer level of the Chern-Simons action.\footnote{The interpretation in terms of $N$ M2-branes can also be found from the `novel Higgs mechanism' for ABJM \cite{Li:2008ya,Pang:2008hw}.} One advantage of the ABJM construction is that it is possible to define a 't Hooft coupling parameter, $\lambda = N/k$. In the limit in which both the number of branes and the Chern-Simons level are large, with $\lambda$ fixed, the theory admits a dual geometric description given by $AdS_4 \times S^7/\mathbb{Z}_k$.  

As the $U(N)\times U(N)$ ABJM theory describes $N$ M2-branes it should exhibit the famous $N^{3/2}$ scaling behaviour for large $N$. By considering so-called localisation techniques the authors of \cite{Drukker:2010nc} were able compute the free energy of an ABJM matrix model in the large 't Hooft limit. This free energy was found to be proportional to $N^2/\sqrt{\lambda}$ which indeed scales like $N^{3/2}$ for large $N$.

The ABJM theory as originally conceived did not use 3-algebras. However in \cite{Aharony:2008ug} it was also argued that for $k=1,2$ the manifest $\mathcal{N}=6$ supersymmetry is enhanced to $\mathcal{N}=8$. For the case of two M2-branes the ABJM theory at levels $k=1,2$ is then equivalent to the BLG theory as written in \eqref{BLG_bifundamental}, also at $k=1,2$. We can then reverse the process at the beginning of this section and write this instance of the ABJM theory as a 3-algebra theory. 
Given this connection, it is of interest to generalise the construction of the BLG model, based on 3-algebras, to the case of $\mathcal{N} = 6$ supersymmetry for arbitrary number of M2-branes. 

Reduced (super-)symmetry implies that there are fewer constraints placed on the theory and for a 3-algebra this can manifest itself as a relaxation of the total antisymmetry condition on the triple product. Another distinction between the 3-algebra BLG and ABJM models is that in the former the fields took values in a real vector space whilst in the latter theory the fields are complex matrices. In \cite{Bagger:2008se}, Bagger and Lambert introduced the concept of a complex 3-algebra which they defined as follows. A complex 3-algebra is a complex vector space with basis $T^a$, $a=1,\ldots,N$, endowed with a triple product,
\begin{equation}
[T^a,T^b;\overline{T}^{\bar c}] = f^{ab\bar c}{}_{d}\,T^d \, .
\end{equation}
The notation for the 3-bracket reflects that it need only be antisymmetric in the first two indices (alternatively one can use the notation $[T^a,T^b;\overline{T}_c] = f^{ab}{}_{cd}\,T^d$). Furthermore, the structure constants, which are now complex, are required to satisfy the following fundamental identity,
\begin{equation}\label{FI_complex}
f^{ef\bar g}{}_bf^{cb\bar a}{}_d +f^{fe\bar a}{}_bf^{cb\bar g}{}_d+
f^{*\bar g\bar a f}{}_{\bar b} f^{ce\bar b}{}_d+f^{*\bar a\bar g e}{}_{\bar b} f^{cf\bar b}{}_d=0 \, .
\end{equation}
There is also an inner product on the complex 3-algebra
\begin{equation}
h^{\bar a b} = {\rm Tr} ({\overline T}^{\bar a} T^b) \, ,
\end{equation}
that is linear in the second entry and complex antilinear in the first and acts as a metric on the 3-algebra indices. We take this metric to have Euclidean signature. Requiring that the inner product is invariant under a gauge transformation generated by the complex 3-bracket leads to the condition
\begin{equation}
f^{ab\bar c\bar d}= {f}^{*\bar c\bar d ab} \, . \label{complex_gauge_inv}
\end{equation}

We use a complex notation in which the $SO(8)$ R-symmetry of the $\mathcal{N}=8$ theory is broken to the subgroup $SO(6) \times SO(2) \cong SU(4)\times U(1)$. The supercharges transform under the $SO(6) \cong SU(4)$ R-symmetry; the $SO(2) \cong U(1)$ provides an additional global symmetry.  We introduce four complex 3-algebra valued scalar fields $Z^A_a$, $A=1,2,3,4$, as well as their complex conjugates $\bar{Z}_{A\bar a}$.  Similarly, we denote the four complex two-component fermions by $\psi_{Aa}$ and their complex conjugates by $\psi^A_{\bar a}$.
A raised $A$ index indicates that the field is in the fundamental $\mathbf{4}$ of $SU(4)$; a lowered index transforms in the antifundamental $\bar{\mathbf{4}}$. We assign $Z^A_a$ and $\psi_{Aa}$ a $U(1)$ charge of
1. Complex conjugation raises or lowers the $A$ index, flips
the sign of the $U(1)$ charge, and interchanges $a \leftrightarrow \bar{a}$.
The supersymmetry generators $\epsilon_{AB}$ are in the antisymmetric $\mathbf{6}$ of
$SU(4)$ and satisfy the reality condition 
\begin{equation}
( \epsilon_{AB} )^* = \epsilon^{AB} = \frac{1}{2}\varepsilon^{ABCD}\epsilon_{CD} \, .
\end{equation}
The gauge field and the supersymmetry generators are not charged under the global $U(1)$. Supersymmetry transformations that preserve the $SU(4)$, $U(1)$ and scale symmetries are\footnote{In chapter \ref{c_M2s_and_background_fields} we will add terms to $\delta \psi_{Bd}$ that are linear in the scalar fields and which lead to a mass deformation.}
\begin{eqnarray}
\delta Z^A_d &=& i\bar\epsilon^{AB}\psi_{Bd} \, , \\[10pt]
\delta \psi_{Bd} &=& \gamma^\mu D_\mu Z^A_d\epsilon_{AB} +
  f^{ab\bar c}{}_dZ^C_a Z^A_b \bar{Z}_{C\bar c} \epsilon_{AB}+
  f^{ab\bar c}{}_d Z^C_a Z^D_{b} \bar{Z}_{B\bar c}\epsilon_{CD} \, , \\[10pt]
\delta \tilde A_\mu{}^c{}_d &=&
-i\bar{\epsilon}_{AB}\gamma_\mu Z^A_a\psi^B_{\bar b} f^{ca\bar b}{}_d +i\bar{\epsilon}^{AB}\gamma_\mu \bar{Z}_{A\bar b}\psi_{Ba} f^{ca\bar b}{}_d \, .
\end{eqnarray}
In \cite{Bagger:2008se} the commutator of these supervariations on each of the fields was shown to give
\begin{eqnarray}
[\delta_1,\delta_2]Z^A_d &=&+ v^\mu D_\mu Z^A_d + \tilde{\Lambda}^a{}_d Z^A_a \, , \\[10pt]
%
%
[\delta_1,\delta_2]\tilde A_\mu{}^c{}_d &=& + v^\nu ( \tilde F_{\mu\nu}{}^c{}_d + \varepsilon_{\mu\nu\lambda} E_{\tilde A^{\lambda c}{}_d} ) + D_\mu( \tilde{\Lambda}^c{}_d ) \, , \\[10pt]
%
%
\nonumber [\delta_1,\delta_2]\psi_{Dd} &=&+ v^\mu D_\mu \psi_{Dd} +
\tilde{\Lambda}^c{}_d \psi_{Dc}\\
\nonumber &&-\frac{i}{2}(\bar\epsilon_1^{AC}\epsilon_{2AD}-\bar\epsilon_2^{AC}\epsilon_{1AD}) E_{\psi_{Cd}} \\
 &&
+\frac{i}{4}(\bar\epsilon^{AB}_1\gamma_\nu\epsilon_{2AB})\gamma^\nu E_{\psi_{Dd}} \, ,
\end{eqnarray}
where
\begin{equation}
v^\mu = \frac{i}{2}\bar\epsilon_2^{CD}\gamma^\mu\epsilon_{1CD} \, ,\qquad
\tilde{\Lambda}^a{}_d=i(\bar\epsilon^{DE}_2\epsilon_{1CE}-
\bar\epsilon^{DE}_1\epsilon_{2CE})\bar{Z}_{D\bar c}Z^C_b f^{ab \bar c}{}_d \, ,
\end{equation}
are a translation and gauge transformation respectively. The gauge field and fermion equations of motion are
\begin{align}
E_{\tilde A^{\lambda c}{}_d} =& + \frac{1}{2} \varepsilon^{\mu\nu\lambda} \tilde F_{\mu\nu}{}^c{}_d +\left(D^\lambda Z^A_a \bar{Z}_{A\bar b}-
Z^A_a D^\lambda \bar{Z}_{A\bar b}-i\bar\psi^A_{\bar b}\gamma^\lambda\psi_{Aa}\right)f^{ca\bar b}{}_d \, , \\[10pt]
E_{\psi_{Cd}} =&+ \gamma^\mu D_\mu\psi_{Cd} +f^{ab\bar c}{}_d \psi_{Ca} Z^D_b\bar{Z}_{D\bar c}-2f^{ab\bar c}{}_d\psi_{Da}Z^D_b\bar{Z}_{C\bar c}-\varepsilon_{CDEF}f^{ab\bar c}{}_d\psi^D_{\bar c} Z^E_aZ^F_b \, .
\end{align}
Hence we see that the supersymmetry algebra closes if we impose the on-shell conditions $E_{\tilde A^{\lambda c}{}_d}=0$ and $E_{\psi_{Cd}}=0$. The scalar field equation can then be identified by taking the supervariation of the fermion equation of motion. Armed with all the field equations we can then `integrate' them to yield a Lagrangian which is automatically $\mathcal{N}=6$ supersymmetric and gauge invariant. That Lagrangian is
\begin{eqnarray}
\nonumber \mathcal{L} &=& - D^\mu \bar{Z}_A^a D_\mu Z^A_a -
i\bar\psi^{Aa}\gamma^\mu D_\mu\psi_{Aa} -V+\mathcal{L}_{CS}\\[5pt]
\nonumber
&& -i f^{ab\bar c\bar d}\bar\psi^A_{\bar d} \psi_{Aa}
Z^B_b\bar{Z}_{B\bar c}+2if^{ab\bar c\bar d}\bar\psi^A_{\bar d}\psi_{Ba}Z^B_b\bar{Z}_{A\bar c}\\[6pt]
&&+\frac{i}{2}\varepsilon_{ABCD}f^{ab\bar c\bar d}\bar{\psi}^A_{\bar d}\psi^B_{\bar c} Z^C_a Z^D_b
-\frac{i}{2}\varepsilon^{ABCD}f^{cd\bar a\bar b}\bar{\psi}_{Ac}\psi_{Bd}\bar{Z}_{C\bar a}\bar{Z}_{D\bar b}\, ,
\end{eqnarray}
where the potential is
\begin{equation}
V = + \frac{2}{3}\Upsilon^{CD}_{Bd}\bar\Upsilon_{CD}^{Bd} \, ,
\end{equation}
and
\begin{equation}
\Upsilon^{CD}_{Bd} = f^{ab\bar c}{}_dZ^C_aZ^D_b\bar{Z}_{B\bar c}
-\frac{1}{2}\delta^C_Bf^{ab\bar c}{}_dZ^E_aZ^D_b\bar{Z}_{E\bar c}+\frac{1}{2}\delta^D_Bf^{ab\bar c}{}_dZ^E_aZ^C_b\bar{Z}_{E\bar c} \, .
\end{equation}
Of course this can equally well be written as
\begin{eqnarray}
\nonumber \mathcal{L} &=& -{\rm Tr}(D^\mu \bar{Z}_A D_\mu Z^A) -
i{\rm Tr}(\bar\psi^A \gamma^\mu D_\mu\psi_A) -\frac{2}{3}{\rm Tr}(\Upsilon^{CD}_B \bar{\Upsilon}^B_{CD})+\mathcal{L}_{CS}\\[5pt]
\nonumber
&& -i{\rm Tr}(\bar\psi^A [\psi_{A},Z^B;\bar{Z}_{B}])+2i{\rm Tr}(\bar\psi^A [\psi_{B},Z^B;\bar{Z}_{A}])\\[6pt]
&&+\frac{i}{2}\varepsilon_{ABCD}{\rm Tr}(\bar\psi^A [Z^C,Z^D;\psi^B]) -\frac{i}{2}\varepsilon^{ABCD}{\rm Tr}(\bar{Z}_D [\bar
\psi_{A},\psi_B;\bar{Z}_{C}])\, , \label{ABJM_act}
\end{eqnarray}
with $\Upsilon^{CD}_B =[Z^C,Z^D;\bar{Z}_B]-\frac{1}{2}\delta^C_B[Z^E,Z^D;\bar{Z}_E]+\frac{1}{2}\delta^D_B[Z^E,Z^C;\bar{Z}_E]$.
As in the $\mathcal{N}=8$ theory the gauge field enters through covariant derivatives and a Chern-Simons action $\mathcal{L}_{CS}$ which is now given by
\begin{equation}
\mathcal{L}_{CS}=\frac{1}{2}\varepsilon^{\mu\nu\lambda}\left(f^{ab\bar c\bar d}A_{\mu
\bar c b}\partial_\nu A_{\lambda \bar d a} +\frac{2}{3}f^{ac\bar d}{}_gf^{ge\bar f \; \bar b}
A_{\mu \bar b a}A_{\nu \bar d c}A_{\lambda \bar f e}\right) \, .
\end{equation}
Consequently the gauge field does not contribute any propagating degrees of freedom and the complex 3-algebra structure constants are quantised.

The gauge symmetries permissible in $\mathcal{N}=6$ Chern-Simons-matter theories have been classified in \cite{Schnabl:2008wj}. The possible choices are $su(n) \oplus su(n)$, $su(m) \oplus su(n) \oplus u(1)$ and $sp(2n) \oplus u(1)$ with matter in the bifundamental representation. We will now see how these algebras arise from complex 3-algebras. The gauge algebra is generated by the parameters $\tilde{\Lambda}^a{}_d = \Lambda_{\bar c d} f^{ab \bar c}{}_d$ and from the metric invariance condition \eref{complex_gauge_inv} we find 
\begin{equation}
( \tilde{\Lambda}^a{}_d )^* = - \tilde{\Lambda}^d{}_a \, ,
\end{equation}
so that the gauge parameters are elements of $u(N)$ ($N$ is the dimension of the 3-algebra and not the number of M2-branes). In addition, coupling $\Lambda_{\bar c d}$ to the complex fundamental identity in \eref{FI_complex} shows that the $\tilde{\Lambda}$ are closed with respect to the matrix commutator and consequently form a Lie subalgebra of $u(N)$. To start we choose the 3-algebra structure constants to be given by
\begin{equation}
f^{ab}{}_{cd} = \frac{2\pi}{k} \left(J^{ab} J_{cd} + (\delta^a{}_c \delta^b{}_d - \delta^a{}_d \delta^b{}_c )\right)\, ,
\end{equation}
where $J^{ab}$ is the invariant antisymmetric tensor of $Sp(2n)$.  These structure constants obey the fundamental identity and have the correct symmetries.  The gauge symmetry can be determined from the gauge transformation on a generic matter field $X_d$,
\begin{equation}
\delta X_{d} = \tilde{\Lambda}^{a}{}_{d} \, X_{a} = f^{ab}{}_{cd} \, \Lambda^{c}{}_b  \, X_{a} = \frac{2\pi}{k} \big( ( \Lambda^{a}{}_d + \Lambda_{d}{}^a ) - \delta^a{}_d \Lambda^{b}{}_b \big) X_a\, .
\end{equation}
This transformation contains two parts:  the first is of the form $\delta' X_d = \tilde\Lambda'^a{}_d X_a$; the second is a phase. It can be verified that $J_{ab}\Lambda'^b{}_c J^{cd} = \Lambda'^d{}_a $ and so the gauge algebra is $sp(2n) \oplus u(1)$.

Perhaps the simplest example of a complex 3-algebra is the vector space of $m \times n$ complex matrices with the triple product of three elements $X$, $Y$, $Z$ given by
\begin{equation}\label{ohhh}
[X,Y;Z]= \frac{2\pi}{k} (XZ^\dagger Y -YZ^\dagger X) \, .
\end{equation}
Here $\dagger$ denotes the matrix Hermitian conjugate and $k$ is the integer level of the Chern-Simons action. It is trivial to show that this definition of the 3-bracket satisfies the $\mathcal{N}=6$ fundamental identity. If we introduce the inner product
\begin{equation}\label{ohhtrace}
{\rm Tr}(\bar{X},Y) = {\rm tr} (X^\dagger Y) \, ,
\end{equation}
where ${\rm tr}$ denotes the ordinary matrix trace, then the structure constants of this 3-algebra satisfy the required symmetry properties outlined earlier in this section. We are free to choose any integer value for $m$ and $n$ and so we actually have an infinite class of 3-algebras in sharp contrast to the Euclidean $\mathcal{N}=8$ theory. For this choice of 3-algebra the gauge transformation of a field is
\begin{equation}
\delta X = \Lambda^L X - X \Lambda^R \, ,
\end{equation}
where $\Lambda^L \in u(m)$ and $\Lambda^R \in u(n)$. Hence the gauge algebra generated by this 3-algebra is a Lie subalgebra of $u(m) \oplus u(n)$. To be more specific, the fields are in the bifundamental representation and consequently carry two indices, $X_{ai}$ with $1 \leq a \leq m$ and $1 \leq i \leq n$. The 3-algebra completely determines the gauge transformation of $X_{dl}$, which is now
\begin{equation}
\nonumber \delta X_{dl} = \tilde{\Lambda}^{ai}{}_{dl} \, X_{ai} = f^{aibj}{}_{ckdl}\,\Lambda^{ck}{}_{bj}\,X_{ai} \, .
\end{equation}
We may choose the 3-algebra structure constants to be,
\begin{equation}\label{UcrossU}
f^{aibj}{}_{ckdl} = \frac{2\pi}{k}\left(\delta^a{}_d \delta^b{}_c\delta^i{}_k \delta^j{}_l -  \delta^a{}_c \delta^b{}_d\delta^i{}_l \delta^j{}_k \right)\, .
\end{equation}
The $f^{aibj}{}_{ckdl}$ have the correct symmetries and satisfy the ${\cal N}=6$ fundamental identity. With this choice of structure constants we find the gauge transformation to be
\begin{equation}
\delta X _{dl} = \tilde{\Lambda}^{ai}{}_{dl} X_{ai} = \frac{2\pi}{k}\left( \delta^a{}_d \Lambda^{bi}{}_{bl} - \delta^i{}_l \Lambda^{aj}{}_{dj} \right)X_{ai}\, .
\end{equation}
When $m=n$ the matrix $\tilde{\Lambda}^{ai}{}_{dl}$ is traceless and the gauge algebra is $su(n) \oplus su(n)$. This lifts to a $SU(n) \times SU(n)$ gauge group. For the case $m \neq n$ the matrix $\tilde{\Lambda}^{ai}{}_{dl}$ has a nonvanishing trace and the gauge algebra is $u(m) \oplus u(n) \cong su(m) \oplus su(n) \oplus u(1)$ which lifts to $SU(m) \times SU(n) \times U(1)$. It has been shown that the $SU(n) \times SU(n)$ theory is related to the $U(n) \times U(n)$ ABJM theory \cite{Lambert:2010ji}, so the 3-algebraic approach describes the complete set of $\mathcal{N} = 6$ ABJM theories. 

We note that in addition to real and complex 3-algebras one can also define so-called symplectic 3-algebras \cite{Chen:2009cwa} where the 3-bracket is symmetric in its first two entries. These algebras are useful in describing three-dimensional CFTs with $\mathcal{N}=5$ supersymmetry \cite{Chen:2009cwa,Bagger:2010zq}. The gauge groups associated with these $\mathcal{N}=5$ theories are again of product type i.e.\ $\mathcal{G} \times \mathcal{H}$ and are generically distinct from the $\mathcal{N}=6,8$ groups. This illustrates one of the fascinating aspects of three-dimensional Chern-Simons-matter theories which is fundamentally different to Yang-Mills theories - the choice of gauge group determines the amount of supersymmetry of the system. There are also models with $\mathcal{N}=4$ supersymmetry \cite{Gaiotto:2008sd,Fuji:2008yj} in addition to those with $\mathcal{N} \leq 3$ which have been known for some time.

\newpage
\section{M2-Branes and Background Fields} \label{c_M2s_and_background_fields}

For a single M2-brane propagating in an eleven-dimensional spacetime with
coordinates $x^m$ the full nonlinear effective action including
fermions and $\kappa$-symmetry was obtained in \cite{Bergshoeff:1987cm}. The bosonic part of the effective action is
\begin{equation}\label{BSTaction}
S = -T_{M2}\int d^3\sigma  \sqrt{-{\rm det}(\partial_\mu
x^m\partial_\nu x^n g_{mn})} + \frac{T_{M2}}{3!}\int
d^3\sigma \, \epsilon^{\mu\nu\lambda}\partial_\mu x^m \partial_\nu
x^n\partial_\lambda x^p C_{mnp}\, .
\end{equation}
Here $C_{mnp}$ is the M-theory 3-form potential,  $g_{mn}$ the
eleven-dimensional metric and $T_{M2}\propto M_{pl}^3$ is the
M2-brane tension.

If we go to static gauge, $\sigma^\mu = x^\mu$, $\mu=0,1,2$ then the
M2-brane has worldvolume coordinates $x^\mu$  and the $x^I$,
$I=3,4,\ldots,10$ become eight scalar fields. In this chapter we will be
interested in the lowest order terms in an expansion in the
eleven-dimensional Planck scale $M_{pl}$. In this case the
canonically normalised scalars are $X^I = x^I \sqrt{T_{M2}}$. These
have mass dimension $1/2$ whereas $g_{mn}$ and $C_{mnp}$ are
dimensionless.

We next seek a generalisation of this action to lowest order in $M_{pl}$ but for multiple M2-branes. The
generalisation of the first term in (\ref{BSTaction}) has been described in chapter \ref{c_Multiple_M2s_and_3_algebras} and is given by the BLG model \cite{Bagger:2006sk,Bagger:2007jr,Bagger:2007vi,Gustavsson:2007vu} for maximal $\mathcal{N}=8$ supersymmetry and the ABJM/ABJ models for $\mathcal{N}=6$ \cite{Aharony:2008ug,Aharony:2008gk}.
In this chapter we will obtain the generalisation of the second term (i.e.\ the Wess-Zumino term) which gives the
coupling of the M2-branes to background gauge fields. In the well studied case of D-branes, where the low
energy effective theory is a maximally supersymmetric Yang-Mills gauge theory with fields in the adjoint
representation, the appropriate generalisation was given by Myers \cite{Myers:1999ps}. In the case of multiple
M2-branes  the scalar fields $X^I$ and fermions now take values in a 3-algebra which carries a bifundamental
representation of the gauge group. Thus we wish to adapt the Myers construction to M2-branes. For alternative
discussions of the coupling of multiple M2-branes to background fields see
\cite{Li:2008eza,Ganjali:2009kt,Kim:2009nc}.

The rest of this chapter is as follows. 
In section \ref{M2_bgf_N8} we will discuss the relevant couplings, to lowest order in $M_{pl}$, for the ${\cal N}=8$ Lagrangian detailed in chapter \ref{c_Multiple_M2s_and_3_algebras} and demonstrate that, by an appropriate choice of terms, the action is local and gauge invariant. We will also supersymmetrise the case where the
background field $G_{IJKL}$ is nonvanishing and demonstrate that this leads to the mass-deformed theories first proposed in \cite{Gomis:2008cv,Hosomichi:2008qk}. 
In section \ref{M2_bgf_N6} we will repeat our analysis for the case of ${\cal N}=6$ supersymmetry
leading to the mass-deformed models of \cite{Hosomichi:2008jb,Gomis:2008vc}. 
In section \ref{M2_bgf_background_curv} we will discuss the physical origin of the flux-squared term that arises by supersymmetry. In particular we will demonstrate that this term arises via back-reaction of the fluxes which leads to a curvature of spacetime. 
Section \ref{s_background_fields_conc} will conclude with a discussion of our results.

\subsection{${\cal N}=8$ Theories}\label{M2_bgf_N8}

Let us first consider the maximally supersymmetric case. 
 Although this case has only been concretely identified with the effective action of two M2-branes 
it is simpler to handle and hence the presentation is clearer. In the next section we will repeat our analysis for the case of ${\cal N}=6$.

\subsubsection{Non-Abelian Couplings to Background Fluxes}

 The scalars
$X^I$ live in a 3-algebra with totally antisymmetric triple product $[X^I,X^J,X^K]$ and invariant inner
product ${\rm Tr}(X^I X^J)$ subject to a quadratic fundamental identity and the condition that ${\rm
Tr}(X^I [X^J,X^K,X^L])$ is totally antisymmetric in $I,J,K,L$ \cite{Bagger:2007jr}. An important distinction
with the usual case of D-branes based on Lie-algebras is that Tr is an inner product and not a map from the
Lie-algebra to the real numbers. In particular there is no gauge invariant object such as ${\rm Tr}(X^I)$. Thus
the only gauge-invariant terms that we can construct involve an even number of scalar fields.

In this chapter we wish to consider the decoupling limit
$T_{M2}\to\infty$ since, unlike string theory, there are no other
parameters that we can tune to turn off the coupling to gravity. In
particular it is not clear to what extent finite $T_{M2}$ effects
can be consistently dealt with in the absence of the full
eleven-dimensional dynamics.

Assuming that there is no metric dependence
we start with the most general form for a non-Abelian pullback of the background gauge fields
to the M2-brane worldvolume:
\begin{align}
\nonumber  S_{C} = \frac{1}{3!}\epsilon^{\mu\nu\lambda}\int
d^3 x \Big( &+a \, T_{M2} C_{\mu\nu\lambda}+3b \, C_{\mu IJ} \, {\rm Tr}
  (D_\nu X^I D_\lambda X^J)  \\
\nonumber     & + 12c\, C_{\mu\nu IJKL} \, {\rm Tr}(D_\lambda
   X^I [X^J,X^K,X^L])\\
   & + 12d\,C_{[\mu IJ}C_{\nu KL]} \, {\rm Tr}(D_\lambda
   X^I [X^J,X^K,X^L])+\ldots 
 \Big) \, ,
\end{align}
where $a,b,c,d$ are dimensionless constants that we
have included for generality and the ellipsis denotes terms that are
proportional to negative powers of $T_{M2}$ and hence vanish in the
limit $T_{M2}\to \infty$.

Let us make several comments. First note that we have allowed the
possibility of higher powers of the background fields. In D-branes
the Myers terms are linear in the R-R fields however they also
include nonlinear couplings to the NS-NS 2-form. Since all these
fields come from the M-theory 3-form or 6-form this suggests that we
allow for a nonlinear dependence in the M2-brane action.

Note that gauge invariance has ruled out any terms where the
$C$-fields have an odd number of indices that are transverse to the
M2-branes (although the last term could have a part of the form $C_{\mu\nu I }C_{JKL}$).
This is consistent with the observation that the ${\cal
N}=8$ theory describes M2-branes in an ${\mathbb R}^8/{\mathbb Z}_2$
orbifold and hence we must set to zero any components of $C_3$ or
$C_6$ with an odd number of $I,J$ indices.

The first term is the ordinary coupling of an M2-brane to the
background 3-form and hence we should take $a=N$ for $N$ M2's. The
second term leads to a non-Lorentz invariant modification of the
effective three-dimensional kinetic terms. It is also present in the
case of a single M2-brane action (\ref{BSTaction}) where we find
$b=1$ which we will assume to be the case in the non-Abelian theory.\footnote{This is an assumption since  the overall centre of mass
zero mode $x^\mu$ that appears in (\ref{BSTaction}) is absent in the
non-Abelian generalisations.} The final
term proportional to $d$ in fact vanishes as ${\rm Tr}(D_\lambda
   X^{[I} [X^J,X^K,X^{L]}])= \frac{1}{4}\partial_\lambda{\rm Tr}(X^I [X^J,X^K,X^L])$ which is symmetric under
   $I,J \leftrightarrow K,L$. Thus we can set
$d=0$.

Finally note that we have allowed the M2-brane to couple to both the
3-form gauge field and  its electromagnetic 6-form dual defined by
$G_4=dC_3$, $G_7=dC_6$ where
\begin{equation}
G_7  = \star \, G_4 - \frac{1}{2}C_3\wedge G_4\, .
\end{equation}
The equations of motion of eleven-dimensional supergravity imply
that $dG_7=0$. However $G_7$ is not gauge invariant under $\delta
C_3 = d\Lambda_2$. Thus $S_{C}$ is not obviously gauge invariant or
even local as a functional of the
 eleven-dimensional gauge fields. As such one should integrate by
parts whenever possible and seek  to find an expression which is
manifestly gauge invariant.

To discuss the gauge invariance under $\delta C_3=d\Lambda_2$ we
first integrate by parts and discard all boundary terms
\begin{align}
\nonumber  S_{C} = \frac{1}{3!}\epsilon^{\mu\nu\lambda}\int d^3 x
\Big(
\nonumber &+NT_{M2} C_{\mu\nu\lambda}\\
\nonumber &+\frac{3}{2} G_{\mu\nu IJ} \, {\rm Tr}
  (X^I D_\lambda X^J) -\frac{3}{2}C_{\mu IJ} \, {\rm Tr}
  ( X^I \tilde F_{\nu\lambda}X^J)\\
   & - c \, G_{\mu\nu\lambda IJKL} \, {\rm Tr}(
   X^I [X^J,X^K,X^L])\Big)\, . \label{N_8_S_C}
\end{align}
Here we have used the fact that $C_{\mu\nu I}$ and $C_{\mu\nu\lambda
IJK}$ have been projected out by the orbifold and hence $G_{\mu\nu
IJ} =2
\partial_{[\mu} C_{\nu ]IJ}$ and $G_{\mu\nu\lambda IJKL} = 3\partial_{[\mu}C_{\nu\lambda] IJKL}$.

We find a coupling to the worldvolume gauge field strength $\tilde
F_{\nu\lambda}$ but this term is not invariant under the gauge
transformation $\delta C_3 = d\Lambda_2$. However it can be cancelled
by adding the term
\begin{equation}
S_{F}=\frac{1}{4}\epsilon^{\mu\nu\lambda}\int d^3 x \, {\rm Tr}(X^I \tilde F_{\mu\nu} X^J)
C_{\lambda
IJ}\, ,
\end{equation}
to $S_{C}$. Such terms involving the worldvolume gauge field
strength also arise in the  action of multiple D-branes.

Next consider the terms on the third line of \eqref{N_8_S_C}. Although $G_7$ is not
gauge invariant $G_7+\frac{1}{2}C_3\wedge G_4$ is. Thus we also add the term
\begin{equation}
S_{CG}=-\frac{c}{2\cdot 3!}\epsilon^{\mu\nu\lambda}\int d^3 x \, {\rm Tr}(X^I [X^J,X^K,X^L])(C_3\wedge G_4)_{\mu\nu\lambda IJKL}\, .
\end{equation}
and obtain a gauge invariant action.


To summarise we find that the total flux terms are, in the limit $T_{M2}\to \infty$,
\begin{eqnarray}
 S_{flux} &=& S_C+S_F+S_{CG} \, , \\[10pt]
\nonumber   &=& \frac{1}{3!}\epsilon^{\mu\nu\lambda}\int d^3
x \Big( +NT_{M2} \, C_{\mu\nu\lambda}+\frac{3}{2} G_{\mu\nu IJ} \, {\rm Tr}
  (X^I D_\lambda X^J)\\
   &&\phantom{\frac{1}{3!}\epsilon^{\mu\nu\lambda}\int d^3
x \Big(}- c \, (G_7+\sfrac{1}{2}C_3\wedge G_4)_{\mu\nu\lambda
IJKL} \, {\rm Tr}(
   X^I [X^J,X^K,X^L])\Big)\, .
\end{eqnarray}
In section \ref{M2_bgf_background_curv} we will argue that $c=2$.

\subsubsection{Supersymmetry}

In this section we wish to supersymmetrise the flux term $S_{flux}$
that we found above. There are also similar calculations in \cite{Grana:2002tu,Marolf:2003vf,Camara:2003ku}
where the flux-induced fermion masses on D-branes were obtained. Here we will be
interested in the final term since only it preserves
three-dimensional Lorentz invariance (the first term is just a constant
if it is Lorentz invariant). Thus for the rest of this section we will
consider backgrounds where
\begin{equation}\label{Sflux}
\mathcal{L}_{flux} = c \, \tilde G_{IJKL} \, {\rm Tr}(X^I [X^J,X^K,X^L])\, ,
\end{equation}
with
\begin{eqnarray}
\tilde G_{IJKL} &=&-
\frac{1}{3!}\epsilon^{\mu\nu\lambda}(G_7+\frac{1}{2}C_3\wedge G_{4})_{\mu\nu\lambda
IJKL}\\[10pt] &=&
+\frac{1}{4!}\epsilon_{IJKLMNPQ}G^{MNPQ}
\end{eqnarray}
and $G_{IJKL}$ is assumed to be constant.

To proceed we take the ansatz for the Lagrangian in the presence of background fields
to be
    \begin{equation}
        \mathcal{L} = \mathcal{L}_{\mathcal{N}=8} + \mathcal{L}_{mass} + \mathcal{L}_{flux}\, ,
    \end{equation}
where $\mathcal{L}_{\mathcal{N}=8}$ is the BLG Lagrangian in \eref{BLG_act_Tr},
%
    \begin{align}
        \mathcal{L}_{mass} &= -\frac{1}{2}m^2\delta_{IJ} \, {\rm Tr} (X^{I} X^{J} ) + b \, {\rm Tr} ( \bar{\psi} \Gamma^{IJKL} \psi ) \tilde{G}_{IJKL} 
    \end{align}
and $m^2$ and $b$ are constants. 
%
As shown in \cite{Bagger:2007jr}, $\mathcal{L}_{\mathcal{N}=8}$ is invariant
under the supersymmetry transformations
    \begin{align}
        \delta X^{I}_{a} &= i \bar{\epsilon} \Gamma^{I} \psi_{a} \, , \\[10pt]
        \delta \tilde{A}_{\mu}{}^{b}{}_{a} &= i \bar{\epsilon} \Gamma_{\mu} \Gamma_{I} X^{I}_{c} \psi_{d} f^{cdb}{}_{a} \label{M2_bgf_N8_susys} \, , \\[10pt]
        \delta \psi_{a} &= D_{\mu}X^{I}_{a}\Gamma^{\mu} \Gamma^{I} \epsilon - \frac{1}{6} X^{I}_{b} X^{J}_{c} X^{K}_{d} f^{bcd}{}_{a} \Gamma^{IJK} \epsilon \, .
    \end{align}
We propose additional supersymmetry transformations of the following
form
    \begin{align}
        \delta ' X^{I}_{a} &= 0 \label{M2_bgf_N8_addn_susys_s} \, , \\[10pt]
        \delta ' \tilde{A}_{\mu}{}^{b}{}_{a} &= 0 \label{M2_bgf_N8_addn_susys_f} \, , \\[10pt]
        \delta ' \psi_{a} &= \omega \Gamma^{IJKL} \Gamma^{M} \epsilon X^{M}_{a} \tilde{G}_{IJKL} \label{M2_bgf_N8_addn_susys_g} \, ,
    \end{align}
where $\omega$ is a real dimensionless parameter.

Applying the supersymmetry transformations to the mass-deformed
Lagrangian gives
    \begin{eqnarray}
        \tilde{\delta} \mathcal{L} &=& ( \delta ' + \delta ) ( \mathcal{L}_{\mathcal{N}=8} + \mathcal{L}_{mass} + \mathcal{L}_{flux} ) \\[10pt]
        &=& ( i\omega + 2b ) \, {\rm Tr} ( \bar{\psi} \Gamma^{\mu} \Gamma^{MNOP} \Gamma^{I} \epsilon D_{\mu} X^{I} ) \tilde{G}_{MNOP} \nonumber \\
        && + \frac{i \omega}{2} \, {\rm Tr} ( \bar{\psi} \Gamma^{IJ} \Gamma^{MNOP} \Gamma^{K} \epsilon [ X^{I} , X^{J} , X^{K} ] ) \tilde{G}_{MNOP} \nonumber \\
        && - \frac{2b}{6} \, {\rm Tr} ( \bar{\psi} \Gamma^{MNOP} \Gamma^{IJK} \epsilon [ X^{I} , X^{J} , X^{K} ] ) \tilde{G}_{MNOP} \nonumber \\
        && + 4ic \, {\rm Tr} ( \bar{\psi} \Gamma^{I} \epsilon [ X^{J} , X^{K} , X^{L} ] ) \tilde{G}_{IJKL} \nonumber \\
        && + i m^{2} \delta_{IJ} \, {\rm Tr} ( \bar{\psi} \Gamma^{I} \epsilon X^{J} ) \nonumber \\
        && + 2b\omega \, {\rm Tr} ( \bar{\psi} \Gamma^{IJKL} \Gamma^{MNOP} \Gamma^{Q} \epsilon X^{Q} ) \tilde{G}_{IJKL} \tilde{G}_{MNOP}\, .
    \end{eqnarray}
To eliminate the term involving the covariant derivative we must set
$b=-i\omega/2$. Substituting for $b$, expanding out the
$\Gamma$-matrices and using antisymmetry of the indices yields
    \begin{eqnarray}
        \tilde{\delta} \mathcal{L} &=& \frac{2i\omega}{3} \, {\rm Tr} ( \bar{\psi} \Gamma^{IJKMNOP} \epsilon [ X^{I} , X^{J} , X^{K} ] ) \tilde{G}_{MNOP} \nonumber \\
        && + (4ic-16i\omega) \, {\rm Tr} ( \bar{\psi} \Gamma^{L} \epsilon [ X^{I} , X^{J} , X^{K} ] ) \tilde{G}_{LIJK} \nonumber \\
        && + i m^{2} \delta_{IJ} \, {\rm Tr} ( \bar{\psi} \Gamma^{I} \epsilon X^{J} ) \nonumber \\
        && - i \omega^{2} \, {\rm Tr} ( \bar{\psi} \Gamma^{JKLM} \Gamma^{NOPQ} \Gamma^{I} \epsilon X^{I} ) \tilde{G}_{JKLM} \tilde{G}_{NOPQ}\, .
    \end{eqnarray}
Defining $\slashed{\tilde{G}} = \tilde{G}_{JKLM}\Gamma^{JKLM} $ and using
Hodge duality of the $\Gamma$-matrices leads to
    \begin{eqnarray}
        \tilde{\delta} \mathcal{L} &=& \frac{96i\omega}{6} \left(-1 + \frac{c}{4\omega} - \star \right) \tilde{G}_{LIJK} \, {\rm Tr} ( \bar{\psi} \Gamma^{L} \epsilon [ X^{I} , X^{J} , X^{K} ] ) \nonumber \\
        && + i \, {\rm Tr} \left( \bar{\psi} \left( m^{2} - \omega^{2} \slashed{\tilde{G}} \slashed{\tilde{G}} \right) \Gamma^{I} \epsilon X^{I} \right)\, .
    \end{eqnarray}
Invariance then follows if the following equations hold
    \begin{equation} \label{1stOrderCondition}
        \left(-1 + \frac{c}{4\omega} - \star \right) \tilde{G}_{LIJK} = 0
\qquad{\rm and}\qquad        \left( m^{2} - \omega^{2}
\slashed{\tilde G} \slashed{\tilde G} \right) \Gamma^{I} \epsilon = 0\, .
    \end{equation}
Since we assume that $c\ne 0$, the first equation implies
$\omega=c/8$ and $\tilde{G}$ is self-dual. It follows from the
result $\Gamma^{3456789(10)} \slashed{\tilde G} = \slashed{\tilde G}$
that the second equation is satisfied by
    \begin{equation}
        \slashed{\tilde G} \slashed{\tilde G} =  \frac{32m^2}{c^{2}} \left( 1 + \Gamma^{3456789(10)} \right)\, .
    \end{equation}
Expanding out the left hand side and using the self-duality of
$\tilde{G}$ one sees that this is equivalent to the two conditions
    \begin{equation}\label{Gcon}
        m^2 = \frac{c^2}{32\cdot 4!} G^2  \qquad {\rm and}\qquad  G_{MN[IJ}
        G_{KL]}{}^{MN}=0\, ,
    \end{equation}
where $G^2 = G_{IJKL}G^{IJKL}$.

The superalgebra can be shown to close on-shell. We first consider
the gauge field and find that the transformations close into the
same translation and gauge transformation as in the un-deformed
theory;
    \begin{align}
        [ \tilde{\delta}_{1} , \tilde{\delta}_{2} ] \tilde{A}_{\mu}{}^{b}{}_{a} &= [ \delta_{1} + \delta_{1}' , \delta_{2} + \delta_{2}' ] \tilde{A}_{\mu}{}^{b}{}_{a} \\[10pt]
            &= v^{\nu} \tilde{F}_{\mu\nu}{}^{b}{}_{a} + D_{\mu}
            \tilde{\Lambda}^{b}{}_{a}\, ,
    \end{align}
where $v^{\nu} = -2i \bar{\epsilon}_{2} \Gamma^{\nu} \epsilon_{1}$ and $\tilde{\Lambda}^{b}{}_{a} = i \bar{\epsilon}_{2} \Gamma_{JK} \epsilon_{1} X^{J}_{c} X^{K}_{d} f^{cdb}{}_{a}$. 

In considering the scalars we find a term, $2 i \omega
\bar{\epsilon}_{2} \Gamma^{MNOPIJ} \epsilon_{1} X^{J}_{a} \tilde{G}_{MNOP}$,
which can be transformed into an object with two $\Gamma$-matrix
indices by utilizing the self-duality of the flux. We find that the
scalars close into a translation plus a gauge transformation and an
SO(8) R-symmetry,
    \begin{align}
   [ \tilde{\delta}_{1} , \tilde{\delta}_{2} ] X^{I}_{a} &= [ \delta_{1} + \delta_{1}' , \delta_{2} + \delta_{2}' ] X^{I}_{a} \\[10pt]
        &= v^{\mu} D_{\mu} X^{I}_{a} + \tilde{\Lambda}^{b}{}_{a} X^{I}_{b}  + i R^{I}{}_{J}
        X^{J}_{a}\, ,
    \end{align}
where $R^{I}{}_{J} = 48 \omega \bar{\epsilon}_{2} \Gamma^{MN}
\epsilon_{1} \tilde{G}_{MNIJ}$ is the $R$-symmetry.

Finally we examine the closure of the fermions. We find again a term
incorporating $\Gamma^{(6)}$ which can be converted to $\Gamma^{(2)}$
using self-duality of $\tilde{G}$. Continuing, we find
    \begin{align}
        [ \tilde{\delta}_{1} , \tilde{\delta}_{2} ] \psi_{a} =& \ [ \delta_{1} + \delta_{1}' , \delta_{2} + \delta_{2}' ] \psi_{a} \\[10pt]
        =& \ v^{\mu} D_{\mu} \psi_{a} + \tilde{\Lambda}^{b}{}_{a} \psi_{b} + i ( \bar{\epsilon}_{2} \Gamma_{\mu} \epsilon_{1} ) \Gamma^{\mu} E_{\psi}' - \frac{i}{4} ( \bar{\epsilon}_{2} \Gamma_{JK} \epsilon_{1} ) \Gamma^{JK} E_{\psi}' \nonumber \\
        & + \frac{i}{4} R_{MN} \Gamma^{MN} \psi_{a}\, .
    \end{align}
Here $E'_{\psi}$ is the mass-deformed fermionic equation of motion,
    \begin{equation} \label{fermionEOM}
        E_{\psi}' = \Gamma^{\nu} D_{\nu} \psi_{a} + \frac{1}{2} \Gamma_{IJ} X^{I}_{c} X^{J}_{d} \psi_{b} f^{cdb}{}_{a} - \omega \Gamma^{MNOP} \psi_{a} \tilde{G}_{MNOP}\, .
    \end{equation}
Consequently, we find that on-shell
    \begin{equation}
        [ \tilde{\delta}_{1} , \tilde{\delta}_{2} ] \psi_{a} = v^{\mu} D_{\mu} \psi_{a} + \tilde{\Lambda}^{b}{}_{a} \psi_{b} + \frac{i}{4} R_{MN} \Gamma^{MN} \psi_{a}\, .
    \end{equation}

We also verify that the fermionic equation of motion maps to the
bosonic equations of motion under the supersymmetry transformations.
From the proposed mass-deformed Lagrangian the scalar equation of
motion is
    \begin{equation}
        E_{X}' = D^{2} X^{I}_{a} - \frac{i}{2} \bar{\psi}_{c} \Gamma^{IJ}   X^{J}_{d} \psi_{b} f^{cdb}{}_{a} - \frac{\partial V}{\partial X^{Ia}} - m^{2} X^{I}_{a} -4c X^{J}_{c} X^{K}_{d} X^{L}_{b} f^{cdb}{}_{a} \tilde{G}_{IJKL} \, .
    \end{equation}
The equation of motion for the gauge field is unchanged and is given
by
    \begin{equation}
        E'_{\tilde{A}} = \tilde{F}_{\mu \nu}{}^{b}{}_{a} + \varepsilon_{\mu \nu \lambda} ( X^{J}_{c} D^{\lambda} X^{J}_{d} + \frac{i}{2} \bar{\psi}_{c} \Gamma^{\lambda} \psi_{d} ) f^{cdb}{}_{a} \, .
    \end{equation}
 Taking the variation of the fermionic equation of motion \eqref{fermionEOM} gives
    \begin{eqnarray}
        0 &=& \Gamma^{I} \Gamma_{\lambda} X^{I}_{b} E'_{\tilde{A}} \epsilon + \Gamma^{I} E_{X}' \epsilon \nonumber \\
        && + \frac{96i\omega}{6} \left(-1 + \frac{c}{4\omega} - \star \right) \tilde{G}_{LIJK} \Gamma^{L} \epsilon X^{I}_{c} X^{J}_{d} X^{K}_{b} f^{cdb}{}_{a} \nonumber \\
        && + \left( m^{2} - \omega^{2} \Gamma^{MNOP} \Gamma^{WXYZ} \tilde{G}_{WXYZ} \tilde{G}_{MNOP} \right) \Gamma^{I} \epsilon X^{I}_{a}\, .
    \end{eqnarray}
Therefore consistency of the equations of motion under supersymmetry
again implies that the conditions \eqref{1stOrderCondition} must be satisfied.

Let us summarise our results. The Lagrangian
    \begin{align}
        \mathcal{L} = {\rm Tr} \bigg( &-\frac{1}{2} D_\mu X^{I} D^\mu X^I + \frac{i}{2} \bar \psi \Gamma^\mu D_\mu \psi + \frac{i}{4} \bar \psi \Gamma^{IJ} [ X^I , X^J , \psi ] - \frac{1}{12} [X^I,X^J,X^K][X^I,X^J,X^K] \nonumber \\
\nonumber
&- \frac{1}{2} m^{2} \delta_{IJ} X^{I} X^{J} - \frac{i c}{16} \, \bar{\psi} \Gamma^{IJKL} \psi \,  \tilde{G}_{IJKL} + c \,  [ X^{I} , X^{J} , X^{K} ] X^{L} \, \tilde{G}_{IJKL} \bigg) \\
&+ \frac{1}{2} \varepsilon^{\mu \nu \lambda} \left(f^{abcd} A_{\mu ab} \partial_\nu A_{\lambda cd} + \frac{2}{3} f^{cda}{}_{g} f^{efgb} A_{\mu ab} A_{\nu cd} A_{\lambda ef} \right) \, ,
    \end{align}
is invariant under the supersymmetries
    \begin{align}
        \delta X^{I}_{a} &= i \bar{\epsilon} \Gamma^{I} \psi_{a} \, , \\[10pt]
        \delta \tilde{A}_{\mu}{}^{b}{}_{a} &= i \bar{\epsilon} \Gamma_{\mu} \Gamma_{I} X^{I}_{c} \psi_{d} f^{cdb}{}_{a} \, ,  \\[10pt]
        \delta \psi_{a} &= D_{\mu}X^{I}_{a}\Gamma^{\mu} \Gamma^{I} \epsilon - \frac{1}{6} X^{I}_{b} X^{J}_{c} X^{K}_{d} f^{bcd}{}_{a} \Gamma^{IJK} \epsilon + \frac{c}{8}\Gamma^{IJKL} \Gamma^{M} \epsilon X^{M}_{a} \tilde{G}_{IJKL} \, , 
    \end{align}
provided $\tilde{G}_{IJKL}$ is self-dual and satisfies the conditions in (\ref{Gcon}).
Moreover the supersymmetry algebra closes according to
    \begin{align}
        [ \delta_{1} , \delta_{2} ] \tilde{A}_{\mu}{}^{b}{}_{a} &= v^{\nu} \tilde{F}_{\mu\nu}{}^{b}{}_{a} + D_{\mu} \tilde{\Lambda}^{b}{}_{a} \, ,  \\[10pt]
        [ \delta_{1} , \delta_{2} ] X^{I}_{a} &= v^{\mu} D_{\mu} X^{I}_{a} + \tilde{\Lambda}^{b}{}_{a} X^{I}_{b}  + i R^{I}{}_{J} X^{J}_{a} \, , \\[10pt]
        [ \delta_{1} , \delta_{2} ] \psi_{a} &= v^{\mu} D_{\mu} \psi_{a} + \tilde{\Lambda}^{b}{}_{a} \psi_{b} + \frac{i}{4} R_{MN} \Gamma^{MN} \psi_{a}\, . 
    \end{align}
Taking %
\begin{equation}
G = \mu (dx^3\wedge dx^4\wedge dx^5\wedge dx^6+dx^7\wedge dx^8\wedge dx^9\wedge dx^{10}) \, ,
\end{equation}
readily leads
to the mass-deformed Lagrangian of \cite{Gomis:2008cv,Hosomichi:2008qk}.

\subsection{${\cal N}=6$ Theories}\label{M2_bgf_N6}

Let us now consider the more general case of ${\cal N}=6$
supersymmetry and in particular the ABJM \cite{Aharony:2008ug} and
ABJ \cite{Aharony:2008gk} models which describe an arbitrary number
of M2-branes in a ${\mathbb C}^4/{\mathbb Z}_k$ orbifold. 
Since the
discussion is similar in spirit to the ${\cal N}=8$ case we will
shorten our discussion and largely just present the results of our calculations.

\subsubsection{Non-Abelian Couplings to Background Fluxes}

In the ${\cal N}=6$ theories there are four complex scalars $Z^A$ and
their complex conjugates $\bar{Z}_A$. These are defined in terms of the
spacetime coordinates through
\begin{eqnarray}
    Z^1&=& \frac{1}{\sqrt{2T_{M2}}}(x^3+ix^4)\qquad
  Z^2 =  \frac{1}{\sqrt{2T_{M2}}}(x^5+ix^6) \, , \\[10pt]
   Z^3 &=&  \frac{1}{\sqrt{2T_{M2}}}(x^7-ix^9)\qquad
 Z^4 = \frac{1}{\sqrt{2T_{M2}}}(x^8-ix^{10})\, .
\end{eqnarray}
In particular we will take the formulation in \cite{Bagger:2008se}.
The scalars and fermions are endowed with a
triple product $[Z^A,Z^B;\bar{Z}_C]$ or $[\bar{Z}_A,\bar{Z}_B;Z^C]$  and an inner product
${\rm Tr}(\bar{Z}_A Z^B)$ subject to a quadratic fundamental identity as well as the
condition
${\rm Tr}(\bar{Z}_D [Z^A,Z^B;\bar{Z}_C])^\star =-{\rm Tr}(\bar{Z}_A [Z^C,Z^D;\bar{Z}_B])$.
As we have seen in chapter \ref{c_Multiple_M2s_and_3_algebras} to obtain the ABJM/ABJ models \cite{Aharony:2008gk,Aharony:2008ug} one should let the fields be $m\times n$ matrices and define
\begin{equation}
[Z^A,Z^B;\bar{Z}_C] = \lambda(Z^A \bar{Z}_C^\dag Z^B-Z^B \bar{Z}_C^\dag Z^A)\, .
\end{equation}
where $\lambda$ is an arbitrary (but quantised) coupling constant. As such the gauge invariant terms
always involve an equal number of $Z$ and $\bar{Z}$ coordinates.
Again this is consistent with the interpretation that the M2-branes
are in an ${\mathbb C}^4/{\mathbb Z}_k$ orbifold which acts as $Z^A \to e^{\frac{2\pi i}{k}}
Z^A$.

Following the discussion of the previous section we start with
\begin{align}
\nonumber   S_{C} = \frac{1}{3!}\epsilon^{\mu\nu\lambda}\int d^3 \, x
\Big(
&+NT_{M2} C_{\mu\nu\lambda}+\frac{3}{2} C_{\mu}{}^A{}_B \, {\rm Tr}
  (D_\nu \bar{Z}_A D_\lambda Z^B)\\
\nonumber & +\frac{3}{2} C_{\mu}{}_A{}^B \, {\rm Tr}
  (D_\nu Z^A D_\lambda \bar{Z}_B)\\
\nonumber & +\frac{3c}{2}C_{\mu\nu AB}{}^{CD} \, {\rm Tr}(
  D_\lambda \bar{Z}_D [Z^A,Z^B;\bar{Z}_C])\\
  & +\frac{3c}{2}C_{\mu\nu}{}^{AB}{}_{CD} \, {\rm Tr}(
  D_\lambda Z^D [\bar{Z}_A,\bar{Z}_B;Z^C])
\Big)\, .
\end{align}
Integrating by parts we again find a non-gauge invariant term proportional to $\epsilon^{\mu\nu\lambda }\tilde
F_{\nu\lambda}C_{\mu}{}^A{}_B$ which is cancelled by adding
\begin{equation}
S_F = \frac{1}{8}\epsilon^{\mu\nu\lambda} \int d^3 x \, C_{\mu}{}^A{}_B \, {\rm Tr}
  (\bar{Z}_A \tilde F_{\nu\lambda} Z^B)+C_{\mu}{}_A{}^B \, {\rm Tr}
  ({Z}^A \tilde F_{\nu\lambda} \bar Z_B)\, .
\end{equation}
As with the case above we also must add
\begin{eqnarray}
S_{CG} &=& -\frac{c}{8\cdot 3!}\epsilon^{\mu\nu\lambda}\int d^3 x \, (C_3\wedge G_4)_{\mu\nu
AB}{}^{CD} \, {\rm Tr}(
  \bar{Z}_D [Z^A,Z^B;\bar{Z}_C])
\end{eqnarray}
to ensure that the last term is gauge invariant. Thus in total we
have
\begin{align}
S_{flux} = S_C+S_F+S_{CG}\\[10pt]
\nonumber =
\frac{1}{3!}\epsilon^{\mu\nu\lambda}\int d^3 x \,
\Big(
&+NT_{M2} C_{\mu\nu\lambda}\\
\nonumber
&+\frac{3}{4} G_{\mu\nu}{}^A{}_B \, {\rm Tr}
  (\bar{Z}_A D_\lambda Z^B)+\frac{3}{4} G_{\mu\nu}{}_A{}^B \, {\rm Tr}
  ({Z}^A D_\lambda \bar Z_B)\\
& -\frac{c}{4}(G_7+\frac{1}{2}C_3\wedge G_4)_{\mu\nu\lambda AB}{}^{CD}\, {\rm Tr}(
  \bar{Z}_D [Z^A,Z^B;\bar{Z}_C])
\Big) \, .
\end{align}

\subsubsection{Supersymmetry}

Following on as before we wish to supersymmetrise the action
    \begin{equation}
        \mathcal{L} = \mathcal{L}_{{\cal N}=6} + \mathcal{L}_{mass} + \mathcal{L}_{flux}\, ,
    \end{equation}
where ${\cal L}_{{\cal N}=6}$ is the ${\cal N}=6$ Lagrangian
given in \eref{ABJM_act}.
We restrict to backgrounds where
\begin{equation}\label{Sflux2}
\mathcal{L}_{flux} = \frac{c}{4} \, {\rm Tr}( \bar Z_D [ Z^A, Z^B;\bar Z_C]) \tilde G_{AB}{}^{CD} \, ,
\end{equation}
with
\begin{eqnarray}
 \tilde G_{AB}{}^{CD} &=&-
\frac{1}{3!}\epsilon^{\mu\nu\lambda}(G_7+\frac{1}{2}C_3\wedge G_{4})_{\mu\nu\lambda
AB}{}^{CD}\\[10pt] &=&
+\frac{1}{4}\epsilon_{ABEF}\epsilon^{CDGH}G^{EF}{}_{GH}\, .
\end{eqnarray}
Finally we take the ansatz for $\mathcal{L}_{mass}$ to be
\begin{equation}
\mathcal{L}_{mass} = - m^2 \, {\rm Tr}( \bar Z_A Z^A ) + b \, {\rm Tr}(\bar \psi^A \psi_F ) \tilde G_{AE}{}^{EF}\, .
\end{equation}
We propose the following modification to the fermion supersymmetry variation
\begin{equation}
\delta ' \psi_{Ad} = \omega \epsilon_{DF} Z^F_d \tilde G_{AE}{}^{ED}\, ,
\end{equation}
where $\omega$ is a real parameter.

After applying the supersymmetry transformations to $\mathcal{L}$ we find that taking $b=-i \omega$ eliminates the covariant derivative terms. The terms that are second order in $\tilde{G}$ must vanish separately and this gives the condition
\begin{equation}\label{Gcon2}
\tilde{G}_{AE}{}^{EB} \tilde{G}_{BF}{}^{FC} = \frac{m^2}{\omega^2} \delta_A^C\, .
\end{equation}

The remaining terms in the variation are
\begin{eqnarray}
\nonumber \delta \mathcal{L} &=& + 2 i \omega \, {\rm Tr} ( \bar Z_D [ \bar \psi_F \epsilon^{DA} , Z^Q ; \bar Z_Q ] ) \tilde{G}_{AE}{}^{EF}\\
\nonumber   && + i \omega \, {\rm Tr} ( \bar Z_D [ \bar \psi_F \epsilon^{QD} , Z^A ; \bar Z_Q ] ) \tilde{G}_{AE}{}^{EF} \\
\nonumber    && + 2 i \omega \, {\rm Tr}( \bar Z_D [ \bar \psi_K \epsilon^{AD}, Z^K ; \bar Z_F ] ) \tilde{G}_{AE}{}^{EF}\\
\nonumber   && + \frac{ic}{2} \, {\rm Tr}( \bar Z_D [ \bar \psi_K \epsilon^{AK}, Z^B ; \bar Z_C ] ) \tilde{G}_{AB}{}^{CD}\\
\nonumber   && + \frac{i \omega}{2} \varepsilon^{AKQD} \varepsilon_{IJFP} \, {\rm Tr}( \bar Z_D [ \bar \psi_K \epsilon^{IJ},Z^P ; \bar Z_Q ] ) \tilde{G}_{AE}{}^{EF}\\
   && + \text{c.c.}\, ,
\end{eqnarray}
where we have made use of the reality condition $\epsilon_{FP} = \frac{1}{2} \varepsilon_{IJFP} \,
\epsilon^{IJ}$. To proceed we need to restrict $\tilde{G}$ to have the form
\begin{equation}\label{Gid}
\tilde{G}_{AB}{}^{CD} = \frac{1}{2} \delta^C_B \tilde{G}_{AE}{}^{ED} - \frac{1}{2} \delta^C_A
\tilde{G}_{BE}{}^{ED} - \frac{1}{2} \delta^D_B \tilde{G}_{AE}{}^{EC} + \frac{1}{2} \delta^D_A
\tilde{G}_{BE}{}^{EC}\, ,
\end{equation}
with $\tilde{G}_{AE}{}^{EA}=0$. Substituting for $\tilde{G}_{AB}{}^{CD}$ allows us to factor out the common
term \\ ${\rm Tr} ( \bar Z_D [ \bar \psi_K \epsilon^{IJ} , Z^P ; \bar Z_Q ] ) \tilde{G}_{AE}{}^{EF}$. This factor is
separately antisymmetric in $IJ$ and $DQ$ so after expanding out $\varepsilon^{AQKD} \varepsilon_{IJFP} = 4!
\delta^{[ AQKD ]}_{\phantom{[} IJFP}$ we have
\begin{eqnarray}
\nonumber \delta \mathcal{L} &=& i \omega \left( \frac{c}{2\omega} -2 \right) ( \delta^A_I \delta^K_J \delta ^D_F \delta^Q_P + \delta^K_I \delta^Q_J \delta ^D_F \delta^A_P ) {\rm Tr} ( \bar Z_D [ \bar \psi_K \epsilon^{IJ}, Z^P ; \bar Z_Q ] ) \tilde{G}_{AE}{}^{EF}\\
&& + \text{c.c.}
\end{eqnarray}
Therefore the Lagrangian is invariant under supersymmetry if $\omega=c/4$. Taking the trace of \eref{Gcon2}
allows us to deduce that
\begin{equation}
m^2 = \frac{1}{32\cdot 4!}c^2 G^2 \, ,
\end{equation}
where $G^2 = 6G_{AB}{}^{CD} G^{AB}{}_{CD}=12 G_{AE}{}^{EB}G_{BF}{}^{FA}$.

In examining the closure of the superalgebra we find
\begin{eqnarray}
\left[ \delta_1,\delta_2 \right] \tilde{A}_{\mu}{}^c{}_d &=& v^{\nu} \tilde{F}_{\mu\nu}{}^c{}_d +D_{\mu} (\Lambda_{\bar a b} f^{cb \bar a}{}_d) \, , \\[10pt]
\left[ \delta_1,\delta_2 \right] Z^A_d &=& v^{\mu} D_{\mu} Z^A_d + \Lambda_{\bar c b} f^{ab \bar c}{}_d Z^A_a -
i R^A{}_B Z^B_d - i Y Z^A_d \, ,
\end{eqnarray}
where
\begin{eqnarray}
v^{\mu} &=& \frac{i}{2} \bar \epsilon_2^{CD} \gamma^{\mu} \epsilon_{1CD} \, , \\[10pt]
\Lambda_{\bar c b} &=& i ( \bar \epsilon_2^{DE} \epsilon_{1CE} - \bar \epsilon_1^{DE} \epsilon_{2CE} ) \bar Z_{D \bar c} Z^C_b \, , \\[10pt]
R^A{}_B &=& \omega \left( ( \bar \epsilon_1^{AC} \epsilon_{2DB} - \bar \epsilon_2^{AC} \epsilon_{1DB} ) - \frac{1}{4} ( \bar \epsilon_1^{EC} \epsilon_{2DE} - \bar \epsilon_2^{EC} \epsilon_{1DE} ) \delta^A_B \right) \tilde{G}_{CM}{}^{MD} \, , \\[10pt]
Y &=& \frac{\omega}{4} ( \bar \epsilon_1^{EC} \epsilon_{2DE} - \bar \epsilon_2^{EC} \epsilon_{1DE} ) \tilde{G}_{CM}{}^{MD} \, .
\end{eqnarray}
Acting with the commutator on the fermions gives
\begin{eqnarray}
\nonumber [ \delta_1 , \delta_2 ] \psi_{Dd} &=& v^\mu D_\mu \psi_{Dd} +
\Lambda_{\bar a b}f^{cb \bar a}{}_d\psi_{Dc}\\
\nonumber &&-\frac{i}{2}(\bar\epsilon_1^{AC}\epsilon_{2AD}-\bar\epsilon_2^{AC}\epsilon_{1AD})E'_{Cd}\\
\nonumber && +\frac{i}{4}(\bar\epsilon^{AB}_1\gamma_\nu\epsilon_{2AB})\gamma^\nu
E'_{Dd}\\
&& + i R^A{}_D \psi_{Ad} - i Y \psi_{Dd} \, ,
\end{eqnarray}
provided the 4-form satisfies $\tilde{G}_{AE}{}^{EA}=0$. The new fermionic equation of motion is
\begin{eqnarray}
\nonumber E'_{Cd} &=& \gamma^{\mu} D_{\mu} \psi_{Cd} +f^{ab \bar c}{}_d \psi_{Ca}
Z^D_b \bar Z_{D \bar c} -2 f^{ab \bar c}{}_d \psi_{Da} Z^D_b \bar Z_{C \bar c} \\
&& - \varepsilon_{CDEF} f^{ab \bar c}{}_d\psi^D_{\bar c} Z^E_a Z^F_b + \frac{c}{4} \tilde{G}_{CE}{}^{EB}
\psi_{Bd}\, .
\end{eqnarray}
Consistency of the bosonic and fermionic equations of motion under supersymmetry requires that $\tilde{G}_{AE}{}^{EB} \tilde{G}_{BF}{}^{FC} = \frac{m^2}{\omega^2} \delta^C_A$, which is the same condition as found in demonstrating invariance of the action.

Choosing $\tilde G_{AB}{}^{CD}$ to have the form (\ref{Gid}) with
\begin{equation}\label{ex2}
\tilde G_{AB}{}^{BC} = \left(
                  \begin{array}{cccc}
                    \mu & 0 &0&0\\
                    0 & \mu &0&0\\
                    0 & 0 &-\mu&0\\
                     0& 0 &0&-\mu\\
                  \end{array}
                \right)\, ,
\end{equation}
gives the mass-deformed Lagrangian of \cite{Hosomichi:2008jb,Gomis:2008vc}.

\subsection{Background Curvature}\label{M2_bgf_background_curv}

Our final point is to understand the physical origin of the mass-squared term in the effective action which is
quadratic in the masses. Note that this term is a simple, $SO(8)$-invariant mass term for all the scalar
fields. Furthermore it does not depend on any non-Abelian features of the theory. Therefore we can derive this
term by simply considering a single M2-brane and compute the unknown constant $c$.

We can understand the origin of this term as follows. We have seen
that it arises as a consequence of supersymmetry. For a single
M2-brane supersymmetry arises as a consequence of $\kappa$-symmetry
and $\kappa$-symmetry is valid whenever an M2-brane is propagating
in a background that satisfies the equations of motion of
eleven-dimensional supergravity \cite{Bergshoeff:1987cm}.

The multiple M2-brane actions implicitly assume that the background
is simply flat space or an orbifold thereof. However the inclusion
of a nontrivial flux  implies that there is now a source for the
eleven-dimensional metric which is of order flux-squared. Thus for
there to be $\kappa$-supersymmetry and hence supersymmetry it
follows that the background must be curved. This in turn will lead
to a potential in the effective action of an M2-brane. In particular
given a 4-form flux $G_4$ the bosonic equations of
eleven-dimensional supergravity are
\begin{eqnarray}
  R_{mn} - \frac{1}{2}g_{mn}R &=& \frac{1}{2\cdot 3!}G_{mpqr}G_n{}^{pqr} - \frac{1}{4\cdot 4!}g_{mn}G^2 \, , \\[10pt]
   d\star G_4 -\frac{1}{2}G_4\wedge G_4 &=&0\, .
\end{eqnarray}
At lowest order in fluxes we see that $g_{mn}=\eta_{mn}$ and $G_4$ is constant. However at second order there
are source terms. To start with we will assume that, at lowest order, only $G_{IJKL}$ is nonvanishing. To
solve these equations we introduce a nontrivial metric of the form
\begin{equation}
g_{mn} = \left(
           \begin{array}{cc}
             e^{2\omega}\eta_{\mu\nu} & 0 \\
             0 & g_{IJ} \\
           \end{array}
         \right)\, ,
\end{equation}
where $\omega = \omega(x^I) = \omega( X^I/T_{M2}^{\frac{1}{2}})$ and
$g_{IJ} = g_{IJ}(x^I) = g_{IJ}( X^I/T_{M2}^{\frac{1}{2}})$.

Let us look at an M2-brane in this background. The first term in the action (\ref{BSTaction}) is
\begin{eqnarray}
 S_1 &=& -T_{M2}\int d^3x  \sqrt{-{\rm det}(e^{2\omega}\eta_{\mu\nu}+\partial_\mu x^I\partial_\nu x^J g_{IJ})} \\[10pt]
   &=&-T_{M2}\int d^3x \, e^{3\omega}\left(1 + \frac{1}{2}e^{-2\omega}\partial_\mu x^I\partial^\mu
   x^Jg_{IJ}+\ldots\right)\\[10pt]
    &=& -\int d^3x \left(T_{M2}e^{3\omega} + \frac{1}{2}e^{\omega}\partial_\mu X^I\partial^\mu
   X^Jg_{IJ}+\ldots\right) \, .
\end{eqnarray}
Next we note that, in the decoupling limit $T_{M2}\to \infty$, we
can expand
\begin{equation}
e^{2\omega(x)} = e^{2\omega(X^I/\sqrt{T_{M2}})} =
1+\frac{2}{T_{M2}}\omega_{IJ}X^IX^J+\ldots
\end{equation}
and
\begin{equation}
g_{IJ}(x) = g_{IJ}( X^I/\sqrt{T_{M2}}) = \delta_{IJ}+\ldots\, ,
\end{equation}
so that
\begin{equation}
S_1 =-\int d^3x \left(T_{M2} + 3\omega_{IJ}X^IX^J + \frac{1}{2}\partial_\mu X^I\partial^\mu
   X^J\delta_{IJ}+\ldots\right)\, ,
\end{equation}
where the ellipsis denotes terms that vanish as $T_{M2}\to \infty$. Thus we see that in the decoupling limit we
obtain the mass term for the scalars. Similar mass terms for M2-branes were also studied in
\cite{Skenderis:2003da}  for pp-waves.

To compute the warp-factor $\omega$ we can expand $g_{mn} =
\eta_{mn} + h_{mn}$, where $h_{mn}$ is second order in the fluxes,
and linearise the Einstein equation. If we impose the gauge
$\partial^m h_{mn} -\frac{1}{2}\partial_n h^p{}_p=0$ then Einstein's
equation becomes
\begin{equation}
-\frac{1}{2}\partial_p\partial^p \left(h_{mn} -\frac{1}{2}\eta_{mn}
h_q{}^q \right)= \frac{1}{2\cdot 3!}G_{mpqr}G_n{}^{pqr} -
\frac{1}{4\cdot 4!}g_{mn}G^2\, .
\end{equation}
This reduces to two coupled sets of equations corresponding to
choosing  indices $(m,n)=(\mu,\nu)$ and $(m,n)=(I,J)$. Contracting
the latter with $\delta^{IJ}$ one finds that $h_I{}^I = 4h_p{}^p$
and hence $h_p{}^p = -\frac{1}{3}h_\mu{}^\mu$. With this in hand the
$(m,n)=(\mu,\nu)$ terms in Einstein's equation reduce to
\begin{equation}
\partial_I\partial^I e^{2\omega} = \frac{1}{3\cdot 4!} G^2 
\end{equation}
and hence, to leading order in the fluxes,
\begin{equation}
e^{2\omega} = 1+\frac{1}{48\cdot4!}G^2 \delta_{IJ}x^Ix^J\, ,
\end{equation}
so that $S_1$ contributes the term
\begin{equation}
S_1 = -\int d^3 x \, \frac{1}{32\cdot4!}G^2X^2
\end{equation}
to the potential.

Next we must look at the second, Wess-Zumino term, in \eqref{BSTaction};
\begin{equation}
S_2 = \frac{T_{M2}}{3!}\int d^3 x \, \epsilon^{\mu\nu\lambda} C_{\mu\nu\lambda}\, .
\end{equation}
Although we have assumed that $C_{\mu\nu\lambda}=0$ at leading order, the $C$-field equation of motion implies
that $G_{I \mu\nu\lambda }=\partial_ I C_{\mu\nu\lambda}$  is second order in $G_{IJKL}$. In particular if we
write $C_{\mu\nu\lambda} = C_0\epsilon_{\mu\nu\lambda}$ we find, assuming $G_{IJKL}$ is self-dual,  the
equation
\begin{equation}
\partial_I\partial^I C_0 = \frac{1}{2\cdot 4!}G^2\, ,
\end{equation}
where $G^2 = G_{IJKL}G^{IJKL}$. The solution is
\begin{equation}
C_0=\frac{1}{32\cdot 4!}G^2 \delta_{IJ}x^Ix^J\, .
\end{equation}
Thus we find that $S_2$ gives a second contribution to the scalar potential
\begin{equation}
S_2 = -\int d^3 x \, \frac{1}{32\cdot4!}G^2X^2\, .
\end{equation}
Note that this is equal to the scalar potential derived from $ S_1$. Therefore if we were to break
supersymmetry and consider anti-M2-branes, where the sign of the Wess-Zumino term changes, we would not find a
mass for the scalars.

In total we find the mass-squared
\begin{equation}
m^2= \frac{1}{8\cdot 4!}G^2 \, .
\end{equation}
Comparing with \eref{Gcon} we see that $c^2=4$, e.g.\ $c=2$. Note that we have performed this
calculation using the notation of the ${\cal N}=8$ theory, however a similar calculation also holds in the
${\cal N}=6$ case with the same result.

\subsection{Conclusions}\label{s_background_fields_conc}

In this chapter we discussed the coupling of multiple M2-branes with ${\cal N}=6,8$ supersymmetry to the
background gauge fields of eleven-dimensional supergravity. In particular we gave
a local and gauge invariant form for the `Myers terms' in the limit $M_{pl}\to \infty$. We supersymmetrised
these flux terms in the case
where the fluxes preserve the supersymmetry and Lorentz symmetry of
M2-branes to obtain the  massive models of \cite{Gomis:2008cv,Hosomichi:2008qk,Hosomichi:2008jb,Gomis:2008vc}. We also showed how the
flux-squared term in the effective action, which arises as a
mass term for the scalar fields, is generated through a back-reaction of the fluxes on the eleven-dimensional geometry.

The results we have found using gauge invariance fit naturally with the ${\mathbb R}^8/{\mathbb Z}_k$ orbifold
interpretation of the background. However for the ${\cal N}=6$ theories with $k=1,2$ the orbifold action is
less restrictive and this allows for additional terms. In particular for $k=2$ we expect terms where the total
number of $Z^A$ and $\bar Z_B$ fields are even (but not necessarily equal). In addition for $k=1$ there should
be terms with any number of $Z^A$ and $\bar Z_B$ fields. Such terms are not gauge invariant on their own but
presumably can be made so by including monopole operators which, for $k=1,2$, are local.

\newpage
\section{Higher Derivative BLG}\label{c_Higher_Derivative_BLG}


The bosonic effective action for a single M2-brane \cite{Bergshoeff:1987cm} in static gauge and in a flat background with zero flux, is given by the Abelian DBI action
\begin{align}
S_{M2} =& - T_{M2} \int d^3 x \ \sqrt{ - {\rm det} ( \eta_{\mu \nu} + \partial_\mu x^I \partial_\nu x^I ) } \, .
\end{align}
The terms in the integral can be expanded as a power series in $(\partial x)^2$ that is, a higher derivative expansion. 
After canonically renormalising the eight scalars so that $X^I = x^I \sqrt{T_{M2}}$, the leading order and next to leading order terms in the expansion are
\begin{align}
\nonumber
S_{M2} =\phantom{+} &\int d^3 x \ - \frac{1}{2} \partial_\mu X^I \partial^\mu X^I \\
+ &\int d^3 x \ \frac{1}{T_{M2}} \left( + \frac{1}{4} \partial_\mu X^I \partial^\mu X^J \partial_\nu X^I \partial^\nu X^J - \frac{1}{8} \partial_\mu X^I \partial^\mu X^I \partial_\nu X^J \partial^\nu X^J \right) \label{M2_action} \\
\nonumber
+ &\ldots \, , \phantom{\int} 
\end{align}
where we have ignored a constant and the ellipsis denotes terms $\mathcal{O}\big( ( 1/T_{M2})^2 \big)$ and higher.

The generalisation of the supersymmetric leading order M2-brane action to multiple M2-branes is given by either the BLG or ABJM model.
There have been several papers which aim to determine the next to leading order i.e.\ the $1/T_{M2}$ higher derivative corrections to multiple M2-branes. It is known \cite{Nicolai:2003bp,deWit:2003ja,deWit:2004yr} that in three dimensions a non-Abelian 2-form is dual to a scalar field. In \cite{Ezhuthachan:2008ch} this dualisation was applied to three-dimensional maximally supersymmetric Yang-Mills (the effective worldvolume theory of multiple D2-branes) and it was shown that it could be rewritten as an $SO(8)$ invariant Lorentzian 3-algebra theory. Three-dimensional maximally supersymmetric Yang-Mills (3D-SYM) arises simply by the appropriate dimensional reduction of 10D-SYM and the higher derivative corrections to this have been uniquely determined (including quartic fermions in the Lagrangian) by superspace considerations in \cite{Cederwall:2001bt,Cederwall:2001td} and independently in \cite{Bergshoeff:2001dc} by calculating open-string scattering amplitudes. The first higher derivative corrections to the 10D-SYM Lagrangian and supersymmetry transformations arise at order $\alpha'^2$ and the same is true in the reduction to three dimensions. In \cite{Alishahiha:2008rs} the authors applied the analysis of \cite{Ezhuthachan:2008ch} to the $\alpha'^2$ corrections of the 3D-SYM Lagrangian. The resulting $SO(8)$ invariant, Lorentzian 3-algebra formulation features only 3-brackets and covariant derivatives of the scalar and fermion fields. This lead to the conjecture that higher derivative corrections to the Euclidean BLG theory would be structurally identical to the Lorentzian theory and only feature 3-brackets and covariant derivatives.

A different approach was considered in \cite{Ezhuthachan:2009sr}. Here the most general $1/T_{M2}$ higher derivative M2-brane Lagrangian with arbitrary coefficients was considered. Then, using the `novel Higgs mechanism' \cite{Mukhi:2008ux} this was reduced uniquely to the four-derivative order correction of the D2-brane effective worldvolume theory. The results of \cite{Ezhuthachan:2009sr} applied both to the Euclidean BLG theory and Lorentzian 3-algebra theories and confirmed the conjecture of \cite{Alishahiha:2008rs}. Other attempts to construct the full nonlinear action for multiple M2-branes include \cite{Li:2008ya,Pang:2008hw,Iengo:2008cq}.

The higher derivative corrected 3-algebra Lagrangians of \cite{Alishahiha:2008rs,Ezhuthachan:2009sr} are expected to possess maximal supersymmetry although this was not verified in either case. An attempt to determine next order corrections to the supersymmetry transformations was made in \cite{Low:2010ie}. Here Low applied the analysis of \cite{Ezhuthachan:2008ch} and \cite{Alishahiha:2008rs} at the level of the multiple D2-brane supersymmetry transformations. It was found that the $\alpha'^2$ corrections to the fermion supersymmetry could be written in an $SO(8)$ fashion but that the scalar transformation could not be. The gauge field supervariation was not considered. By taking an Abelian truncation of the higher derivative Lorentzian 3-algebra action and showing it was supersymmetric, Low was able to partially determine the higher derivative scalar supersymmetry transformation. 

As it is not possible to derive higher derivative $SO(8)$ invariant 3-algebra valued supersymmetries from the multiple D2-brane ones by the 2-form/scalar dualisation approach, it seems the only way to unequivocally determine them is to examine the full supervariation of the higher derivative Lagrangian and by closing the superalgebra.  This is the approach we will take here, focusing solely on the Euclidean BLG theory of \cite{Ezhuthachan:2009sr}. 

The rest of the chapter is as follows. 
In section \ref{HDer_BLG} we revisit the higher derivative action of \cite{Ezhuthachan:2009sr} and introduce our ansatz for the $1/T_{M2}$ corrections to the Euclidean BLG supersymmetry transformations. 
We determine all arbitrary coefficients in the system in section \ref{Invariance} by examining the supervariation of the higher derivative Lagrangian for Euclidean BLG. In addition, in \ref{Closure}, the supersymmetry algebra is shown to close on the scalar and gauge fields for the coefficients we find.
In section \ref{s_Summary_of_results} we collect together our results and in section \ref{s_HDer_conc} we will offer some concluding remarks.
%
\subsection{Higher Derivative Lagrangian and Supersymmetries} \label{HDer_BLG}

We begin with the most general four-derivative order Lagrangian as considered in \cite{Ezhuthachan:2009sr}, to lowest nontrivial order in fermions
\begin{align}
\nonumber
\mathcal{L}_{T^{-1}_{M2}} = \sfrac{1}{T_{M2}} {\rm STr} \Big\{ &+ \bold{a} \, D^\mu X^I D_\mu X^J  D^\nu X^J D_\nu X^I + \bold{b} \, D^\mu X^I D_\mu X^I D^\nu X^J D_\nu X^J \\
\nonumber
&+ \bold{c} \, \varepsilon^{\mu\nu\lambda}\, X^{IJK} D_\mu X^I D_\nu X^J D_\lambda X^K \\
\nonumber
&+ \bold{d} \, X^{IJK} X^{IJL} D^\mu X^K D_\mu X^L + \bold{e} \, X^{IJK} X^{IJK} D^\mu X^L  D_\mu X^L \\
\nonumber
&+ \bold{f} \, X^{IJK} X^{IJK} X^{LMN} X^{LMN} + \bold{g} \, X^{IJK} X^{IJL} X^{KMN} X^{LMN} \\
%
%
\nonumber
&+ i \hat{\bold{d}} \, \bar{\psi} \Gamma^\mu \Gamma^{IJ} D^\nu\psi {D}_\mu X^{I} {D}_\nu X^J + i \hat{\bold{e}} \, \bar{\psi}\Gamma^\mu D^\nu\psi  {D}_\mu X^{I} {D}_\nu X^I \\
\nonumber
&+ i \hat{\bold{f}} \, \bar{\psi}\Gamma^{IJKL} D^\nu\psi\; X^{IJK}  {D}_\nu X^L + i \hat{\bold{g}} \, \bar{\psi}\Gamma^{IJ}  D^\nu\psi\; X^{IJK}{D}_\nu X^K \\
\nonumber
&+ i \hat{\bold{h}} \, \bar{\psi}\Gamma^{IJ}[X^J,X^{K},\psi] {D}^\mu X^{I} {D}_\mu X^K \\
\nonumber
&+ i \hat{\bold{i}} \, \bar{\psi}\Gamma^{\mu\nu}[X^I,X^{J},\psi] {D}_\mu X^{I}{D}_\nu X^J + i \hat{\bold{j}} \, \bar{\psi}\Gamma^{\mu\nu}\Gamma^{IJ}[X^J,X^{K},\psi] {D}_\mu X^{I}{D}_\nu X^K \\
\nonumber
&+ i \hat{\bold{k}} \, \bar{\psi}\Gamma^\mu\Gamma^{IJ}[X^K,X^{L},\psi] {D}_\mu X^{I}X^{JKL} + i \hat{\bold{l}} \, \bar{\psi}\Gamma^\mu[X^I,X^{J},\psi] {D}_\mu X^{K}X^{IJK} \\
\nonumber
&+ i \hat{\bold{m}} \, \bar{\psi}\Gamma^\mu\Gamma^{IJKL}[X^L,X^M,\psi] {D}_\mu X^M X^{IJK} + i \hat{\bold{n}} \, \bar{\psi}\Gamma^\mu\Gamma^{IJ}[X^K,X^{L},\psi] {D}_\mu X^L X^{IJK} \\
&+ i \hat{\bold{o}} \, \bar{\psi}\Gamma^{IJKL}[X^M,X^N,\psi] X^{IJL}X^{KMN} + i \hat{\bold{p}} \, \bar{\psi}\Gamma^{IJ}[X^K,X^{L},\psi] X^{IJM}X^{KLM} \Big\} \, . \label{L_Higher_Starting_Point}
\end{align}
We have adopted the notation $X^{IJK} := [ X^I , X^J , X^K ]$ which we will use to save space wherever possible. Let us make some comments on this Lagrangian. The symmetrised trace of four basis elements of the $\mathcal{A}_4$ 3-algebra is given by ${\rm STr} \big\{ T^a T^b T^c T^d \big\} = d^{abcd}$ and is totally symmetric and linear in its four entries. To be specific we could take $d^{abcd} = \frac{1}{4} h^{(ab} h^{cd)}$ as in \cite{Ezhuthachan:2009sr}. Next, we require that each term within this higher derivative Lagrangian is gauge invariant. As we have mentioned in chapter \ref{c_Multiple_M2s_and_3_algebras}, the 3-bracket generates a gauge symmetry whose action on an arbitrary 3-algebra element $Y= Y_a T^a$ is 
\begin{equation}
\delta Y = [ \alpha , \beta , Y ] \, , \label{gauge_trans}
\end{equation}
where $\alpha$ and $\beta$ are two other elements of the 3-algebra. Acting on a generic four-derivative order term with the gauge transformation \eqref{gauge_trans} we see that the requirement of gauge invariance leads to 
\begin{align}
{\rm STr} \big\{ [ \alpha , \beta, Y_1 ] Y_2 Y_3 Y_4 + Y_1 [ \alpha , \beta, Y_2 ] Y_3 Y_4 + Y_1 Y_2 [ \alpha , \beta, Y_3 ] Y_4 + Y_1 Y_2 Y_3 [ \alpha , \beta, Y_4 ] \big\} =0 \, , \label{Gauge_Inv}
\end{align}
where $Y_1, \ldots, Y_4$ are arbitrary fields. In basis form this symmetrised trace invariance condition reads
\begin{equation}
d^{abcd} f^{efg}{}_a + d^{aecd} f^{bfg}{}_a + d^{abed} f^{cfg}{}_a + d^{abce} f^{dfg}{}_a =0 \ \ \text{ i.e.} \ \ d^{a(bcd} f^{e)fg}{}_a = 0 \, , \label{Gauge_Inv_basis}
\end{equation}
and can be seen as a generalisation of the trace invariance property: $ h^{a(b} f^{e)fg}{}_a = 0$.

There are further identities we can construct using the symmetrised trace. To start with we note that due to their simple nature the structure constants of the $\mathcal{A}_4$ 3-algebra satisfy 
\begin{equation}
f^{[abcd} f^{e] fgh} =0 \ \ \text{ i.e.} \ \ f^{abcd} f^{efgh} = + f^{bced} f^{afgh} - f^{cead} f^{bfgh} + f^{eabd} f^{cfgh} - f^{eabc} f^{dfgh} \, . \label{A4_Id}
\end{equation}
We can combine this identity with the symmetrised trace to find
\begin{align}
{\rm STr} \big\{ T_d T_h T_i T_j \big\} f^{abcd} f^{efgh} =& {\rm STr} \big\{ T_d T_h T_i T_j \big\} ( f^{bced} f^{afgh} - f^{cead} f^{bfgh} + f^{eabd} f^{cfgh} ) \, ,
\end{align}
where the final term, ${\rm STr} \big\{ T_d T_h T_i T_j \big\} f^{eabc} f^{dfgh} $, vanishes because of symmetry/antisymmetry under $d \leftrightarrow h$. Contracting the gauge indices with the fields leads to the following identities
\begin{align}
{\rm STr} \Big\{ \alpha \beta \, X^{I_1 I_2 I_3} X^{J_1 J_2 J_3} \Big\} = {\rm STr} \Big\{ \alpha \beta \left(  X^{J_1 I_2 I_3} X^{I_1 J_2 J_3} + X^{I_1 J_1 I_3} X^{I_2 J_2 J_3} + X^{I_1 I_2 J_1} X^{I_3 J_2 J_3} \right) \Big\} \, , \label{Useful_Id} 
\end{align}
\begin{align}
\nonumber
{\rm STr} \Big\{ \alpha \beta \, [ X^{I_1} , X^{I_2} , \gamma ] X^{J_1 J_2 J_3} \Big\} = {\rm STr} \Big\{ \alpha \beta \, \big( [ X^{J_1} , X^{I_2} , \gamma ] X^{I_1 J_2 J_3} + &[ X^{I_1} , X^{J_1} , \gamma ] X^{I_2 J_2 J_3} \\
+ &[ X^{J_2} , X^{J_3} , \gamma ] X^{I_1 I_2 J_1} \big) \Big\} \, , \label{Useful_Id2}
\end{align}
where $\alpha$, $\beta$ and $\gamma$ are arbitrary fields and $I_1 , J_1 , \ldots$ are transverse Lorentz indices. 

The starting ansatz for the four-derivative order Lagrangian can be simplified using these identities. Equation \eqref{Useful_Id} shows that the $\bold{f}$ and $\bold{g}$ terms in $\mathcal{L}_{1/T_{M2}}$ are proportional to each other. The same equation, together with antisymmetry in the $\Gamma$-matrix indices, tells us that the term in $\mathcal{L}_{1/T_{M2}}$ with coefficient $\hat{\bold{o}}$ is identically zero. Similarly, the term with coefficient $\hat{\bold{m}}$ is identically zero through the use of \eref{Useful_Id2}. We subsequently drop the terms with coefficients $\bold{g}$, $\hat{\bold{m}}$ and $\hat{\bold{o}}$ to leave
\begin{align}
\nonumber
\mathcal{L}_{T^{-1}_{M2}} = \sfrac{1}{T_{M2}} {\rm STr} \Big\{ &+ \bold{a} \, D^\mu X^I D_\mu X^J  D^\nu X^J D_\nu X^I + \bold{b} \, D^\mu X^I D_\mu X^I D^\nu X^J D_\nu X^J \\
\nonumber
&+ \bold{c} \, \varepsilon^{\mu\nu\lambda}\, X^{IJK} D_\mu X^I D_\nu X^J D_\lambda X^K \\
\nonumber
&+ \bold{d} \, X^{IJK} X^{IJL} D^\mu X^K D_\mu X^L + \bold{e} \, X^{IJK} X^{IJK} D^\mu X^L  D_\mu X^L \\
\nonumber
&+ \bold{f} \, X^{IJK} X^{IJK} X^{LMN} X^{LMN} \\
%
%
\nonumber
&+ i \hat{\bold{d}} \, \bar{\psi} \Gamma^\mu \Gamma^{IJ} D^\nu\psi {D}_\mu X^{I} {D}_\nu X^J + i \hat{\bold{e}} \, \bar{\psi}\Gamma^\mu D^\nu\psi  {D}_\mu X^{I} {D}_\nu X^I \\
\nonumber
&+ i \hat{\bold{f}} \, \bar{\psi}\Gamma^{IJKL} D^\nu\psi\; X^{IJK}  {D}_\nu X^L + i \hat{\bold{g}} \, \bar{\psi}\Gamma^{IJ}  D^\nu\psi\; X^{IJK}{D}_\nu X^K \\
\nonumber
&+ i \hat{\bold{h}} \, \bar{\psi}\Gamma^{IJ}[X^J,X^{K},\psi] {D}^\mu X^{I} {D}_\mu X^K \\
\nonumber
&+ i \hat{\bold{i}} \, \bar{\psi}\Gamma^{\mu\nu}[X^I,X^{J},\psi] {D}_\mu X^{I}{D}_\nu X^J + i \hat{\bold{j}} \, \bar{\psi}\Gamma^{\mu\nu}\Gamma^{IJ}[X^J,X^{K},\psi] {D}_\mu X^{I}{D}_\nu X^K \\
\nonumber
&+ i \hat{\bold{k}} \, \bar{\psi}\Gamma^\mu\Gamma^{IJ}[X^K,X^{L},\psi] {D}_\mu X^{I}X^{JKL} + i \hat{\bold{l}} \, \bar{\psi}\Gamma^\mu[X^I,X^{J},\psi] {D}_\mu X^{K}X^{IJK} \\
\nonumber
&+ i \hat{\bold{n}} \, \bar{\psi}\Gamma^\mu\Gamma^{IJ}[X^K,X^{L},\psi] {D}_\mu X^L X^{IJK} \\
&+ i \hat{\bold{p}} \, \bar{\psi}\Gamma^{IJ}[X^K,X^{L},\psi] X^{IJM}X^{KLM} \Big\} . \label{L_Higher}
\end{align}

We now give the general starting point for the $1/T_{M2}$ higher derivative corrections to the $\mathcal{N}=8$ supersymmetry transformations which are consistent with mass dimension, 3-algebra index structure, parity under $\Gamma_{012}$ and Lorentz invariance. We assume that the higher derivative scalar and fermion supersymmetry transformations are built out of $\psi$, $DX$ and $[X,X,X]$ only. In particular, as the Chern-Simons term in the $\mathcal{A}_4$ BLG Lagrangian does not receive higher derivative corrections, the gauge field strength is not present in the $1/T_{M2}$ supersymmetries. The gauge field variation additionally requires the presence of a `bare' scalar field. 

Our ansatz for the scalar supersymmetry transformation, to lowest order in fermions, is
\begin{align}
\delta' X^I_a &= \sfrac{1}{T_{M2}} \left( \delta_{2  DX}' X^I_a + \delta_{1  DX}' X^I_a + \delta_{0  DX}' X^I_a \right) \, ,
\end{align}
where
\begin{align}
\nonumber
\delta_{2  DX}' X^I_a =&+ i s_1 ( \bar{\epsilon} \Gamma^{IJK} \Gamma^{\mu \nu} \psi_b ) D_{\mu} X^{J}_c D_{\nu} X^{K}_d \ d^{bcd}{}_{a} \\
\nonumber
&+  i s_2 ( \bar{\epsilon} \Gamma^{J} \Gamma^{\mu \nu} \psi_b ) D_{\mu} X^{I}_c D_{\nu} X^{J}_d \ d^{bcd}{}_{a} \\
\nonumber
&+  i s_3 ( \bar{\epsilon} \Gamma^{J} \psi_b ) D_{\mu} X^{I}_c D^{\mu} X^{J}_d \ d^{bcd}{}_{a} \\
&+  i s_4 ( \bar{\epsilon} \Gamma^{I} \psi_b ) D_{\mu} X^{J}_c D^{\mu} X^{J}_d \ d^{bcd}{}_{a} \label{S1} \, , \\ 
\nonumber \\
\nonumber
\delta_{1  DX}' X^I_a =&+ i s_5 ( \bar{\epsilon} \Gamma^{IJKLM} \Gamma^{\mu} \psi_b ) D_{\mu} X^{J}_c X^{KLM}_d \ d^{bcd}{}_{a} \\
\nonumber
&+ i s_6	( \bar{\epsilon} \Gamma^{KLM} \Gamma^{\mu} \psi_b ) D_{\mu} X^{I}_c X^{KLM}_d \ d^{bcd}{}_{a} \\
\nonumber
&+ i s_7	( \bar{\epsilon} \Gamma^{JLM} \Gamma^{\mu} \psi_b ) D_{\mu} X^{J}_c X^{ILM}_d \ d^{bcd}{}_{a} \\
\nonumber
&+ i s_8	( \bar{\epsilon} \Gamma^{ILM} \Gamma^{\mu} \psi_b ) D_{\mu} X^{J}_c X^{JLM}_d \ d^{bcd}{}_{a} \\
&+ i s_9	( \bar{\epsilon} \Gamma^{M} \Gamma^{\mu} \psi_b ) D_{\mu} X^{J}_c X^{IJM}_d \ d^{bcd}{}_{a} \label{S2} \, , \\
\nonumber \\
\nonumber
\delta_{0  DX}' X^I_a =&+ i s_{10} ( \bar{\epsilon} \Gamma^{I} \psi_b ) X^{JKL}_c X^{JKL}_d \ d^{bcd}{}_{a} \\
&+ i s_{11} ( \bar{\epsilon} \Gamma^{L} \psi_b ) X^{JKL}_c X^{JKI}_d \ d^{bcd}{}_{a} \label{S3} \, .
\end{align}
The ansatz for the fermion supersymmetry transformation is
\begin{align}
\delta' \psi_a &= \sfrac{1}{T_{M2}} \left( \delta_{3  DX}' \psi_a + \delta_{2  DX}' \psi_a + \delta_{1  DX}' \psi_a + \delta_{0  DX}' \psi_a \right) \, ,
\end{align}	
where
\begin{align}
\nonumber
\delta_{3  DX}' \psi_a =&+ f_1 \Gamma^{JKL} \Gamma^{\mu \nu \lambda} \epsilon D_{\mu} X^{J}_b D_{\nu} X^{K}_c D_{\lambda} X^{L}_d \ d^{bcd}{}_{a} \\
\nonumber
&+ f_2 \Gamma^{K} \Gamma^{\mu} \epsilon D_{\mu} X^{J}_b D_{\nu} X^{J}_c D^{\nu} X^{K}_d \ d^{bcd}{}_{a} \\
&+ f_3 \Gamma^{K} \Gamma^{\mu} \epsilon D_{\mu} X^{K}_b D_{\nu} X^{J}_c D^{\nu} X^{J}_d \ d^{bcd}{}_{a} \, , \\
\nonumber \\
\nonumber
\delta_{2 DX}' \psi_a =&+ f_4 \Gamma^{JKLMN} \Gamma^{\mu \nu} \epsilon  D_\mu X^J_b D_\nu X^K_c X^{LMN}_d \, d^{bcd}{}_a \\
\nonumber
&+ f_5 \Gamma^{KLM} \Gamma^{\mu \nu} \epsilon D_\mu X^J_b D_\nu X^K_c X^{JLM}_d \, d^{bcd}{}_a \\
\nonumber
&+ f_6 \Gamma^M \Gamma^{\mu \nu} \epsilon D_\mu X^J_b D_\nu X^K_c X^{JKM}_d \, d^{bcd}{}_a \\
\nonumber
&+ f_7 \Gamma^{KLM} \epsilon D_\mu X^J_b D^\mu X^J_c X^{KLM}_d \, d^{bcd}{}_a \\
&+ f_8 \Gamma^{KLM} \epsilon D_\mu X^J_b D^\mu X^K_c X^{JLM}_d \, d^{bcd}{}_a \, , \\
\nonumber \\
\nonumber
\delta_{1 DX}' \psi_a =&+ f_9 \Gamma^J \Gamma^\mu \epsilon D_\mu X^J_b X^{KLM}_c X^{KLM}_d \, d^{bcd}{}_a \\
&+ f_{10} \Gamma^M \Gamma^\mu \epsilon D_\mu X^J_b X^{JKL}_c X^{KLM}_d \, d^{bcd}{}_a \, , \\
\nonumber \\
\delta_{0 DX}' \psi_a =&+ f_{11} \Gamma^{NOP} \epsilon X^{JKL}_b X^{JKL}_c X^{NOP}_d \, d^{bcd}{}_a \, .
\end{align}
Finally, the ansatz for the gauge field variation, again to lowest order in fermions, is 
\begin{equation}
\delta' \tilde{A}_{\mu}{}^b{}_a = \sfrac{1}{T_{M2}} \left( \delta_{2  DX}' \tilde{A}_{\mu}{}^b{}_a + \delta_{1  DX}' \tilde{A}_{\mu}{}^b{}_a + \delta_{0  DX}' \tilde{A}_{\mu}{}^b{}_a \right) \, ,
\end{equation}
where
\begin{align}
\nonumber
\delta_{2  DX}' \tilde{A}_{\mu}{}^b{}_a =&+ i g_1 ( \bar{\epsilon} \Gamma_\mu \Gamma^I \psi_e ) D_\nu X^J_f D^\nu X^J_g X^I_c \, d^{efg}{}_d f^{cdb}{}_a \\
\nonumber
&+ i g_2 ( \bar{\epsilon} \Gamma^\nu \Gamma^I \psi_e ) D_\mu X^J_f D_\nu X^J_g X^I_c \, d^{efg}{}_d f^{cdb}{}_a \\
\nonumber
&+ i g_3 ( \bar{\epsilon} \Gamma^\nu \Gamma^J \psi_e ) D_\mu X^J_f D_\nu X^I_g X^I_c \, d^{efg}{}_d f^{cdb}{}_a \\
\nonumber
&+ i g_4 ( \bar{\epsilon} \Gamma^\nu \Gamma^J \psi_e ) D_\mu X^I_f D^\nu X^J_g X^I_c \, d^{efg}{}_d f^{cdb}{}_a \\
\nonumber
&+ i g_5 ( \bar{\epsilon} \Gamma_\mu \Gamma^J \psi_e ) D_\nu X^J_f D^\nu X^I_g X^I_c \, d^{efg}{}_d f^{cdb}{}_a \\
\nonumber
&+ i g_6 ( \bar{\epsilon} \Gamma_{\mu \nu \lambda} \Gamma^J \psi_e ) D^\nu X^J_f D^\lambda X^I_g X^I_c \, d^{efg}{}_d f^{cdb}{}_a \\
\nonumber
&+ i g_7 ( \bar{\epsilon} \Gamma_{\mu \nu \lambda} \Gamma^{IJK} \psi_e ) D^\nu X^J_f D^\lambda X^K_g X^I_c \, d^{efg}{}_d f^{cdb}{}_a \\
&+ i g_8 ( \bar{\epsilon} \Gamma^{\nu} \Gamma^{IJK} \psi_e ) D_\mu X^J_f D^\nu X^K_g X^I_c \, d^{efg}{}_d f^{cdb}{}_a \label{G1} \, , \\
\nonumber \\
\nonumber
\delta_{1  DX}' \tilde{A}_{\mu}{}^b{}_a =&+i g_9 ( \bar{\epsilon} \Gamma_{\mu \nu} \Gamma^{KLM} \psi_e ) D^\nu X^I_f X^{KLM}_g X^I_c \, d^{efg}{}_d f^{cdb}{}_a \\
\nonumber
&+i g_{10} ( \bar{\epsilon} \Gamma_{\mu \nu} \Gamma^{JLM} \psi_e ) D^\nu X^J_f X^{ILM}_g X^I_c \, d^{efg}{}_d f^{cdb}{}_a \\
\nonumber
&+i g_{11} ( \bar{\epsilon} \Gamma_{\mu \nu} \Gamma^{M} \psi_e ) D^\nu X^J_f X^{IJM}_g X^I_c \, d^{efg}{}_d f^{cdb}{}_a \\
\nonumber
&+i g_{12} ( \bar{\epsilon} \Gamma^{KLM} \psi_e ) D_\mu X^I_f X^{KLM}_g X^I_c \, d^{efg}{}_d f^{cdb}{}_a \\	
\nonumber
&+i g_{13} ( \bar{\epsilon} \Gamma^{JLM} \psi_e ) D_\mu X^J_f X^{ILM}_g X^I_c \, d^{efg}{}_d f^{cdb}{}_a \\	
&+i g_{14} ( \bar{\epsilon} \Gamma^{M} \psi_e ) D_\mu X^J_f X^{IJM}_g X^I_c \, d^{efg}{}_d f^{cdb}{}_a \label{G2} \, , \\
\nonumber \\
\delta_{0  DX}' \tilde{A}_{\mu}{}^b{}_a =&+ i g_{15} ( \bar{\epsilon} \Gamma_{\mu} \Gamma^{I} \psi_e ) X^{JKL}_f X^{JKL}_g X^I_c \, d^{efg}{}_d f^{cdb}{}_a \label{G3} \, .
\end{align}
There are other terms which are consistent with mass dimensions etc.\ that could be added to the $\delta'$ variations however, we can apply the $\mathcal{A}_4$ identity $f^{[abcd} f^{e]fgh} =0$ at the level of the supersymmetry transformations to find
\begin{equation}
\alpha_b X^{I_1 I_2 I_3}_c X^{J_1 J_2 J_3}_d \, d^{bcd}{}_a = \alpha_b \big( X^{J_1 I_2 I_3}_c X^{I_1 J_2 J_3}_d + X^{I_1 J_1 I_3}_c X^{I_2 J_2 J_3}_d + X^{I_1 I_2 J_1}_c X^{I_3 J_2 J_3}_d \big) \, d^{bcd}{}_a \, , \label{Useful_Id_5} \\[5pt]
\end{equation}
\begin{equation}
\alpha_e \beta_f X^{J_1 J_2 J_3}_g X^I_c \, d^{efg}{}_d f^{cdb}{}_a = \alpha_e \beta_f \big( X^{I J_2 J_3}_g X^{J_1}_c + X^{J_1 I J_3}_g X^{J_2}_c + X^{J_1 J_2 I}_g X^{J_3}_c \big) \, d^{efg}{}_d f^{cdb}{}_a \, , \label{Useful_Id_4}
\end{equation}
where $\alpha$ and $\beta$ are either $\psi$, $DX$ or $[X,X,X]$. Using these identities it is possible to show that the additional terms are either identically zero or proportional to terms we have already listed.

\subsection{Invariance of the Lagrangian} \label{Invariance}

We want to determine the coefficients for which the BLG Lagrangian, given in \eref{BLG_act_Tr}, together with its $1/T_{M2}$ correction given in \eref{L_Higher} is maximally supersymmetric. As the BLG Lagrangian is invariant under the lowest order supersymmetries given in Eqs.\,(\ref{BLG_susy_s})\,-\,\eqref{BLG_susy_g} i.e.\ $\delta \mathcal{L}_{BLG}=0$, the full corrected Lagrangian varies into 
\begin{equation}
\tilde{\delta} \mathcal{L} = \delta' \mathcal{L}_{BLG} + \delta \mathcal{L}_{1/T_{M2}} + \mathcal{O} \left( \sfrac{1}{(T_{M2})^2} \right) = \tilde{\delta} \mathcal{L}_4 + \tilde{\delta} \mathcal{L}_3 + \tilde{\delta} \mathcal{L}_2 + \tilde{\delta} \mathcal{L}_1 + \tilde{\delta} \mathcal{L}_0 \, ,
\end{equation}
where we ignore $\mathcal{O} \left( 1/(T_{M2})^2 \right)$ terms. The subscript in $\tilde{\delta} \mathcal{L}_n$ enumerates the total number of covariant derivatives acting on the fields and because the terms in $\tilde{\delta} \mathcal{L}_n$ are independent of those in any other $\tilde{\delta} \mathcal{L}_m$, invariance of the full Lagrangian means each $\tilde{\delta} \mathcal{L}_n$ must be invariant up to total derivatives.\footnote{There is the possibility that $\tilde{\delta} \mathcal{L}=0$ only after terms are removed using $1/T_{M2} \ \times$ lowest order equations of motion (which are $\mathcal{O} ( 1/T^2_{M2} ))$, in which case the different $\tilde{\delta} \mathcal{L}_n$ are not independent. However, we find for the Euclidean theory that invariance does not require use of the lowest order field equations.}

When we insert the higher derivative supersymmetries which are of the form $\delta' \chi_a = \alpha_b \beta_c \gamma_d \, d^{bcd}{}_a$, into the varied kinetic terms in $\delta' \mathcal{L}_{BLG}$ we find Tr is promoted to STr because
\begin{align}
{\rm Tr} ( \phi \delta' \chi ) = \phi^a \delta' \chi_a = \phi^a \alpha_b \beta_c \gamma_d \, d^{bcd}{}_a = \phi_a \alpha_b \beta_c \gamma_d \, d^{abcd} = {\rm STr} \{ \phi \alpha \beta \gamma \} \, .
\end{align}
Inserting the higher derivative supersymmetries into the varied bosonic potential and Yukawa terms in $\delta' \mathcal{L}_{BLG}$ requires more manipulation:
\begin{align}
{\rm Tr} ( \phi [ \lambda , \varphi , \delta' \chi ] ) = \phi_g \lambda_e \varphi_f \delta' \chi_a f^{efag} = \phi_g \lambda_e \varphi_f \alpha_b \beta_c \gamma_d \, d^{bcd}{}_a f^{efag} = - \phi_g \lambda_e \varphi_f \alpha_b \beta_c \gamma_d \, d^{abcd} f^{efg}{}_a \, .
\end{align}
Using the gauge invariance condition in \eref{Gauge_Inv} we can write this as
\begin{align}
{\rm Tr} ( \phi [ \lambda , \varphi , \delta' \chi ] ) = {\rm STr} \{ \phi \beta \gamma [ \lambda , \varphi , \alpha ] + \phi \alpha \gamma [ \lambda , \varphi , \beta ] + \phi \alpha \beta [ \lambda , \varphi , \gamma ] \} =: {\rm STr} \{ \phi [ \lambda , \varphi , \alpha \beta \gamma ] \} \, .
\end{align}

We are now in a position where we can proceed to compute $\tilde{\delta} \mathcal{L}$. We start by investigating the terms in the variation of the full corrected Lagrangian which contain four covariant derivatives. These come from
\begin{align}
\nonumber
\tilde{\delta} \mathcal{L}_{4} =\sfrac{1}{T_{M2}} {\rm STr} \Big\{ &- D_\mu ( \delta'_{2 DX} X^I ) D^\mu X^I + \sfrac{i}{2} \delta'_{3 DX} \bar{\psi} \Gamma^\mu D_\mu \psi + \sfrac{i}{2} \bar{\psi} \Gamma^\mu D_\mu ( \delta'_{3 DX} \psi ) + \sfrac{1}{2} \varepsilon^{\mu\nu\lambda} F_{\nu \lambda} \delta'_{2DX} \tilde{A}_\mu \\
\nonumber
&+ 4 \bold{a} \, D^\mu ( \delta X^I ) D_\mu X^J  D^\nu X^J D_\nu X^I + 4 \bold{b} \, D^\mu ( \delta X^I ) D_\mu X^I D^\nu X^J D_\nu X^J \\
\nonumber
&+ i \hat{\bold{d}} \, \delta_{1 DX} \bar{\psi} \Gamma^\mu \Gamma^{IJ} D_\nu\psi {D}_\mu X^{I} {D}^\nu X^J + i \hat{\bold{d}} \, \bar{\psi} \Gamma^\mu \Gamma^{IJ} D_\nu ( \delta_{1 DX}  \psi ) {D}_\mu X^{I} {D}^\nu X^J \\
&+ i \hat{\bold{e}} \, \delta_{1 DX}  \bar{\psi}\Gamma^\mu D^\nu\psi {D}_\mu X^{I} {D}_\nu X^I + i \hat{\bold{e}} \, \bar{\psi}\Gamma^\mu D^\nu ( \delta_{1 DX}  \psi ) {D}_\mu X^{I} {D}_\nu X^I \Big\} \, .
\end{align}
Note that the gauge field strength contributes two derivatives through its definition as the commutator of covariant derivatives. We have also split the lowest order fermion supersymmetry into $\delta \psi = \delta_{0DX} \psi + \delta_{1DX} \psi$ with $\delta_{0DX} \psi = \Gamma^\mu \Gamma^I \epsilon \, D_\mu X^I$ and $\delta_{0DX} \psi = - \sfrac{1}{6} \Gamma^{IJK} \epsilon \, X^{IJK}$. The next steps in the calculation are to insert the appropriate supersymmetry transformations, canonically reorder $\bar{\psi}$ and $\epsilon$ using the spinor flip condition \eref{Spinor_Flip} and then commute the worldvolume $\Gamma$-matrices through the transverse ones. Doing all this gives
\begin{align}
\nonumber
\tilde{\delta} \mathcal{L}_{4} = \sfrac{1}{T_{M2}} {\rm STr} \Big\{ &- i s_1 \bar{\epsilon} \Gamma^{IJK} \Gamma^{\lambda \mu} D_\nu ( \psi D_\lambda X^J D_\mu X^K ) D^\nu X^I - i s_2 \bar{\epsilon} \Gamma^I \Gamma^{\lambda \mu} D_\nu ( \psi D_\lambda X^J D_\mu X^I ) D^\nu X^J \\
\nonumber
&- i s_3 \bar{\epsilon} \Gamma^I D_\mu ( \psi D_\nu X^J D^\nu X^I ) D^\mu X^J - i s_4 \bar{\epsilon} \Gamma^I D_\mu ( \psi D_\nu X^J D^\nu X^J ) D^\mu X^I \\
\nonumber
&- \sfrac{i}{2} f_1 \bar{\epsilon} \Gamma^{IJK} \Gamma^{\nu \lambda \rho} \Gamma^\mu D_\mu \psi D_\nu X^I D_\lambda X^J D_\rho X^K \\
\nonumber
&+ \sfrac{i}{2} f_1 \bar{\epsilon} \Gamma^{IJK} \Gamma^{\nu \lambda \rho} \Gamma^\mu \psi D_\mu ( D_\nu X^I D_\lambda X^J D_\rho X^K ) \\
\nonumber
&- \sfrac{i}{2} f_2 \bar{\epsilon} \Gamma^I \Gamma^\lambda \Gamma^\mu D_\mu \psi D_\lambda X^J D_\nu X^J D^\nu X^I + \sfrac{i}{2} f_2 \bar{\epsilon} \Gamma^I \Gamma^\lambda \Gamma^\mu \psi D_\mu ( D_\lambda X^J D_\nu X^J D^\nu X^I ) \\
\nonumber
&- \sfrac{i}{2} f_3 \bar{\epsilon} \Gamma^I \Gamma^\lambda \Gamma^\mu D_\mu \psi D_\lambda X^I D_\nu X^J D^\nu X^J + \sfrac{i}{2} f_3 \bar{\epsilon} \Gamma^I \Gamma^\lambda \Gamma^\mu \psi D_\mu ( D_\lambda X^I D_\nu X^J D^\nu X^J ) \\
\nonumber
&- \sfrac{i}{2} g_1 \varepsilon^{\mu \rho \sigma} \bar{\epsilon} \Gamma^I \Gamma_\mu \psi D_\nu X^J D^\nu X^J ( \tilde{F}_{\rho \sigma} X^I ) - \sfrac{i}{2} g_2 \varepsilon^{\mu \rho \sigma} \bar{\epsilon} \Gamma^I \Gamma^\nu \psi D_\mu X^J D_\nu X^J ( \tilde{F}_{\rho \sigma} X^I ) \\
\nonumber
&- \sfrac{i}{2} g_3 \varepsilon^{\mu \rho \sigma} \bar{\epsilon} \Gamma^J \Gamma^\nu \psi D_\mu X^J D_\nu X^I ( \tilde{F}_{\rho \sigma} X^I ) - \sfrac{i}{2} g_4 \varepsilon^{\mu \rho \sigma} \bar{\epsilon} \Gamma^J \Gamma^\nu \psi D_\mu X^I D_\nu X^J ( \tilde{F}_{\rho \sigma} X^I ) \\
\nonumber
&- \sfrac{i}{2} g_5 \varepsilon^{\mu \rho \sigma} \bar{\epsilon} \Gamma^J \Gamma_\mu \psi D_\nu X^J D^\nu X^I ( \tilde{F}_{\rho \sigma} X^I ) - \sfrac{i}{2} g_6 \varepsilon^{\mu \rho \sigma} \bar{\epsilon} \Gamma^J \Gamma_{\mu \nu \lambda} \psi D^\nu X^J D^\lambda X^I ( \tilde{F}_{\rho \sigma} X^I ) \\
\nonumber
&- \sfrac{i}{2} g_7 \varepsilon^{\mu \rho \sigma} \bar{\epsilon} \Gamma^{IJK} \Gamma_{\mu \nu \lambda} \psi D^\nu X^J D^\lambda X^K ( \tilde{F}_{\rho \sigma} X^I ) \\
\nonumber
&- \sfrac{i}{2} g_8 \varepsilon^{\mu \rho \sigma} \bar{\epsilon} \Gamma^{IJK} \Gamma^{\nu} \psi D_\mu X^J D_\nu X^K ( \tilde{F}_{\rho \sigma} X^I ) \\
\nonumber
&+ 4 i \bold{a} \, \bar{\epsilon} \Gamma^I D^\mu \psi D_\mu X^J  D^\nu X^J D_\nu X^I + 4 i \bold{b} \, \bar{\epsilon} \Gamma^I D^\mu \psi D_\mu X^I D^\nu X^J D_\nu X^J \\
\nonumber
&+ i \hat{\bold{d}} \, \bar{\epsilon} \Gamma^K \Gamma^{IJ} \Gamma^\lambda \Gamma^\mu D_\nu \psi D_\mu X^I D^\nu X^J D_\lambda X^K \\
\nonumber
&+ i \hat{\bold{d}} \, \bar{\epsilon} \Gamma^K \Gamma^{IJ} \Gamma^\lambda \Gamma^\mu \psi D^\nu ( D_\lambda X^K ) D_\mu X^I D_\nu X^J \\
&+ i \hat{\bold{e}} \, \bar{\epsilon} \Gamma^J \Gamma^\lambda \Gamma^\mu D^\nu \psi D_\mu X^I D_\nu X^I D_\lambda X^J - i \hat{\bold{e}}  \bar{\epsilon} \Gamma^J \Gamma^\lambda \Gamma^\mu \psi D^\nu ( D_\lambda X^J ) D_\mu X^I D_\nu X^I \Big\}.
\end{align}
After using worldvolume $\Gamma$-matrix duality \eqref{M2_wvol_duality_f} wherever $\varepsilon^{\mu \rho \sigma}$ occurs and then expanding out the $\Gamma$-matrices (this has been aided by use of Cadabra \cite{Peeters:2006kp,Peeters:2007wn}) we find the appearance of four distinct and independent types of $\Gamma$-matrix terms;  $\Gamma^{IJK} \Gamma^{\lambda \mu}$, $\Gamma^{IJK}$, $\Gamma^I \Gamma^{\lambda \mu}$ and $\Gamma^I$. We consider each of these types in turn.

We find the $\Gamma^{IJK} \Gamma^{\lambda \mu}$ terms to be
\begin{align}
\nonumber
\sfrac{1}{T_{M2}} {\rm STr} \Big\{ &+ i \left( - \sfrac{3}{2} f_1 - s_1 + \hat{\bold{d}} \right) \bar{\epsilon} \Gamma^{IJK} \Gamma^{\lambda \mu} D^\nu \psi D_\mu X^I D_\nu X^J D_\lambda X^K \\
\nonumber
&+ i \left( + \sfrac{3}{2} f_1 \right) \bar{\epsilon} \Gamma^{IJK} \Gamma^{\lambda \mu} \psi D^\nu ( D_\mu X^I D_\nu X^J D_\lambda X^K ) \\
\nonumber
&+ i \left(  - 2 s_1 +  \hat{\bold{d}} \right) \bar{\epsilon} \Gamma^{IJK} \Gamma^{\lambda \mu} \psi D^\nu ( D_\mu X^I ) D_\nu X^J D_\lambda X^K \\
&+ i \left( - g_8 \right) \bar{\epsilon} \Gamma^{IJK} \Gamma^{\lambda\mu} \psi D_\mu X^I D^\nu X^J ( \tilde{F}_{\nu \lambda} X^K ) \Big\} \, .
\end{align}
The first two lines combine to form a total derivative if they share the same coefficient. Hence we require $- \sfrac{3}{2} f_1 - s_1 + \hat{\bold{d}} = + \sfrac{3}{2} f_1$. The two remaining terms are invariant if $s_1 = + \sfrac{1}{2} \hat{\bold{d}}$ and $g_8=0$. The value for $s_1$ allows us to identify $f_1 = + \sfrac{1}{6} \hat{\bold{d}}$. 

The $\Gamma^{IJK}$ terms are
\begin{align}
\nonumber
\sfrac{1}{T_{M2}} {\rm STr} \Big\{ &- i \hat{\bold{d}} \, \bar{\epsilon} \Gamma^{IJK} \psi D^\mu ( D^\nu X^K ) D_\mu X^I D_\nu X^J + i g_7 \, \bar{\epsilon} \Gamma^{IJK} \psi D^\nu X^J D^\lambda X^K ( \tilde{F}_{\nu \lambda} X^I ) \Big\} \\
= \sfrac{1}{T_{M2}} {\rm STr} \Big\{ &+ i ( - \hat{\bold{d}} + 2 g_7 ) \bar{\epsilon} \Gamma^{IJK} \psi D^\mu ( D^\nu X^K ) D_\mu X^I D_\nu X^J \Big\} \, ,
\end{align}
where we have made use of the definition $\tilde{F}_{\mu \nu} X = [ D_\mu , D_\nu ] X$ and relabelled dummy Lorentz indices. Invariance of the $\Gamma^{IJK}$ terms then follows if $g_7 = + \sfrac{1}{2} \hat{\bold{d}}$.

After some manipulation the $\Gamma^I \Gamma^{\lambda \mu}$ terms are
\begin{align}
\nonumber
\sfrac{1}{T_{M2}} {\rm STr} \Big\{ &+ i \left( - \sfrac{1}{2} f_2 \right) \bar{\epsilon} \Gamma^I \Gamma^{\lambda \mu} D_\mu \psi D_\lambda X^J D_\nu X^J D^\nu X^I \\
\nonumber
&+ i \left( - \sfrac{1}{2} f_2 \right) \bar{\epsilon} \Gamma^I \Gamma^{\lambda \mu} \psi D_\mu ( D_\lambda X^J ) D_\nu X^J D^\nu X^I \\
\nonumber
&+ i \left( + \sfrac{1}{2} f_2 - \hat{\bold{d}} \right) \bar{\epsilon} \Gamma^I \Gamma^{\lambda \mu} \psi D_\lambda X^J D_\mu ( D_\nu X^J ) D^\nu X^I \\
\nonumber
&+ i \left( + \sfrac{1}{2} f_2 + \hat{\bold{e}} - s_2 \right) \bar{\epsilon} \Gamma^I \Gamma^{\lambda \mu} \psi D_\lambda X^J D_\nu X^J D_\mu ( D^\nu X^I ) \\[10pt]
\nonumber
&+ i \left( - \sfrac{1}{2} f_3 \right) \bar{\epsilon} \Gamma^I \Gamma^{\lambda \mu} D_\mu \psi D_\lambda X^I D_\nu X^J D^\nu X^J \\
\nonumber
&+ i \left( - \sfrac{1}{2} f_3 \right) \bar{\epsilon} \Gamma^I \Gamma^{\lambda \mu} \psi D_\mu ( D_\lambda X^I ) D_\nu X^J D^\nu X^J \\
\nonumber
&+ i \left( + \sfrac{1}{2} f_3 + \sfrac{1}{2} \hat{\bold{d}} + \sfrac{1}{2} s_2 \right) \bar{\epsilon} \Gamma^I \Gamma^{\lambda \mu} \psi D_\lambda X^I D_\mu ( D_\nu X^J D^\nu X^J ) \\[10pt]
\nonumber
&+ i \left( - \hat{\bold{d}} - \hat{\bold{e}} - s_2 \right) \bar{\epsilon} \Gamma^I \Gamma^{\lambda \mu} D^\nu \psi D_\mu X^I D_\nu X^J D_\lambda X^J \\
\nonumber
&+ i \left( + \sfrac{1}{2} f_2 + \sfrac{1}{2} g_3 + \sfrac{1}{2} g_4 + \sfrac{1}{2} g_5 \right) \bar{\epsilon} \Gamma^I \Gamma^{\lambda \mu} \psi ( \tilde{F}_{\mu \lambda} X^J ) D^\nu X^I D_\nu X^J \\
\nonumber
&+ i \left( + \sfrac{1}{2} f_3 + \sfrac{1}{2} g_1 + \sfrac{1}{2} g_2 \right) \bar{\epsilon} \Gamma^I \Gamma^{\lambda \mu} \psi ( \tilde{F}_{\mu \lambda} X^I ) D_\nu X^J D^\nu X^J \\
\nonumber
&+ i \left( - \hat{\bold{d}} - s_2 - g_3 \right) \bar{\epsilon} \Gamma^I \Gamma^{\lambda \mu} \psi ( \tilde{F}_{\mu \nu} X^J ) D_\lambda X^I D^\nu X^J \\
\nonumber
&+ i \left( - \hat{\bold{e}} + s_2 - g_2 \right) \bar{\epsilon} \Gamma^I \Gamma^{\lambda \mu} \psi ( \tilde{F}_{\mu \nu} X^I ) D_\lambda X^J D^\nu X^J \\
&+ i \left( + \hat{\bold{d}} - g_4 \right) \bar{\epsilon} \Gamma^I \Gamma^{\lambda \mu} \psi ( \tilde{F}_{\mu \nu} X^J ) D^\nu X^I D_\lambda X^J \Big\} \, .
\end{align}
The first seven lines can be written as two distinct total derivatives provided
\begin{align}
&- \sfrac{1}{2} f_2 = + \sfrac{1}{2} f_2 - \hat{\bold{d}} = + \sfrac{1}{2} f_2 + \hat{\bold{e}} - s_2 \, , \\[6pt]
&- \sfrac{1}{2} f_3 = + \sfrac{1}{2} f_3 + \sfrac{1}{2} \hat{\bold{d}} + \sfrac{1}{2} s_2 \, .
\end{align}
The remaining terms vanish if 
\begin{align}
0=&- \hat{\bold{d}} - \hat{\bold{e}} - s_2 \, , \\
0=&+ \sfrac{1}{2} f_2 + \sfrac{1}{2} g_3 + \sfrac{1}{2} g_4 + \sfrac{1}{2} g_5 \, , \\
0=&+ \sfrac{1}{2} f_3 + \sfrac{1}{2} g_1 + \sfrac{1}{2} g_2 \, , \\
0=&- \hat{\bold{d}} - s_2 - g_3 \, , \\
0=&- \hat{\bold{e}} + s_2 - g_2 \, , \\
0=&+ \hat{\bold{d}} - g_4 \, .
\end{align}
The solution to these simultaneous equations is
\begin{equation}
f_2 = + \hat{\bold{d}} \, , \quad f_3 = - \sfrac{1}{2} \hat{\bold{d}} \, , \quad s_2 = 0 \, , \quad \hat{\bold{e}} = - \hat{\bold{d}} \, , \label{Intermediate_1}
\end{equation}
\begin{equation}
g_1 = - \sfrac{1}{2} \hat{\bold{d}} \, , \quad g_2 = + \hat{\bold{d}} \, , \quad g_3 = - \hat{\bold{d}} \, , \quad g_4 = + \hat{\bold{d}} \, , \quad g_5 = - \hat{\bold{d}} \, .
\end{equation}
Finally, the $\Gamma^I$ terms can be manipulated to arrive at
\begin{align}
\nonumber
\sfrac{1}{T_{M2}} {\rm STr} \Big\{ &+ i \left( - \sfrac{1}{2} f_2 - s_3 + 4 \bold{a} - \hat{\bold{d}} + \hat{\bold{e}} \right) \bar{\epsilon} \Gamma^I D^\mu \psi D^\nu X^I D_\mu X^J D_\nu X^J \\
\nonumber
&+ i \left( + \sfrac{1}{2} f_2 - s_3 - \hat{\bold{e}} \right) \bar{\epsilon} \Gamma^I \psi D^\mu ( D^\nu X^I ) D_\mu X^J D_\nu X^J \\
\nonumber
&+ i \left( + \sfrac{1}{2} f_2 \right) \bar{\epsilon} \Gamma^I \psi D^\nu X^I D^\mu ( D_\mu X^J ) D_\nu X^J \\
\nonumber
&+ i \left( + \sfrac{1}{2} f_2 - s_3 - g_6 - \hat{\bold{d}} \right) \bar{\epsilon} \Gamma^I \psi D^\nu X^I D_\mu X^J D^\mu ( D_\nu X^J ) \\[10pt]
\nonumber
&+i \left( - \sfrac{1}{2} f_3 - s_4 + 4 \bold{b} + \hat{\bold{d}} \right) \bar{\epsilon} \Gamma^I D^\mu \psi D_\mu X^I D_\nu X^J D^\nu X^J \\
\nonumber
&+ i \left( + \sfrac{1}{2} f_3 \right) \bar{\epsilon} \Gamma^I \psi D^\mu ( D_\mu X^I ) D_\nu X^J D^\nu X^J \\
&+ i \left( + \sfrac{1}{2} f_3 - s_4 + \sfrac{1}{2} g_6 + \sfrac{1}{2} \hat{\bold{d}} \right) \bar{\epsilon} \Gamma^I \psi D_\mu X^I D^\mu ( D_\nu X^J D^\nu X^J ) \Big\} \, . \label{4DX_Gamma_I}
\end{align}
We see that the first four lines combine to form a total derivative if
\begin{equation}
- \sfrac{1}{2} f_2 - s_3 + 4 \bold{a} - \hat{\bold{d}} + \hat{\bold{e}} = + \sfrac{1}{2} f_2 - s_3 - \hat{\bold{e}} = + \sfrac{1}{2} f_2 = + \sfrac{1}{2} f_2 - s_3 - g_6 - \hat{\bold{d}} \, .
\end{equation}
The last three lines form another total derivative provided
\begin{equation}
- \sfrac{1}{2} f_3 - s_4 + 4 \bold{b} + \hat{\bold{d}} = + \sfrac{1}{2} f_3 = + \sfrac{1}{2} f_3 - s_4 + \sfrac{1}{2} g_6 + \sfrac{1}{2} \hat{\bold{d}} \, .
\end{equation}
Using the values for $f_2$, $f_3$ and $\hat{\bold{e}}$ in \eref{Intermediate_1} we can solve these latest simultaneous equations to discover
\begin{equation}
s_3 = + \hat{\bold{d}} \, , \quad s_4 = - \sfrac{1}{2} \hat{\bold{d}} \, , \quad g_6 = - 2 \hat{\bold{d}} \, , \quad \bold{a} = + \hat{\bold{d}} \, , \quad \bold{b} = - \sfrac{1}{2} \hat{\bold{d}} \, .
\end{equation}

To summarise, the four covariant derivative terms $\tilde{\mathcal{L}}_4$ are invariant up to boundary terms if the coefficients in the Lagrangian and supersymmetry transformations are given by
\begin{equation}
\bold{a} = + \hat{\bold{d}} \, , \quad \bold{b} = - \sfrac{1}{2} \hat{\bold{d}} \, , \quad \hat{\bold{e}} = - \hat{\bold{d}} \, , \label{4DX_L}
\end{equation}
\begin{equation}
f_1 = + \sfrac{1}{6} \hat{\bold{d}} \, , \quad f_2 = + \hat{\bold{d}} \, , \quad f_3 = - \sfrac{1}{2} \hat{\bold{d}} \, , \label{4DX_f}
\end{equation}
\begin{equation}
s_1 = + \sfrac{1}{2} \hat{\bold{d}} \, , \quad s_2 = 0 \, , \quad s_3 = + \hat{\bold{d}} \, , \quad s_4 = - \sfrac{1}{2} \hat{\bold{d}} \, , \label{4DX_s}
\end{equation}
\begin{equation}
\nonumber
g_1 = - \sfrac{1}{2} \hat{\bold{d}} \, , \quad g_2 = + \hat{\bold{d}} \, , \quad g_3 = - \hat{\bold{d}} \, , \quad g_4 = + \hat{\bold{d}} \, , 
\end{equation}
\begin{equation}
g_5 = - \hat{\bold{d}} \, , \quad g_6 = - 2 \hat{\bold{d}} \, , \quad g_7 = + \sfrac{1}{2} \hat{\bold{d}} \, , \quad g_8 = 0 \, . \label{4DX_g}
\end{equation}
The coefficients in Eqs.\,(\ref{4DX_f}) and (\ref{4DX_s}) satisfy the relations previously found by Low \cite{Low:2010ie}.

We now consider the terms in $\delta' \mathcal{L}_{BLG} + \delta \mathcal{L}_{1/T_{M2}}$ which contain a total of three covariant derivatives. These are,
\begin{align}
\nonumber
\tilde{\delta} \mathcal{L}_3 = \sfrac{1}{T_{M2}} {\rm STr} \Big\{ &- D_\mu ( \delta'_{1 DX} X^I ) D^\mu X^I + \delta'_{2 DX} \tilde{A}_\mu X^I D^\mu X^I \\
\nonumber
&+ \sfrac{i}{2} \delta'_{2 DX} \bar{\psi} \Gamma^\mu D_\mu \psi + \sfrac{i}{2} \bar{\psi} \Gamma^\mu D_\mu ( \delta'_{2 DX} \psi ) \\
\nonumber
&+ \sfrac{i}{4} \delta'_{3 DX} \bar{\psi} \Gamma_{IJ} [ X^I , X^J , \psi ] + \sfrac{i}{4} \bar{\psi} \Gamma_{IJ} [ X^I , X^J , \delta'_{3 DX} \psi ] \\
\nonumber
&+ \sfrac{1}{2} \varepsilon^{\mu \rho \sigma} F_{\rho \sigma} \delta'_{1DX} \tilde{A}_\mu \\
\nonumber
&-4 \bold{a} \, \delta \tilde{A}^\mu X^I D_\mu X^J  D^\nu X^J D_\nu X^I - 4 \bold{b} \, \delta \tilde{A}^\mu X^I D_\mu X^I D^\nu X^J D_\nu X^J \\
\nonumber
&+ 3 \bold{c} \, \varepsilon^{\mu\nu\lambda} \, [ \delta X^I , X^J , X^K ] D_\mu X^I D_\nu X^J D_\lambda X^K + 3 \bold{c} \, \varepsilon^{\mu\nu\lambda}\, X^{IJK} D_\mu ( \delta X^I ) D_\nu X^J D_\lambda X^K \\
\nonumber
&+ i \hat{\bold{d}} \, \delta_{0 DX} \bar{\psi} \Gamma^\mu \Gamma^{IJ} D_\nu\psi {D}_\mu X^{I} {D}^\nu X^J + i \hat{\bold{d}} \, \bar{\psi} \Gamma^\mu \Gamma^{IJ} D_\nu ( \delta_{0 DX} \psi ) {D}_\mu X^{I} {D}^\nu X^J \\
\nonumber
&+ i \hat{\bold{e}} \, \delta_{0 DX} \bar{\psi} \Gamma^\mu D^\nu\psi {D}_\mu X^{I} {D}_\nu X^I + i \hat{\bold{e}} \, \bar{\psi} \Gamma^\mu D^\nu ( \delta_{0 DX} \psi ) {D}_\mu X^{I} {D}_\nu X^I \\
\nonumber
&+ i \hat{\bold{f}} \, \delta_{1 DX} \bar{\psi} \Gamma^{IJKL} D_\nu\psi\; X^{IJK} {D}^\nu X^L + i \hat{\bold{f}} \, \bar{\psi} \Gamma^{IJKL} D_\nu ( \delta_{1 DX} \psi ) X^{IJK}  {D}^\nu X^L \\
\nonumber
&+ i \hat{\bold{g}} \, \delta_{1 DX} \bar{\psi} \Gamma^{IJ} D_\nu\psi\; X^{IJK}{D}^\nu X^K + i \hat{\bold{g}} \, \bar{\psi} \Gamma^{IJ} D_\nu ( \delta_{1 DX} \psi ) X^{IJK}{D}^\nu X^K \\
\nonumber
&+ i \hat{\bold{h}} \, \delta_{1 DX} \bar{\psi}\Gamma^{IJ}[X^J,X^{K},\psi] {D}^\mu X^{I} {D}_\mu X^K + i \hat{\bold{h}} \, \bar{\psi}\Gamma^{IJ}[X^J,X^{K}, \delta_{1 DX} \psi] {D}^\mu X^{I} {D}_\mu X^K \\
\nonumber
&+ i \hat{\bold{i}} \, \delta_{1 DX} \bar{\psi}\Gamma^{\mu\nu}[X^I,X^{J},\psi] {D}_\mu X^{I}{D}_\nu X^J + i \hat{\bold{i}} \, \bar{\psi}\Gamma^{\mu\nu}[X^I,X^{J}, \delta_{1 DX} \psi] {D}_\mu X^{I}{D}_\nu X^J \\
\nonumber
&+ i \hat{\bold{j}} \, \delta_{1 DX} \bar{\psi}\Gamma^{\mu\nu} \Gamma^{IJ} [X^J,X^{K},\psi] {D}_\mu X^{I}{D}_\nu X^K \\
&+ i \hat{\bold{j}} \, \bar{\psi}\Gamma^{\mu\nu} \Gamma^{IJ} [X^J,X^{K},\delta_{1 DX} \psi] {D}_\mu X^{I}{D}_\nu X^K \Big\} \, .
\end{align}
Once again we insert the appropriate supersymmetry transformations, canonically reorder $\bar{\psi}$ and $\epsilon$ using the spinor flip condition \eref{Spinor_Flip} and then commute the worldvolume $\Gamma$-matrices through the transverse ones. The result is
\begin{align}
\nonumber
{\rm STr} \Big\{ &- i s_5 \bar{\epsilon} \Gamma^{IJKLM} \Gamma^\mu D^\nu ( \psi D_\mu X^J X^{KLM} ) D_\nu X^I - i s_6 \bar{\epsilon} \Gamma^{KLM} \Gamma^\mu D^\nu ( \psi D_\mu X^I X^{KLM} ) D_\nu X^I \\
\nonumber
&- i s_7 \bar{\epsilon} \Gamma^{JLM} \Gamma^\mu D^\nu ( \psi D_\mu X^J X^{ILM} ) D_\nu X^I - i s_8 \bar{\epsilon} \Gamma^{ILM} \Gamma^\mu D^\nu ( \psi D_\mu X^J X^{JLM} ) D_\nu X^I \\
\nonumber
&- i s_9 \bar{\epsilon} \Gamma^{M} \Gamma^\mu D^\nu ( \psi D_\mu X^J X^{IJM} ) D_\nu X^I \\
\nonumber
&- i g_1 \bar{\epsilon} \Gamma^J \Gamma^\mu \psi [ X^J ,  X^I , D^\mu X^I ] D_\nu X^K D^\nu X^K - i g_2 \bar{\epsilon} \Gamma^J \Gamma^\nu \psi [ X^J , X^I , D^\mu X^I ] D_\mu X^K D_\nu X^K \\
\nonumber
&- i g_3 \bar{\epsilon} \Gamma^J \Gamma^\nu \psi [ X^K , X^I , D^\mu X^I ] D_\mu X^J D_\nu X^K - i g_4 \bar{\epsilon} \Gamma^J \Gamma^\nu \psi [ X^K , X^I , D^\mu X^I ] D_\mu X^K D_\nu X^J \\
\nonumber
&- i g_5 \bar{\epsilon} \Gamma^J \Gamma^\mu \psi [ X^K , X^I , D_\mu X^I ] D_\nu X^J D^\nu X^K - i g_6 \bar{\epsilon} \Gamma^J \Gamma^{\mu \nu \lambda} \psi [ X^K , X^I , D_\mu X^I ] D_\nu X^J D_\lambda X^K \\
\nonumber
&- i g_7 \bar{\epsilon} \Gamma^{JKL} \Gamma^{\mu \nu \lambda} \psi [ X^J , X^I , D_\mu X^I ] D_\nu X^K D_\lambda X^L - i g_8 \bar{\epsilon} \Gamma^{JKL} \Gamma^\nu \psi [ X^J , X^I , D^\mu X^I ] D_\mu X^K D_\nu X^L \\
\nonumber
&+ \sfrac{i}{2} f_4 \bar{\epsilon} \Gamma^{IJKLM} \Gamma^{\nu \lambda} \Gamma^\mu D_\mu \psi D_\nu X^I D_\lambda X^J X^{KLM} - \sfrac{i}{2} f_4 \bar{\epsilon} \Gamma^{IJKLM} \Gamma^{\nu \lambda} \Gamma^\mu \psi D_\mu ( D_\nu X^I D_\lambda X^J X^{KLM} ) \\
\nonumber
&- \sfrac{i}{2} f_5 \bar{\epsilon} \Gamma^{KLM} \Gamma^{\nu \lambda} \Gamma^\mu D_\mu \psi D_\nu X^J D_\lambda X^K X^{JLM} + \sfrac{i}{2} f_5 \bar{\epsilon} \Gamma^{KLM} \Gamma^{\nu \lambda} \Gamma^\mu \psi D_\mu ( D_\nu X^J D_\lambda X^K X^{JLM} ) \\
\nonumber
&+ \sfrac{i}{2} f_6 \bar{\epsilon} \Gamma^M \Gamma^{\nu \lambda} \Gamma^\mu D_\mu \psi D_\nu X^J D_\lambda X^K X^{JKM} - \sfrac{i}{2} f_6 \bar{\epsilon} \Gamma^M \Gamma^{\nu \lambda} \Gamma^\mu \psi D_\mu ( D_\nu X^J D_\lambda X^K X^{JKM} ) \\
\nonumber
&+ \sfrac{i}{2} f_7 \bar{\epsilon} \Gamma^{KLM} \Gamma^\mu D_\mu \psi D_\nu X^J D^\nu X^J X^{KLM} - \sfrac{i}{2} f_7 \bar{\epsilon} \Gamma^{KLM} \Gamma^\mu \psi D_\mu ( D_\nu X^J D^\nu X^J X^{KLM} ) \\
\nonumber
&+ \sfrac{i}{2} f_8 \bar{\epsilon} \Gamma^{KLM} \Gamma^\mu D_\mu \psi D_\nu X^J D^\nu X^K X^{JLM} - \sfrac{i}{2} f_8 \bar{\epsilon} \Gamma^{KLM} \Gamma^\mu \psi D_\mu ( D_\nu X^J D^\nu X^K X^{JLM} ) \\
\nonumber
&- \sfrac{i}{4} f_1 \bar{\epsilon} \Gamma^{KLM} \Gamma^{IJ} \Gamma^{\mu \nu \lambda} [ X^I , X^J , \psi ] D_\mu X^K D_\nu X^L D_\lambda X^M \\
\nonumber
&+ \sfrac{i}{4} f_1 \bar{\epsilon} \Gamma^{KLM} \Gamma^{IJ} \Gamma^{\mu \nu \lambda} \psi [ X^I , X^J , D_\mu X^K D_\nu X^L D_\lambda X^M ] \\
\nonumber
&- \sfrac{i}{4} f_2 \bar{\epsilon} \Gamma^K \Gamma^{IJ} \Gamma^\mu [ X^I , X^J , \psi ] D_\mu X^L D_\nu X^L D^\nu X^K \\
\nonumber
&+ \sfrac{i}{4} f_2 \bar{\epsilon} \Gamma^K \Gamma^{IJ} \Gamma^\mu \psi [ X^I , X^J , D_\mu X^L D_\nu X^L D^\nu X^K ] \\
\nonumber
&- \sfrac{i}{4} f_3 \bar{\epsilon} \Gamma^K \Gamma^{IJ} \Gamma^\mu [ X^I , X^J , \psi ] D_\mu X^K D_\nu X^L D^\nu X^L \\
\nonumber
&+ \sfrac{i}{4} f_3 \bar{\epsilon} \Gamma^K \Gamma^{IJ} \Gamma^\mu \psi [ X^I , X^J , D_\mu X^K D_\nu X^L D^\nu X^L ] \\
\nonumber
&+ \sfrac{i}{2} g_9 \varepsilon^{\mu \rho \sigma} \bar{\epsilon} \Gamma^{KLM} \Gamma_{\mu \nu} \psi D^\nu X^I X^{KLM} ( \tilde{F}_{\rho \sigma} X^I ) + \sfrac{i}{2} g_{10} \varepsilon^{\mu \rho \sigma} \bar{\epsilon} \Gamma^{JLM} \Gamma_{\mu \nu} \psi D^\nu X^J X^{ILM} ( \tilde{F}_{\rho \sigma} X^I ) \\
\nonumber
&+ \sfrac{i}{2} g_{11} \varepsilon^{\mu \rho \sigma} \bar{\epsilon} \Gamma^{M} \Gamma_{\mu \nu} \psi D^\nu X^J X^{IJM} ( \tilde{F}_{\rho \sigma} X^I ) + \sfrac{i}{2} g_{12} \varepsilon^{\mu \rho \sigma} \bar{\epsilon} \Gamma^{KLM} \psi D_\mu X^I X^{KLM} ( \tilde{F}_{\rho \sigma} X^I ) \\	
\nonumber
&+ \sfrac{i}{2} g_{13} \varepsilon^{\mu \rho \sigma} \bar{\epsilon} \Gamma^{JLM} \psi D_\mu X^J X^{ILM} ( \tilde{F}_{\rho \sigma} X^I ) + \sfrac{i}{2} g_{14} \varepsilon^{\mu \rho \sigma} \bar{\epsilon} \Gamma^{M} \psi D_\mu X^J X^{IJM} ( \tilde{F}_{\rho \sigma} X^I ) \\
\nonumber
&- 4 i \bold{a} \, \bar{\epsilon} \Gamma^K \Gamma^\mu [X^K, X^I ,\psi ] D_\mu X^J D^\nu X^J D_\nu X^I - 4 i \bold{b} \, \bar{\epsilon} \Gamma^K \Gamma^\mu [X^K, X^I ,\psi ] D_\mu X^I D^\nu X^J D_\nu X^J \\
\nonumber
&+ 3i \bold{c} \, \varepsilon^{\mu\nu\lambda}\, \bar{\epsilon} \Gamma^I [ \psi , X^J , X^K ] D_\mu X^I D_\nu X^J D_\lambda X^K + 3i \bold{c} \, \varepsilon^{\mu\nu\lambda}\, \bar{\epsilon} \Gamma^I D_\mu \psi D_\nu X^J D_\lambda X^K X^{IJK} \\
\nonumber
&- \sfrac{i}{6} \hat{\bold{d}} \, \bar{\epsilon} \Gamma^{KLM} \Gamma^{IJ} \Gamma^\mu D_\nu \psi D_\mu X^I D^\nu X^J X^{KLM} - \sfrac{i}{6} \hat{\bold{d}} \, \bar{\epsilon} \Gamma^{KLM} \Gamma^{IJ} \Gamma^\mu \psi D_\nu ( X^{KLM} ) D_\mu X^I D^\nu X^J \\
\nonumber
&- \sfrac{i}{6} \hat{\bold{e}} \, \bar{\epsilon} \Gamma^{KLM} \Gamma^\mu D^\nu \psi D_\mu X^I D_\nu X^I X^{KLM} + \sfrac{i}{6} \hat{\bold{e}} \, \bar{\epsilon} \Gamma^{KLM} \Gamma^\mu \psi D^\nu ( X^{KLM}) D_\mu X^I D_\nu X^I \\
\nonumber
&+ i \hat{\bold{f}} \, \bar{\epsilon} \Gamma^M \Gamma^{IJKL} \Gamma^\mu D_\nu \psi X^{IJK} D^\nu X^L D_\mu X^M + i \hat{\bold{f}} \, \bar{\epsilon} \Gamma^M \Gamma^{IJKL} \Gamma^\mu \psi D_\nu ( D_\mu X^M ) X^{IJK} D^\nu X^L \\
\nonumber
&+ i \hat{\bold{g}} \, \bar{\epsilon} \Gamma^L \Gamma^{IJ} \Gamma^\mu D_\nu \psi X^{IJK} D^\nu X^K D_\mu X^L - i \hat{\bold{g}} \, \bar{\epsilon} \Gamma^L \Gamma^{IJ} \Gamma^\mu \psi D_\nu ( D_\mu X^L ) X^{IJK} D^\nu X^K \\
\nonumber
&+ i \hat{\bold{h}} \, \bar{\epsilon} \Gamma^L \Gamma^{IJ} \Gamma^\mu [X^J,X^{K},\psi] D^\nu X^I D_\nu X^K D_\mu X^L - i \hat{\bold{h}} \, \bar{\epsilon} \Gamma^L \Gamma^{IJ} \Gamma^\mu \psi [X^J,X^K, D_\mu X^L ] D^\nu X^I D_\nu X^K \\
\nonumber
&+ i \hat{\bold{i}} \, \bar{\epsilon} \Gamma^K \Gamma^\lambda \Gamma^{\mu\nu} [X^I,X^J,\psi] D_\mu X^ID_\nu X^J D_\lambda X^K - i \hat{\bold{i}} \, \bar{\epsilon} \Gamma^K \Gamma^\lambda \Gamma^{\mu\nu} \psi [X^I,X^J, D_\lambda X^K ] D_\mu X^I D_\nu X^J \\
\nonumber
&+ i \hat{\bold{j}} \, \bar{\epsilon} \Gamma^L \Gamma^{IJ} \Gamma^\lambda \Gamma^{\mu\nu} [X^J,X^{K},\psi] D_\mu X^I D_\nu X^K D_\lambda X^L \\
&+ i \hat{\bold{j}} \, \bar{\epsilon} \Gamma^L \Gamma^{IJ} \Gamma^\lambda \Gamma^{\mu\nu} \psi [X^J,X^K,D_\lambda X^L ] D_\mu X^I D_\nu X^K \Big\} \, .
\end{align}
We have omitted the calculations due to their length (they can be found in appendix \ref{a_HDer_invariance}) however, after using worldvolume dualisation and performing the $\Gamma$-matrix algebra we find all the terms in $\tilde{\delta} \mathcal{L}_3$ can be assembled into total derivatives or made to vanish through the gauge invariance condition in \eref{Gauge_Inv}. As in $\tilde{\delta} \mathcal{L}_4$, this requires the coefficients to satisfy certain constraints. Using the coefficient data from $\tilde{\delta} \mathcal{L}_4$ we can solve these additional simultaneous equations to find that $\tilde{\delta} \mathcal{L}_3$ is invariant if
\begin{equation}
\bold{c} = + \sfrac{2}{3} \hat{\bold{d}} \, , \quad \hat{\bold{f}} = + \sfrac{1}{6} \hat{\bold{d}} \, , \quad \hat{\bold{g}} = - \sfrac{1}{2} \hat{\bold{d}} \, , \quad \hat{\bold{h}} = + \hat{\bold{d}} \, , \quad \hat{\bold{i}} = - \hat{\bold{d}} \, , \quad \hat{\bold{j}} = - \hat{\bold{d}} \, , \label{3DX_L}
\end{equation}
\begin{equation}
f_4 = + \sfrac{1}{12} \hat{\bold{d}} \, , \quad f_5 = 0 \, , \quad f_6 = - \sfrac{3}{2} \hat{\bold{d}} \, , \quad f_7 = - \sfrac{1}{12} \hat{\bold{d}} \, , \quad f_8 = + \sfrac{1}{2} \hat{\bold{d}} \, , \label{3DX_f}
\end{equation}
\begin{equation}
s_5 = + \sfrac{1}{6} \hat{\bold{d}} \, , \quad s_6 = 0 \, , \quad s_7 = 0 \, , \quad s_8 = 0 \, , \quad s_9 = + \hat{\bold{d}} \, , \label{3DX_s}
\end{equation}
\begin{equation}
g_9 = 0 \, , \quad g_{10} = + \sfrac{1}{2} \hat{\bold{d}} \, , \quad g_{11} = 0 \, , \quad g_{12} = 0 \, , \quad g_{13} = 0 \, , \quad g_{14} = + \hat{\bold{d}} \, . \label{3DX_g}
\end{equation}

Demonstrating invariance of the terms $\tilde{\delta} \mathcal{L}_2$, $\tilde{\delta} \mathcal{L}_1$ and $\tilde{\delta} \mathcal{L}_0$ proceeds analogously to $\tilde{\delta} \mathcal{L}_4$ and $\tilde{\delta} \mathcal{L}_3$ only now the presence of two or more 3-brackets means we can manipulate terms using the fundamental identity in \eref{real_FI} as well as the $\mathcal{A}_4$ identities in Eqs.\,(\ref{Useful_Id}) and (\ref{Useful_Id2}). We find invariance of $\tilde{\delta} \mathcal{L}_2$ is achieved if,
\begin{equation}
\bold{d} = + \hat{\bold{d}} \, , \quad \bold{e} = - \sfrac{1}{6} \hat{\bold{d}} \, , \quad \hat{\bold{k}} = + \sfrac{1}{2} \hat{\bold{d}} \, , \quad \hat{\bold{l}} = - \sfrac{1}{2} \hat{\bold{d}} \, , \quad \hat{\bold{n}} = - \sfrac{1}{2} \hat{\bold{d}} \, , \label{2DX_L}
\end{equation}
\begin{equation}
f_9 = - \sfrac{1}{12} \hat{\bold{d}} \, , \quad f_{10} = + \sfrac{1}{2} \hat{\bold{d}} \, , \label{2DX_f}
\end{equation}
\begin{equation}
s_{10} = - \sfrac{1}{12} \hat{\bold{d}} \, , \quad s_{11} = + \sfrac{1}{2} \hat{\bold{d}} \, , \label{2DX_s}
\end{equation}
\begin{equation}
g_{15} = - \sfrac{1}{12} \hat{\bold{d}} \, . \label{2DX_g}
\end{equation}

The additional constraints from invariance of the $\tilde{\delta} \mathcal{L}_1$ terms are $\hat{\bold{p}} = - \sfrac{1}{4} \hat{\bold{d}}$ and $f_{11} = + \sfrac{1}{72} \hat{\bold{d}}$ whilst the $\tilde{\delta} \mathcal{L}_0$ terms require $\bold{f} = + \sfrac{1}{72} \hat{\bold{d}}$.

We have been able to determine all the arbitrary coefficients in the order $1/T_{M2}$ Lagrangian and supersymmetry transformations up to a scale factor parametrised by $\hat{\bold{d}}$. The numerical value for $\hat{\bold{d}}$ can be fixed by reference to the action for a single M2-brane in \eref{M2_action}. We have seen in moving from a single M2-brane to multiple M2-branes the lowest order scalar kinetic terms are generalised as 
\begin{equation}
- \sfrac{1}{2} \partial_\mu X^I \partial^\mu X^I \rightarrow{\rm Tr} \big( - \sfrac{1}{2} D_\mu X^I D^\mu X^I \big) \, .
\end{equation}
It seems reasonable that the $1/T_{M2}$ corrections in \eref{M2_action} have a similar generalisation so that
\begin{align}
\nonumber
&\sfrac{1}{T_{M2}} \Big( + \sfrac{1}{4} \partial_\mu X^I  \partial^\mu X^J \partial_\nu X^I \partial^\nu X^J - \sfrac{1}{8} \partial_\mu X^I \partial^\mu X^I \partial_\nu X^J \partial^\nu X^J \Big) \\
\rightarrow \ &\sfrac{1}{T_{M2}} {\rm STr} \Big( + \sfrac{1}{4} D_\mu X^I D^\mu X^J D_\nu X^I D^\nu X^J - \sfrac{1}{8} D_\mu X^I D^\mu X^I D_\nu X^J D^\nu X^J \Big) \, .
\end{align}
Thus, comparing with the $1/T_{M2}$ ansatz in \eref{L_Higher} we find $\bold{a} = + \hat{\bold{d}} = + \sfrac{1}{4}$ and $\bold{b} = -\sfrac{1}{2} \hat{\bold{d}} = - \sfrac{1}{8}$ which implies $\hat{\bold{d}} = + \sfrac{1}{4}$. Having fixed the scale parameter the remaining numerical values of the coefficients are\footnote{In comparing our results for the Lagrangian coefficients to those of \cite{Ezhuthachan:2009sr} we find some differences: although the values for the coefficients $\bold{a}$, $\bold{b}$, $\bold{d}$-$\bold{f}$ and $\hat{\bold{d}}$-$\hat{\bold{g}}$ match, in \cite{Ezhuthachan:2009sr} nonzero values are assigned to $\hat{\bold{m}}$ and $\hat{\bold{o}}$ whereas we have found they should be dropped from the $\mathcal{A}_4$ Lagrangian. For the remaining coefficients, $\bold{c}$ and $\hat{\bold{h}}$-$\hat{\bold{p}}$, we find the absolute values match but that there is disagreement over signs.},
\begin{align}
\nonumber
s_1 &= + \sfrac{1}{8} \, , & &f_1 = + \sfrac{1}{24} \, , & &g_1 = - \sfrac{1}{8} \, , & &\bold{a} = + \sfrac{1}{4} \, , \\
\nonumber
 s_2 &=  0 \, , & &f_2 = + \sfrac{1}{4} \, , & &g_2 = + \sfrac{1}{4} \, , & &\bold{b} = - \sfrac{1}{8} \, , \\
\nonumber
 s_3 &= + \sfrac{1}{4} \, , & &f_3 = - \sfrac{1}{8} \, , & &g_3 = - \sfrac{1}{4} \, , & &\bold{c} = + \sfrac{1}{6} \, , \\
\nonumber
 s_4 &= - \sfrac{1}{8} \, , & &f_4 = + \sfrac{1}{48} \, , & &g_4 = + \sfrac{1}{4} \, , & &\bold{d} = + \sfrac{1}{4} \, , \\
\nonumber
 s_5 &= + \sfrac{1}{24} \, , & &f_5 = 0 \, , & &g_5 = - \sfrac{1}{4} \, , & &\bold{e} = - \sfrac{1}{24} \, , \\
\nonumber
 s_6 &= 0 \, , & &f_6 = - \sfrac{3}{8} \, , & &g_6 = - \sfrac{1}{2} \, , & &\bold{f} = + \sfrac{1}{288} \, , \\
\nonumber
 s_7 &= 0 \, , & &f_7 = - \sfrac{1}{48} \, , & &g_7 = + \sfrac{1}{8} \, , & &\hat{\bold{d}} = + \sfrac{1}{4} \, , \\
\nonumber
 s_8 &= 0 \, , & &f_8 = + \sfrac{1}{8} \, , & &g_8 = 0 \, , & &\hat{\bold{e}} = - \sfrac{1}{4} \, , \\
%
 s_9 &= + \sfrac{1}{4} \, , & &f_9 = - \sfrac{1}{48} \, , & &g_9 =0 \, & &\hat{\bold{f}} = + \sfrac{1}{24} \, , \label{Numerical_Results} \\ 
\nonumber
 s_{10} &= - \sfrac{1}{48} \, , & &f_{10} = + \sfrac{1}{8} \, , & &g_{10} = + \sfrac{1}{8} \, , & &\hat{\bold{g}} = - \sfrac{1}{8} \, , \\
\nonumber
 s_{11} &= + \sfrac{1}{8} \, , & &f_{11} = + \sfrac{1}{288} \, , & &g_{11} = 0 \, , & &\hat{\bold{h}} = + \sfrac{1}{4} \, , \\
\nonumber
 && && &g_{12} = 0 \, , & &\hat{\bold{i}} = - \sfrac{1}{4} \, , \\
\nonumber
 && && &g_{13} = 0 \, , & &\hat{\bold{j}} = - \sfrac{1}{4} \, , \\
\nonumber
 && && &g_{14} = + \sfrac{1}{4} \, , & &\hat{\bold{k}} = + \sfrac{1}{8} \, , \\
\nonumber
 && && &g_{15} = - \sfrac{1}{48} \, , & &\hat{\bold{l}} = - \sfrac{1}{8} \, , \\
 %
%
\nonumber
 && && && &\hat{\bold{n}} = - \sfrac{1}{8} \, , \\  
%
%
\nonumber
 && && && &\hat{\bold{p}} = - \sfrac{1}{16} \, .
\end{align}
%

\subsection{Closure of the Superalgebra} \label{Closure}
We have seen that the higher derivative corrected Euclidean BLG theory is invariant under our supersymmetry ansatz. However, for a truly supersymmetric theory the supersymmetry transformations must close on-shell on to translations and gauge transformations. In this section we show the superalgebra does indeed close for the coefficients listed in \eqref{Numerical_Results}. In the absence of cubic fermion terms in $\delta ' X^I_a$ and $\delta ' \tilde{A}_\mu{}^b{}_a$ and quadratic fermions in $\delta ' \psi_a$ we are unable to close on the fermion field. 

We present only our results here as the detailed calculations are long (more details are given in appendix \ref{a_HDer_closure}). Our methodology in the closure calculations is the same for both the scalar and gauge fields and we detail it here. We first separate out certain terms according to their number of covariant derivatives and then insert the relevant supersymmetry transformations. Next, we use the relation $\{ \Gamma^\mu , \Gamma^I \} = 0$ to group all worldvolume $\Gamma$-matrices together and then expand them out using the Clifford algebra relation. Following this, we perform the $( 1 \leftrightarrow 2 )$ antisymmetrisation in the supersymmetry parameters making heavy use of \eref{Commutator_Relation}. The transverse $\Gamma$-matrix algebra is performed next and our calculations have again been helped by using the symbolic computer package Cadabra \cite{Peeters:2006kp,Peeters:2007wn}. Finally, we simplify the remaining expressions wherever possible using the identities in Eqs.\,(\ref{Useful_Id_5}) and (\ref{Useful_Id_4}).
%
\subsubsection{Closure on the Scalar Fields} \label{Scalar_Closure}
The full supersymmetry transformations can be written as $\tilde{\delta} = \delta + \delta'$ where $\delta$ are the lowest order variations and $\delta'$ are the $1/T_{M2}$ corrections. Closure on the scalars then takes the form
\begin{align}
[ \tilde{\delta}_1 , \tilde{\delta}_2 ] X^I_a = [ \delta_1 , \delta_2 ] X^I_a + (\delta_1 \delta_2' + \delta'_1 \delta_2 ) X^I_a - (\delta_2 \delta_1' + \delta'_2 \delta_1 ) X^I_a + [ \delta'_1 , \delta'_2 ] X^I_a \, .
\end{align}
The lowest order commutator, $[ \delta_1 , \delta_2 ] X^I_a$, closes on to translations and gauge transformations \cite{Bagger:2007jr} as we have seen previously. The commutator $[ \delta'_1 , \delta'_2 ] X^I_a$ is $\mathcal{O} (T_{M2}^{-2})$ and can be ignored because we are not considering the $T_{M2}^{-2}$ corrections to the supersymmetry transformations. The remaining mixed terms, $(\delta_1 \delta_2' + \delta'_1 \delta_2 ) X^I_a - (\delta_2 \delta_1' + \delta'_2 \delta_1 ) X^I_a$, are the focus of this section and must be zero for the algebra to close. As closing on the scalar field does not involve use of the equation of motion the mixed terms must be zero either through symmetry arguments or by constraining the coefficients to be zero. Performing the supervariations we find that the resulting terms can be grouped according to the number of covariant derivatives they contain.
To begin, we consider terms which involve three covariant derivatives,
%
\begin{align}
\nonumber
T_{M2} \, (\delta_1 \delta_2' X^I_a + \delta'_1 \delta_2 X^I_a  )_{3DX} &- (1 \leftrightarrow 2) \\
\nonumber
	=& +i ( 6 f_1 - 2 s_1 - 2 s_2 ) ( \bar{\epsilon}_2 \Gamma^{JK} \Gamma^{\mu \nu \lambda} \epsilon_1) D_{\mu} X^{I}_b D_{\nu} X^{J}_c D_{\lambda} X^{K}_d \ d^{bcd}{}_{a} \\
\nonumber	
	&+ i ( 2 f_2 + 2 s_2 - 2 s_3 ) ( \bar{\epsilon}_2 \Gamma^\mu \epsilon_1) D_{\mu} X^{J}_b D_{\nu} X^{I}_c D^{\nu} X^{J}_d \ d^{bcd}{}_{a} \\
	&+ i ( 2 f_3 -  2 s_2 - 2 s_4) ( \bar{\epsilon}_2 \Gamma^{\mu} \epsilon_1) D_{\mu} X^{I}_b D_{\nu} X^{J}_c D^{\nu} X^{J}_d \ d^{bcd}{}_{a} \, .
\end{align}
Closure requires each of these terms is zero. Hence,
\begin{align}
f_1 =& + \sfrac{1}{3} s_1 + \sfrac{1}{3} s_2 \, , \qquad f_2 = - s_2 + s_3 \, , \qquad f_3 = + s_2 + s_4 \, .
\end{align}
%
Next, we consider terms which involve two covariant derivatives,
\begin{align}
\nonumber
T_{M2} \, (\delta_1 \delta_2' X^I_a + \delta'_1 \delta_2 X^I_a  )_{2DX} &- (1 \leftrightarrow 2) \\
\nonumber
=&+ i ( 6 f_4 - s_1 + 2 s_7 ) ( \bar{\epsilon}_2 \Gamma^{JKLM} \Gamma^{\mu \nu} \epsilon_1 ) D_{\mu} X^{J}_b D_{\nu} X^{K}_c X^{ILM}_d \ d^{bcd}{}_{a} \\
\nonumber
&+ i ( 4 f_4 - \sfrac{1}{3} s_2 - 2 s_5 - 2 s_6 ) ( \bar{\epsilon}_2 \Gamma^{JKLM} \Gamma^{\mu \nu} \epsilon_1 ) D_{\mu} X^{I}_b D_\nu X^J_c X^{KLM}_d \ d^{bcd}{}_{a} \\		
\nonumber
&+ i ( 2 f_5 + 2 s_1 - 6 s_5 + 2 s_8 ) ( \bar{\epsilon}_2 \Gamma^{IKLM} \Gamma^{\mu \nu} \epsilon_1 ) D_{\mu} X^{J}_b D_{\nu} X^{K}_c X^{JLM}_d \ d^{bcd}{}_{a} \\		
\nonumber
&+ i ( 2 f_6 + 2 s_1 + 2 s_9 ) ( \bar{\epsilon}_2 \Gamma^{\mu \nu} \epsilon_1 ) D_{\mu} X^{J}_b D_{\nu} X^{K}_c X^{IJK}_d \ d^{bcd}{}_{a} \\
\nonumber
&+ i ( 6 f_7 - s_4 + 2 s_7 ) ( \bar{\epsilon}_2 \Gamma^{JK} \epsilon_1 ) D_{\mu} X^{L}_b D^{\mu} X^{L}_c X^{IJK}_d \ d^{bcd}{}_{a} \\
\nonumber
&+ i ( 2 f_8 - s_3 + 6 s_6 + 2 s_8 ) ( \bar{\epsilon}_2 \Gamma^{KL} \epsilon_1 ) D_{\mu} X^{I}_b D^{\mu} X^{J}_c X^{JKL}_d \ d^{bcd}{}_{a} \\	
&+ i ( - 4 f_8 + 4 s_7 + 2 s_9 ) ( \bar{\epsilon}_2 \Gamma^{KL} \epsilon_1 ) D^\mu X^J_b D_{\mu} X^{L}_c X^{IJK}_d \ d^{bcd}{}_{a} \, .
\end{align}
These two derivatives terms are then zero if
\begin{equation}
f_4 = \sfrac{1}{6} s_1 - \sfrac{1}{3} s_7 \, , \qquad f_4 = \sfrac{1}{12} s_2 + \sfrac{1}{2} s_5 + \sfrac{1}{2} s_6 \, , 
\end{equation}
\begin{equation}
f_5 = - s_1 + 3 s_5 - s_8 \, ,  \qquad f_6 = - s_1 - s_9 \, , \qquad f_7 = \sfrac{1}{6} s_4 - \sfrac{1}{3} s_7 \, ,
\end{equation}
\begin{equation}
f_8 = \sfrac{1}{2} s_3 - 3 s_6 - s_8 \, , \qquad f_8 = s_7 + \sfrac{1}{2} s_9 \, .
\end{equation}
The terms which involve a single covariant derivative are
%
\begin{align}
\nonumber
T_{M2} \, (\delta_1 \delta_2' X^I_a + \delta'_1 \delta_2 X^I_a & )_{1DX} - (1 \leftrightarrow 2) \\
\nonumber
=&+ i ( 2 f_9 - 2 s_6 - 2 s_{10} ) ( \bar{\epsilon}_2 \Gamma^\mu \epsilon_1 ) D_\mu X^I_b X^{JKL}_c X^{JKL}_d \, d^{bcd}{}_a \\
&+ i ( 2 f_{10} - 2 s_7 - 2 s_8 - 2 s_{11} ) ( \bar{\epsilon}_2 \Gamma^\mu \epsilon_1 ) D_\mu X^J_b X^{IKL}_c X^{JKL}_d \, d^{bcd}{}_a \, .
\end{align}
Closure requires 
\begin{equation}
f_9 = s_6 + s_{10} \, ,  \qquad f_{10} = s_7 + s_8 + s_{11} \, .
\end{equation}
Finally, we consider those terms which contain no covariant derivatives,
%
\begin{align}
\nonumber
T_{M2} \, (\delta_1 \delta_2' X^I_a + \delta'_1 \delta_2 X^I_a  )_{0DX} &- (1 \leftrightarrow 2) \\
=&+ i \left( 6 f_{11} - s_{10} - \sfrac{1}{3} s_{11} \right) ( \bar{\epsilon}_2 \Gamma^{JK} \epsilon_1 ) X^{IJK}_b X^{LMN}_c X^{LMN}_d \, d^{bcd}{}_a \, ,
\end{align}
and we require
\begin{equation}
f_{11} = \sfrac{1}{6} s_{10} + \sfrac{1}{18} s_{11} \, .
\end{equation}

It is easily verified that the conditions for closure are satisfied when the $f$ and $s$ coefficients take the values found in \eqref{Numerical_Results}.

\subsubsection{Closure on the Gauge Fields} \label{Gauge_Closure}
Closing the algebra on $\tilde{A}_\mu$ gives
 \begin{align}
[ \tilde{\delta}_1 , \tilde{\delta}_2 ] \tilde{A}_\mu{}^b{}_a = [ \delta_1 , \delta_2 ] \tilde{A}_\mu{}^b{}_a + (\delta_1 \delta_2' + \delta'_1 \delta_2 ) \tilde{A}_\mu{}^b{}_a - (\delta_2 \delta_1' + \delta'_2 \delta_1 ) \tilde{A}_\mu{}^b{}_a + [ \delta'_1 , \delta'_2 ] \tilde{A}_\mu{}^b{}_a \, .
\end{align}
As for the scalar field, the terms in $[ \delta'_1 , \delta'_2 ] \tilde{A}_\mu{}^b{}_a$ can be ignored. The lowest order terms may be written as
\begin{align}
[\delta_1,\delta_2] \tilde A_\mu{}^b{}_a =& + 2 i ( \bar{\epsilon}_2 \Gamma^\nu \epsilon_1 ) \varepsilon_{\mu\nu\lambda} \left( \sfrac{1}{2} \varepsilon^{\rho \sigma \lambda} \tilde{F}_{\rho \sigma}{}^b{}_a + E_{A_\lambda{}^a{}_b} \right) - 2i ( \bar{\epsilon}_2 \Gamma_{IJ} \epsilon_1 ) X^I_c D_\mu X^J_d f^{cdb}{}_a \, ,
\end{align}
where $E_{A_\lambda{}^a{}_b}$ is the lowest order gauge field equation of motion. From the presence of covariant derivatives in the higher derivative Lagrangian it follows that the gauge field equation of motion picks up $1/T_{M2}$ corrections. Hence for on-shell closure we require the mixed terms make the following contribution to the higher order equation of motion
\begin{align}
(\delta_1 \delta_2' + \delta'_1 \delta_2 ) \tilde{A}_\mu{}^b{}_a - ( 1 \leftrightarrow 2 ) =& + \sfrac{2 i}{T_{M2}} ( \bar{\epsilon}_2 \Gamma^\nu \epsilon_1 ) \varepsilon_{\mu\nu\lambda} E'_{A_\lambda{}^a{}_b} \\[10pt]
%
%
\nonumber
=&+ \sfrac{2 i}{T_{M2}} ( \bar{\epsilon}_2 \Gamma^\nu \epsilon_1 ) \varepsilon_{\mu\nu\lambda} \Big( + 4 \bold{a} \, D^\lambda X^J_e D^\rho X^J_f D_\rho X^I_g X^I_c \Big) \, d^{efg}{}_d f^{cdb}{}_a \\
\nonumber
&+ \sfrac{2 i}{T_{M2}} ( \bar{\epsilon}_2 \Gamma^\nu \epsilon_1 ) \varepsilon_{\mu\nu\lambda} \Big( + 4 \bold{b} D^\lambda X^I_e D^\rho X^J_f D_\rho X^J_g X^I_c \Big) \, d^{efg}{}_d f^{cdb}{}_a \\
\nonumber
&+ \sfrac{2 i}{T_{M2}} ( \bar{\epsilon}_2 \Gamma^\nu \epsilon_1 ) \varepsilon_{\mu\nu\lambda} \Big( + 3 \bold{c} \, \varepsilon^{\rho \sigma \lambda} D_\rho X^J_e D_\sigma X^K_f X^{IJK}_g X^I_c \Big) \, d^{efg}{}_d f^{cdb}{}_a \\
\nonumber
&+ \sfrac{2 i}{T_{M2}} ( \bar{\epsilon}_2 \Gamma^\nu \epsilon_1 ) \varepsilon_{\mu\nu\lambda} \Big( + \left( \sfrac{2}{3} \bold{d} + 2 \bold{e} \right) D^\lambda X^I_e X^{JKL}_f X^{JKL}_g X^I_c \Big) \, d^{efg}{}_d f^{cdb}{}_a \\
&+ \mathcal{O} ( \psi^2 ) \label{Gauge_ReqEOM} \, ,
\end{align}
with all others terms in $(\delta_1 \delta_2' + \delta'_1 \delta_2 ) \tilde{A}_\mu{}^b{}_a - ( 1 \leftrightarrow 2 )$ being zero.
Once again, the closure terms can be neatly split according to their number of covariant derivatives. We first consider terms which involve three covariant derivatives,
%
\begin{align}
\nonumber
T_{M2} \, (\delta_1 \delta_2' \tilde{A}_\mu{}^b{}_a &+ \delta'_1 \delta_2 \tilde{A}_\mu{}^b{}_a )_{3DX} - (1 \leftrightarrow 2) \\
\nonumber
=&+ i ( 2 f_2 - 2 g_5 - 2 g_6 ) \varepsilon_{\mu \nu \lambda} ( \bar{\epsilon}_2 \Gamma^{\nu} \epsilon_1 ) X^I_c D^{\rho} X^{I}_e D^{\lambda} X^{J}_f D_{\rho} X^{J}_g \, d^{efg}{}_{d} f^{cdb}{}_a \\
\nonumber
&+ i ( 2 f_3 - 2 g_1 + 2 g_6 ) \varepsilon_{\mu \nu \lambda} ( \bar{\epsilon}_2 \Gamma^{\nu} \epsilon_1 ) X^I_c D^{\lambda} X^{I}_e D_{\rho} X^{J}_f D^{\rho} X^{J}_g \, d^{efg}{}_{d} f^{cdb}{}_a \\
\nonumber
&+ i ( - 2 g_2 + 2 g_3 - 2 g_6 ) \varepsilon_{\nu \lambda \rho} ( \bar{\epsilon}_2 \Gamma^\nu \epsilon_1 ) X^I_c D^\lambda X^I_e D_\mu X^J_f D^\rho X^J_g \, d^{efg}{}_d f^{cdb}{}_a \\
\nonumber
&+ i ( - 2 g_3 + 2 g_5 - 2 g_8 ) ( \bar{\epsilon}_2 \Gamma^{JK} \epsilon_1 ) D_\mu X^J_e D_\nu X^K_f D^\nu X^I_g X^I_c \, d^{efg}{}_d f^{cdb}{}_a \\			
\nonumber
&+ i ( - 6 f_1 + 2 g_7 + 2 g_8 ) \varepsilon^{\nu \lambda \rho} ( \bar{\epsilon}_2 \Gamma^{IJKL} \Gamma_\rho \epsilon_1 ) X^I_c D_{\mu} X^J_e D_{\nu} X^K_f D_{\lambda} X^L_g \, d^{efg}{}_{d} f^{cdb}{}_a \\	
\nonumber
&+ i ( 2 f_3 - 2 g_1 - 2 g_8 ) ( \bar{\epsilon}_2 \Gamma^{IJ} \epsilon_1 ) X^I_c D_{\mu} X^{J}_e D_{\nu} X^{K}_f D^{\nu} X^{K}_g \, d^{efg}{}_{d} f^{cdb}{}_a \\			
&+ i ( 2 f_2 - 2 g_2 + 2 g_8 ) ( \bar{\epsilon}_2 \Gamma^{IK} \epsilon_1 ) X^I_c D_{\mu} X^{J}_e D_{\nu} X^{J}_f D^{\nu} X^{K}_g \, d^{efg}{}_{d} f^{cdb}{}_a \, .
\end{align}
The first and second terms form part of the higher derivative equation of motion and after comparing with \eref{Gauge_ReqEOM} we find
\begin{equation}
2 f_2 - 2 g_5 - 2 g_6 = 8 \bold{a} \, , \qquad 2 f_3 - 2 g_1 + 2 g_6 = 8 \bold{b} \, .
\end{equation}
The remaining coefficients must be zero for closure of the superalgebra. Hence,
\begin{equation}
f_1 = \sfrac{1}{3} g_7 + \sfrac{1}{3} g_8 \, , \qquad f_2 = g_2 - g_8 \, , \qquad f_3 = g_1 + g_8 \, ,
\end{equation}
\begin{equation}
g_2 - g_3 + g_6 = 0 \, , \qquad  g_3 - g_5 + g_8 =0 \, .
\end{equation}
Next we consider terms which involve two covariant derivatives,
%
%
\begin{align}
\nonumber
T_{M2} \, (\delta_1 &\delta_2' \tilde{A}_\mu{}^b{}_a + \delta'_1 \delta_2 \tilde{A}_\mu{}^b{}_a )_{2DX} - (1 \leftrightarrow 2) \\
\nonumber
=&+ i ( 4 f_6 + 2 g_8 - 2 g_{11} - 2 g_{14} ) ( \bar{\epsilon}_2 \Gamma^{\nu} \epsilon_1 ) X^I_c D_\mu X^J_e D_\nu X^K_f X^{IJK}_g \, d^{efg}{}_d f^{cdb}{}_a \\	
\nonumber
&+ i ( 2 f_5 + 2 f_6 - g_6 + 2 g_7 + 6 g_9 ) \varepsilon_{\mu \nu \lambda} ( \bar{\epsilon}_2 \Gamma^{KL} \epsilon_1 ) X^I_c D^\nu X^I_e D^\lambda X^J_f X^{JKL}_g \, d^{efg}{}_d f^{cdb}{}_a \\	
\nonumber
&+ i ( 4 f_5 + 4 g_7 - 4 g_{10} + 2 g_{11} ) \varepsilon_{\mu \nu \lambda} ( \bar{\epsilon}_2 \Gamma^{KL} \epsilon_1 ) X^I_c D^\nu X^J_e D^\lambda X^L_f X^{IJK}_g \, d^{efg}{}_d f^{cdb}{}_a \\
\nonumber
&+ i ( 2 f_8 + g_5 - 6 g_9 ) ( \bar{\epsilon}_2 \Gamma_\mu \Gamma^{IJKL} \epsilon_1 ) X^I_c D^\nu X^J_e D_\nu X^M_f X^{KLM}_g \, d^{efg}{}_d f^{cdb}{}_a \\	
\nonumber
&+ i ( 12 f_4 - 2 f_5 + g_3 - g_8 + 6 g_9 ) ( \bar{\epsilon}_2 \Gamma^{\nu} \Gamma^{IJLM} \epsilon_1 ) X^I_c D_\mu X^J_e D_\nu X^K_f X^{KLM}_g \, d^{efg}{}_d f^{cdb}{}_a \\		
\nonumber
&+ i ( - 12 f_4 + 2 f_5 + g_4 + g_8 - 6 g_{12} ) ( \bar{\epsilon}_2 \Gamma^{\nu} \Gamma^{IJKL} \epsilon_1 ) X^I_c D_\nu X^J_e D_\mu X^M_f X^{KLM}_g \, d^{efg}{}_d f^{cdb}{}_a \\	
&+ i ( 12 f_4 - g_8 - 2 g_{10} - 2 g_{13} ) ( \bar{\epsilon}_2 \Gamma^{\nu} \Gamma^{JKLM} \epsilon_1 ) X^I_c D_\mu X^J_e D_\nu X^K_f X^{ILM}_g \, d^{efg}{}_d f^{cdb}{}_a \, .		
\end{align}
The first term contributes to the gauge field equation of motion. After multiplying out the $\varepsilon$-tensors in \eref{Gauge_ReqEOM} we find that closure on-shell requires
\begin{equation}
4 f_6 + 2 g_8 - 2 g_{11} - 2 g_{14} = -12 \bold{c} \, . 
\end{equation}
The remaining terms are zero provided 
\begin{equation}
f_4 = \sfrac{1}{12} g_8 + \sfrac{1}{6} g_{10} + \sfrac{1}{6} g_{13} \, , 
\end{equation}
\begin{equation}
6 f_4 - f_5 = - \sfrac{1}{2} g_3 + \sfrac{1}{2} g_8 - 3 g_9 \, , \qquad 6 f_4 - f_5 = \sfrac{1}{2} g_4 + \sfrac{1}{2} g_8 - 3 g_{12} \, ,
\end{equation}
\begin{equation}
f_5 = - g_7 + g_{10} - \sfrac{1}{2} g_{11} \, ,
\end{equation}
\begin{equation}
f_5 + f_6 = \sfrac{1}{2} g_6 - g_7 - 3 g_9  \, , \qquad f_8 = - \sfrac{1}{2} g_5 + 3 g_9 \, .
\end{equation}
The terms which involve a single covariant derivative are
%
%
\begin{align}
\nonumber
T_{M2} \, (\delta_1 \delta_2' \tilde{A}_\mu{}^b{}_a &+ \delta'_1 \delta_2 \tilde{A}_\mu{}^b{}_a )_{1DX} - (1 \leftrightarrow 2) \\
\nonumber
=&+ i ( 2 f_9 + \sfrac{2}{3} f_{10} + 2 g_9 + \sfrac{2}{3} g_{10} - 2 g_{15} ) ( \bar{\epsilon}_2 \Gamma^{\nu} \epsilon_1 ) \varepsilon_{\mu \nu \lambda} D^\lambda X^I_e X^{JKL}_f X^{JKL}_g X^I_c \, d^{efg}{}_d f^{cdb}{}_a \\	
\nonumber
&+ i ( 2 f_9 - \sfrac{2}{3} g_{13} - 2 g_{15} ) ( \bar{\epsilon}_2 \Gamma^{IJ} \epsilon_1 ) D_\mu X^J_e X^{KLM}_f X^{KLM}_g X^I_c \, d^{efg}{}_d f^{cdb}{}_a \\
&+ i ( 2 f_{10} + 2 g_{13} - g_{14} ) ( \bar{\epsilon}_2 \Gamma^{LM} \epsilon_1 ) D_\mu X^J_e X^{IJK}_f X^{KLM}_g X^I_c \, d^{efg}{}_d f^{cdb}{}_a \, .
\end{align}
The first term forms part of the gauge field equation of motion. Comparing with \eref{Gauge_ReqEOM} we see that
\begin{align}
2 f_9 + \sfrac{2}{3} f_{10} + 2 g_9 + \sfrac{2}{3} g_{10} - 2 g_{15} = + \sfrac{4}{3} \bold{d} + 4 \bold{e} \, .
\end{align}
The remaining coefficients must be zero hence,
\begin{equation}
f_9 = \sfrac{1}{3} g_{13} + g_{15} \, , \qquad f_{10} = - g_{13} + \sfrac{1}{2} g_{14}  \, .
\end{equation}
Next we consider terms which involve no covariant derivatives,
%
\begin{align}
\nonumber
T_{M2} \, (\delta_1 \delta_2' \tilde{A}_\mu{}^b{}_a &+ \delta'_1 \delta_2 \tilde{A}_\mu{}^b{}_a )_{0DX} - (1 \leftrightarrow 2) \\
=&+ i \left( 2 f_{11} - \sfrac{1}{3} g_{15} \right) ( \bar{\epsilon}_2 \Gamma_\mu \Gamma^{IJKL} \epsilon_1 ) X^{JKL}_e X^{MNO}_f X^{MNO}_g X^I_c \, d^{efg}{}_d f^{cdb}{}_a \, .
\end{align}
At first sight we should take the coefficient to be zero however, using the identity \eqref{Useful_Id_4} we can show that the term is zero independently of its coefficient and consequently this part of algebra closes automatically.

Once more it is easy to verify that all the gauge field closure conditions are satisfied by the coefficients listed in \eqref{Numerical_Results}.

\subsection{Summary of Results} \label{s_Summary_of_results}
In summary, we have found that the maximally supersymmetric higher derivative corrected Lagrangian of the $\mathcal{A}_4$ BLG theory, to lowest nontrivial order in fermions, is
\begin{align}
\nonumber
\mathcal{L} = \mathcal{L}_{BLG} + \sfrac{1}{T_{M2}} {\rm STr} \Big\{ &+ \sfrac{1}{4} \, D^\mu X^I D_\mu X^J  D^\nu X^J D_\nu X^I - \sfrac{1}{8} \, D^\mu X^I D_\mu X^I D^\nu X^J D_\nu X^J \\
\nonumber
&+ \sfrac{1}{6} \, \varepsilon^{\mu\nu\lambda}\, X^{IJK} D_\mu X^I D_\nu X^J D_\lambda X^K \\
\nonumber
&+ \sfrac{1}{4} \, X^{IJK}X^{IJL} D^\mu X^K D_\mu X^L - \sfrac{1}{24} \, X^{IJK} X^{IJK} D^\mu X^L  D_\mu X^L \\
\nonumber
&+ \sfrac{1}{288} \, X^{IJK}X^{IJK} X^{LMN}X^{LMN} \\
%
%
\nonumber
&+ \sfrac{i}{4} \, \bar{\psi} \Gamma^\mu \Gamma^{IJ} D^\nu\psi {D}_\mu X^{I} {D}_\nu X^J - \sfrac{i}{4} \, \bar{\psi}\Gamma^\mu D^\nu\psi  {D}_\mu X^{I} {D}_\nu X^I \\
\nonumber
&+ \sfrac{i}{24} \, \bar{\psi}\Gamma^{IJKL} D^\nu\psi\; X^{IJK}  {D}_\nu X^L - \sfrac{i}{8} \, \bar{\psi}\Gamma^{IJ}  D^\nu\psi\; X^{IJK}{D}_\nu X^K \\
\nonumber
&+ \sfrac{i}{4} \, \bar{\psi}\Gamma^{IJ}[X^J,X^{K},\psi] {D}^\mu X^{I} {D}_\mu X^K \\
\nonumber
&- \sfrac{i}{4} \, \bar{\psi}\Gamma^{\mu\nu}[X^I,X^{J},\psi] {D}_\mu X^{I}{D}_\nu X^J - \sfrac{i}{4} \, \bar{\psi}\Gamma^{\mu\nu}\Gamma^{IJ}[X^J,X^{K},\psi] {D}_\mu X^{I}{D}_\nu X^K \\
\nonumber
&+ \sfrac{i}{8} \, \bar{\psi}\Gamma^\mu\Gamma^{IJ}[X^K,X^{L},\psi] {D}_\mu X^{I}X^{JKL} - \sfrac{i}{8} \, \bar{\psi}\Gamma^\mu[X^I,X^{J},\psi] {D}_\mu X^{K}X^{IJK} \\
\nonumber
&- \sfrac{i}{8} \, \bar{\psi}\Gamma^\mu\Gamma^{IJ}[X^K,X^{L},\psi] {D}_\mu X^L X^{IJK} \\
&- \sfrac{i}{16} \, \bar{\psi}\Gamma^{IJ} [X^K,X^{L},\psi] X^{IJM}X^{KLM} \Big\} \, ,
\end{align}
where $\mathcal{L}_{BLG}$ was given in \eref{BLG_act_Tr}. The preceding Lagrangian is invariant under the following $\mathcal{N}=8$ supersymmetry transformations;
\begin{align}
%
%
\nonumber
\delta X^I_a = i ( \bar{\epsilon} \Gamma^{I} \psi_a ) + \sfrac{1}{T_{M2}} \big\{ &+\sfrac{i}{8} ( \bar{\epsilon} \Gamma^{IJK} \Gamma^{\mu \nu} \psi_b ) D_{\mu} X^{J}_c D_{\nu} X^{K}_d \ d^{bcd}{}_{a} \\
\nonumber
&+ \sfrac{i}{4} ( \bar{\epsilon} \Gamma^{J} \psi_b ) D_{\mu} X^{I}_c D^{\mu} X^{J}_d \ d^{bcd}{}_{a} \\
\nonumber
&- \sfrac{i}{8} ( \bar{\epsilon} \Gamma^{I} \psi_b ) D_{\mu} X^{J}_c D^{\mu} X^{J}_d \ d^{bcd}{}_{a} \\ 
\nonumber
&+ \sfrac{i}{24} ( \bar{\epsilon} \Gamma^{IJKLM} \Gamma^{\mu} \psi_b ) D_{\mu} X^{J}_c X^{KLM}_d \ d^{bcd}{}_{a} \\
\nonumber
&+ \sfrac{i}{4} ( \bar{\epsilon} \Gamma^{M} \Gamma^{\mu} \psi_b ) D_{\mu} X^{J}_c X^{IJM}_d \ d^{bcd}{}_{a} \\
\nonumber
&- \sfrac{i}{48} ( \bar{\epsilon} \Gamma^{I} \psi_b ) X^{JKL}_c X^{JKL}_d \ d^{bcd}{}_{a} \\
&+ \sfrac{i}{8} ( \bar{\epsilon} \Gamma^{L} \psi_b ) X^{JKL}_c X^{JKI}_d \ d^{bcd}{}_{a} \big\} \, , \\
\nonumber \\
%
%
%
%
\nonumber
\delta \psi_a =\Gamma^\mu \Gamma^I \epsilon \, D_\mu X^I_a - \sfrac{1}{6} \Gamma^{IJK} \epsilon \, X^{IJK}_a + \sfrac{1}{T_{M2}} \big\{ &+ \sfrac{1}{24} \Gamma^{JKL} \Gamma^{\mu \nu \lambda} \epsilon \, D_{\mu} X^{J}_b D_{\nu} X^{K}_c D_{\lambda} X^{L}_d \ d^{bcd}{}_{a} \\
\nonumber
&+ \sfrac{1}{4} \Gamma^{K} \Gamma^{\mu} \epsilon \, D_{\mu} X^{J}_b D_{\nu} X^{J}_c D^{\nu} X^{K}_d \ d^{bcd}{}_{a} \\
\nonumber
&- \sfrac{1}{8} \Gamma^{K} \Gamma^{\mu} \epsilon \, D_{\mu} X^{K}_b D_{\nu} X^{J}_c D^{\nu} X^{J}_d \ d^{bcd}{}_{a} \\
\nonumber
&+ \sfrac{1}{48} \Gamma^{JKLMN} \Gamma^{\mu \nu} \epsilon \, D_\mu X^J_b D_\nu X^K_c X^{LMN}_d \, d^{bcd}{}_a \\
\nonumber
&- \sfrac{3}{8} \Gamma^M \Gamma^{\mu \nu} \epsilon \, D_\mu X^J_b D_\nu X^K_c X^{JKM}_d \, d^{bcd}{}_a \\
\nonumber
&- \sfrac{1}{48} \Gamma^{KLM} \epsilon \, D_\mu X^J_b D^\mu X^J_c X^{KLM}_d \, d^{bcd}{}_a \\
\nonumber
&+ \sfrac{1}{8} \Gamma^{KLM} \epsilon \, D_\mu X^J_b D^\mu X^K_c X^{JLM}_d \, d^{bcd}{}_a \\
\nonumber
&- \sfrac{1}{48} \Gamma^J \Gamma^\mu \epsilon \, D_\mu X^J_b X^{KLM}_c X^{KLM}_d \, d^{bcd}{}_a \\
\nonumber
&+ \sfrac{1}{8} \Gamma^M \Gamma^\mu \epsilon \, D_\mu X^J_b X^{JKL}_c X^{KLM}_d \, d^{bcd}{}_a \\
&+ \sfrac{1}{288} \Gamma^{NOP} \epsilon \, X^{JKL}_b X^{JKL}_c X^{NOP}_d \, d^{bcd}{}_a \big\} \, , \\
\nonumber \\
%
%
%
%
\nonumber
\delta \tilde{A}_{\mu}{}^b{}_a =i \bar{\epsilon} \Gamma_\mu \Gamma_I X^I_c \psi_d f^{cdb}{}_a + \sfrac{1}{T_{M2}} \big\{ &- \sfrac{i}{8} ( \bar{\epsilon} \Gamma_\mu \Gamma^I \psi_e ) D_\nu X^J_f D^\nu X^J_g X^I_c \, d^{efg}{}_d f^{cdb}{}_a \\
\nonumber
&+ \sfrac{i}{4} ( \bar{\epsilon} \Gamma^\nu \Gamma^I \psi_e ) D_\mu X^J_f D_\nu X^J_g X^I_c \, d^{efg}{}_d f^{cdb}{}_a \\
\nonumber
&- \sfrac{i}{4} ( \bar{\epsilon} \Gamma^\nu \Gamma^J \psi_e ) D_\mu X^J_f D_\nu X^I_g X^I_c \, d^{efg}{}_d f^{cdb}{}_a \\
\nonumber
&+ \sfrac{i}{4} ( \bar{\epsilon} \Gamma^\nu \Gamma^J \psi_e ) D_\mu X^I_f D^\nu X^J_g X^I_c \, d^{efg}{}_d f^{cdb}{}_a \\
\nonumber
&- \sfrac{i}{4} ( \bar{\epsilon} \Gamma_\mu \Gamma^J \psi_e ) D_\nu X^J_f D^\nu X^I_g X^I_c \, d^{efg}{}_d f^{cdb}{}_a \\
\nonumber
&- \sfrac{i}{2} ( \bar{\epsilon} \Gamma_{\mu \nu \lambda} \Gamma^J \psi_e ) D^\nu X^J_f D^\lambda X^I_g X^I_c \, d^{efg}{}_d f^{cdb}{}_a \\
\nonumber
&+ \sfrac{i}{8} ( \bar{\epsilon} \Gamma_{\mu \nu \lambda} \Gamma^{IJK} \psi_e ) D^\nu X^J_f D^\lambda X^K_g X^I_c \, d^{efg}{}_d f^{cdb}{}_a \\
\nonumber
&+ \sfrac{i}{8} ( \bar{\epsilon} \Gamma_{\mu \nu} \Gamma^{JLM} \psi_e ) D^\nu X^J_f X^{ILM}_g X^I_c \, d^{efg}{}_d f^{cdb}{}_a \\
\nonumber
&+ \sfrac{i}{4} ( \bar{\epsilon} \Gamma^{M} \psi_e ) D_\mu X^J_f X^{IJM}_g X^I_c \, d^{efg}{}_d f^{cdb}{}_a \\
&- \sfrac{i}{48} ( \bar{\epsilon} \Gamma_{\mu} \Gamma^{I} \psi_e ) X^{JKL}_f X^{JKL}_g X^I_c \, d^{efg}{}_d f^{cdb}{}_a \big\} \, .
\end{align}

\subsection{Conclusions}\label{s_HDer_conc}

In this chapter we have determined the four-derivative order corrections to both the supersymmetry transformations and Lagrangian of the $\mathcal{A}_4$ Bagger-Lambert-Gustavsson theory. Supersymmetric invariance of the Lagrangian requires that the arbitrary coefficients in the system are fixed up to an overall scale parameter and by reference to the Abelian DBI action for a single M2-brane, the scale parameter is itself fixed leading to definite numerical values for all the coefficients. We have also shown that the supersymmetry algebra closes on-shell on the scalar and gauge fields at linear order in the fermions. With the coefficients we have determined, it can also be demonstrated that the presence of higher derivative corrections in the fermion supersymmetry does not modify the BPS equation.

\newpage
\section{Periodic Arrays of M2-branes} \label{c_Periodic_arrays_of_M2s}

One of the early results of the M2-brane theories \cite{Bagger:2006sk,Bagger:2007jr,Gustavsson:2007vu,Bagger:2007vi,Aharony:2008ug}\footnote{For a review see \cite{Bagger:2012jb}.} was that their relation to D2-branes   arises by a `novel Higgs mechanism' \cite{Mukhi:2008ux} where, far out on the Coulomb branch, the nondynamical gauge fields `eat' a scalar so that the theory is described at low energy - low compared to the vacuum expectation value (vev) on the Coulomb branch -  by three-dimensional maximally supersymmetric Yang-Mills (3D-SYM). On the other hand, at least
naively, the most straightforward way to reduce from M2-branes to D2-branes is to compactify one transverse dimension on a circle. This can be done by considering an infinite array of M2-branes with equal spacing between them along some direction. Such  arrays of D-branes were considered in \cite{Taylor:1996ik} within the context of T-duality and therefore it is of interest to extend this discussion to the case of M2-branes.

At first this would seem to be a clear-cut and well-defined goal. After all the ABJM model \cite{Aharony:2008ug} allows us to consider an arbitrary number of M2-branes located in any configuration in ${\mathbb C}^4/{\mathbb Z}_k$. We can therefore use this to describe an infinite periodic array. In particular the vacuum is described by the scalar field vev:
 \be\label{config}
\langle Z^{A'}\rangle =0\, ,\qquad \langle Z^4\rangle = 2\pi i R \left(\begin{array}{ccccc} \ddots && && \\  &1&&& \\ &&0&&\\ &&&-1& \\ && && \ddots   \end{array}\right) \, ,
 \ee
where $A'=1,2,3$.  Note that each entry should be viewed as multiplying an $M\times M$ identity matrix corresponding to  $M$ M2-branes located at each site.  This configuration is illustrated in Figure 1, where we have also indicated the action of the inherent ${\mathbb Z}_k$ orbifold. 
\begin{figure}[h]
\centering
\includegraphics[width=0.8\textwidth]{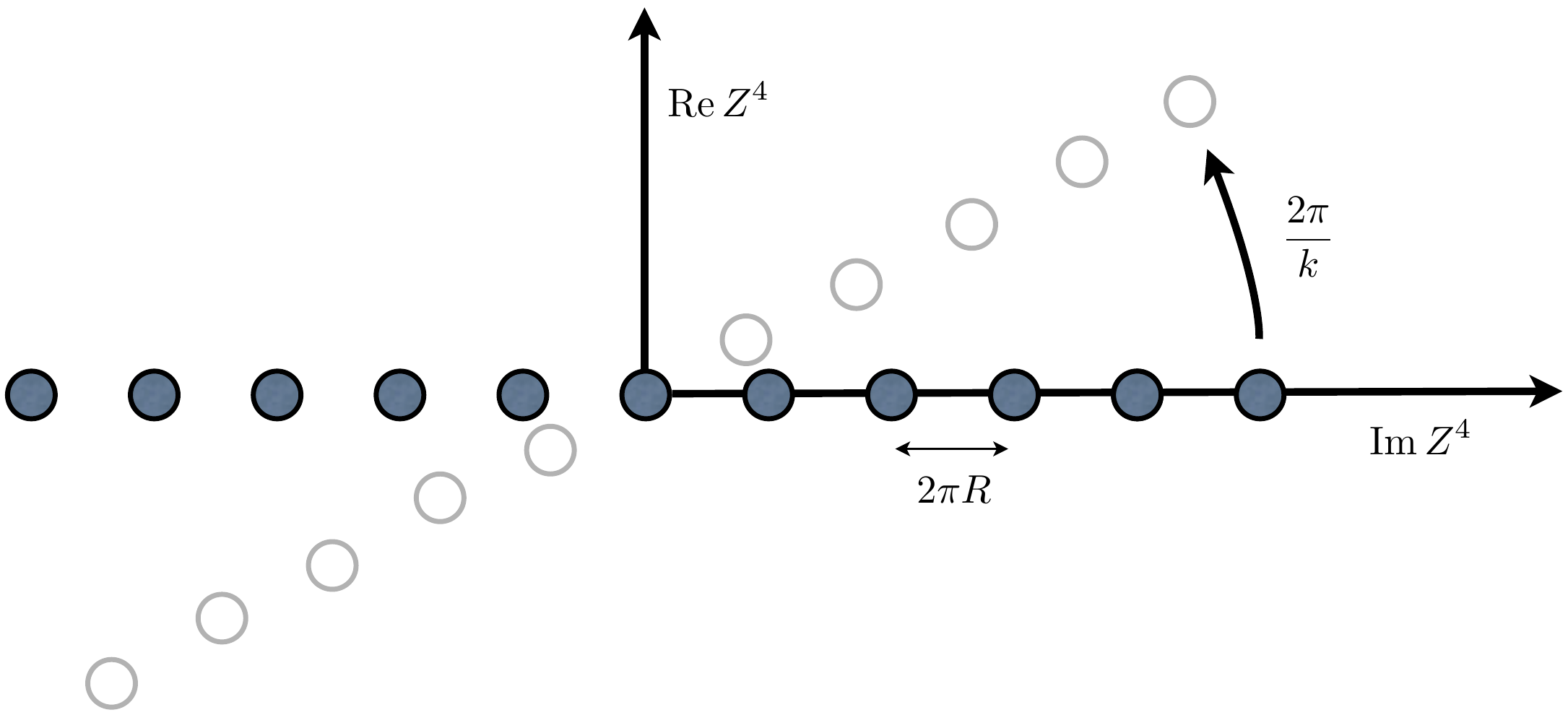}
\caption{The array of M2-branes\label{figure1}}
\end{figure}
One might worry about the effect of this orbifold  however this could in principle be avoided by taking $k=1$. Although this is strongly coupled we might expect to recover  a weak coupling expansion by taking the periodicity $2\pi R$ small and thus reducing to type IIA string theory.

On further reflection more serious difficulties present themselves. Although the vacuum configuration of an infinite periodic array is readily accommodated for in the ABJM model, the dynamics of the array come from considering an additional orbifold that imposes a discrete translational invariance along the array, as was done for D-branes in \cite{Taylor:1996ik}. But in the ABJM model this translational invariance is broken and, even for $k=1$, it is not a symmetry of the Lagrangian.  Rather, the restoration of this shift symmetry at $k=1$   is through nonperturbative effects involving monopole ('t Hooft) operators \cite{Bashkirov:2010kz}. Another issue is that the classical Lagrangian analysis gives spurious massless fluctuations whenever two M2-branes lie at the same distance from the origin, which are expected to be lifted by nonperturbative effects \cite{Martinec:2011nb}\footnote{Although in the case at hand this problem seems to be washed-away by the sum over the infinite array.}.

Another puzzle is that in the D-brane analysis taking a periodic array leads to an infinite tower of massive states. These have a natural interpretation in string theory as the Kaluza-Klein (KK) modes of the T-dual D-brane that is wrapped on a circle. But when taking such an array in M-theory one would not expect to find an extra tower of KK-like states of D2-branes. What happens to these modes?

Nevertheless, even with all these difficulties, since the ABJM theory  is supposed to describe an arbitrary number of M2-branes, and at least for large $k$ it is weakly coupled and perturbatively reliable, there ought to be some prescription for studying the periodic array and obtaining a suitable description of D2-branes, and more generally D$p$-branes,  from M-theory. The aim of this chapter then is to do just that. We  note that there are also other papers that relate M2-branes to D$p$-branes \cite{Ho:2009nk,Kobo:2009gz,Honma:2009bx,Nastase:2010ft}. 

Another motivation for studying arrays of M2-branes is that one might expect that a cubic periodic array of M2-branes could somehow be related via an M-theory version of T-duality, to M5-branes wrapped on ${\mathbb T}^3$.

The rest of this chapter is organised as follows. In section \ref{s_Array_setup} we discuss the M2-brane array and the way in which we impose discrete translational invariance on the ABJM theory and the regularisation method that we use. In section \ref{s_Array_reduced_lagrangian} we then evaluate the Lagrangian to obtain the Lagrangian for the periodic M2-brane array. In section \ref{s_Array_3dSYM} we show how this Lagrangian is related to that of three-dimensional maximally supersymmetric Yang-Mills and hence D2-branes in string theory. In section \ref{s_Array_torus} we consider a further T-duality along a transverse torus which maps our result to a non-manifest-Lorentz invariant five-dimensional Lagrangian which is similarly related to five-dimensional maximally supersymmetric Yang-Mills. Finally in section \ref{s_Array_conc} we give our conclusions.


\subsection{Set-up}\label{s_Array_setup}

The ABJM Lagrangian is
\be
{\cL}_{ABJM}=-{\rm Tr}(D_{\mu}Z^{A}D^{\mu}\bar{Z}_{A})-i{\rm Tr} (\bar\Psi^{{A}}\gamma^\mu D_\mu\Psi_{{A}})+\cL_{Yukawa}-V+\cL_{CS}\,,
\ee
where
\begin{align}
D_{\mu}Z^{A}=&\partial_{\mu}Z^{A}-iA^{L}_{\mu}Z^{A}+iZ^{A}A^{R}_{\mu} \, , \\[10pt]
V=&-\frac{2}{3}{\rm Tr}\left([Z^{A},Z^{B};\bar{Z}_{C}][\bar{Z}_{A},\bar{Z}_{B};Z^{C}]-\frac{1}{2}[Z^{A},Z^{B};\bar{Z}_{A}][\bar{Z}_{C},\bar{Z}_{B};Z^{C}]\right) \, , \\[10pt]
\nonumber \cL_{Yukawa} =&  - i{\rm Tr}(\bar\Psi^A[\Psi_A,Z^B;\bar Z_B])+ 2i{\rm Tr}(\bar\Psi^A[\Psi_B,Z^B;\bar Z_A])\\
&+\frac{i}{2}\varepsilon_{ABCD}{\rm Tr}(\bar\Psi^A[ Z^B,  Z^C;\Psi^D])-\frac{i}{2}\varepsilon^{ABCD}{\rm Tr}(\bar Z_D [\Psi_A,\Psi_B;\bar Z_C]) \, ,\\[10pt]
\cL_{CS}=&\frac{k}{4\pi}\varepsilon^{\mu\nu\lambda}\left({\rm Tr} \left( A^{L}_{\mu}\partial_{\nu} A^{L}_{\lambda}-\frac{2i}{3} A^{L}_{\mu} A^{L}_{\nu} A^{L}_{\lambda} \right) -{\rm Tr} \left( A^{R}_{\mu}\partial_{\nu}A^{R}_{\lambda}-\frac{2i}{3} A^{R}_{\mu} A^{R}_{\nu} A^{R}_{\lambda} \right) \right)\, ,
\end{align}
and
\be
[Z^A,Z^B;\bar Z_C] = \frac{2\pi}{k}(Z^A\bar Z_CZ^B -Z^B\bar Z_CZ^A )\, .
\ee

At this point we should mention our
conventions. Firstly $A,B=1,2,3,4$, $A',B'=1,2,3$, $\mu = 0,1,2$ and $\bar\Psi_A = \Psi^T_A\gamma_0$, where $\gamma_\mu$ are a real basis of the three-dimensional Clifford algebra.  We raise/lower the $SU(4)$ $A,B$ indices when taking a Hermitian conjugate. To describe an infinite array of M2-branes we need to consider infinite matrices $Z^A_{mn}$, where $m,n\in \mathbb Z$ and,  for each $m,n$, $Z^A_{mn}$ is itself an $M\times M$ matrix. We use  $\bar Z_A$ to denote the Hermitian matrix conjugate of the full infinite-dimensional system: in components $(\bar Z_A)_{mn}=Z^\dag_{A \, nm}$, where $\dag$ denotes the matrix Hermitian conjugate of the internal $M\times M$ matrix.

The maximally supersymmetric vacua of this Lagrangian consist of commuting scalars. Hence the configuration (\ref{config}) is indeed a good vacuum and describes $M$ M2-branes located at ${{\rm Im} \, Z^4 = 2\pi i n}$ for every $n\in{\mathbb Z}$. The infinite array is invariant under the shift symmetry $Z^4\to Z^4 + 2\pi i R $:
\begin{eqnarray}
\langle Z^4_{mn}\rangle &\to& \langle Z^4_{mn}\rangle + 2\pi i R \delta_{mn}  \\[10pt]
&=& 2\pi i R (n+1)\delta_{mn} \\[10pt]
&=& \langle Z^4_{m+1 \, n+1}\rangle \, .
\end{eqnarray}
Next we need to impose the above finite shift symmetry on the whole theory, including the fluctuations.
 We can think of this as an orbifold action on M2-branes where the orbifold group is $\Gamma = \mathbb Z$ acting by
 \be\label{orbifold}
  Z^A_{mn}\to  Z^A_{m+1 \, n+1}\, ,
 \ee
 and similarly for the other fields.
 We must then consider configurations of M2-branes that are invariant under the action of $\Gamma$ , with the exception of $Z^4$ which is allowed to
carry integer `winding number' along the array:
\begin{eqnarray}
Z^{A}_{mn} &=& 2\pi i R\delta^{A}_{4}\delta_{mn}+Z^{A}_{(m-1)(n-1)}\, ,\\[10pt]
A^{L}_{\mu\,mn} &=& A^{L}_{\mu\,(m-1)(n-1)}\,, \qquad A^{R}_{\mu\,mn}=A^{R}_{\mu\,(m-1)(n-1)}\, ,\\[10pt]
\Psi_{A \, mn} &=& \Psi_{A \, (m-1)(n-1)}\, .
\end{eqnarray}
As mentioned above the problem with this group action is that it is not a symmetry of the Lagrangian. Imposing it leads to additional constraints. Furthermore it is not consistent with the  supersymmetry transformations. Nevertheless we simply proceed and  consider the theory in this case.

We first note that the infinite size of the array leads to divergent terms in the Lagrangian. For example consider the kinetic term for the scalars
\begin{eqnarray}
\sum_{m,n} \tr (\partial_\mu Z^A_{mn} \partial^\mu \bar Z_{A \,nm} )&=&
\sum_{m,n} \tr(\partial_\mu Z^A_{m-n 0 }\partial^\mu Z^\dag_{A \, m-n 0}) \\[10pt]
&=& \sum_q \sum_p \tr(\partial_\mu Z^A_{p 0}\partial^\mu Z^\dag_{A\,p 0}) \\[10pt]
&=& |\Gamma| \sum_p \tr(\partial_\mu Z^A_{p 0}\partial^\mu Z^\dag_{A\,p 0}) \, , 
\end{eqnarray}
where
\be
|\Gamma| = \sum_q 1\, .
\ee
In the D-brane case \cite{Taylor:1996ik} the effect of this divergence is harmless as each term in the Lagrangian comes with the same overall factor of $|\Gamma|$. In our case however,
the fact that the shift invariance we impose is not a symmetry of the Lagrangian, causes other divergences to appear. We therefore need a way to regulate and compare divergences.

To do this we simply consider a very large but finite array consisting of M2-branes located at $Z^4 = 2\pi in R$ with $n=-N,\ldots,N$. We then always impose the limit $N\to \infty$ in any final expressions and therefore only consider the leading large $N$ terms. Using this regulator we see that
\begin{equation}
|\Gamma| = \sum_q 1  = 2N+1 \sim 2N\, ,
\end{equation}
where $\sim$ denotes the leading order behaviour as $N\to\infty$.
We will also be cavalier about ignoring possible boundary effects that occur when $N$ is finite.  Thus our starting point is a ${U((2N+1)M)\times U((2N+1)M)}$ ABJM model with $N>>1$. Note that in such a theory the 't Hooft coupling constant grows as $NM/k$. Furthermore, when taking the limit $N\to \infty$, we will allow for both $k$ and  $R$  to scale in appropriate ways with $N$.

With this in mind we note that we can solve the shift symmetry condition in terms of  the $M\times M$ matrix valued fields $\phi^A_{n}$, $\psi_{A\,n}$, $a^{L/R}_{\mu\, n}$:
\begin{equation}
Z^{4}_{mn}:= 2\pi i R n 1_{M\times M}\delta_{mn}+\frac{1}{\sqrt{2N}}\phi^4_{n-m} \, , \quad  Z^{A'}_{mn}:=\frac{1}{\sqrt{2N}}\phi^{A'}_{n-m} \, , 
\end{equation}
\begin{equation}
A^{L/R}_{\mu\,mn}:=a^{L/R}_{\mu\,n-m} \, , \quad\Psi_{A \, mn} := \frac{1}{\sqrt{2N}}\psi_{A \, n-m}\, .
\end{equation}
Here we have included factors of $(2N)^{-\frac{1}{2}}$ so that the fields $\phi^A_p$ and $\psi_{A \, p}$ have canonical kinetic terms.
Note that since $A^{L/R}_\mu$ are Hermitian we require that  $(a^{L/R }_{\mu\, n})^\dag = a^{L/R }_{\mu \,-n}$. We have not rescaled the gauge fields by $(2N)^{-\frac{1}{2}}$ since their role in covariant derivatives and gauge field strengths does not readily allow for this.


\subsection{Reduced Lagrangian} \label{s_Array_reduced_lagrangian}

Having set up our configuration we can now construct the reduced action for the infinite array. Let us start with the kinetic terms:
\begin{align}
\nonumber
D_{\mu}Z^{4}_{mn}=&\frac{1}{\sqrt{2N}}\nabla_{\mu}\phi^{4}_{n-m} -\frac{i}{\sqrt{2N}}\sum_{p\ne 0}[\,a^{+}_{\mu\,p}\,,\phi^{4}_{n-m-p}\,]-\frac{i}{\sqrt{2N}}\sum_{p}\{a^{-}_{\mu\,p}\,,\phi^{4}_{n-m-p}\} \\
%
&+2\pi R(n-m)a^{+}_{\mu\,n-m}+2\pi R(n+m) a^{-}_{\mu\,n-m}\, , \label{mass} \\[10pt]
D_{\mu}Z^{A'}_{mn}=&\frac{1}{\sqrt{2N}}\nabla_{\mu}\phi^{A'}_{n-m}-\frac{i}{\sqrt{2N}}\sum_{p\ne 0}[\,a^{+}_{\mu\,p}\,,\phi^{A'}_{n-m-p}\,]-\frac{i}{\sqrt{2N}}\sum_{p}\{a^{-}_{\mu\,p}\,,\phi^{A'}_{n-m-p}\}\, , \\[10pt]
%
D_{\mu}\Psi_{{A}\,mn}=&\frac{1}{\sqrt{2N}}\nabla_{\mu}\psi_{{A}\,n-m}-\frac{i}{\sqrt{2N}}\sum_{p\ne 0 }[\,a^{+}_{\mu\,p}\,,\psi_{{A}\,n-m-p}\,]-\frac{i}{\sqrt{2N}}\sum_{p}\{a^{-}_{\mu\,p}\,,\psi_{{A}\,n-m-p}\} \, ,
\end{align}
where $a^{\pm}_{\mu\,n}:=\frac{1}{2}(a^{L}_{\mu\, n}\pm a^{R}_{\mu\,n})$ and
\be
\nabla_\mu \phi^A_p = \partial_\mu\phi^A_p - i[a^+_{\mu \,0},\phi^A_p]\, .
\ee
Note the appearance of terms involving $m + n$ on the right-hand-side of \eref{mass}. These arise because the Lagrangian is not invariant under our orbifold action (\ref{orbifold}). This leads to a divergent term of the form
\begin{eqnarray}
(2\pi R)^2\sum_{m,n}(m+n)^2 \tr ( a^{-}_{\mu\,n-m}a^{\mu -}_{\,m-n} ) &=& (2\pi R)^2\sum_q \sum_p (p+2q)^2\tr ( a^{-}_{\mu\,p}a^{\mu -}_{\,-p} ) \\[10pt] 
&\sim &\frac{8}{3}N^3(2\pi R)^2 \sum_p  \tr ( a^{-}_{\mu\,p}a^{\mu -}_{\,-p} ) \label{div1} \, .
\end{eqnarray}
This diverges (unless $R$ is taken to vanish at least as fast as $N^{-\frac{3}{2}}$, which we will not consider here) and is not cancelled by anything else in the Lagrangian. We therefore conclude that, to obtain finite energy configurations we must have
\be\label{aconstraint}
0 = a^{-}_{\mu\,p} =\frac{1}{2} (a^{L}_{\mu\,p}-a^{R}_{\mu\,p}) \, .
\ee
Note that one might be tempted to simply rescale $a^{-}_{\mu \,p}$ by a factor proportional to $N^{-3/2}$ so as to render \eref{div1} finite. However one would then simply find that, in the limit $N\to\infty$, $a^-_{\mu \,p}$ drops out from   the covariant derivative and hence the  Lagrangian.
Thus \eref{aconstraint} should be viewed as a constraint on the system that breaks the gauge group to $U(M)$.

In particular the gauge group associated to the zero mode is just $U(M)$. This should be viewed as a constraint on the system.

Next we look at the quadratic terms that come from expanding the potential:
\begin{eqnarray}
V_{\phi^2} &=&  \frac{1}{2N}\left(\frac{2\pi}{k}\right)^2(2\pi R)^4\sum_{p,q}p^{2}(p+2q)^{2} \tr(\phi^{A'}_{p}{\phi}^\dag_{A' \,p}) \\[10pt]
&\sim &  M_b^2\sum_{p}p^{2} \tr(\phi^{A'}_{p}{\phi}^\dag_{A' \,p}) \, ,
\end{eqnarray}
where
\be
M_b^2 = \frac{1}{2N}\left(\frac{2\pi}{k}\right)^2(2\pi R)^4 \sum_q (p+2q)^2\sim \frac{4}{3}(2\pi)^6\frac{N^2R^4}{k^2}\, .
\ee
Although there could be cases where it is finite if $R\to 0$ sufficiently quickly, we will consider the case that $M_b\to \infty$ as $N\to \infty$.
Here we see that the mysterious KK-like tower is lifted to infinite mass, resolving one of the puzzles raised in the introduction. In particular
we must impose the constraint:
\be
\phi^{A'}_p =0\, , \qquad p\ne 0\, .
\ee
Note that the masslessness of $\phi^4_p$ does  not seem related to the problem mentioned in the introduction, where spurious massless states arise when pairs of M2-branes are at equal distance from the origin, since that degeneracy applies to all four scalars in the same way.

Let us next examine the quadratic fermion term:
\begin{align}
{\cal L}_{\psi^2}=& \ \frac{i}{2N}\left(\frac{2\pi}{k}\right)(2\pi R)^2\sum_{p,q} (p^2+2pq) \tr ( \bar\psi^{A'}_p   \psi_{A'\,p} ) -\frac{i}{2N} \left(\frac{2\pi}{k}\right)(2\pi R)^2\sum_{p,q} (p^2+2pq) \tr (\bar\psi^{4}_p   \psi_{4\,p} ) \\[10pt]
\sim & \ iM_f\sum_p p \tr ( \bar\psi^{A'}_p \psi_{A'\,p} ) -iM_f\sum_p p \tr ( \bar\psi^{4}_p   \psi_{4\,p} ) \label{fmass} \, ,
\end{align}
where
\be
M_f = \frac{1}{2N}\left(\frac{2\pi}{k}\right)(2\pi R)^2\sum_q (p+2q)  \sim \Omega(2\pi)^3\frac{R^2N}{k}\, .
\ee
Here we have used the regularisation
\be
\sum_q q \sim \Omega N^2\, ,
\ee
where $\Omega$ is an undetermined constant of order 1. In particular we note that this  sum is ill-defined. To determine how to treat it we will  use supersymmetry.\footnote{It is conceivable that this ambiguity can be avoided by performing our calculations with a superspace formalism.} This suggests that the fermion masses should be the same as the bosons, i.e.\ $M_f=M_b$, and hence gives
\be
\Omega = {\frac{2}{\sqrt3}}\, ,
\ee
however we will keep $\Omega$ general in our calculations in this section.
Assuming $\Omega\ne 0$ we conclude that there is also a fermionic constraint
\be
\psi^A_{p}=0\, , \qquad p\ne 0 \, .
\ee

Thus  we see that in order to avoid divergences in the Lagrangian (and correspondingly Hamiltonian) we must impose the constraints
\be
\phi^{A'}_p=0\, , \qquad \psi_{A\,p}=0\, ,\qquad a^{-}_{\mu \,0 }=0\, ,\qquad a^{-}_{\mu \,p }=0\, ,\qquad p\ne 0\, ,
\ee
where $A'=1,2,3$. This leaves us with the zero-modes
\be\label{zeromodes}
\phi^A_0\, ,\qquad \psi_{A\,0}\, ,\qquad a^+_{\mu\, 0}\, ,
\ee
as well as three infinite towers of fields:
\be
a^+_{\mu \,p}\, ,\qquad \chi_p = \frac{1}{2} \phi^4_p + \frac{1}{2} \phi^\dag_{4\,-p}\, ,\qquad \omega_p = -\frac{i}{2}\phi^4_p +  \frac{i}{2}\phi^\dag_{4\,-p} \, ,\qquad p\ne 0 \, ,
\ee
which satisfy  $\omega^\dag_p = \omega_{-p}$, $\chi^\dag_p = \chi_{-p}$ and $(a^+_{\mu \,p})^\dag = a^+_{\mu \,-p}$.

Once we have set these infinitely massive fields to zero we must also ensure that there are no source terms for them in the action. Classically this is a clear requirement to solve the equations of motion. Quantum mechanically it follows from the general procedure for quantisation with a constraint. In particular if we have a Hamiltonian $H(\vec q,\vec p)$ on some phase space with coordinates $(\vec q,\vec p)$ and impose a constraint $C(\vec q,\vec p)=0$ then we require that $\{H,C\}=0$ so that the constraint is consistent with time evolution. In general this leads to a new set of constraints $\{H,C\} = C_1$, $\{H,C_1\}=C_2$ etc. In our case the original constraints simply set $\phi^{A'}_p=a^-_{\mu \,p}=\psi_{A\,p}=0$. One then finds that the resulting additional constraints are simply the vanishing of the sources for $\phi^{A'}_p, a^-_{\mu\, p}$ and $\psi_{A\,p}$, $p\ne 0$. We take the view here that, in order to ensure a smooth large $N$ limit,  such sources must be set to zero even for finite, but large $N$. In addition this means that sources that scale differently with $N$ must be made to vanish separately. However we don't expect that our results depend significantly on this.

Let us examine such sources. First  we look at the kinetic terms. Here we see that there will be a source for $\phi^{A'}_p$, $p\ne 0$ arising from $a^+_{\mu\, p}$:
\be
D^2 Z^{A'}_{mn} = \frac{1}{\sqrt{2N}}\partial^2 \phi^{A'}_0\delta_{mn}- \frac{i}{\sqrt{2N}}[\partial_\mu a^{\mu +}_{n-m},\phi^{A'}_{0}]- \frac{2i}{\sqrt{2N}}[ a^{\mu +}_{n-m},\partial_\mu \phi^{A'}_{0}] - \frac{1}{\sqrt{2N}}\sum_{p}[a^+_{\mu\, p-m},[a^{\mu +}_{n-p},\phi^{A'}_{0}]] .
\ee
Thus we also require that $a^+_{\mu\, p}$ is proportional to the $M\times M$ identity matrix if $p\ne 0$. This means that $a^+_{\mu\, p}$, $p\ne 0$, does not appear in $D_\mu Z^{A'}_{mn} = \frac{1}{\sqrt{2N}}\nabla_\mu \phi^{A'}_0\delta_{mn}$.

We can also expand the potential to cubic order in $\phi^A_p$. Although this vanishes if $\phi^{A'}_p=0$, $p\ne 0$, one does find a source term for $\phi^{A'}_p$, $p\ne 0$:
\begin{align}
V_{\phi^3} =&i\frac{(2 \pi R)^{3}}{(2N)^{3/2}}\left(\frac{2\pi}{k}\right)^2 \sum_{p,q}\tr\left[\left(p^2(p+2q)\{\chi_{p},\phi^{A'}_{0}\}+ip(p+2q)^2[\omega_{p},\phi^{A'}_{0}]\right)\phi_{A'\, p}\right]+h.c. +\ldots \\[10pt]
\sim & i\frac{(2 \pi )^{5}}{\sqrt{2}}\sum_{p}\tr\left[\left(\Omega\frac{N^{1/2}R^3}{k^2}p^2\{\chi_{p},\phi^{A'}_{0}\}+i\frac{4N^{3/2}R^3}{3k^2}p[\omega_{p},\phi^{A'}_{0}]\right)\phi_{A'\, p}\right]+h.c.+\ldots\, ,
\end{align}
where the ellipsis denote further cubic terms that  are not linear in $\phi^{A'}_p$ and hence not sources. Requiring that this vanishes tells us that
\be
\chi_p=0\, ,\qquad \omega_p \propto 1_{M\times M} \, , \qquad  p\ne 0\, .
\ee

There is also a source for $a^-_{\mu\, p}$. Setting this to zero leads to the constraint\footnote{{We denote $[\bar\Psi^{A},\gamma_\mu\Psi_{A}]:=\bar\Psi^{Aa}\gamma_\mu\Psi_{Ab}[T^\dagger_a,T^b]$.}}
\be
[Z^A,D_\mu  \bar Z_A] +  [\bar Z_A,D_\mu  Z^A] - i [\bar\Psi^A,\gamma_\mu\Psi_A]=0\, . 
\ee
The nonzero mode part of this constraint leads to
\be
2\pi i R p a^{+}_{\mu\, p} =\frac{1}{\sqrt{2N}}\nabla_{\mu}\omega_{p} \, . \label{efg}
\ee
The zero mode part of the constraint is the $a^+_{\mu\, 0}$ equation of motion and will be dealt with later. One then finds that, substituting back into the Lagrangian,  (\ref{efg}) simply removes both $\omega_p$ and $a^+_{\mu\, p}$  all together.

Thus, once all the constraints are fully considered we are effectively left with just the zero-modes (\ref{zeromodes}).
We can now evaluate the Lagrangian. In the limit that $N\to\infty$ the sixth order terms in the potential and fourth order terms in the Yukawa interaction vanish. The final result for the Lagrangian evaluated on an infinite M2-brane array is
\begin{align}\label{M2act}
{\cal L}_{array} =& -\tr(\nabla_\mu\phi^{A'}_0\nabla^\mu\phi^\dag_{A'\, 0}) -\tr(\nabla_\mu{\rm Re} \, \phi^4_0\nabla^\mu{\rm Re} \, \phi^4_0) -\tr(\nabla_\mu{\rm Im} \, \phi^4_0\nabla^\mu{\rm Im} \, \phi^4_0)- i\tr(\bar\psi^A_0\gamma^\mu\nabla_\mu \psi_{A\,0})\nonumber\\ &+{\cal L}_{Yukawa} -V\, ,
\end{align}
where ${\rm Re} \, \phi^4_0 = \frac{1}{2}(\phi^4_0+\phi^\dag_{4\,0})$, ${\rm Im} \, \phi^4_0 = -\frac{i}{2}(\phi^4_0-\phi^\dag_{4\,0})$. The potential and Yukawa terms are
\begin{align}
V =& - \frac{g^2_{YM}}{2} \tr\left( [\phi^{A'}_0,\phi^\dag_{B'\,0}][\phi^\dag_{A'\,0},\phi^{B'}_{0}]+[\phi^{A'}_0,\phi^ {B'}_{0}][\phi^\dag_{A'\,0},\phi^\dag_{B'\,0}]+4[\phi^{A'}_0,{\rm Im} \, \phi^4_0][\phi^\dag_{A'\,0},{\rm Im} \, \phi^4_0]\right) \, , \\[10pt]
\nonumber
{\cal L}_{Yukawa} =& \phantom{-} g'_{YM} \tr \left( 2i\bar\psi^{A'}_0[{\rm Im} \, \phi^4_0,\psi_{A'\,0}]-2i\bar\psi^{4}_0[{\rm Im} \, \phi^4_0,\psi_{4\,0}]+2\bar\psi^{A'}_0[\phi^\dag_{A'\,0},\psi_{4\,0}]+2\bar\psi_{A'\,0}[\phi^{A'}_0,\psi^{4}_0]\right. \\
&\left. \ \qquad \qquad+\varepsilon_{A'B'C'}\bar{\psi}^{A'}_0[ \phi^{B'}_0,\psi^{C'}_0]+\varepsilon^{A'B'C'}\bar{\psi}_{A'\,0}[ \phi_{B'\,0},\psi_{C'\,0}]\right)\, ,
\end{align}
with
\begin{eqnarray}
g^2_{YM}&=& \frac{(2\pi R)^2}{2N^2} \left(\frac{2\pi}{k}\right)^2 \sum_q q^2  \sim   \frac{1}{3}(2\pi)^4 \frac{R^2N}{k^2}\, , \\[10pt]
g'_{YM} &=& \frac{ 2\pi R }{(2 N)^{3/2}}\left(\frac{2\pi}{k}\right) \sum_{q}q \sim \frac{\Omega}{2\sqrt{2}} (2\pi)^2 \frac{R \sqrt{N}}{k} =\sqrt{\frac{3}{8}}\Omega g_{YM}\, .
\end{eqnarray}
Thus to obtain an interesting theory, with Lagrangian (\ref{M2act}), we require $M_b\to\infty$ with $g_{YM}$ finite in the limit $N\to \infty$.


\subsection{Comparing to Three-Dimensional Maximally Supersymmetric Yang-Mills} \label{s_Array_3dSYM}

The theory we have obtained looks rather strange as there is no kinetic term for the gauge field. Therefore we should consider comparing our result to three-dimensional maximally supersymmetric Yang-Mills that is obtained from the open string description of D2-branes:
\begin{eqnarray}
{\cal L}_{3DSYM} &=& -\frac{1}{4 g^2_{YM}} \tr ( F_{\mu\nu}F^{\mu\nu} ) -  \frac{1}{2}\tr ( \nabla_\mu X^I\nabla^\mu X^I ) - \frac{i}{2}\tr ( \bar\Lambda\Gamma^\mu \nabla_\mu \Lambda ) \nonumber \\
&& +\frac{g_{YM}}{2}\tr ( \bar\Lambda\Gamma^{11}\Gamma^I[X^I,\Lambda] ) +\frac{g^2_{YM}}{4}\tr \sum_{I,J}([X^I,X^J])^2\, ,
\end{eqnarray}
where $\bar \Lambda = \Lambda^T\Gamma_0$ and $\nabla_\mu X^I =\partial_\mu X^I - i[A_\mu, X^I]$.
Here there are seven scalars $X^I$, $I=3,4,\ldots,9$, a gauge field $A_\mu$ and fermions $\Lambda$ which, as written, form a real 32-component $SO(1,9)$ spinor that satisfies $\Gamma_{012}\Lambda = -\Lambda$. Furthermore  $\Gamma_\mu$, $\Gamma_I$ are real $32\times 32$ $\Gamma$-matrices and $\Gamma^{11} = \Gamma_0 \cdots \Gamma_9$.

To compare with our results we need to break the manifest $SO(7)$ symmetry to $SU(3)$. To this end we let (we will consider the fermions shortly)
\begin{equation}
X^{A'+2} = \frac{1}{\sqrt{2}}(\phi^{A'}_0+\phi^\dag_{A' \,0}) \, , \qquad X^{A'+5} = \frac{1}{\sqrt{2i}}(\phi^{A'}_0-\phi^\dag_{A' \,0}) \, , \qquad X^9 = \sqrt{2}{\rm Im} \, \phi^4_0\, ,
\end{equation}
where
\begin{equation}
{\rm Re} \, \phi^A_0 = \frac{1}{2}(\phi^A_0+\phi^\dag_{A\,0}) \, , \qquad {\rm Im} \, \phi^A_0 = - \frac{i}{2}(\phi^A_0-\phi^\dag_{A\,0})\, .
\end{equation}
The bosonic part of ${\cal L}_{3DSYM}$ can now be written as
\begin{align}
\nonumber
{\cal L}^{(b)}_{3DSYM} =& -\frac{1}{4g^2_{YM}} \tr ( F_{\mu\nu}F^{\mu\nu} ) -  \tr ( \nabla_\mu \phi^{A'}_0\nabla^\mu \phi^\dag_{A'\,0} ) -  \tr ( \nabla_\mu {\rm Im} \, \phi^{4}_0\nabla^\mu {\rm Im} \, \phi^4_{0} ) \\
& +\frac{g^2_{YM}}{2}\tr \left([\phi^{A'}_0,\phi^{B'}_0][\phi^\dag_{A'\,0},\phi^\dag_{B'\,0}]+[\phi^{A'}_0,\phi^\dag_{B\,0}][\phi^\dag_{A\,0},\phi^{B'}_0]+4[{\rm Im} \, \phi^4_0,\phi^{A'}_0][{\rm Im} \, \phi^4_0,\phi^\dag_{A'\,0}]\right)\, .
\end{align}

Next we consider the fermions. Here we need to reduce $\Lambda$ to four, complex, two-component spinors $\psi^A$. To do this we note that the Clifford algebra can be reduced as
\be
\Gamma_\mu = \gamma_\mu\otimes \rho_\star\qquad \Gamma^{I} = 1\otimes \rho^{I-2}\, ,
\ee
where $\rho^1,\ldots,\rho^8$ are a real, $16\times 16$-matrix representation of the Euclidean eight-dimensional Clifford algebra and $\rho_\star = \rho^1\rho^2 \cdots \rho^8$. In this formulation $\Gamma^{11} = 1\otimes\rho^8$ and $\Gamma_{012} = 1\otimes\rho_\star$. We can therefore decompose
\be
\Lambda = \lambda_\Sigma\otimes \eta^\Sigma = \frac{1}{\sqrt{2}}\left(\lambda_A\otimes \eta^A+\lambda^A\otimes \eta_A\right)\, ,
\ee
where $\Sigma=1,2,\ldots,16$ and $\rho_\star\eta^\Sigma=-\eta^\Sigma$ (so that there are just eight independent $\eta^\Sigma$) which we take to be normalised such that $(\eta^{\Sigma})^T \eta^\Pi=\delta^{\Sigma\Pi}$. We have also introduced a complex basis of spinors $\eta^A$, along with a suitable complex basis of $\rho^A$-matrices,  that will be useful later.
The fermion terms are now
\begin{align}
{\cal L}^{(f)}_{3DSYM} =& -\frac{i}{2} \tr ( \bar\lambda_\Sigma \gamma^\mu \nabla_\mu \lambda_\Pi ) ( \delta^{\Sigma\Pi} ) \nonumber \\
& -\frac{g_{YM}}{2\sqrt{2}}\tr(\bar \lambda_\Sigma [\phi^{A'}_0,\lambda_\Pi])\left((\eta^{\Sigma})^T\rho_8\rho_{A'}\eta^\Pi\right)-\frac{g_{YM}}{2\sqrt{2}}\tr(\bar \lambda_\Sigma [\phi_{A'\,0},\lambda_\Pi])((\eta^{\Sigma})^T\rho_8\rho^{A'}\eta^\Pi)\nonumber\\
& + \frac{ig_{YM}}{2\sqrt{2}}\tr(\bar \lambda_\Sigma [\phi^4_0,\lambda_\Pi])\left((\eta^{\Sigma})^Ti \rho_8\rho_{7}\eta^\Pi\right) + \frac{ig_{YM}}{2\sqrt{2}}\tr(\bar \lambda_\Sigma [\phi^\dag_{4\,0},\lambda_\Pi])\left((\eta^{\Sigma})^T i\rho_8\rho_{7}\eta^\Pi\right) \, ,
\end{align}
where $\bar\lambda_\Sigma  = \lambda^T_\Sigma\gamma_0$.

Let us first consider the last line. If we consider complex fermions then we can diagonalise $i\rho_8\rho_7$ with
\be
i\rho_8\rho_7\eta^{A'} = \eta^{A'} \, , \qquad i\rho_8\rho_7\eta_4 = \eta_4 \, .
\ee
It then follows that the complex conjugates satisfy
\be
i\rho_8\rho_7\eta_{A'} = -\eta_{A'} \, , \qquad i\rho_8\rho_7\eta^4 = - \eta^4 \, .
\ee
We can choose to normalise this basis such that
\be
(\eta^A)^T\eta_B = 2\delta^A_B\, ,\qquad (\eta_A)^T\eta_B = 0\, .
\ee
Next we need to deduce the action of $\rho_8\rho_{A'}$, $\rho_8\rho^{A'}$ on $\eta^{B'}$,  $\eta_4$ and their complex conjugates. We note that the Clifford algebra is equivalent to
\be\label{blah}
\{\rho_8\rho_{A'},\rho_8\rho_{B'}\} =0\, ,\qquad \{\rho_8\rho_{A'},\rho_8\rho^{B'}\} =-4\delta_{A'}^{B'}\, .
\ee
Since we have not been very precise about the exact definition of $\eta^{A'}$ and $\eta_4$ it is enough to observe that the choice
\begin{eqnarray}
\rho_8\rho_{A'}\eta^{B'} &=& 2\delta_{A'}^{B'}\eta^4\, ,\qquad\rho_8\rho_{A'}\eta_{B'} = -2\varepsilon_{A'B'C'}\eta^{C'} \, , \nonumber\\[10pt]
 \rho_8\rho_{A'}\eta_{4} &=& -2\eta_{A'}\, ,\qquad\rho_8\rho_{A'}\eta^{4} = 0\, ,
\end{eqnarray}
and similarly for the complex conjugates,
satisfies the algebra (\ref{blah}). In this basis the fermion terms become
\begin{eqnarray}
{\cal L}^{(f)}_{3DSYM} &=& -i \tr \left( \bar\lambda^{A} \gamma^\mu \nabla_\mu \lambda_{A} \right)   \nonumber \\
&& +\frac{g_{YM}}{\sqrt{2}}\tr\left(2i\bar \lambda^{A'} [{\rm Im} \, \phi^4_{0},\lambda_{A'}]-2i\bar \lambda^{4} [{\rm Im} \, \phi^4_{0},\lambda_{4}] + 2\bar \lambda^{A'}[\phi^\dag_{A'\,0},\lambda_{4}] - 2\bar \lambda^4 [\phi^{A'}_{0},\lambda_{A'}]\right.\nonumber\\
&&\left.  +  \varepsilon_{A'B'C'}\bar \lambda^{A'} [\phi^{B'}_0,\lambda^{C'}] + \varepsilon^{A'B'C'} \bar \lambda_{A'} [\phi_{B'\,0},\lambda_{C'}]\right)\, .
\end{eqnarray}
 In particular we see that, if we identify $A_\mu = a^+_{\mu \,0}$, $\lambda^A=\psi^A_{0}$ and take $g'_{YM}=g_{YM}/\sqrt{2}$, corresponding to $\Omega = 2/\sqrt{3}$ as before, then (\ref{M2act}) can be  written as
\begin{eqnarray}
{\cal L}_{array}&=&
-\frac{1}{2} \tr \left( \nabla_\mu Y \nabla^\mu Y \right) -  \frac{1}{2}\tr \left( \nabla_\mu X^I\nabla^\mu X^I \right) - \frac{i}{2}\tr \left( \bar\Lambda\Gamma^\mu \nabla_\mu \Lambda \right) \nonumber \\
&&  +\frac{g_{YM}}{2}\tr \left( \bar\Lambda \Gamma^{11}\Gamma^I[X^I,\Lambda] \right) +\frac{g^2_{YM}}{4} \sum_{I,J} \tr ([X^I,X^J])^2\, ,
\end{eqnarray}
where
\be
Y = {\rm Re} \, \phi^4_0\, .
\ee
Thus we find that the three-dimensional maximally supersymmetric Yang-Mills Lagrangian is in agreement with the M2-brane Lagrangian, with the exception of the kinetic term of the gauge field, which is absent, along with an additional scalar field $Y$ which does not enter in the potential. In particular we see that the M2-brane Lagrangian from the infinite array has an $SO(7)$ symmetry, which is enhanced from the manifest $SU(3)$ symmetry that we started with.

Since our action has no kinetic term for the gauge field, its equation of motion imposes a constraint:
\be\label{noJ}
i[Y,\nabla_\mu  Y] +   i[X^I,\nabla_\mu  X^I]+\frac{1}{2}[\bar\Lambda , \Gamma_\mu\Lambda]=0\, .
\ee
Furthermore the scalar $Y$ couples to the gauge field but does not enter into the potential. Its equation of motion is
\be
\nabla^2Y=0\, .
\ee
A solution to this equation is
\be\label{dualisation}
\nabla_\mu Y = -\frac{1}{2 g_{YM}}\varepsilon_{\mu\nu\lambda}F^{\nu\lambda}\, .
\ee
From this we deduce that
\begin{eqnarray}
[Y,\nabla_\mu  Y] &=&- \frac{1}{2 g_{YM}}\varepsilon_{\mu\nu\lambda}[Y,F^{\nu\lambda}] \\[10pt]
&=& \frac{i}{ g_{YM}}\varepsilon_{\mu\nu\lambda}\nabla^\nu\nabla^\lambda Y \\[10pt]
&=&\frac{i}{ g^2_{YM}}\nabla^\nu F_{\mu\nu}\, .
\end{eqnarray}
In which case we find that the  constraint (\ref{noJ}) can be written as
\be
\frac{ 1}{ g^2_{YM}}\nabla^\nu F_{\mu\nu} = i[X^I,\nabla_\mu  X^I]+\frac{1}{2}[\bar\Lambda, \Gamma_\mu\Lambda]\, ,
\ee
which is precisely the equation for three-dimensional maximally supersymmetric Yang-Mills. In particular we have recovered all 16 supersymmetries in addition to the $SO(7)$ R-symmetry. Note that this is an on-shell dualisation of the scalar field into a gauge field. Without this dualisation ${\cal L}_{array}$ is not supersymmetric, however one may conjecture that it secretly enjoys a hidden quantum supersymmetry, much like the case of the enhanced maximal supersymmetry in the ABJM models at $k=1,2$ (although here it would seem to appear even at weak coupling).

Thus the system we obtained is related classically to 3D-SYM. In particular, to be more precise, every solution to 3D-SYM solves our system. However our system is slightly more general. In particular consider a pure-gauge configuration
\be
A_{\mu  } = i g\partial_\mu g^{-1}\, ,
\ee
where $g\in U(M)$.
The solution to (\ref{dualisation}) is
\be
Y = gY_0 g^{-1}\, ,
\ee
where $Y_0$ is any constant Hermitian matrix. Thus there is an additional, nondynamical, `modulus' that appears in the M2-brane description of D2-branes. This is not surprising and should be thought of as the positions of the D2-branes in the eleventh dimension. In particular it is possible to break the gauge group while keeping all the D2-branes at the origin of the string theory Coulomb branch $X^I=0$. Note that the classical vacuum moduli space condition does not require that the vevs of $Y$ and $X^I$ commute.

Let us now discuss some curiosities of our results. We see that $X^9=\sqrt{2}{\rm Im} \, \phi^4_0 $ appears in the potential whereas in the original array ${\rm Im} \, Z^4$ represents the direction along the array, i.e.\ the M-theory direction, and is subject to a discrete shift symmetry: ${\rm Im} \, Z^4 \to {\rm Im} \, Z^4 +2\pi  R $. This symmetry is still present in our system at finite $N$,  although due to the field normalisation it is rescaled to ${\rm Im} \, \phi^4_0 \to {\rm Im} \, \phi^4_0 +2\pi \sqrt{2N} R$ which diverges when we take $N\to\infty$. However in the D-brane interpretation $X^9$ is not the M-theory direction.

As mentioned above we need to take a limit $N\to\infty$ such that $M_b\to\infty$ and $g_{YM}$ finite. This can be done in a variety of ways. In particular since $M_b\propto g^2_{YM} k$ all  we require is that $N,k\to \infty$ with $R\propto k/\sqrt{N}$. We could achieve this by keeping $R$ fixed and $N\propto k^2$. Since the scalar fields have canonical dimensions of $(mass)^{1/2}$ the physical radius of the array is
\be
R_{11} = RT^{-1/2}_{M2}\, .
\ee
Since $T_{M2} = (2\pi)^{-2} l_p^{-3}$, $l_s=g_s^{-1/3}l_p$ and $R_{11}=g_sl_s$ we see that
\be
g^2_{YM} = \frac{(2\pi)^2}{3} \frac{N}{ k^2}\frac{g_s}{l_s}\, .
\ee
whereas the precise relationship for a D2-brane is $g^2_{YM} = g_s/l_s$. We can arrange for this by taking $N=ab$, $k=b$ where $a/b$ is a rational approximation to $3/(2\pi)^{2}$. However this seems very ad hoc.

One way to avoid any conflicts with these issues is to consider a scaling limit where $R\propto \sqrt{N}$, $k\propto N$. In this way we can remain at weak 't Hooft coupling $NM/k$ throughout. The distance between the M2-branes then diverges so that the fluctuations of ${\cal L}_{array}$ really just describe an isolated block of $M$ M2-branes, inside the array. (In fact this is true more generally as the normalisation of $\phi^A$ ensures that  fluctuations of $\phi^A_0$ do not correspond to finite fluctuations of $Z^A$.) In this case the reduction to type IIA string theory arises because of $k\to\infty$, and the associated spacetime ${\mathbb C}^4/{\mathbb Z}_k$ orbifold, in addition to the periodicity imposed by the array.

Finally we note that it is not clear how to quantise the action we have found. Although we have derived it from the ABJM model which does have a well-defined quantisation. One way is to map it to an equivalent classical Lagrangian which is more suitable to quantisation, i.e.\ one which admits a simple Hamiltonian without constraints, or with constraints that can be readily solved. Ignoring the subtlety that we have mentioned above this would lead to three-dimensional maximally supersymmetric Yang-Mills and its familiar quantisation. Another approach would be to use Dirac quantisation applied to the constraint induced from the $A_{\mu }$ equation of motion. We will not address this problem in this thesis. Assuming that there is a suitable quantum theory involving $Y$ we can consider operators such as
\be
{\cal M} = e^{iY }\, ,
\ee
which correspond to monopole (or 't Hooft) operators. Thus we have arrived at a more refined version of three-dimensional maximally supersymmetric Yang-Mills as the description of D2-branes in type IIA string theory.

\subsection{Further Compactification on ${\mathbb T}^2$ and M5-branes} \label{s_Array_torus}

Let us now consider a doubly periodic array in the $X^3$ and $X^4$ directions. For simplicity we will only consider the bosonic part of the action in this section. The extension to include the fermions is straightforward. Firstly let us rescale the scalars by a factor of $g^{-1}_{YM}$ to cast the (bosonic) action as
\be\label{bM2act}
S^{(b)}_{array}=-\frac{1}{g^2_{YM}}\tr\, \int d^3 x\
 \frac{1}{2}  \nabla_\mu Y \nabla^\mu Y  +  \frac{1}{2}  \nabla_\mu X^I\nabla^\mu X^I -\frac{1}{4} \sum_{I,J}([X^I,X^J])^2\, .
\ee
If we impose the on-shell dualisation discussed above then we arrive at the familiar three-dimensional maximally supersymmetric Yang-Mills theory of D2-branes. In this case imposing a further periodic array along $X^3$, $X^4$ was studied some time ago in \cite{Taylor:1996ik} and leads to the same action as five-dimensional maximally supersymmetric Yang-Mills compactified on ${\mathbb T}^2$. We follow the same steps here but without the dualisation.

We first consider an infinite parallel array along the $X^3$ direction by imposing the shift symmetry:
\begin{align}
X^I_{mn} \ =& \ 2\pi R' n\delta_{mn}\delta^I_3 + X^I_{n-m} \, , \\[10pt]
Y_{mn} \ =& \ Y_{n-m} \, , \\[10pt]
A_{\mu\, m n} \ =& \ A_{\mu\, n-m}
\end{align}
here $m,n\in \mathbb Z$ and as before each field is an $M\times M$ Hermitian matrix. Note that $R'$ has dimensions of mass. (In the interests of not introducing more symbols we are being rather brief in our notation.)  We can then repackage these fields in terms of a higher dimensional gauge theory on ${\mathbb R}^{3}\times S^1$:
\begin{eqnarray}
X^I &=&\sum_n e^{i n x^3R'}X^I_{n}\, , \quad I = 4,5,\ldots,9 \, , \\[10pt]
Y &=&  \sum_n e^{i n x^3R'}Y_{n} \, , \\[10pt]
A_{\mu} &=& \sum_n e^{i n x^3R'}A_{\mu\, n} \, , \\[10pt]
A_{3}&=&\sum_n e^{i n x^3R'}X^3_{n}\, ,
\end{eqnarray}
where $x^3$ is periodic with period   $2\pi/R'$.
Next we repeat this procedure for an array along $X^4$, with the same periodicity. In this way we construct five-dimensional Hermitian matrix valued fields $Y,A_{\mu'}, X^{I'}$ where $\mu'=0,1,\ldots,4$ and $I' =5,6,\ldots,9$.

Following the analysis of \cite{Taylor:1996ik} leads to the action\footnote{We could also use our regularisation technique of introducing a large but finite array of size $N>>1$. This would simply result in an additional factor of $4N^2$ in front of the action which we would then remove by an appropriate rescaling of $g_{YM}$ and $R'$.}
\begin{align}\label{M5act}
\nonumber
S^{(b)}_{cubic\ array}=-\frac{R'^2}{(2\pi)^2 g^2_{YM}}\tr\, \int d^5  x  \ \frac{1}{2}&  \nabla_{\mu'} X^{I'}\nabla^{\mu'} X^{I'} -\frac{1}{4} \sum_{I',J'}([X^{I'},X^{J'}])^2
 \\ 
+\frac{1}{2}& \nabla_{\mu} Y \nabla^{\mu} Y  + \frac{1}{2}F_{\mu \alpha}F^{\mu\alpha }+ \frac{1}{4}F_{\alpha\beta}F^{\alpha\beta}\, .
\end{align}
where $\alpha,\beta=3,4$ and $\mu,\nu = 0,1,2$.
This is not five-dimensional Lorentz invariant. In particular there is no kinetic term for $Y$ along the torus directions and no $F_{\mu\nu}$ terms. The first issue  arises because there is no $[X^I,Y]$ term in (\ref{bM2act}) whereas  the second  arises because there was no $F_
{\mu\nu}$ term to start with.
Nevertheless we note that the $A_\alpha$ equation of motion is that of five-dimensional maximally supersymmetric Yang-Mills:
\be
\nabla^{\mu'} F_{\alpha\mu'}  = i[X^I,\nabla_\alpha X^I]\, .
\ee
One the other hand the $A_\mu$ equation of motion is similar to before but with an extra term:
\be
\nabla^\alpha F_{\nu\alpha} =  i[Y,\nabla_\nu Y] + i[X^I,\nabla_\nu X^I]\, .
\ee
Once again we can consider the on-shell dualisation and choose:
\be
\nabla_\mu Y =  -\frac{1}{2}\varepsilon_{\mu\nu\lambda}F^{\nu\lambda}\, ,
\ee
which is still consistent with the $Y$ equation of motion $\nabla_\mu \nabla^\mu Y =0$. In this way we obtain
\be
\nabla^{\mu'} F_{\nu\mu'} =   i[X^I,\nabla_\nu X^I]\, ,
\ee
so that our equations are those of (the bosonic part of) five-dimensional maximally supersymmetric Yang-Mills, which restores five-dimensional Lorentz symmetry and 16 supersymmetries.

We can also consider another on-shell dualisation, which is more naturally associated with the broken Lorentz symmetry, and take
\be\label{Yagain}
\nabla_\mu Y= \frac{1}{2}\varepsilon^{\alpha\beta}H_{\mu \alpha\beta} \, .
\ee
We can then also write $F_{\mu \alpha}=H_{\mu\alpha 5}$ and $F_{\alpha\beta}=H_{\alpha\beta 5}$ which is sufficient to determine all the components of a self-dual six-dimensional 3-form $H$. In this language the `moduli' $Y_0$ associated to solving (\ref{Yagain}) can be thought of as the  period of a 2-form potential $B_{34}$ integrated over the two-cycle of the torus.

\subsection{Conclusions}\label{s_Array_conc}

In this chapter we have investigated periodic arrays of M2-branes using the ABJM model. By introducing a regularisation method, where we consider a large but finite array with $2N+1$ sites, imposing a discrete translational symmetry, computing the action, and then letting $N\to\infty$. The Chern-Simons level $k$ and array radius $R$ were also allowed to scale with $N$ and to obtain a suitable theory of D2-branes we required that  $k\to\infty$ with $R\propto k/ \sqrt{N}$. Our result is a curious variation of three-dimensional maximally supersymmetric Yang-Mills where the gluon kinetic term is replaced by that of a dual scalar field. All solutions of three-dimensional maximally supersymmetric Yang-Mills can be made solutions of our Lagrangian but in addition we find  nondynamical moduli. We also considered further doubly-periodic arrays that map the D2-branes to D4-branes. This led to a non-Lorentz invariant version of five-dimensional maximally supersymmetric Yang-Mills. This in turn can be viewed as the M-theory description of a cubic array of M2-branes, which should therefore also describe M5-branes wrapped on ${\mathbb T}^3$. 

However it is not entirely clear how much our results should be trusted at strong coupling. In particular the M2-brane physics relies crucially on 't Hooft (monopole) operators which we have not addressed. Although these are not expected to be   important at large $k$ our analysis is largely justified by the fact that our results can be mapped to the known open string description obtained at weak $g_{YM}$. Therefore it remains  possible that 't Hooft operators  are important here at large $g_{YM}$.

\newpage
\section{M5-branes and 3-algebras} \label{c_Multiple_M5s_and_3_algebras}

In this chapter we move our focus to the other extended object in M-theory namely the M5-brane. In section \ref{s_The_M5_brane} we will discuss the properties of M5-branes. For various reasons that will be mentioned, the M5-brane is a more difficult object to study than the M2-brane. We have seen in chapter \ref{c_Multiple_M2s_and_3_algebras} that the key to unlocking the worldvolume description of multiple M2-branes was the introduction of 3-algebras. In section \ref{s_A_nonabelian_2_0_tensor_multiplet} we describe work to formulate M5-branes using 3-algebra language \cite{Lambert:2010wm}. 

\subsection{The M5-brane}\label{s_The_M5_brane}

The worldvolume of the M5-brane is a six-dimensional hypersurface propagating in an eleven-dimensional background. The low energy dynamics of a single M5-brane is described by the Abelian $\mathcal{N}=(2,0)$ tensor multiplet \cite{Gibbons:1993sv,Kaplan:1995cp,Adawi:1998ta}. The field content of this multiplet can be determined as follows. There are five scalar fields which arise as Goldstone bosons from spontaneous breaking of eleven-dimensional translational invariance by the brane. The M5-brane is a $\sfrac{1}{2}$-BPS object and therefore preserves 16 of the 32 supersymmetries of the eleven-dimensional background. Consequently there are 16 Goldstone fermions which contribute eight fermionic degrees of freedom. For a supersymmetric theory the number of bosonic degrees of freedom must match the number of fermionic ones. So far we are short of three bosonic degrees of freedom. These additional degrees of freedom are provided by an Abelian 2-form gauge field $B_{\mu \nu}$ which has a self-dual field strength, $H_{\mu \nu \lambda}$. The 2-form originates from breaking of the gauge symmetry of the 3-form potential which exists in M-theory.

For this field content one would then construct a Lorentz invariant action for the M5-brane in a flat Minkowski background
\begin{equation}
S_{M5} = \int d^6 x \ - \frac{1}{2} \partial^\mu X^I \partial_\mu X^I + \frac{i}{2} \bar{\psi} \Gamma^\mu \partial_\mu \psi + \frac{1}{12} H^{\mu \nu \lambda} H_{\mu \nu \lambda} \, .
\end{equation}
However there is immediately a problem with this action: the kinetic term for the 2-form gauge field is identically zero because of the self-duality condition. Due to this a standard Lagrangian description of the gauge sector of the M5-brane does not seem possible. Various proposals for dealing with this problem have been suggested.
One approach to finding an M5-brane action is to drop manifest Lorentz invariance. In the literature this is often done at the expense of introducing an auxiliary field as in the PST action \cite{Pasti:1997gx}.
Another viewpoint is that, at least classically, the action is simply a tool which gives the equations of motion. If we have the field equations then an action is unnecessary. The equations of motion are simpler to deal with because they are linear in their respective field so in the case of the 2-form its field strength appears only once and the self-duality problem which occurs in the action does not appear. The covariant field equations for a single M5-brane are known \cite{Howe:1983fr,Howe:1996yn,Howe:1997fb}.

 
We now imagine that $N$ parallel copies of the M5-brane become coincident. The low energy theory which describes this system is an interacting six-dimensional CFT with $(2,0)$ supersymmetry. Unlike the case of the three-dimensional worldvolume theories of M2-branes very little is known about six-dimensional UV complete quantum field theories, let alone those with maximal supersymmetry. Famously, the degrees of freedom of the worldvolume theory of $N$ M5-brane exhibit $N^3$ scaling although the microscopic origin of this scaling is not currently understood. The gauge sector of the interacting $(2,0)$ theory should now consist of a non-Abelian 2-form with self-dual field strength. As for M2-branes there is no obvious parameter which controls the coupling strength. In the ABJM theory we have seen that the coupling is controlled by the quantity $N/k$ where $k$ is associated with the three-dimensional Chern-Simons action. The worldvolume of M5-branes is even-dimensional and consequently does not admit a Chern-Simons action from which to construct an analogous coupling parameter. There have been several attempts to understand the $(2,0)$ theory in the literature. Some time ago a light-cone formulation was proposed in \cite{Aharony:1997th,Aharony:1997an}, for the case of light-like compactification of the M5-brane as well as related constructions from Matrix Theory \cite{Berkooz:1996is,Maldacena:2002rb}. In addition a four-dimensional `deconstruction' was presented in \cite{ArkaniHamed:2001ie}. 
In the next section we will review the construction of a non-Abelian $(2,0)$ tensor multiplet in which the fields take values in a 3-algebra \cite{Lambert:2010wm}. Other, even more recent, discussions on formulating the dynamics of the non-Abelian $(2,0)$ theory are \cite{Ho:2011ni,Bolognesi:2011nh,Samtleben:2011fj,Chu:2011fd}.

\subsection{A Non-Abelian $(2,0)$ Tensor Multiplet}\label{s_A_nonabelian_2_0_tensor_multiplet}

We proceed in a similar manner to constructing the BLG theory in chapter \ref{c_Multiple_M2s_and_3_algebras}, starting with the covariant supersymmetry transformations of a free six-dimensional $(2,0)$ tensor multiplet \cite{Howe:1997fb} which are
\begin{align}
\delta X^I \ =& \ i \bar \epsilon \Gamma^I \psi \, , \\[10pt]
\delta  \psi \ =& \ \Gamma^\mu \Gamma^I \partial_\mu X^I \epsilon + \frac{1}{3!}\frac{1}{2} \Gamma^{\mu\nu\lambda}H_{\mu\nu\lambda}\epsilon \, ,  \\[10pt]
\delta B_{\mu\nu} \ =& \ i \bar \epsilon \Gamma_{\mu\nu} \psi \, .
\end{align}
Here $\mu = 0,1,\ldots,5$, $I = 6,7,\ldots,10$ and $H_{\mu\nu\lambda} = 3\partial_{[\mu}B_{\nu\lambda]}$ is self-dual. The supersymmetry generator $\epsilon$ and fermion $\psi$ satisfy the projection conditions
\begin{align}
\Gamma_{012345} \epsilon \ =& \ +\epsilon \, , \\[10pt]
\Gamma_{012345} \psi \ =& \ -\psi \, .
\end{align}
The commutator of the supersymmetry transformations on each of the fields closes on to a translation provided the following equations of motion hold 
\begin{align}
\partial^\mu\partial_\mu X^I \ =& \ 0 \, , \\[10pt]
\Gamma^\mu \partial_\mu \psi \ =&  \ 0 \, . 
\end{align}
We note that, from the point of view of supersymmetry, it is sufficient to write the algebra purely in terms of $H_{\mu\nu\lambda}$ and not mention $B_{\mu\nu}$:
\begin{align}
\delta X^I \ =& \ i \bar \epsilon \Gamma^I \psi \, , \\[10pt]
\delta \psi \ =& \ \Gamma^\mu \Gamma^I \partial_\mu X^I \epsilon + \frac{1}{3!}\frac{1}{2} \Gamma^{\mu\nu\lambda}H_{\mu\nu\lambda}\epsilon\; \label{Habelian} \, , \\[10pt]
\delta H_{\mu\nu\lambda} \ =& \ 3i \bar \epsilon \Gamma_{[\mu\nu}\partial_{\lambda]} \psi \, ,
\end{align}
in which case the equations of motion have to be supplemented by $\partial_{[\mu}H_{\nu\lambda\rho]}=0$.

We wish to try and generalise the free M5-brane superalgebra to allow for non-Abelian fields and interactions. In analogy with M2-branes we assume the fields take values in a vector space with a basis $T^a$. Derivatives are also promoted to covariant derivatives through a coupling to a gauge field $\tilde A_\mu{}^b{}_a$. For example the covariant derivative of the scalar field is
\begin{equation}\label{deriv}
D_\mu X^I_a = \partial_\mu X^I_a - \tilde A_\mu{}^b{}_a X^I_a \, .
\end{equation}

Upon reduction on a circle one expects that the six-dimensional  $(2,0)$ transformation rules reduce to those of five-dimensional supersymmetric Yang-Mills, which are given by
\begin{align}
\delta X^I \ =& \ i \bar \epsilon \Gamma^I \psi \label{SYM_s} \, , \\[10pt]
\delta  \psi \ =& \ \Gamma^\alpha \Gamma^I D_\alpha X^I \epsilon + \frac{1}{2} \Gamma^{\alpha\beta}\Gamma^{5}F_{\alpha\beta}\epsilon -\frac{i}{2} g_{YM} \Gamma^{IJ}\Gamma^{5} [X^I,X^J]\epsilon \label{SYM_f} \, , \\[10pt]
\delta A_{\alpha} \ =& \ i \bar \epsilon \Gamma_{\alpha}\Gamma_{5} \psi \label{SYM_g} \, ,
\end{align}
where $D_\alpha X^I = \partial_\alpha X^I - i g_{YM} [ A_\alpha , X^I]$ and $\alpha = 0,1,\ldots,4$.

In order to obtain a term analogous to the $[X^I,X^J]$ for $\delta\psi$ in \eref{Habelian} we need to introduce a $\Gamma_\mu$ matrix to account for the fact that $\epsilon$ and $\psi$ have opposite chirality. Thus a natural guess is to  propose the existence of a new algebra-valued field $C^\mu_a$ so that we can consider the  ansatz:
\begin{align}\label{ansatz1}
\delta X^I_a \ &=\ i \bar{\epsilon} \Gamma^I \psi_a \, , \\[10pt]
\delta \psi_a \ &= \ \Gamma^\mu \Gamma^I \epsilon D_\mu X^I_a + \frac{1}{3!} \frac{1}{2} \Gamma_{\mu \nu \lambda} \epsilon H^{\mu \nu \lambda}_a - \frac{1}{2} \Gamma_\lambda \Gamma^{IJ} \epsilon C^\lambda_b X^I_c X^J_d f^{cdb}{}_a \, , \\[10pt]
\delta H_{\mu \nu \lambda \; a} \ &= \ 3i \bar{\epsilon } \Gamma_{[\mu \nu} D_{\lambda]} \psi_a + i \bar{\epsilon} \Gamma^I \Gamma_{\mu \nu \lambda \kappa} C^\kappa_b X^I_c \psi_d f^{cdb}{}_a \, , \\[10pt]
\delta \tilde{A}_\mu{}^b{}_a \ &= \ i \bar{\epsilon} \Gamma_{\mu \lambda} C^\lambda_c \psi_d f^{cdb}{}_a \, , \\[10pt]
\delta C^\mu_a \ &= \ 0 \, .
\end{align}

As in the Abelian case the 3-form is self-dual
\begin{equation}
H_{\mu\nu\lambda \; a} = \frac{1}{3!}\epsilon_{\mu\nu\lambda\tau\sigma\rho}H^{\tau\sigma\rho}{}_a \, .
\end{equation}
In \cite{Lambert:2010wm} it was shown that there is no 2-form $B_{\mu \nu}$ such that $H_{\mu\nu\lambda}$ is its field strength.
Demanding that the self-duality relation is preserved under supersymmetry gives rise to the fermion equation:
\begin{equation}\label{fermionanticipated}
\Gamma^\mu D_\mu \psi_a + C^\mu_b X^I_c\Gamma_\mu\Gamma^I \psi_d f^{cdb}{}_a  =0\, .
\end{equation}

Note that consistency of the above set of equations with respect to their scaling dimensions gives
\begin{align}
\nonumber [H] = [X]+1 \, ,\qquad &\qquad [\tilde A] = 1\, ,\qquad\qquad [C] = 1-[X] \, , \\[10pt]
[\epsilon] = -\frac{1}{2}\, ,\qquad &\qquad [\psi] = [X] +\frac{1}{2}\, ,\qquad\qquad [X] \, ,
\end{align}
so one could still make this work with some other assignment  that are all related to the choice of $[X]$. However the canonical choice is
$[X]=2,[H]=3,[\psi]=\frac{5}{2},[C]=-1$. In particular we see that the new field $C^\mu$ has scaling dimension $-1$. Therefore, if we compactify the theory on a circle of radius $R$ we expect the expectation value of $C^\mu$ to be proportional to $R$. 

Provided the $f^{cdb}{}_a$ satisfy 
\begin{equation}
f^{bcd}{}_a = f^{[bcd]}{}_a \, , \qquad f^{efg}{}_d f^{abc}{}_g = f^{efa}{}_g f^{bcg}{}_d + f^{efb}{}_g f^{cag}{}_d + f^{efc}{}_g f^{abg}{}_d \, ,
\end{equation}
which are the conditions that define a real 3-algebra in chapter \ref{c_Multiple_M2s_and_3_algebras}, then the proposed non-Abelian $(2,0)$ supersymmetry transformations close on-shell on to a translation and a gauge transformation. The $(2,0)$ equations of motion are \cite{Lambert:2010wm} 
\begin{align}
0 &= \Gamma^\mu D_\mu\psi_a+X^I_c C^\nu_b \Gamma_\nu\Gamma^I\psi_d f^{cdb}{}_a \label{fermion_EOM_2_0} \, , \\[10pt]
0 &=D^2 X_a^I -\frac{i}{2}\bar\psi_c C^\nu_b \Gamma_\nu\Gamma^I \psi_d f^{cdb}{}_a + C^\nu_b C_{\nu g} X^J_c X^J_e X^I_f f^{efg}{}_{d}f^{cdb}{}_a \label{scalar_EOM_2_0} \, , \\[10pt]
0 &= D_{[\mu}H_{\nu\lambda\rho]\;a}+\frac{1}{4}\epsilon_{\mu\nu\lambda\rho\sigma\tau}C^\sigma_b X^I_c D^\tau X^I_d f^{cdb}{}_a + \frac{i}{8}\epsilon_{\mu\nu\lambda\rho\sigma\tau}C^\sigma_b \bar\psi_c \Gamma^\tau \psi_d f^{cdb}{}_a \label{H_EOM_2_0} \, , \\[10pt]
0&= \tilde F_{\mu\nu}{}^b{}_a + C^\lambda_c H_{\mu\nu\lambda\; d}f^{cdb}{}_a \label{F_EOM_2_0} \, , \\[10pt]
0 &= D_\mu C^\nu_a = C^\mu_c C^\nu_d f^{bcd}{}_a \label{C_EOM_2_0} \, , \\[10pt]
0 &= C^\rho_c D_\rho X^I_d f^{cdb}{}_a = C^\rho_c D_\rho \psi_d f^{cdb}{}_a =C^\rho_c D_\rho H_{\mu\nu\lambda\; a} f^{cdb}{}_a \label{pconstraint} \, . 
\end{align}
Where the scalar equation of motion is identified by taking the supervariation of the fermion field equation.
In all these equations $\tilde F_{\mu\nu}{}^b{}_a$ is the field strength of the gauge connection $\tilde A_\mu{}^b{}_a$ and as usual satisfies the Bianchi identity $D_{[\lambda} F_{\mu \nu]}{}^b{}_a=0$. Equation \eqref{H_EOM_2_0} can be dualised to give the alternative $H$ equation of motion 
\begin{equation}
0 = D_\mu H^{\mu \nu \lambda}{}_a - C^{[\nu}_b D^{\lambda]} X^I_d X^I_c f^{cdb}{}_a - \frac{i}{2} C^{[\nu}_b \bar\psi_c \Gamma^{\lambda]} \psi_d f^{cdb}{}_a \, . \label{H_EOM_2_0_dual}
\end{equation}
Note that the second to last equation \eqref{C_EOM_2_0} implies that $C^\mu_a$ is constant and hence selects a preferred direction in spacetime and in the 3-algebra. The final equations \eqref{pconstraint} imply that the non-Abelian components of the fields can only propagate in the five dimensions orthogonal to $C^\mu_a$.


All the calculations have been performed at the level of the supersymmetries and equations of motion and there has been no need to use the 3-algebra inner product to construct gauge invariant quantities. As we do not require an inner product, we are free to use any real 3-algebra we like i.e.\ Euclidean, Lorentzian or multiple time-like. We will now investigate the vacuum solutions of this 3-algebra formulation of the $(2,0)$ theory.

Our first choice will be a Lorentzian 3-algebra built from the Lie-algebra $su(N)$. We look for vacua of this theory by expanding around a particular point
\begin{equation}\label{vev}
\langle C^\lambda_a \rangle = g \delta_5^\lambda \delta^+_a \, ,
\end{equation}
where $g$ is dimensionful. With this vacuum expectation value the gauge field equation of motion \eqref{F_EOM_2_0} gives
\begin{equation}\label{FvH}
\tilde F_{\alpha\beta}{}^b{}_a =  - g H_{\alpha\beta 5\; d}f^{db}{}_a\, ,
\end{equation}
with $\mu = \alpha, 5$ and all other components of $\tilde F_{\mu\nu}{}^b{}_a$ are zero. As a result the latter correspond to flat connections that can be set to zero up to gauge transformations and \eref{C_EOM_2_0} reduces to $\partial_\mu g = 0$ so that $g$ is a dimensionful constant. 

The remaining $(2,0)$ equations of motion become:
\begin{eqnarray}
0 &=&\tilde  D^\alpha \tilde{D}_\alpha X_a^I - \frac{i}{2} g \bar{\psi}_c\Gamma_5\Gamma^I \psi_d f^{cd}{}_a - g^2 X^J_c X^J_e X^I_f f^{ef}{}_{d} f^{cd}{}_a  \, , \\[10pt]
0 &=& \tilde{D}_{[\alpha}H_{\beta\gamma] 5\;a} \, ,  \\[10pt]
0 &=& \tilde{D}^\alpha H_{\alpha\beta 5\; a}+\frac{1}{2} g \left( X^I_c \tilde{D}_\beta X^I_d + \frac{i}{2} \bar{\psi}_c \Gamma_\beta \psi_d \right) f^{cd}{}_a \, , \\[10pt]
0 &=& \Gamma^\mu \tilde{D}_\mu \psi_a+ g X^I_c \Gamma_5 \Gamma^I \psi_d f^{cd}{}_a  \, , \\[10pt]
0 &=& \partial_5 X^I_d = \partial_5 \psi_d = \partial_5 H_{\mu\nu\lambda\;d} \, ,
\end{eqnarray}
where $\tilde  D_\alpha X^I_a= \partial_\alpha X^I_a - \tilde A_{\alpha}{}^b{}_a X^I_b$, while the $(2,0)$ supersymmetry transformations become
\begin{eqnarray}
\delta X^I_a &=& i \bar{\epsilon} \Gamma^I \psi_a \, , \\[10pt]
\delta \psi_a &=& \Gamma^\alpha \Gamma^I \tilde{D}_\alpha X_a^I \epsilon + \frac{1}{2} \Gamma_{\alpha\beta} \Gamma_5 H_a^{\alpha\beta 5}\epsilon- \frac{1}{2}\Gamma_5 \Gamma^{IJ} X^I_c X^J_d {f^{cd}}_a\epsilon \, , \\[10pt]
\delta  \tilde{A}_{\alpha}{}^{b}{}_a &=& i \bar{\epsilon} \Gamma_{\alpha} \Gamma_5 \psi_d {f^{db}}_a \, .
\end{eqnarray} 
We immediately see that with the identifications
\begin{equation}
g= g_{YM}^2\, , \qquad H_{\alpha\beta 5 \; a} = \frac{1}{g_{YM}^2} F_{\alpha\beta \; a} \, , \qquad \tilde A_{\alpha}{}^{b}{}_a =  A_{\alpha\;c} {f^{cb}}_a\, ,
\end{equation}
we recover the equations of motion, Bianchi identity and supersymmetry transformations of five-dimensional supersymmetric $SU(N)$ Yang-Mills theory. In particular since $g$ has scaling dimension $-1$, we see that $g_{YM}$ also has the correct scaling dimension. Furthermore the fundamental identity reduces to the Jacobi identity for the structure constants of $su(N)$. 
However from the time-like generators in the 3-algebra we also get the additional equations
\begin{eqnarray}
0 &=&\partial^\mu \partial_\mu X_\pm^I  \, , \\[10pt]
0 &=& \partial_{[\mu}H_{\nu\lambda\rho]\; \pm} \, , \\[10pt]
0 &=& \Gamma^\mu \partial_\mu \psi_\pm \, ,
\end{eqnarray}
with transformations 
\begin{eqnarray}
\delta X^I_\pm &=& i \bar{\epsilon} \Gamma^I \psi_\pm \, ,  \\[10pt]
\delta \psi_\pm &=& \Gamma^\mu \Gamma^I \partial_\mu X_\pm^I \epsilon + \frac{1}{3!} \frac{1}{2} \Gamma_{\mu\nu\lambda} H_\pm^{\mu\nu\lambda} \epsilon \, , \\[10pt]
\delta H_{\mu\nu\lambda\; \pm} &=& 3 i \bar{\epsilon} \Gamma_{[\mu\nu}\partial_{\lambda]} \psi_\pm \, .
\end{eqnarray}
These comprise two free, Abelian $(2,0)$ multiplets which are genuinely six-dimensional. 

If we take the Euclidean 3-algebra, single out one of the gauge directions, say 4, and repeat the analysis around the vacuum given by 
\begin{equation}
\langle C^\lambda_a \rangle = g \delta_5^\lambda \delta^4_a \, ,
\end{equation}
then five-dimensional supersymmetric Yang-Mills is again recovered only now there is a single Abelian $(2,0)$ multiplet. Solutions to the equations of motion in the case of multiple time-like 3-algebras have been studied in \cite{Kawamoto:2011ab,Honma:2011br} and were found to lead to the description of other D$p$-branes in string theory.

A closely related system can essentially be reverse-engineered directly from maximally supersymmetric five-dimensional Yang-Mills \cite{Singh:2011id}. At this point one may argue that the model presented here is nothing more than multiple D4-branes written in 3-algebra language rather like the story with multiple D2-branes and Lorentzian 3-algebras in chapter \ref{c_Multiple_M2s_and_3_algebras}. However, in the next chapter we will see that the 3-algebra $(2,0)$ model is also capable of describing a separate nontrivial multiple M5-brane conjecture.


Recently it has been suggested that five-dimensional maximally supersymmetric Yang-Mills contains all the degrees of freedom of the non-Abelian $(2,0)$ theory, where the Kaluza-Klein states from the $S^1$ are mapped to the instantons of the five-dimensional theory \cite{Douglas:2010iu,Lambert:2010iw}. In chapter \ref{c_Periodic_arrays_of_M2s} we showed that the ABJM model can be used to describe a cubic periodic array of M2-branes and yields an action which is essentially the same as that of five-dimensional maximally supersymmetric Yang-Mills but with additional nondynamical `moduli'. Although we were required to rescale $k\to\infty$ to obtain this action we did keep all eleven-dimensional momentum modes. In particular the on-shell dualisation of $Y$ implies that magnetic flux $F_{12}$ plays the role of the `missing' eleven-dimensional momentum. Furthermore the resulting theory has a coupling constant that we could tune to be small but which we could also take to be large. Therefore our results in chapter \ref{c_Periodic_arrays_of_M2s} appear to be in broad agreement with the proposal of \cite{Douglas:2010iu,Lambert:2010iw}. We also note that if this proposal is true then five-dimensional maximally supersymmetric Yang-Mills should also display the $N^3$ scaling behaviour at strong coupling. Recent work in this direction can be found in \cite{Bolognesi:2011rq,Kim:2011mv,Maxfield:2012aw,Hatefi:2012sy,Kallen:2012zn}.

\newpage
\section{Light-cone Description of Multiple M5-branes} \label{c_Light_cone_M5s}

In this chapter we wish to study the $(2,0)$ system of equations obtained in \cite{Lambert:2010wm} and reviewed in chapter \ref{c_Multiple_M5s_and_3_algebras}, in the case where the auxiliary field $C^\mu_a$ is null.
This chapter is organised as follows. 
In section \ref{s_Conserved_currents} we determine the conserved energy-momentum tensor and supercurrent of the $(2,0)$ system that we outlined in the previous chapter. We also compute the superalgebra including the central charges. 
In section \ref{s_Null_reduction} we consider in detail the resulting dynamical system when the auxiliary vector field $C^\mu_a$ has a null vacuum expectation value. This leads to a curious system of equations with 16 supersymmetries and an $SO(5)$ R-symmetry that propagate in one null and four space directions.  We show how the equations reduce to motion on instanton moduli space, where the instanton number is the null momentum parallel to $C^\mu_a$. 
In section \ref{s_Quantisation} we then quantise the system by using the other null momentum generator as a Hamiltonian. This leads directly to the light-cone quantisation proposal of the $(2,0)$ theory proposed in \cite{Aharony:1997th,Aharony:1997an}, generalised to include a potential when the scalars have a vacuum expectation value and also  couplings to background gauge  and self-dual 2-form fields.
In section \ref{s_Null_reduction_as_infinite_boost} we show that the null reduction can be seen as the limit of an infinite Lorentz boost.
We end with our conclusions in section \ref{s_Light_cone_conc}.

\subsection{Conserved Currents and Superalgebra}\label{s_Conserved_currents}

It will be useful to construct the conserved currents of the $(2,0)$ equations of motion given in chapter \ref{c_Multiple_M5s_and_3_algebras}. In particular we look for an energy-momentum tensor $T_{\mu\nu}$ as well as a supercurrent $J^\mu$.
Simple trial and error leads to the following expressions:
\begin{align}
\nonumber
T_{\mu \nu} =&+ D_\mu X^I_a D_\nu X^{Ia} - \frac{1}{2} \eta_{\mu \nu} D_\lambda X^I_a D^\lambda X^{Ia} \\
\nonumber
	&+ \frac{1}{4} \eta_{\mu \nu} C^\lambda_b X^I_a X^J_c C_{\lambda g} X^I_f X^J_e f^{cdba} f^{efg}{}_d + \frac{1}{4} H_{\mu \lambda \rho \; a} H_{\nu}{}^{\lambda \rho \; a} \\
	&- \frac{i}{2} \bar{\psi}_a \Gamma_\mu D_\nu \psi^a + \frac{i}{2} \eta_{\mu \nu} \bar{\psi}_a \Gamma^\lambda D_\lambda \psi^a - \frac{i}{2} \eta_{\mu \nu} \bar{\psi}_a C^\lambda_b X^I_c \Gamma_\lambda \Gamma^I \psi_d f^{abcd} \, , \label{2_0_energy_momentum} \\
	\nonumber \\
J^\mu =&+ \frac{1}{2} \frac{1}{3!} H_{\nu \lambda \rho \; a} \Gamma^{\nu \lambda \rho} \Gamma^\mu \psi^a - D_\nu X^I_a \Gamma^\nu \Gamma^I \Gamma^\mu \psi^a - \frac{1}{2} C^\nu_b X^I_c X^J_d \Gamma_\nu \Gamma^{IJ} \Gamma^\mu \psi^a f^{bcd}{}_a \, . \label{2_0_supercurrent}
\end{align}
In the Abelian case this agrees with the linearised form of the energy-momentum tensor derived in \cite{Barwald:1999jh}. The associated conserved charges are
\begin{align}
P_\mu = \int d^5 x \ T_{\mu 0}\, ,\qquad \qquad Q = \int d^5 x \ J^0\, ,
\end{align}
where the integrals are over the spatial coordinates, corresponding to the momentum and supercharge respectively.

The superalgebra of the $(2,0)$ theory can then be deduced by evaluating $\delta J^0 = \delta _\epsilon J^{0\alpha} \epsilon_\alpha$ viz:
\begin{align}
\{ Q_\alpha , Q_\beta \} =& - \int d^5x \ ( \delta_\epsilon J^0 C^{-1} )_{\alpha \beta} \\[10pt]
=&-2 ( \Gamma^\mu C^{-1} )_{\alpha \beta} P_\mu + ( \Gamma^\mu \Gamma^I C^{-1} )_{\alpha \beta} Z_\mu^{I} + ( \Gamma^{\mu \nu \lambda} \Gamma^{IJ} C^{-1} )_{\alpha \beta} Z_{\mu \nu \lambda}^{IJ} \, .
\end{align}
The central charges we obtain in this way are (in the case of vanishing fermions):\footnote{The original equations in \cite{Lambert:2011gb} contained errors in the sign of $Z_0^{I}$ and the final term in $Z_i^I$, which we have corrected.}
\begin{align}
Z_0^{I} &= \int  d^5x \ \left( - 4 C^0_{b} X^I_c X^J_d D^0 X^J_a f^{cdba} \right) \, , \\[10pt]
\nonumber 
Z_i^I &= \int d^5x \ \bigg( + H^{0}{}_{j i \; a} D^j X^{I a}  +  \frac{1}{6} \partial_{j} ( H_{klm \; a} X^{Ia} \varepsilon_{0ijklm} ) \\
& \hskip2cm + C_{i b} X^I_c X^J_d D^0 X^J_a f^{cdba} + 2 C^0_{b} X^I_c X^J_d D_i X^J_a f^{cdba} \bigg) \, , \\[10pt]
Z_{0 i j}^{IJ} &= \int d^5x \  \left( + \frac{1}{2} H_{0 i j \; a} C^0_b X^I_c X^J_d f^{cdba} - \partial_i ( X^I_a D_j X^{Ja} ) \right) \, , \\[10pt]
Z_{k l m}^{IJ} &= \int d^5x \  \left( + \frac{1}{12} H_{k l m \; a} C^0_b X^I_c X^J_d f^{cdba} + \frac{1}{36} \partial^i ( C^j_b X^K_c X^L_d X^M_a f^{cdba} \varepsilon_{0ijklm} \varepsilon^{IJKLM} ) \right) \, ,
\end{align}
where here, in this section, $i,j=1,2,3,4,5$.

\subsection{Null Reduction}\label{s_Null_reduction}

Next we wish to consider the above system of equations for the special case where $C^\mu_a$ is a null vector:
\begin{equation}
C^\mu_a = \frac{g^2}{\sqrt{2}}(\delta^\mu_0+\delta^\mu_5)\delta_a^* \, ,
\end{equation}
where $g^2$ has dimensions of length and $*$ denotes some preferred direction in the 3-algebra. We choose to go to light-cone coordinates i.e.
$x^\mu = ( x^{+} , x^{-} , x^i )$ where
\begin{equation}
x^{-} = \frac{1}{\sqrt{2}} ( x^0 - x^5 )\, ,\qquad x^{+}  = \frac{1}{\sqrt{2}} (x^0 + x^5 )\, ,
\end{equation}
so that $C^\mu_a = g^2\delta^\mu_+\delta_a^*$. Note that for the rest of this chapter we have $i,j=1,2,3,4$ (rather than $i,j=1,2,3,4,5$ that was used in the previous section). The constraint (\ref{pconstraint}) now tells us that $D_{+}$ vanishes on all the fields. Furthermore $\tilde F_{i+}{}^b{}_a=0$ so $\tilde A_+{}^b{}_a$ is a flat connection and can be set to zero (at least locally). Thus the fields are essentially just functions of $x^i$ and $x^-$. Here we wish to view these equations of motion as a dynamical system where $x^-$ plays the role of time.

Let us now give the equations of motion that follow from the choice $C^\mu_a = g^2 \delta^\mu_+\delta_a^*$. Fixing the element $T^*$ in the 3-algebra means that the remaining generators behave as an ordinary Lie-algebra with Lie-bracket:
\begin{equation}
i [T^c,T^d]= [T^*,T^c,T^d]=f^{*cd}{}_a T^a\, .
\end{equation}
The components of the fields along the $*$ direction in the 3-algebra decouple and behave as a free six-dimensional tensor multiplet and for the rest of this chapter we simply discard them. Alternatively one could have started from a non-Abelian $(2,0)$ system where the $C$-field does not take values in the algebra, i.e.\ $C^\mu$ instead of $C^\mu_a$, as in the construction of \cite{Singh:2011id}.

For the sake of clarity we will use a notation whereby all the fields are taken to be Lie-algebra valued: e.g.\ $X^I = \sum_{a\ne *}X_a^I T^a$, and the $a$ index is dropped. We also note that the gauge field $\tilde A_\mu{}^b{}_a$ and field strength $\tilde F_{\mu\nu}{}^b{}_a$ also take values in the Lie-algebra and act on the other fields through the commutator. Therefore we drop the $a,b$ indices and tilde on these fields in what follows.

In the $(x^+,x^-,x^i)$ coordinates self-duality of $H_{\mu\nu\lambda}$ implies that $F_{ij} = -g^2 H_{ij+}$ is anti-self-dual, $G_{ij} = -g^2 H_{ij-}$ is self-dual and
\begin{equation}
H_{ijk} = g^{-2} \epsilon_{ijkl}F^l{}_-\, .
\end{equation}
Noting that the constraint implies that only the derivatives $D_-$ and $D_i$ are nonvanishing we find the remaining equations of motion can be written as
\begin{align}
0 &= \Gamma^- D_-\psi +\Gamma^iD_i\psi+ig^2[X^I  ,\Gamma_+\Gamma^I\psi] \, , \\[10pt]
0 &=D_iD^i X^I +\frac{g^2}{2}[\bar\psi, \Gamma_+\Gamma^I \psi] \, , \\[10pt]
0 &= D^iF_{i-} + \frac{g^4}{2} [\bar\psi, \Gamma_+ \psi]  \, , \\[10pt]
\label{secondorder}0 &= D_-F_{i-}-D^jG_{ij} - ig^4[X^I, D_i X^I] + \frac{g^4}{2} [\bar\psi, \Gamma_i \psi] \, ,  \\[10pt]
0 &= D_{[i}F_{j-]}\, .
\end{align}
One sees that the final equation is just the Bianchi identity and automatically satisfied.

Our strategy now is to solve as many of the equations of motion as possible. We will do this by setting the fermions to zero with the understanding that the supersymmetry can be used to generate fermionic solutions. We will see that all but the second order equation (\ref{secondorder}) can be solved and reduced to ADHM data.

To continue we first observe that the gauge field $A_i$ is determined by the ADHM construction \cite{Atiyah:1978ri}. Thus the degrees of freedom of the gauge field are reduced to the finite dimensional instanton moduli space with local coordinates $m^\alpha$. Note that although $G_{ij}$
is self-dual it has no interpretation as the field strength of $A_i$. Therefore $G_{ij}$ is not necessarily the field strength of a gauge field and one cannot solve for it using the ADHM construction. In fact $G_{ij}$ behaves as a nondynamical background field since its $D_-$ derivative never appears.

With vanishing fermions the scalar equation of motion is just $D_iD^i X^I=0$. It is easy to see that there is a unique solution to this equation for any given asymptotic value of $X^I$. In addition for an instanton background there exists smooth solutions. Thus $X^I$ is uniquely determined in terms of the ADHM data of the gauge field $ A_i$ and its asymptotic value:
\begin{equation}
X^I = v^I+{\cal O}\left(\frac{1}{|x|^2}\right)\, ,
\end{equation}
where $v^I$ is an element of the Lie-algebra.

Next we consider the equation $D^i F_{i-}=0$. In terms of gauge fields this is
\begin{equation}
D^iD_i A_- - D^i\partial_-A_i=0\, .
\end{equation}
To solve this equation we need to recall some facts about instanton moduli space, for reviews see \cite{Tong:2005un,Vandoren:2008xg}. In particular the instanton equations are
\begin{equation}
F_{ij} =-\frac{1}{2}\varepsilon_{ijkl}F^{kl}\, .
\end{equation}
Moduli correspond to infinitesimal changes to the gauge fields that preserve this condition:
\begin{equation}
D_i \delta A_{j} - D_j\delta A_{i} = -\varepsilon_{ijkl}D^k\delta A^l\, .
\end{equation}
However gauge transformations $\delta A_{i}= D_i\omega$ will clearly solve these equations and we do not wish to include them in the moduli. To exclude them we require that $\delta A_{i}$ is orthogonal to all gauge modes:
\begin{equation}
{\rm Tr}\int d^4 x \ \delta A_{i} D^i\omega =0\, .
\end{equation}
Integrating by parts, and requiring that $\omega=0$ at infinity,  shows that we therefore impose the gauge fixing condition
\begin{equation}
D^i\delta A_{i}=0\, .
\end{equation}
We have seen that the solution to the equations of motion requires that $A_{i}$ has anti-self-dual field strength. Therefore the $x^-$ dependence comes entirely through the dependence of the moduli on $x^-$ and hence we conclude that
\begin{equation}\label{gauge}
D^i\partial_-  A_{i}=0\, ,
\end{equation}
with $\partial_-A_i = \frac{\partial A_i}{\partial m^\alpha} \partial_-m^\alpha+D_i\omega$ where $\omega$ is chosen to ensure that \eref{gauge} is satisfied.
Thus the $D^i F_{i-} =0 $ equation simply becomes $D^iD_i A_-=0$. This is the same as the  $X^I$ equation and so $A_-$ is also determined in terms of ADHM data and its asymptotic value:
\begin{equation}
A_- = w+{\cal O}\left(\frac{1}{|x|^2}\right)\, ,
\end{equation}
where $w$ is an element of the Lie-algebra.

We are now left with just one equation which is second order in $x^-$:
\begin{equation}\label{dd}
D_-F_{i-}-D^jG_{ij} - ig^4[X^I, D_i X^I]=0\, .
\end{equation}
However as we mentioned above we do not aim to solve this equation - which would amount to a complete solution to all the classical field equations. Rather we now wish to quantise the classical field configurations that we have constructed and use the momentum generator along $x^-$ as the Hamiltonian.

\subsubsection{Conserved Charges}

To proceed we note that we need to use a slightly different definition of the conserved charge. In particular the problem with the standard definition given in section \ref{s_Conserved_currents} is that the integral over all space includes an integral over $x^5$. However one can simply change integration variable from $x^5$ to $x^-$ so that the integral is over all the coordinates. The resulting conserved charge is therefore constant not for dynamical reasons but because we have integrated over all the coordinates upon which the fields depend.

On the other hand we can consider
\begin{align}
{\cal P}_\mu = g^2\int d^4 x \ T_{\mu +}\, ,\qquad \qquad {\cal Q} = g^2\int d^4 x \ J^-\, ,
\end{align}
where we have included a factor of $g^2$ to ensure that they have the canonical dimensions.
Since $D_+=0$, ${\cal P}_\mu$ and ${\cal Q}$ are conserved in the sense that $\partial_-{\cal P}_\mu=\partial_-{\cal Q}=0$. Note that this assumes that the fields vanish sufficiently quickly at infinity so that the boundary terms in the integrals can be discarded. In particular conservation of ${\cal Q}$ requires that $D_-X^I$ and $[X^I,X^J]\to 0$ as $x^i\to\infty$. Therefore, in this chapter, in order to obtain conserved charges that can be used to define the quantum theory we assume that
\begin{equation}\label{vbdry}
[v^I,v^J]=[v^I,w]=0\, .
\end{equation}
That is we require that the scalar fields and gauge field are in a vacuum configuration at infinity.
More explicitly these expressions are (in the case of vanishing fermions):
\begin{align}
{\cal P}_- =& {\rm Tr}\int d^4 x \   \frac{1}{2g^2} F_{i-} F^i{}_- +\frac{g^2}{2} D_i X^ID^i X^{I} \, , \\[10pt]
{\cal P}_+ =& -\frac{1}{8g^2}{\rm Tr}\int d^4 x \    \varepsilon^{ijkl}F_{ij} F_{kl} \, , \\[10pt]
{\cal P}_i =& \frac{1}{2g^2}{\rm Tr}\int d^4 x \   F_{ij} F_-{}^j \, , \\[10pt]
{\cal Q} =& {\rm Tr}\int d^4 x\   F_{i-} \Gamma^i \Gamma^- \psi - \frac{1}{4} F_{ij}\Gamma^{i j} \Gamma^+ \Gamma^- \psi + g^2 D_i X^J \Gamma^J \Gamma^i \Gamma^- \psi \, .
\end{align}
Note that ${\cal P}_+ = -4\pi^2 g^{-2}k $, where $k$ is the instanton number. Thus the ${\cal P}_+$ eigenvalues are discrete. Physically we interpret this as arising because the $x^+$ direction is restricted to lie on a circle with radius $R=g^2/4\pi^2$.

We can further decompose $\cal Q={\cal Q}_++{\cal Q}_-$ where $\Gamma_{-+}{\cal Q}_\pm=\pm{\cal Q}_\pm$. In this case the superalgebra becomes
\begin{align}
\{ \cal Q_{-\alpha} , \cal Q_{-\beta }\}
=& - 2 {\cal P}_- ( \Gamma^- C^{-1} )_{\alpha \beta} + {\cal Z}^I_+ ( \Gamma^- \Gamma^I C^{-1} )_{\alpha \beta} + {\cal Z}^{IJ}_{ij+} ( \Gamma^{ij} \Gamma^- \Gamma^{IJ} C^{-1})_{\alpha \beta} \, , \\[10pt]
%
%
\{ \cal Q_{+\alpha} , \cal Q_{+\beta } \}
=& - 2 {\cal P}_+ ( \Gamma^+ C^{-1} )_{\alpha \beta} \, , \\[10pt]
%
%
\{ \cal Q_{-\alpha} , \cal Q_{+\beta } \}
=& - 2 {\cal P}_i ( \Gamma^i C^{-1} )_{\alpha \beta} + {\cal Z}^I_i  (\Gamma^i \Gamma^I C^{-1} )_{\alpha \beta} \, ,
\end{align}
where $C=\Gamma_0$ is the charge conjugation matrix and the central charges are
\begin{align}
{\cal Z}^I_+=& -2{\rm Tr}\int d^4 x \  F_{-i} D^i X^I \, , \\[10pt]
{\cal Z}_i^I=& -{\rm Tr}\int d^4 x \  G_{ij} D^j X^I \, , \\[10pt]
{\cal Z}^{IJ}_{ij+}=& -g^2{\rm Tr}\int d^4 x \  D_{[i} X^I D_{j]} X^J \, .
\end{align}
Note that although there are 16 supersymmetry charges only 8 of them ($\cal Q_-$) have a nontrivial relation with $\cal P_-$. This is a well-known feature of light-cone gauge (c.f.\ the Green-Schwarz superstring).  Furthermore any state with a nonvanishing ${\cal P}_+$ must break the ${\cal Q}_+$ supersymmetries. 

We also see that $G_{ij}$ only appears through its contribution to the central charge ${\cal Z}_i^I$. Here we take it to be a background, nondynamical field, in which case one only seems to obtain a conserved quantity in the case that $D^jG_{ij}=0$, so that it decouples from (\ref{dd}). In this case  ${\cal Z}_i^I$ is simply a boundary term depending on $v^I$ and $G_{ij}$.

Thus, to summarise, we impose the constraints $D^iG_{ij}=[v^I,v^J]=[v^I,w]=0$ on the fields to ensure that their charges given above are well-defined and conserved. This is necessary in our treatment since we will ultimately quantise the theory and use the Hamiltonian as the generator of time evolution through a Schr\"odinger equation.

\subsection{Quantisation}\label{s_Quantisation}

We have seen above that the classical equations of motion can be solved up to a single second order evolution. We have also constructed the conserved momentum and central charges in the $(2,0)$ algebra. In this section, rather than solve the second order classical evolution equation we instead wish to quantise the system using ${\cal P}_-$ as the Hamiltonian. In particular we see that it can be written as
\begin{equation}
{\cal P}_- = \frac{1}{2g^2}{\rm Tr}\int d^4 x \ \partial_-A_i\partial_-A^i - 2\partial_-A_i D^i A_- + D_iA_-D^iA_-+g^4D_iX^ID^iX^I   \, .
\end{equation}
The first term gives the kinetic energy and can be expressed in terms of the metric $g_{\alpha\beta}$ on instanton moduli space defined by
\begin{equation}
{\rm Tr}\int d^4 x \ \delta A_i \delta A^i = g_{\alpha\beta}\delta m^{\alpha}\delta m^\beta \, .
\end{equation}
Here $\delta A_i = \partial A_i/\partial m^\alpha \delta m^\alpha + D_i\delta \omega$,  with $\delta\omega$ is the gauge transformation required to preserve $D^i\delta A_i=0$.

Next we have a term that is linear in time derivatives:
\begin{equation}
{\rm Tr}\int d^4 x \ \partial_-A_i D^i A_- = {\rm Tr}\oint \partial_-A_r w = L_\alpha \dot m^\alpha\, .
\end{equation}
where $r$ is the radial normal direction to the sphere at infinity,  $\dot m^\alpha = \partial_- m^\alpha$  and $L_\alpha$ is a vector field on the instanton moduli space. We note that it is proportional to $w$,  i.e.\ it is determined by the vacuum expectation value of $A_-$, and can be viewed as a background gauge field.

 The last two terms can be written as a boundary integral and contribute to the potential. Thus we find that the Hamiltonian is
 \begin{equation}
{\cal P}_-= \frac{1}{2g^2}g_{\alpha\beta} (\dot m^\alpha - L^\alpha) (\dot m^\beta-L^\beta )+V\, ,
 \end{equation}
 where
 \begin{equation}\label{V}
 V  = - \frac{1}{2g^2} g_{\alpha\beta}L^\alpha L^\beta+\frac{1}{2g^2}{\rm Tr}\oint g^{4}X^ID_rX^I+A_-D_rA_-\  .
 \end{equation}

 For $w=0$ this Hamiltonian has appeared before \cite{Lambert:1999ua}  and is known to admit 8 supersymmetries, which correspond to the ${\cal Q}_-$ here. In particular it was shown that
 \begin{equation}
 V = \frac{g^2}{2} g_{\alpha\beta} K^\alpha K^\beta\, ,
 \end{equation}
 where $K^\alpha$ is a tri-holomorphic Killing vector on the instanton moduli space which
 can be expressed purely in terms of the asymptotic values of $X^I$ and the ADHM data  \cite{Lambert:1999ua} . By construction the Hamiltonian is also invariant under 8 supersymmetries when $w\ne 0$.

The next step is to decide on a momentum conjugate to the moduli coordinates $m^\alpha$. The obvious choice is
\begin{equation}
p_\alpha = g_{\alpha\beta}  \dot m^\beta \, .
\end{equation}
An alternative quantisation could be $p_\alpha = g_{\alpha\beta} (\dot m^\beta-L^\beta)$ however since $L^\alpha$ depends on $w_a$ this quantisation would then differ in various sectors of the theory.
It would be interesting to obtain a symplectic structure on the entire $(2,0)$ system that leads to this.
Quantisation is now straightforward and we just consider wavefunctions $\Psi(m^\alpha,x^-)$ and define
\begin{equation}
\hat p_\alpha\Psi = -i\frac{\partial \Psi}{\partial m^\alpha}\, , \qquad \qquad \hat m^\alpha \Psi = m^\alpha \Psi \, ,
\end{equation}
where a hat denotes the quantum operator. 

There is  one issue that requires some discussion, namely the moduli space generically contains singularities where the instantons shrink to zero size. These are not curvature singularities but rather more like orbifold singularities. Thus we should either seek to remove them or simply come up with  a suitable prescription on the behaviour of the wavefunction at the singularities. Methods for pursuing the first approach were considered in \cite{Aharony:1997th}. For the second approach one could simply assume that physical wavefunctions need to be even under the orbifold action at each singularity.

\subsubsection{One Instanton Example}

For concreteness we now give the expressions above for the case of a single instanton i.e.\ 
\begin{equation}
{\cal P}_+=-4\pi^2/g^2 \, ,
 \end{equation}
 with gauge group $SU(2)$, including all the moduli. In this case we have ($\eta^a_{ij}$ are the self-dual 't Hooft matrices)
\begin{eqnarray}
A_i &=& \frac{1}{(x-y)^2}\frac{\rho^2}{(x-y)^2+\rho^2}\eta^a_{ij}(x-y)^j U\sigma_a U^{-1} \, , \\[10pt]
X^I &=& \frac{(x-y)^2}{(x-y)^2+\rho^2}{v^I_aU\sigma_a U^{-1}} \, , \\[10pt]
 A_- &=&\frac{(x-y)^2}{(x-y)^2+\rho^2}{w_aU\sigma_a U^{-1} }\, .
 \end{eqnarray}
Here there are eight moduli represented by the instanton size $\rho$, position $y^i$ and gauge embedding $U\in SU(2)\equiv S^3$. Therefore, in total the moduli space is eight-dimensional.

Our first task is to compute the metric. To do this we note that to ensure $D^i\partial_-A_i=0$ we find that $\omega$ is given by
\begin{equation}
\omega =  \frac{1}{(x-y)^2}\frac{\rho^2}{(x-y)^2+\rho^2} \eta^a_{ij}\dot y ^i (x-y)^jU\sigma_aU^{-1}- \frac{\rho^2}{(x-y)^2+\rho^2} \dot u^a U\sigma_a U^{-1}\, ,
\end{equation}
where we have introduced
\begin{equation}
U^{-1}\dot U = i\dot u^a\sigma_a\, . 
\end{equation}
We can now compute the metric and find
\begin{equation}
ds^2 = 8\pi^2 (d\rho^2+\rho^2du^adu^a)+4\pi^2dy^kdy^k\, .
\end{equation}
This is just the flat metric on ${\mathbb R}^4\times {\mathbb R}^4$ ($u^a$ are the left-invariant $SU(2)$ forms of the unit $S^3$). However we note that, by construction, $U$ is indistinguishable from $-U$ and therefore  the actual moduli space is obtained by identifying $U\cong -U$ and hence is the quotient ${\mathbb R}^4/{\mathbb Z}_2\times {\mathbb R}^4$.

Next we evaluate
\begin{equation}\label{bt}
\oint  \partial_-A_r = \oint \frac{\partial A_r}{\partial m^\alpha} \dot m^\alpha + D_r \omega\, ,
\end{equation}
where $r$ is the normal direction to the boundary. The only contributions to this come from the ${\cal O}(1/r^3)$ term in $\partial_-A_r$.  To evaluate (\ref{bt}) one notes that  $\partial A_i /\partial y^k = {\cal O}(r^{-4})$ and, although the $\partial A_i /\partial \rho$  and $\partial A_i/\partial U$ terms are ${\cal O}(r^{-3})$, their $\partial A_r /\partial \rho$  and $\partial A_r/\partial U$ components vanish. Thus we have
\begin{equation}\label{btone}
\oint  \partial_-A_r = \oint  D_r \omega =  4\pi^2 \rho^2U\dot u^a\sigma_aU^{-1} 
\end{equation}
and hence
\begin{equation}
L_\alpha \dot m^\alpha=   8\pi^2 \rho^2 w_a\dot u^a\, ,
\end{equation}
or equivalently $L^\alpha=w^a\delta_a^\alpha$. If we consider gauge transformations of the form $U(x^-)$ then $L^\alpha$ will transform as a gauge field.
For $V$ we find
\begin{equation}
V =  4\pi^2g^2 v^I_av^I_a  \rho^2\, .
\end{equation}
Note that the first and last terms in (\ref{V}) have completely cancelled each other and we expect that this is generically the case. Thus we have found that
\begin{equation}
{\cal P}_- = \frac{4\pi^2} {g^{2}}\left(\dot \rho^2 +  \rho^2(\dot u^a - w^a)(\dot u^a - w^a)+\frac{1}{2}\dot y^k\dot y^k\right)\ +  4\pi^2g^2v^I_av^I_a  \rho^2 .
\end{equation}
It is also straightforward to show that the conserved momentum is
\begin{equation}
{\cal P}_i = -2\pi^2g^{-2} \dot y_i\, .
\end{equation}
 More generally, for the case of  point-like multi-instantons   (i.e.\ widely separated compared to their individual scale sizes), one finds  ${\cal P}_i\sim -2\pi^2g^{-2}\sum\dot y_i$ is just the centre of mass momentum.

Let us now discuss the central charges. First consider ${\cal Z}^I_+$;
\begin{eqnarray}
{\cal Z}^I_+ &=& - 2{\rm Tr} \int d^4x (\partial_- A_i-D_iA_-) D^iX^I\\[10pt]
&=& - 2{\rm Tr} \oint (\partial_- A_r-D_rA_-)X^I\\[10pt]
&=&  -16\pi^2 \rho^2 v^I_a( \dot u^a -   w^a )\, .
\end{eqnarray}
This is  the angular momentum associated to the action of $SU(2)$ on the moduli space.

In the one-instanton case the unique solution to $D^iG_{ij}=0$ is given by 
\begin{equation} 
G_{ij} = G_0 {(x^2+\rho^2)^2}{x^{-4}}\eta^a_{ij}\sigma_a \, ,
\end{equation}
where $G_0$ is a constant. However conservation of ${\cal Q}$ and ${\cal P}_\mu$ requires that all fields vanish at infinity (and are not too singular at the origin) and hence we must take $G_0=0$ so that $Z_i^I=0$. We  expect that   any states that carry $Z_i^I$ charge are string-like states extended along some direction say $x^4$. In this case
the total  ${\cal P}_+$ momentum is infinite but the ${\cal P}_+$ per unit length should be finite. Therefore the quantum mechanical system reduces to motion on the monopole moduli space determined by the Nahm construction \cite{Nahm:1979yw}.

In addition we find that  ${\cal Z}^{IJ}_{ij+}$ is given by
\begin{equation}
{\cal Z}^{IJ}_{ij+} = -2\pi^2\rho^2\epsilon^{abc}\eta^a_{ij} v^I_bv^J_c\, .
\end{equation}
However this vanishes since we demand that $[v^I,v^J]=0$ in order that $\cal Q$ is conserved. More generally we expect that any  state with nonvanishing ${\cal Z}^{IJ}_{ij+}$ should have co-dimension two, corresponding to 3-brane states of the M5-brane. In this case we need to consider states with finite  ${\cal P}_+$
per unit area and the quantum mechanical system should then be reduced to the vortex moduli space.

\subsection{Null Reduction as Infinite Boost}\label{s_Null_reduction_as_infinite_boost}

Finally it is instructive to see how the null reduction of the $(2,0)$ system above can be viewed as the limit of an infinite boost. This is in agreement with the general arguments for matrix models and light-cone quantisation given in \cite{Seiberg:1997ad}. In particular let us return to the general discussion for arbitrary $C^\mu$ and set
$$
C^\mu =  \frac{g^2}{\sqrt{1+\beta^2}}(\beta\delta^\mu_0+\delta^\mu_5)\, ,
$$
where $\beta$ is real. For any $|\beta|<1$, $C^\mu$ is space-like and after a suitable Lorentz transformation could be taken to simply be $C^\mu = g^2\delta^\mu_5$ and one reproduces five-dimensional maximally supersymmetric Yang-Mills. Taking $\beta\to \pm1$ corresponds to an infinite boost of the system along $x^5$ and leads to the null reduction we have discussed.

Let us see how this works in the $(2,0)$ system. We introduce coordinates
\begin{align}
u = \frac{x^0-\beta x^5}{\sqrt{1+\beta^2}} \, , \qquad \qquad v=\frac{x^5+\beta x^0}{\sqrt{1+\beta^2}} \, ,
\end{align}
so that $C^\mu=g^2\delta^\mu_v$ (again we are cavalier about the 3-algebra indices for the sake of clarity).
We now find that if we let
\begin{align}
F_{ij} &= -g^2 H_{ijv} \, , \\[10pt]
F_{iu} &= -g^2 H_{iuv} \, , \\[10pt]
G_{ij} &= -g^2 H_{iju} \, ,
\end{align}
then self-duality of $H$ implies that $H_{ijk} = g^{-2} \varepsilon_{ijkl} F^l{}_u$ and also:
\begin{align}
\frac{1}{2}\varepsilon_{ijkl} F^{kl} =\frac{2\beta}{1+\beta^2} F_{ij}+ \frac{1-\beta^2}{1+\beta^2} G_{ij}\, .
\end{align}
In the limit that $\beta = 1-\varepsilon$ with $\epsilon <<1$ we see that
\begin{align}
\frac{1}{2}\varepsilon_{ijkl} F^{kl} =F_{ij}+\varepsilon G_{ij}+{\cal O}(\varepsilon^2)\, ,
\end{align}
and therefore the non-self-dual part of $F_{ij}$ is boosted away. However for any $\beta \ne \pm 1$ the gauge fields are not required to be self-dual and the reduction to quantum mechanics that we found above will not occur.

\subsection{Conclusions}\label{s_Light_cone_conc}

In this chapter we have constructed the conserved energy-momentum tensor and supercurrent for the $(2,0)$ system we reviewed in chapter \ref{c_Multiple_M5s_and_3_algebras}.
We then considered in detail the case of a null reduction to a novel dynamical system with 16 supersymmetries and an $SO(5)$ R-symmetry in one null and four space dimensions. In particular we showed how the classical equations can be reduced to motion on the instanton moduli space. This allows us to quantise the system. In so doing we obtained the light-cone quantisation proposal of \cite{Aharony:1997th,Aharony:1997an}, generalised to include a potential that arises when the scalars (or gauge field $A_-$) have a nonvanishing vacuum expectation value, corresponding to the Coulomb branch where the M5-branes are separated. We were also able to obtain expressions for the six-dimensional supersymmetry and Poincar\'e algebras in terms of ADHM data of the instanton moduli space. This clarifies the relation of the quantum mechanical system to the full six-dimensional one.

In our opinion the work in this chapter presents evidence that the $(2,0)$ system of \cite{Lambert:2010wm} represents a complete Lorentz covariant picture of the M5-brane on a six-dimensional spacetime of the form ${\cal M}\times S^1$. In particular it is capable of including and interpolating between two conjectures on the dynamics of M5-branes: namely the recent suggestions that the $(2,0)$ theory on a space-like circle is precisely five-dimensional maximally supersymmetric Yang-Mills \cite{Douglas:2010iu,Lambert:2010iw} and also the older light-cone proposal of  \cite{Aharony:1997th}. In particular the latter can now be seen to arise as a space-like boost of the former in accordance with the general prescription of \cite{Seiberg:1997ad}. Nevertheless it remains to be seen if these conjectures can be made to lead to a more robust and complete description of the $(2,0)$ theory and hence the M5-brane, particularly on uncompactified spacetimes.


\newpage
\section{Summary and Further Work} \label{c_Conclusions}

\subsection{Thesis Summary}

Let us summarise what we have done in this thesis. We began in chapter \ref{c_Multiple_M2s_and_3_algebras} by describing the expected properties of the worldvolume theory of multiple M2-branes. This was followed by a review of the construction of the Bagger-Lambert-Gustavsson model which is superficially the long sought for theory of arbitrary multiple M2-branes. The key to this construction was the introduction of a novel algebraic structure called a real 3-algebra. The unique gauge algebra of the Euclidean BLG theory restricts it to describing at most two M2-branes and we provided details of this interpretation by examining its moduli space and outlining the `novel Higgs mechanism'. We then mentioned work to circumvent the uniqueness theorem for the $\mathcal{A}_4$ 3-algebra by relaxing the assumption of a positive definite 3-algebra metric and noted that the resulting ghost free theories seem to describe D2-branes or other D$p$-branes wrapped on tori. After this we gave some details on how the Euclidean BLG can be rewritten as an ordinary gauge theory albeit with matter fields in the bifundamental representation of $SU(2) \times SU(2)$. We then introduced the ABJM theory which generalises the $SU(2) \times SU(2)$ model to gauge groups of the form $U(N) \times U(N)$ and describes $N$ M2-branes in a $\mathbb{C}^4/\mathbb{Z}_k$ orbifold but has only manifest $\mathcal{N}=6$ supersymmetry. In the final part of the chapter we discussed complex 3-algebras and ultimately showed that for a particular choice of structure constants the ABJM theory is recovered.

The $\mathcal{N}=6,8$ supersymmetric 3-algebra Lagrangians detailed in chapter \ref{c_Multiple_M2s_and_3_algebras} are only leading order terms. The full actions involve an infinite expansion in powers of $1/T_{M2}$ and also include the coupling to the M-theory background 3-form gauge field. These full actions are expected to remain supersymmetric (with $\mathcal{N}=6,8$ as appropriate) and consequently the lowest order supersymmetry transformations must also be modified. We have extended the Lagrangians of chapter \ref{c_Multiple_M2s_and_3_algebras} in the following ways.

In chapter \ref{c_M2s_and_background_fields} we were able to construct the coupling, in the infinite tension limit and at linear order, of both the $\mathcal{N}=8$ and $\mathcal{N}=6$ Lagrangians to the background 3-form gauge field under the assumption that its 4-form field strength was constant. These coupling terms are gauge invariant and we showed explicitly that they were invariant under suitably modified supersymmetry transformations. We were also able to understand the origin of the mass terms on the worldvolume in terms of the back-reaction of the geometry.  Subsequent work by other groups extending our results beyond linear order and for non-constant field strength can be found in \cite{Kim:2009nc,Kim:2010hj,Allen:2011pm}.

In chapter \ref{c_Higher_Derivative_BLG} we tackled the next order Lagrangian and supersymmetry transformations for the Euclidean $\mathcal{N}=8$ theory. We started by positing an ansatz for both the $1/T_{M2}$ corrections to the Lagrangian and supersymmetry transformations (both to lowest order in fermions). The Lagrangian was then varied and we showed that it was invariant if the coefficients in our ansatz satisfied certain relations. We were able to solve the simultaneous equations relating the coefficients up to an overall parameter. By looking to the Abelian DBI action for a single M2-brane the overall parameter was itself fixed. As a consistency check we computed the commutator of two supersymmetries on the scalar and gauge fields (again to lowest order in fermions). Pleasingly, the resulting simultaneous equations are solved by the numerical values we identified in the invariance calculations and consequently the superalgebra closes. We have provided the full backing calculations, which were not included in \cite{Richmond:2012by}, in appendix \ref{a_HDer_invariance} and \ref{a_HDer_closure}.

By M-theory/IIA string theory duality, circle compactification of a direction transverse to the worldvolume of $N$ M2-branes should lead to $N$ D2-branes in ten dimensions. One method of constructing a circle of radius $R$ is to take the infinite real line and quotient by a discrete symmetry group generated by the integers i.e.\ $S^1 \cong \mathbb{R}/(2\pi R \mathbb{Z})$. In chapter \ref{c_Periodic_arrays_of_M2s} we considered such a compactification by taking an infinite array of M2-branes in the ABJM model and imposing translational invariance on all the fields except for the imaginary component of one of the scalars. We then evaluated the effect on the ABJM Lagrangian of imposing translational invariance.  We found that the resulting theory incorporates three-dimensional maximally supersymmetric Yang-Mills but is slightly more general as it allows for the appearance of nondynamical moduli. We also considered a cubic array of M2-branes and found a non-Lorentz invariant version of five-dimensional maximally supersymmetric Yang-Mills.

Our focus switched from M2-branes to M5-branes in chapter \ref{c_Multiple_M5s_and_3_algebras}. Here we reviewed the field content of the worldvolume theory of a single M5-brane. We built on this by explaining the expected properties of a theory of multiple M5-branes and some of the obstacles in constructing such a theory. The focus of the early part of this thesis means we are primarily interested in a 3-algebra formulation of the non-Abelian $(2,0)$ theory which describes the dynamics of M5-branes. Such a formulation is known in the literature and we outlined its construction. We saw that the 3-algebra $(2,0)$ theory involves an auxiliary vector field whose equation of motion restricts the nontrivial dynamics of the system to be five-dimensional. We also showed that five-dimensional supersymmetric Yang-Mills could be found from this non-Abelian 3-algebra system by choosing a space-like vacuum expectation value for the auxiliary field.

In the penultimate chapter we constructed the conserved energy-momentum tensor and supercurrent associated with the 3-algebra formulation of the $(2,0)$ system. Both of these quantities are genuinely six-dimensional. We then investigated the $(2,0)$ equations of motion when the auxiliary vector field was given a null vacuum expectation value. Importantly, in light-cone coordinates the scalar potential vanishes and the equation of motion for the self-dual 3-form identifies a 2-form gauge field strength which is self-dual in Euclidean space. This allows us to solve all but one (the time evolution equation) of the null reduced $(2,0)$ equations of motion in terms of ADHM data. In light-cone coordinates the momentum parallel to the null auxiliary field is quantised and this is interpreted as compactification of the M5-branes on a null circle. We were also able to quantise the null reduced system and connect with much earlier work involving the light-cone quantisation proposal of the $(2,0)$ theory.  The work in chapter \ref{c_Light_cone_M5s} together with the reduction to five-dimensional supersymmetric Yang-Mills in chapter \ref{c_Multiple_M5s_and_3_algebras} suggests that the 3-algebra formulation of the non-Abelian $(2,0)$ system describes multiple M5-branes whose worldvolume is of the form $\mathcal{M} \times S^1$.

The original construction of the BLG model is a wonderful example of the power of supersymmetry. Armed with only a few basic properties of the field content and a willingness to not be limited by any preconceived notion of gauge symmetry, the requirements of supersymmetry determine the structure of the theory. The most striking aspect of the BLG theory is the appearance of a 3-algebra. That they can also appear in the description of M5-branes suggests a deeper role in M-theory. However, this is tempered by the fact that there exist dual formulations of the three-dimensional 3-algebra theories in terms of bifundamental gauge theories. It is therefore debatable whether 3-algebras are fundamental in M-theory or not. Whatever the outcome of that debate, it is the view of this author that the 3-algebra formulations have a certain beauty to them and it would be disappointing if 3-algebras did not play a wider role in M-theory. 


\subsection{Further Work}

Our original research in this thesis has hopefully made a modest but useful contribution to the collective body of M-theory knowledge. To finish, we offer some suggestions for extending our work.
%
%
In establishing the higher derivative results in chapter \ref{c_Higher_Derivative_BLG} we have made use of the identity $f^{[abcd} f^{e] fgh} =0$, which is trivially satisfied by the structure constants of the $\mathcal{A}_4$ 3-algebra as $f^{abcd} \propto \epsilon^{abcd}$ and $a,b,c,d \in \{1,2,3,4\}$. However, the Lorentzian and other non-Euclidean 3-algebras of \cite{Gomis:2008uv,Benvenuti:2008bt,Ho:2008ei,deMedeiros:2008bf,Ho:2009nk,deMedeiros:2009hf} do not necessarily satisfy this identity and it is clear that our results do not hold for these wider classes of theories. Therefore, to extend our results to the non-Euclidean BLG theories we must abandon use of the identities which follow from $f^{[abcd} f^{e] fgh} =0$ i.e.\ Eqs.\,(\ref{Useful_Id}), \eqref{Useful_Id2}, \eqref{Useful_Id_5} and \eqref{Useful_Id_4}. Consequently, we should reinstate the $\bold{g}$, $\hat{\bold{m}}$ and $\hat{\bold{o}}$ terms in $\mathcal{L}_{1/T_{M2}}$ \eqref{L_Higher_Starting_Point} as well as adding terms to the order $1/T_{M2}$ supersymmetry transformations. The coefficients of the new terms would then be determined by repeating the analysis for the $\mathcal{A}_4$ case. 

Our higher derivative work is incomplete in the sense that we have only worked to lowest order in fermions.  The quartic fermion terms in the action, which coincide for the Lorentzian \cite{Alishahiha:2008rs} and Euclidean \cite{Ezhuthachan:2009sr} BLG theories, are known. Incorporating higher fermions in the supersymmetry transformations would, in principle, allow us to verify that the entire theory at $\mathcal{O} ( 1/T_{M2} )$ is maximally supersymmetric and additionally, to close the superalgebra on all the fields. To proceed at this level would require the addition of supersymmetry transformations of the form
\begin{align}
T_{M2} \, \delta' X =&+ ( \bar{\epsilon} \Gamma \psi ) ( \bar{\psi} \Gamma D \psi ) + ( \bar{\epsilon} \Gamma \psi ) \, \bar{\psi} \Gamma [ \psi , X , X ] \, , \label{X_susy_f} \\[10pt]
T_{M2} \, \delta' \tilde{A} =& + ( \bar{\epsilon} \Gamma \psi ) ( \bar{\psi} \Gamma D \psi ) X + ( \bar{\epsilon} \Gamma \psi ) \, \bar{\psi} \Gamma [ \psi , X , X ] X \, , \label{A_susy_f} \\[10pt]
\nonumber
T_{M2} \, \delta' \psi =&+ \Gamma \epsilon ( \bar{\psi} \Gamma D \psi ) DX + \Gamma \epsilon ( \bar{\psi} \Gamma D \psi ) [X,X,X] \\
&+ \Gamma \epsilon \, \bar{\psi} \Gamma [ \psi ,X,X] DX + \Gamma \epsilon \, \bar{\psi} \Gamma [ \psi ,X,X] [X,X,X] \, . \label{psi_susy_f}
\end{align}
The most general starting point would then involve taking all independent Lorentz invariant combinations. However, the presence of two sets of $\Gamma$-matrices in the supersymmetries allows for many ways of contracting Lorentz indices and also brings into play the transverse duality relation $\star \Gamma^{(n)} \propto \varepsilon_{(8)} \Gamma^{(8-n)}$. In addition, the cubic fermions in Eqs.\,(\ref{X_susy_f}) and (\ref{A_susy_f}) can be rearranged using the Fierz relation. The impact of these features is to obscure which terms are independent so that even the starting point is difficult to determine. Moreover, the subsequent invariance and closure calculations would involve heavy use of the Fierz rearrangement and whilst tractable would represent a formidable computational challenge.

%
With the exception of the Abelian $U(1) \times U(1)$ theory \cite{Sasaki:2009ij}, the order $1/T_{M2}$ higher derivative extension of the ABJM model has not been examined. Possible methods of approaching the ABJM higher derivative extension have been discussed in \cite{Ezhuthachan:2009sr} and \cite{Low:2010ie}. A separate brute force approach is simply to consider the most general action and supervariations which are consistent with all symmetries of the system and try to demonstrate invariance and closure as we have done for the $\mathcal{A}_4$ BLG theory. It is conceivable that the arbitrary coefficients in the $1/T_{M2}$ extension of ABJM can likewise be determined up to an overall scaling parameter. It would then remain to fix this scale parameter and there are at least two possible ways of doing this. First, we could directly compare against multiple D2-branes written in a suitable complex format by using the `novel Higgs mechanism' for ABJM \cite{Li:2008ya,Pang:2008hw} or perhaps by taking an infinite periodic array of M2-branes as we did in chapter \ref{c_Periodic_arrays_of_M2s}. Secondly we could rewrite the results of chapter \ref{c_Higher_Derivative_BLG} in complex $SU(2) \times SU(2)$ form \cite{VanRaamsdonk:2008ft} and exploit the equivalence, at levels $k=1,2$ of the $U(2) \times U(2)$ ABJM and $\mathcal{A}_4$ BLG theories. More generally it would be interesting to investigate the possibility of a 3-algebra DBI action for M2-branes. 

In chapter \ref{c_Multiple_M5s_and_3_algebras} we saw that multiple D4-branes could be recovered by choosing a space-like vacuum expectation value for the auxiliary field in the 3-algebra formulation of $(2,0)$ theory. Chapter \ref{c_Light_cone_M5s} was concerned with giving the auxiliary field a light-like vev. There is one remaining choice for $C^\mu_a$ which is to give it a time-like vev. The resulting system is very similar to five-dimensional maximally supersymmetric Yang-Mills but in Euclidean signature. There has recently been interest in placing Euclideanised supersymmetric Yang-Mills on compact manifolds, particularly spheres. After taking the time-like reduction of the $(2,0)$ system it may be possible to put the theory on a five-sphere and use localisation techniques to exactly compute the partition function. This would hopefully lead to an improved understanding of the relationship between the six-dimensional $(2,0)$ theory and five-dimensional supersymmetric Yang-Mills.

\begin{appendices}

\newpage
\section{Conventions and Useful Identities}\label{a_Conventions}


Throughout this thesis we have adopted the mostly plus convention for the Minkowski metric. We denote brane worldvolume indices by lower case Greek letters, $\mu , \nu , \ldots $ and directions transverse to the brane worldvolume by upper case Roman letters, $I, J , \ldots$. In eleven dimensions our Clifford algebra matrices are $\Gamma_m , \, m = 0,1, \ldots ,10$.  We always take the Clifford matrices to be real. 

The totally antisymmetric products of $k$ $\Gamma$-matrices have transposes given by;
\begin{align}
\left( \Gamma^{(m_k)} \right)^{\rm T} &= (-1)^{\frac{1}{2} k ( k + 1) } C \Gamma^{(m_k)} C^{-1} = 
\begin{cases} 
- C \Gamma^{(m_k)} C^{-1} \ \text{ if $k=1,2,5,6,9$ or 10,} \, , \\[10pt]
+ C \Gamma^{(m_k)} C^{-1} \ \text{ if $k=0,3,4,7,8$ or 11,} \label{Transpose}
\end{cases}
\end{align}
%
%
%
where $C=\Gamma_0$ is the antisymmetric charge conjugation matrix. We denote by $\Gamma^{(\mu_n)}$ and $\Gamma^{(I_m)}$ the totally antisymmetric product of $n$ worldvolume and $m$ transverse $\Gamma$-matrices respectively. Using the transpose property \eqref{Transpose} together with $\{ \Gamma^I , \Gamma^\mu \} =0$ we find for any two spinors $\chi$ and $\lambda$
%
%
\begin{equation}\label{Spinor_Flip}
\begin{aligned}
\bar{\chi} \Gamma^{(I_m)} \Gamma^{(J_n)} \Gamma^{(\mu_p)} \lambda =& (-1)^{\theta ( m,n,p )} \bar{\lambda} \Gamma^{(J_n)} \Gamma^{(I_m)} \Gamma^{(\mu_p)} \chi \, , \\[10pt]
\bar{\chi} \Gamma^{(\mu_m)} \Gamma^{(\nu_n)} \Gamma^{(I_p)} \lambda =& (-1)^{\theta ( m,n,p )} \bar{\lambda} \Gamma^{(\nu_n)} \Gamma^{(\mu_m)} \Gamma^{(I_p)} \chi \, , \\
\end{aligned}
\end{equation}
where $\bar{\chi} = \chi^{\rm T} C$ and $\theta ( m,n,p ) = p(m+n) + \frac{1}{2} m ( m + 1 ) + \frac{1}{2} n ( n + 1 ) + \frac{1}{2} p ( p + 1 )$. Hence for $\chi=\epsilon_2$ and $\lambda=\epsilon_1$
\begin{align}
\bar{\epsilon}_2 \Gamma^{(I_m)} \Gamma^{(J_n)} \Gamma^{(\mu_p)} \epsilon_1 - ( 1 \leftrightarrow 2 ) =
\begin{cases}
\bar{\epsilon}_2 \left[ \Gamma^{(I_m)} , \Gamma^{(J_n)} \right] \Gamma^{(\mu_p)} \epsilon_1 \ \text{ if $(-1)^{\theta ( m,n,p )}=+1$,} \\[10pt]
\bar{\epsilon}_2 \left\{ \Gamma^{(I_m)} , \Gamma^{(J_n)} \right\} \Gamma^{(\mu_p)} \epsilon_1 \ \text{ if $(-1)^{\theta ( m,n,p )}=-1$.} \label{Commutator_Relation}
\end{cases}
\end{align}
The Fierz identity in eleven dimensions is
\begin{align}
(\bar \epsilon \chi) \phi =- 2^{-[\frac{11}{2}]} \Big( +(\bar \epsilon \phi) \chi+(\bar \epsilon \Gamma_{m}\phi)\Gamma^{m} \chi-\frac{1}{2!}(\bar \epsilon \Gamma_{mn}\phi)\Gamma^{mn} \chi-\frac{1}{3!}(\bar \epsilon \Gamma_{mnp}\phi)\Gamma^{mnp} &\chi \cr
+\frac{1}{4!}(\bar \epsilon \Gamma_{mnpq}\phi)\Gamma^{mnpq} \chi+\frac{1}{5!}(\bar \epsilon \Gamma_{mnpqr}\phi)\Gamma^{mnpqr} &\chi\Big) \, ,
\end{align}
where $\left[ n \right]$ denotes the integer part of $n$. The Fierz relations for eleven-dimensional spinors living on the worldvolume of either M2- or M5-branes are found by splitting the indices into worldvolume and transverse indices as appropriate.

\subsection{M2-branes}

\subsubsection{$\mathcal{N}=8$}

The M2-brane worldvolume indices are $\mu, \nu , \ldots = 0, 1, 2$ and the transverse indices are $I,J, \ldots = 3, 4, \ldots, 10$. The unbroken supersymmetry parameters, which are 32-component Majorana spinors, satisfy the following chirality conditions
\begin{align}
\Gamma_{012} \epsilon =& + \epsilon \, , \\[10pt]
\Gamma_{012} \psi =& - \psi \, .
\end{align}
From the chirality conditions and the choices $\Gamma_{0123456789(10)} = +1$, $\varepsilon_{012} = - 1$, we deduce the following M2-brane $\Gamma$-matrix duality relations
\begin{equation}
\Gamma_{\mu \nu \lambda} \epsilon = - \varepsilon_{\mu \nu \lambda} \epsilon \, , \qquad \Gamma_{\nu \lambda} \epsilon = - \varepsilon_{\mu \nu \lambda} \Gamma^\mu \epsilon \, , \qquad \Gamma_{\lambda} \epsilon = + \frac{1}{2} \varepsilon_{\mu \nu \lambda} \Gamma^{\mu \nu} \epsilon \, , \label{M2_wvol_duality_e} 
\end{equation}
\begin{equation}
\Gamma_{\mu \nu \lambda} \psi = + \varepsilon_{\mu \nu \lambda} \psi \, , \qquad \Gamma_{\nu \lambda} \psi = + \varepsilon_{\mu \nu \lambda} \Gamma^\mu \psi \, , \qquad \Gamma_{\lambda} \psi = - \frac{1}{2} \varepsilon_{\mu \nu \lambda} \Gamma^{\mu \nu} \psi \, , \label{M2_wvol_duality_f} \\[10pt]
\end{equation}
\begin{equation}
\Gamma^{I_1 \ldots I_k} \epsilon = - \, \frac{(-1)^{\frac{1}{2}(8-k-1)(8-k)}}{(8-k)!} \varepsilon^{I_1 \ldots I_k J_{k+1} \ldots J_8} \Gamma_{J_{k+1} \ldots J_8} \epsilon \label{M2_trans_duality_e} \, , \\[10pt]
\end{equation}
\begin{equation}
\Gamma^{I_1 \ldots I_k} \psi = + \, \frac{(-1)^{\frac{1}{2}(8-k-1)(8-k)}}{(8-k)!} \varepsilon^{I_1 \ldots I_k J_{k+1} \ldots J_8} \Gamma_{J_{k+1} \ldots J_8} \psi \, , \label{M2_trans_duality_f}
\end{equation}
where $\varepsilon^{(8)}$ is the totally antisymmetric tensor in eight dimensions.

For the $\mathcal{N}=8$ calculations in chapter \ref{c_M2s_and_background_fields} we have used the following identities found in \cite{Bagger:2007jr}
\begin{align}
\Gamma_M\Gamma^{IJ}\Gamma^M =&+ 4\Gamma^{IJ} \, , \\[10pt]
\Gamma_M\Gamma^{IJKL}\Gamma^M =&0 \, , \\[10pt]
\Gamma^{IJP}\Gamma^{KLMN}\Gamma_P =&-\Gamma^I\Gamma^{KLMN}\Gamma^J +\Gamma^J\Gamma^{KLMN}\Gamma^I \, , \\[10pt]
\nonumber
\Gamma^I\Gamma^{KL}\Gamma^J - \Gamma^J\Gamma^{KL}\Gamma^I =&+ 2\Gamma^{KL}\Gamma^{IJ} - 2\Gamma^{KJ}\delta^{IL} + 2\Gamma^{KI}\delta^{JL}- 2\Gamma^{LI}\delta^{JK} \\
& + 2\Gamma^{LJ}\delta^{IK}- 4\delta^{KJ}\delta^{IL}+4\delta^{KI}\delta^{JL} \, , \\[10pt]
\nonumber
\Gamma^{IJM}\Gamma^{KL}\Gamma_M =&+2\Gamma^{KL}\Gamma^{IJ} - 6\Gamma^{KJ}\delta^{IL} + 6\Gamma^{KI}\delta^{JL}- 6\Gamma^{LI}\delta^{JK} \\
& + 6\Gamma^{LJ}\delta^{IK}+ 4\delta^{KJ}\delta^{IL}-4\delta^{KI}\delta^{JL} \, .
\end{align}
\\
\noindent We have also used the eleven-dimensional Fierz identity, which for three spinors of the same chirality with respect to $\Gamma_{012}$, is
\begin{equation}\label{M2_N8_fierz}
(\bar{\epsilon} \chi) \phi - (\bar{\phi} \chi) \epsilon =-\frac{1}{16}\left(2(\bar\epsilon \Gamma_\mu \phi)\Gamma^\mu \chi -(\bar\epsilon \Gamma_{IJ} \phi) \Gamma^{IJ} \chi +\frac{1}{4!}(\bar\epsilon \Gamma_{\mu} \Gamma_{IJKL} \phi) \Gamma^\mu \Gamma^{IJKL} \chi \right) \, .
\end{equation}

\subsubsection{$\mathcal{N}=6$}
The M2-brane worldvolume indices are $\mu, \nu , \ldots = 0, 1, 2$ and the transverse indices are $A,B, \ldots = 1,2,3,4$. For the ABJM model we use two-component Majorana spinors and a real three-dimensional Clifford algebra with matrices $\gamma_\mu$ such that $\gamma_{012}=1$. 

For the $\mathcal{N}=6$ calculations in chapter \ref{c_M2s_and_background_fields} we have used the following identities found in \cite{Bagger:2008se}
\begin{align}
\frac{1}{2}\bar\epsilon^{CD}_1\gamma_\nu\epsilon_{2CD}\,\delta^A_B =& +\bar\epsilon^{AC}_1\gamma_\nu\epsilon_{2BC} -\bar\epsilon^{AC}_2\gamma_\nu\epsilon_{1BC} \, , \\[10pt]
\nonumber
2\bar\epsilon^{AC}_1\epsilon_{2BD}-2\bar\epsilon^{AC}_2\epsilon_{1BD} =& +\bar\epsilon^{CE}_1\epsilon_{2DE}\delta^A_B-\bar\epsilon^{CE}_2\epsilon_{1DE}\delta^A_B\\
\nonumber
&-\bar\epsilon^{AE}_1\epsilon_{2DE}\delta^C_B+\bar\epsilon^{AE}_2\epsilon_{1DE}\delta^C_B\\
\nonumber
&+\bar\epsilon^{AE}_1\epsilon_{2BE}\delta^C_D-\bar\epsilon^{AE}_2\epsilon_{1BE}\delta^C_D\\
&-\bar\epsilon^{CE}_1\epsilon_{2BE}\delta^A_D+\bar\epsilon^{CE}_2\epsilon_{1BE}\delta^A_D \, , \\[10pt]
\nonumber 
\frac{1}{2}\varepsilon_{ABCD} \, \bar\epsilon^{EF}_1\gamma_\mu\epsilon_{2EF}=&+\bar\epsilon_{1AB}\gamma_\mu\epsilon_{2CD}-\bar\epsilon_{2AB}\gamma_\mu\epsilon_{1CD}\\
\nonumber
&+\bar\epsilon_{1AD}\gamma_\mu\epsilon_{2BC}-\bar\epsilon_{2AD}\gamma_\mu\epsilon_{1BC}\\
&-\bar\epsilon_{1BD}\gamma_\mu\epsilon_{2AC}+\bar\epsilon_{2BD}\gamma_\mu\epsilon_{1AC} \, .
\end{align}
We have also used the three-dimensional Fierz identity, which for three spinors of the same chirality with respect to $\gamma_{012}$, is
\begin{equation}\label{M2_N6_fierz} 
(\bar\lambda\chi)\psi = -\frac{1}{2}(\bar\lambda\psi)\chi -\frac{1}{2} (\bar\lambda\gamma_\nu\psi)\gamma^\nu\chi \, .
\end{equation}

\subsection{M5-branes}
The M5-brane worldvolume indices are $\mu, \nu , \ldots = 0, 1, \ldots , 5$ and the transverse indices are $I,J, \ldots = 6,7, \ldots, 10$. The unbroken supersymmetry parameters, which are 32-component Majorana spinors, satisfy the following chirality conditions
\begin{align}
\Gamma_{012345} \epsilon =& + \epsilon \, , \\[10pt]
\Gamma_{012345} \psi =& - \psi \, .
\end{align}
From the chirality conditions and the choices $\Gamma_{0123456789(10)} = +1$, $\varepsilon_{012345} = - 1$, we deduce the following M5-brane $\Gamma$-matrix duality relations
\begin{equation}
\Gamma^{\mu_1 \ldots \mu_k} \epsilon = + \, \frac{ (-1)^{ \left[ \frac{k}{2} \right] } }{ ( 6 - k )! } \varepsilon^{\mu_1 \ldots \mu_k \nu_1 \ldots \nu_{(6-k)}} \Gamma_{\nu_1 \ldots \nu_{(6-k)}} \epsilon \, , \label{M5_wvol_duality_e}
\end{equation}
\begin{equation}
\Gamma^{\mu_1 \ldots \mu_k} \psi = - \, \frac{ (-1)^{ \left[ \frac{k}{2} \right] } }{ ( 6 - k )! } \varepsilon^{\mu_1 \ldots \mu_k \nu_1 \ldots \nu_{(6-k)}} \Gamma_{\nu_1 \ldots \nu_{(6-k)}} \psi \, , \label{M5_wvol_duality_f}
\end{equation}
\begin{equation}
\Gamma^{I_1 \ldots I_l} \epsilon = + \, \frac{ (-1)^{ \left[ \frac{l}{2} \right] } }{(5-l)!} \varepsilon^{I_1 \ldots I_l J_1 \ldots J_{(5-l)}} \Gamma_{J_1 \ldots J_{(5-l)}} \epsilon \, , \label{M5_trans_duality_e}
\end{equation}
\begin{equation}
\Gamma^{I_1 \ldots I_l} \psi = - \, \frac{ (-1)^{ \left[ \frac{l}{2} \right] } }{(5-l)!} \varepsilon^{I_1 \ldots I_l J_1 \ldots J_{(5-l)}} \Gamma_{J_1 \ldots J_{(5-l)}} \psi \, , \label{M5_trans_duality_f}
\end{equation}
where $\varepsilon^{(5)}$ and $\varepsilon^{(6)}$ are the totally antisymmetric tensors in five and six dimensions. For the calculations in chapter \ref{c_Light_cone_M5s} the following identities are useful 
\begin{align}
\Gamma^K \Gamma^K &= +5 \, , \\[10pt]
\Gamma^K \Gamma^I \Gamma^K &= -3 \Gamma^I \, , \\[10pt]
\Gamma^K \Gamma^{IJ} \Gamma^K &= +\Gamma^{IJ} \, .
\end{align}
We will also need the Fierz identity, which for two spinors $\epsilon$, $\phi$ of the same chirality with respect to $\Gamma_{012345}$ and a spinor $\chi$ of opposite chirality, is
\begin{align}\label{M5_fierz}
&(\bar{\epsilon} \chi) \phi - (\bar{\phi} \chi) \epsilon = -\frac{1}{16} \bigg( 2 ( \bar{\epsilon} \Gamma^\mu \phi ) \Gamma_\mu \chi - 2 ( \bar{\epsilon} \Gamma^\mu \Gamma^I \phi ) \Gamma_\mu \Gamma^I \chi + \frac{1}{3!} \frac{1}{2!} ( \bar{\epsilon} \Gamma^{\mu \nu \lambda} \Gamma^{IJ} \phi ) \Gamma_{\mu \nu \lambda} \Gamma^{IJ} \chi \bigg) \, .
\end{align}

\newpage
\section{Higher Derivative BLG Invariance Calculations}\label{a_HDer_invariance}


In this appendix we show in detail the calculations behind the invariance of the higher derivative corrected Lagrangian of chapter \ref{c_Higher_Derivative_BLG}. The variation of the $1/T_{M2}$ corrected Lagrangian is 
\begin{equation}
\tilde{\delta} \mathcal{L} = \delta' \mathcal{L}_{BLG} + \delta \mathcal{L}_{1/T_{M2}} \, ,
\end{equation}
where (setting $T_{M2}=1$ for convenience) 
\begin{align}
\nonumber
\delta' \mathcal{L}_{BLG} = {\rm Tr} \Big\{ &- D_\mu \big( \delta'_{2 DX} X^I + \delta'_{1 DX} X^I + \delta'_{0 DX} X^I \big) D^\mu X^I \\
\nonumber
&+ \big( \delta'_{2 DX} \tilde{A}_\mu + \delta'_{1 DX} \tilde{A}_\mu + \delta'_{0 DX} \tilde{A}_\mu \big) X^I D^\mu X^I \\
\nonumber
&- \sfrac{1}{2} X^{IJK} [ ( \delta'_{2 DX} X^I + \delta'_{1 DX} X^I + \delta'_{0 DX} X^I ) , X^J , X^K ] \\
\nonumber
&+ \sfrac{i}{2} \big( \delta'_{3 DX} \bar{\psi} + \delta'_{2 DX} \bar{\psi} + \delta'_{1 DX} \bar{\psi} + \delta'_{0 DX} \bar{\psi} \big)\Gamma^\mu D_\mu \psi \\
\nonumber
&+ \sfrac{i}{2} \bar{\psi} \Gamma^\mu D_\mu \big( \delta'_{3 DX} \psi + \delta'_{2 DX} \psi + \delta'_{1 DX} \psi + \delta'_{0 DX} \psi \big) \\
\nonumber
&+ \sfrac{i}{4} \big( \delta'_{3 DX} \bar{\psi} + \delta'_{2 DX} \bar{\psi} + \delta'_{1 DX} \bar{\psi} + \delta'_{0 DX} \bar{\psi} \big) \Gamma^{IJ} [ X^I , X^J , \psi ] \\
\nonumber
&+ \sfrac{i}{4} \bar{\psi} \Gamma^{IJ} [ X^I , X^J , ( \delta'_{3 DX} \psi + \delta'_{2 DX} \psi + \delta'_{1 DX} \psi + \delta'_{0 DX} \psi ) ] \\
\nonumber
&+ \sfrac{1}{2} \varepsilon^{\mu\nu\lambda} F_{\nu \lambda} \, \big( \delta'_{2 DX} \tilde{A}_\mu + \delta'_{1 DX} \tilde{A}_\mu + \delta'_{0 DX} \tilde{A}_\mu \big) \Big\} \\
&+ \mathcal{O} ( \psi^3 ) \, , \label{Varied_L_BLG}
\end{align}
\begin{align}
\nonumber
\delta \mathcal{L}_{T^{-1}_{M2}} = {\rm STr} \Big\{ &+ 4 \bold{a} \, D^\mu ( \delta X^I ) D_\mu X^J  D^\nu X^J D_\nu X^I - 4 \bold{a} \, ( \delta \tilde{A}^\mu X^I ) D_\mu X^J  D^\nu X^J D_\nu X^I \\
\nonumber
&+ 4 \bold{b} \, D^\mu ( \delta X^I ) D_\mu X^I D^\nu X^J D_\nu X^J - 4 \bold{b} \, ( \delta \tilde{A}^\mu X^I ) D_\mu X^I D^\nu X^J D_\nu X^J \\
\nonumber
&+ 3 \bold{c} \, \varepsilon^{\mu\nu\lambda}\, [ \delta X^I , X^J , X^K ] D_\mu X^I D_\nu X^J D_\lambda X^K \\
\nonumber
&+ 3 \bold{c} \, \varepsilon^{\mu\nu\lambda}\, X^{IJK} D_\mu ( \delta X^I ) D_\nu X^J D_\lambda X^K - 3 \bold{c} \, \varepsilon^{\mu\nu\lambda}\, X^{IJK} ( \delta \tilde{A}_\mu X^I ) D_\nu X^J D_\lambda X^K \\
\nonumber
&+ 4 \bold{d} \, [ \delta X^I , X^J , X^K ] X^{IJL} D^\mu X^K D_\mu X^L + 2 \bold{d} \, [ X^I , X^J , \delta X^K ] X^{IJL} D^\mu X^K D_\mu X^L \\
\nonumber
&+ 2 \bold{d} \, X^{IJK}X^{IJL} D^\mu ( \delta X^K ) D_\mu X^L - 2 \bold{d} \, X^{IJK}X^{IJL} ( \delta \tilde{A}^\mu X^K ) D_\mu X^L \\
\nonumber
&+ 6 \bold{e} \, [ \delta X^I , X^J , X^K ] X^{IJK} D^\mu X^L  D_\mu X^L \\
\nonumber
&+ 2 \bold{e} \, X^{IJK} X^{IJK} D^\mu ( \delta X^L ) D_\mu X^L - 2 \bold{e} \, X^{IJK} X^{IJK} ( \delta \tilde{A}^\mu X^L ) D_\mu X^L \\
\nonumber
&+ 12 \bold{f} \, [ \delta X^I , X^J ,X^K ] X^{IJK} X^{LMN}X^{LMN} \\
\nonumber
&+ i \hat{\bold{d}} \delta_{0 DX} \bar{\psi} \Gamma^\mu \Gamma^{IJ} D_\nu\psi  D_\mu X^{I}  D^\nu X^J + i \hat{\bold{d}} \bar{\psi} \Gamma^\mu \Gamma^{IJ} D_\nu ( \delta_{0 DX} \psi )  D_\mu X^{I}  D^\nu X^J \\
\nonumber
&+ i \hat{\bold{d}} \delta_{1 DX} \bar{\psi} \Gamma^\mu \Gamma^{IJ} D_\nu\psi  D_\mu X^{I}  D^\nu X^J + i \hat{\bold{d}} \bar{\psi} \Gamma^\mu \Gamma^{IJ} D_\nu ( \delta_{1 DX}  \psi )  D_\mu X^{I}  D^\nu X^J \\
\nonumber
&+ i \hat{\bold{e}} \delta_{0 DX} \bar{\psi}\Gamma^\mu D_\nu\psi  D_\mu X^{I}  D^\nu X^I + i \hat{\bold{e}} \bar{\psi}\Gamma^\mu D_\nu ( \delta_{0 DX} \psi )  D_\mu X^{I}  D^\nu X^I \\
\nonumber
&+ i \hat{\bold{e}} \delta_{1 DX}  \bar{\psi}\Gamma^\mu D_\nu\psi  D_\mu X^{I}  D^\nu X^I + i \hat{\bold{e}} \bar{\psi}\Gamma^\mu D_\nu ( \delta_{1 DX}  \psi )  D_\mu X^{I}  D^\nu X^I \\
\nonumber
&+ i \hat{\bold{f}} \delta_{0 DX} \bar{\psi}\Gamma^{IJKL} D_\nu\psi X^{IJK}  D^\nu X^L + i \hat{\bold{f}} \bar{\psi}\Gamma^{IJKL} D_\nu ( \delta_{0 DX} \psi ) X^{IJK}   D^\nu X^L \\
\nonumber
&+ i \hat{\bold{f}} \delta_{1 DX} \bar{\psi}\Gamma^{IJKL} D_\nu\psi X^{IJK}  D^\nu X^L + i \hat{\bold{f}} \bar{\psi}\Gamma^{IJKL} D_\nu ( \delta_{1 DX} \psi ) X^{IJK}   D^\nu X^L \\
\nonumber
&+ i \hat{\bold{g}} \delta_{0 DX} \bar{\psi}\Gamma^{IJ} D_\nu\psi X^{IJK} D^\nu X^K + i \hat{\bold{g}} \bar{\psi}\Gamma^{IJ} D_\nu ( \delta_{0 DX} \psi ) X^{IJK} D^\nu X^K \\
\nonumber
&+ i \hat{\bold{g}} \delta_{1 DX} \bar{\psi}\Gamma^{IJ} D_\nu\psi X^{IJK} D^\nu X^K + i \hat{\bold{g}} \bar{\psi}\Gamma^{IJ} D_\nu ( \delta_{1 DX} \psi ) X^{IJK} D^\nu X^K \\
\nonumber
&+ i \hat{\bold{h}} \delta_{0 DX} \bar{\psi}\Gamma^{IJ}[X^J,X^{K},\psi]  D^\mu X^{I}  D_\mu X^K + i \hat{\bold{h}} \bar{\psi}\Gamma^{IJ}[X^J,X^{K}, \delta_{0 DX} \psi]  D^\mu X^{I}  D_\mu X^K \\
\nonumber
&+ i \hat{\bold{h}} \delta_{1 DX} \bar{\psi}\Gamma^{IJ}[X^J,X^{K},\psi]  D^\mu X^{I}  D_\mu X^K + i \hat{\bold{h}} \bar{\psi}\Gamma^{IJ}[X^J,X^{K}, \delta_{1 DX} \psi]  D^\mu X^{I}  D_\mu X^K \\
\nonumber
&+ i \hat{\bold{i}} \delta_{0 DX} \bar{\psi}\Gamma^{\mu\nu}[X^I,X^{J},\psi]  D_\mu X^{I} D_\nu X^J + i \hat{\bold{i}} \bar{\psi}\Gamma^{\mu\nu}[X^I,X^{J}, \delta_{0 DX} \psi]  D_\mu X^{I} D_\nu X^J \\
\nonumber
&+ i \hat{\bold{i}} \delta_{1 DX} \bar{\psi}\Gamma^{\mu\nu}[X^I,X^{J},\psi]  D_\mu X^{I} D_\nu X^J + i \hat{\bold{i}} \bar{\psi}\Gamma^{\mu\nu}[X^I,X^{J}, \delta_{1 DX} \psi]  D_\mu X^{I} D_\nu X^J \\
\nonumber
&+ i \hat{\bold{j}} \delta_{0 DX} \bar{\psi}\Gamma^{\mu\nu} \Gamma^{IJ} [X^J,X^{K},\psi]  D_\mu X^{I} D_\nu X^K + i \hat{\bold{j}} \bar{\psi}\Gamma^{\mu\nu} \Gamma^{IJ} [X^J,X^{K},\delta_{0 DX} \psi]  D_\mu X^{I} D_\nu X^K \\
\nonumber
&+ i \hat{\bold{j}} \delta_{1 DX} \bar{\psi}\Gamma^{\mu\nu} \Gamma^{IJ} [X^J,X^{K},\psi]  D_\mu X^{I} D_\nu X^K + i \hat{\bold{j}} \bar{\psi}\Gamma^{\mu\nu} \Gamma^{IJ} [X^J,X^{K},\delta_{1 DX} \psi]  D_\mu X^I D_\nu X^K \\
\nonumber
&+ i \hat{\bold{k}}  \delta_{0 DX} \bar{\psi}\Gamma^\mu\Gamma^{IJ}[X^K,X^{L},\psi]  D_\mu X^{I}X^{JKL} + i \hat{\bold{k}}  \bar{\psi}\Gamma^\mu\Gamma^{IJ}[X^K,X^{L},\delta_{0 DX} \psi] D_\mu X^{I}X^{JKL} \\
\nonumber
&+ i \hat{\bold{k}}  \delta_{1 DX} \bar{\psi}\Gamma^\mu\Gamma^{IJ}[X^K,X^{L},\psi] D_\mu X^{I}X^{JKL} + i \hat{\bold{k}}  \bar{\psi}\Gamma^\mu\Gamma^{IJ}[X^K,X^{L},\delta_{1 DX} \psi] D_\mu X^{I}X^{JKL} \\
\nonumber
&+ i \hat{\bold{l}}  \delta_{0 DX} \bar{\psi}\Gamma^\mu[X^I,X^{J},\psi] D_\mu X^{K}X^{IJK} + i \hat{\bold{l}}  \bar{\psi}\Gamma^\mu[X^I,X^{J},\delta_{0 DX} \psi] D_\mu X^{K}X^{IJK} \\
\nonumber
&+ i \hat{\bold{l}}  \delta_{1 DX} \bar{\psi}\Gamma^\mu[X^I,X^{J},\psi] D_\mu X^{K}X^{IJK} + i \hat{\bold{l}}  \bar{\psi}\Gamma^\mu[X^I,X^{J},\delta_{1 DX} \psi] D_\mu X^{K}X^{IJK} \\
\nonumber
&+ i \hat{\bold{n}}  \delta_{0 DX} \bar{\psi}\Gamma^\mu\Gamma^{IJ}[X^K,X^{L},\psi] X^{IJK} D_\mu X^L + i \hat{\bold{n}}  \bar{\psi}\Gamma^\mu\Gamma^{IJ}[X^K,X^{L},\delta_{0 DX} \psi] X^{IJK} D_\mu X^L \\
\nonumber
&+ i \hat{\bold{n}}  \delta_{1 DX} \bar{\psi}\Gamma^\mu\Gamma^{IJ}[X^K,X^{L},\psi] X^{IJK} D_\mu X^L + i \hat{\bold{n}}  \bar{\psi}\Gamma^\mu\Gamma^{IJ}[X^K,X^{L},\delta_{1 DX} \psi] X^{IJK} D_\mu X^L \\
\nonumber
&+ i \hat{\bold{p}}  \delta_{0DX} \bar{\psi}\Gamma^{IJ}[X^K,X^{L},\psi] X^{IJM}X^{KLM} + i \hat{\bold{p}}  \bar{\psi}\Gamma^{IJ}[X^K,X^{L},\delta_{0DX} \psi] X^{IJM}X^{KLM} \\
\nonumber
&+ i \hat{\bold{p}}  \delta_{1DX} \bar{\psi}\Gamma^{IJ}[X^K,X^{L},\psi] X^{IJM}X^{KLM} + i \hat{\bold{p}}  \bar{\psi}\Gamma^{IJ}[X^K,X^{L},\delta_{1DX} \psi] X^{IJM}X^{KLM} \Big\} \\
&+ \mathcal{O} ( \psi^3 ) \, , \label{Varied_L_Higher}
\end{align}
and the variations $\delta$ and $\delta'$ have been given in Eqs.\,\eqref{BLG_susy_s}\,-\,\eqref{BLG_susy_f} and \eqref{S1}\,-\,\eqref{G3} in the main body of this thesis. From $\delta' \mathcal{L}_{BLG}$ and $\delta \mathcal{L}_{1/T_{M2}}$ we can pick terms according to the number of derivatives they contain. We will consider these terms separately and in decreasing order of derivatives in the following sections.

After selecting the required terms we perform the following manipulations which are common to each section. Into the chosen terms we insert the appropriate supersymmetry transformations and use the spinor flip condition \eqref{Spinor_Flip} to place the supersymmetry parameter $\epsilon$ to the left of the fermions $\psi$ and $D \psi$. We then commute the worldvolume $\Gamma$-matrices through the transverse ones so that they lie next to the fermions. The next step is to dualise the worldvolume $\Gamma$-matrices using \eqref{M2_wvol_duality_f} whenever they are accompanied by $\varepsilon_{\mu \nu \lambda}$. We are then in a position to expand out all $\Gamma$-matrices using the Clifford algebra. The matrix algebra, which we perform separately in each subsection below, has been aided by use the Cadabra symbolic computer algebra package \cite{Peeters:2006kp,Peeters:2007wn}.

\subsection{Four Derivative Terms - $\tilde{\delta} \mathcal{L}_4$}

These terms have been covered in chapter \ref{c_Higher_Derivative_BLG}.

\subsection{Three Derivative Terms - $\tilde{\delta} \mathcal{L}_3$}

The terms from \eqref{Varied_L_BLG} and \eqref{Varied_L_Higher} which contain a total of three derivatives are
\begin{align}
\nonumber
{\rm STr} \Big\{ &- i s_5 \bar{\epsilon} \Gamma^{IJKLM} \Gamma^\mu D^\nu ( \psi D_\mu X^J X^{KLM} ) D_\nu X^I - i s_6 \bar{\epsilon} \Gamma^{KLM} \Gamma^\mu D^\nu ( \psi D_\mu X^I X^{KLM} ) D_\nu X^I \\
\nonumber
&- i s_7 \bar{\epsilon} \Gamma^{JLM} \Gamma^\mu D^\nu ( \psi D_\mu X^J X^{ILM} ) D_\nu X^I - i s_8 \bar{\epsilon} \Gamma^{ILM} \Gamma^\mu D^\nu ( \psi D_\mu X^J X^{JLM} ) D_\nu X^I \\
\nonumber
&- i s_9 \bar{\epsilon} \Gamma^{M} \Gamma^\mu D^\nu ( \psi D_\mu X^J X^{IJM} ) D_\nu X^I \\
\nonumber
&- i g_1 \bar{\epsilon} \Gamma^J \Gamma^\mu \psi [ X^J ,  X^I , D^\mu X^I ] D_\nu X^K D^\nu X^K - i g_2 \bar{\epsilon} \Gamma^J \Gamma^\nu \psi [ X^J , X^I , D^\mu X^I ] D_\mu X^K D_\nu X^K \\
\nonumber
&- i g_3 \bar{\epsilon} \Gamma^J \Gamma^\nu \psi [ X^K , X^I , D^\mu X^I ] D_\mu X^J D_\nu X^K - i g_4 \bar{\epsilon} \Gamma^J \Gamma^\nu \psi [ X^K , X^I , D^\mu X^I ] D_\mu X^K D_\nu X^J \\
\nonumber
&- i g_5 \bar{\epsilon} \Gamma^J \Gamma^\mu \psi [ X^K , X^I , D_\mu X^I ] D_\nu X^J D^\nu X^K - i g_6 \bar{\epsilon} \Gamma^J \Gamma^{\mu \nu \lambda} \psi [ X^K , X^I , D_\mu X^I ] D_\nu X^J D_\lambda X^K \\
\nonumber
&- i g_7 \bar{\epsilon} \Gamma^{JKL} \Gamma^{\mu \nu \lambda} \psi [ X^J , X^I , D_\mu X^I ] D_\nu X^K D_\lambda X^L - i g_8 \bar{\epsilon} \Gamma^{JKL} \Gamma^\nu \psi [ X^J , X^I , D^\mu X^I ] D_\mu X^K D_\nu X^L \\
\nonumber
&+ \sfrac{i}{2} f_4 \bar{\epsilon} \Gamma^{IJKLM} \Gamma^{\nu \lambda} \Gamma^\mu D_\mu \psi D_\nu X^I D_\lambda X^J X^{KLM} - \sfrac{i}{2} f_4 \bar{\epsilon} \Gamma^{IJKLM} \Gamma^{\nu \lambda} \Gamma^\mu \psi D_\mu ( D_\nu X^I D_\lambda X^J X^{KLM} ) \\
\nonumber
&- \sfrac{i}{2} f_5 \bar{\epsilon} \Gamma^{KLM} \Gamma^{\nu \lambda} \Gamma^\mu D_\mu \psi D_\nu X^J D_\lambda X^K X^{JLM} + \sfrac{i}{2} f_5 \bar{\epsilon} \Gamma^{KLM} \Gamma^{\nu \lambda} \Gamma^\mu \psi D_\mu ( D_\nu X^J D_\lambda X^K X^{JLM} ) \\
\nonumber
&+ \sfrac{i}{2} f_6 \bar{\epsilon} \Gamma^M \Gamma^{\nu \lambda} \Gamma^\mu D_\mu \psi D_\nu X^J D_\lambda X^K X^{JKM} - \sfrac{i}{2} f_6 \bar{\epsilon} \Gamma^M \Gamma^{\nu \lambda} \Gamma^\mu \psi D_\mu ( D_\nu X^J D_\lambda X^K X^{JKM} ) \\
\nonumber
&+ \sfrac{i}{2} f_7 \bar{\epsilon} \Gamma^{KLM} \Gamma^\mu D_\mu \psi D_\nu X^J D^\nu X^J X^{KLM} - \sfrac{i}{2} f_7 \bar{\epsilon} \Gamma^{KLM} \Gamma^\mu \psi D_\mu ( D_\nu X^J D^\nu X^J X^{KLM} ) \\
\nonumber
&+ \sfrac{i}{2} f_8 \bar{\epsilon} \Gamma^{KLM} \Gamma^\mu D_\mu \psi D_\nu X^J D^\nu X^K X^{JLM} - \sfrac{i}{2} f_8 \bar{\epsilon} \Gamma^{KLM} \Gamma^\mu \psi D_\mu ( D_\nu X^J D^\nu X^K X^{JLM} ) \\
\nonumber
&- \sfrac{i}{4} f_1 \bar{\epsilon} \Gamma^{KLM} \Gamma^{IJ} \Gamma^{\mu \nu \lambda} [ X^I , X^J , \psi ] D_\mu X^K D_\nu X^L D_\lambda X^M \\
\nonumber
&+ \sfrac{i}{4} f_1 \bar{\epsilon} \Gamma^{KLM} \Gamma^{IJ} \Gamma^{\mu \nu \lambda} \psi [ X^I , X^J , D_\mu X^K D_\nu X^L D_\lambda X^M ] \\
\nonumber
&- \sfrac{i}{4} f_2 \bar{\epsilon} \Gamma^K \Gamma^{IJ} \Gamma^\mu [ X^I , X^J , \psi ] D_\mu X^L D_\nu X^L D^\nu X^K \\
\nonumber
&+ \sfrac{i}{4} f_2 \bar{\epsilon} \Gamma^K \Gamma^{IJ} \Gamma^\mu \psi [ X^I , X^J , D_\mu X^L D_\nu X^L D^\nu X^K ] \\
\nonumber
&- \sfrac{i}{4} f_3 \bar{\epsilon} \Gamma^K \Gamma^{IJ} \Gamma^\mu [ X^I , X^J , \psi ] D_\mu X^K D_\nu X^L D^\nu X^L \\
\nonumber
&+ \sfrac{i}{4} f_3 \bar{\epsilon} \Gamma^K \Gamma^{IJ} \Gamma^\mu \psi [ X^I , X^J , D_\mu X^K D_\nu X^L D^\nu X^L ] \\
\nonumber
&- i g_9 \bar{\epsilon} \Gamma^{KLM} \Gamma^\lambda \psi D^\nu X^I X^{KLM} ( \tilde{F}_{\nu \lambda} X^I ) - i g_{10} \bar{\epsilon} \Gamma^{JLM} \Gamma^\lambda \psi D^\nu X^J X^{ILM} ( \tilde{F}_{\nu \lambda} X^I ) \\
\nonumber
&- i g_{11} \bar{\epsilon} \Gamma^{M} \Gamma^\lambda \psi D^\nu X^J X^{IJM} ( \tilde{F}_{\nu \lambda} X^I ) + \sfrac{i}{2} g_{12} \bar{\epsilon} \Gamma^{KLM} \Gamma^{\mu \rho \sigma} \psi D_\mu X^I X^{KLM} ( \tilde{F}_{\rho \sigma} X^I ) \\	
\nonumber
&+ \sfrac{i}{2} g_{13} \bar{\epsilon} \Gamma^{JLM} \Gamma^{\mu \rho \sigma} \psi D_\mu X^J X^{ILM} ( \tilde{F}_{\rho \sigma} X^I ) + \sfrac{i}{2} g_{14} \bar{\epsilon} \Gamma^{M} \Gamma^{\mu \rho \sigma} \psi D_\mu X^J X^{IJM} ( \tilde{F}_{\rho \sigma} X^I ) \\
\nonumber
&- 4 i \bold{a} \, \bar{\epsilon} \Gamma^K \Gamma^\mu [X^K, X^I ,\psi ] D_\mu X^J D^\nu X^J D_\nu X^I - 4 i \bold{b} \, \bar{\epsilon} \Gamma^K \Gamma^\mu [X^K, X^I ,\psi ] D_\mu X^I D^\nu X^J D_\nu X^J \\
\nonumber
&+ 3i \bold{c} \, \bar{\epsilon} \Gamma^I \Gamma^{\mu\nu\lambda} [ \psi , X^J , X^K ] D_\mu X^I D_\nu X^J D_\lambda X^K + 3i \bold{c} \, \bar{\epsilon} \Gamma^I \Gamma^{\mu\nu\lambda} D_\mu \psi D_\nu X^J D_\lambda X^K X^{IJK} \\
\nonumber
&- \sfrac{i}{6} \hat{\bold{d}} \bar{\epsilon} \Gamma^{KLM} \Gamma^{IJ} \Gamma^\mu D_\nu \psi D_\mu X^I D^\nu X^J X^{KLM} - \sfrac{i}{6} \hat{\bold{d}} \bar{\epsilon} \Gamma^{KLM} \Gamma^{IJ} \Gamma^\mu \psi D_\nu ( X^{KLM} ) D_\mu X^I D^\nu X^J \\
\nonumber
&- \sfrac{i}{6} \hat{\bold{e}} \bar{\epsilon} \Gamma^{KLM} \Gamma^\mu D^\nu \psi D_\mu X^I D_\nu X^I X^{KLM} + \sfrac{i}{6} \hat{\bold{e}} \bar{\epsilon} \Gamma^{KLM} \Gamma^\mu \psi D^\nu ( X^{KLM}) D_\mu X^I D_\nu X^I \\
\nonumber
&+ i \hat{\bold{f}} \bar{\epsilon} \Gamma^M \Gamma^{IJKL} \Gamma^\mu D_\nu \psi X^{IJK} D^\nu X^L D_\mu X^M + i \hat{\bold{f}} \bar{\epsilon} \Gamma^M \Gamma^{IJKL} \Gamma^\mu \psi D_\nu ( D_\mu X^M ) X^{IJK} D^\nu X^L \\
\nonumber
&+ i \hat{\bold{g}} \bar{\epsilon} \Gamma^L \Gamma^{IJ} \Gamma^\mu D_\nu \psi X^{IJK} D^\nu X^K D_\mu X^L - i \hat{\bold{g}} \bar{\epsilon} \Gamma^L \Gamma^{IJ} \Gamma^\mu \psi D_\nu ( D_\mu X^L ) X^{IJK} D^\nu X^K \\
\nonumber
&+ i \hat{\bold{h}} \bar{\epsilon} \Gamma^L \Gamma^{IJ} \Gamma^\mu [X^J,X^{K},\psi] D^\nu X^I D_\nu X^K D_\mu X^L - i \hat{\bold{h}} \bar{\epsilon} \Gamma^L \Gamma^{IJ} \Gamma^\mu \psi [X^J,X^K, D_\mu X^L ] D^\nu X^I D_\nu X^K \\
\nonumber
&+ i \hat{\bold{i}} \bar{\epsilon} \Gamma^K \Gamma^\lambda \Gamma^{\mu\nu} [X^I,X^J,\psi] D_\mu X^ID_\nu X^J D_\lambda X^K - i \hat{\bold{i}} \bar{\epsilon} \Gamma^K \Gamma^\lambda \Gamma^{\mu\nu} \psi [X^I,X^J, D_\lambda X^K ] D_\mu X^I D_\nu X^J \\
\nonumber
&+ i \hat{\bold{j}} \bar{\epsilon} \Gamma^L \Gamma^{IJ} \Gamma^\lambda \Gamma^{\mu\nu} [X^J,X^{K},\psi] D_\mu X^I D_\nu X^K D_\lambda X^L \\
&+ i \hat{\bold{j}} \bar{\epsilon} \Gamma^L \Gamma^{IJ} \Gamma^\lambda \Gamma^{\mu\nu} \psi [X^J,X^K,D_\lambda X^L ] D_\mu X^I D_\nu X^K \Big\} \, ,
\end{align}
where we have used $\varepsilon^{\mu \rho \sigma} \Gamma_{\mu \nu} \psi \tilde{F}_{\rho \sigma} = - 2 \Gamma^\lambda \psi \tilde{F}_{\nu \lambda}$. Expanding out the $\Gamma$-matrices we find six distinct types of index structure: 
\begin{equation}
\Gamma^{(5)} \Gamma^{\mu \nu \lambda} \, , \qquad \Gamma^{(3)} \Gamma^{\mu \nu \lambda} \, , \qquad \Gamma^{(1)} \Gamma^{\mu \nu \lambda} \, , \qquad  \Gamma^{(5)} \Gamma^{\mu} \, , \qquad \Gamma^{(3)} \Gamma^{\mu} \, , \qquad \Gamma^{(1)} \Gamma^{\mu} \, , 
\end{equation}
where $\Gamma^{(n)}$ represents $\Gamma$-matrices with $n$ antisymmetric transverse Lorentz indices. We now consider each of these in turn.
%
\subsubsection{$\Gamma^{(5)} \Gamma^{\mu \nu \lambda}$ terms}
The $\Gamma^{(5)} \Gamma^{\mu \nu \lambda}$ terms are
\begin{align}
\nonumber
{\rm STr} \Big\{ &+ i \left( - \sfrac{1}{4} f_1 \right) \bar{\epsilon} \Gamma^{IJKLM} \Gamma^{\mu \nu \lambda} [ X^I , X^J , \psi ] D_\mu X^K D_\nu X^L D_\lambda X^M \\
\nonumber
&+ i \left( + \sfrac{3}{4} f_1 - \sfrac{3}{2} f_4 \right) \bar{\epsilon} \Gamma^{IJKLM} \Gamma^{\mu \nu \lambda} \psi [ X^I , X^J , D_\mu X^K ] D_\nu X^L D_\lambda X^M \\
\nonumber
&+ i \left( + \sfrac{1}{2} f_4 \right) \bar{\epsilon} \Gamma^{IJKLM} \Gamma^{\mu \nu \lambda} D_\mu \psi D_\nu X^I D_\lambda X^J X^{KLM} \\
&+ i \left( - \sfrac{1}{2} f_4 \right) \bar{\epsilon} \Gamma^{IJKLM} \Gamma^{\mu \nu \lambda} \psi D_\mu ( D_\nu X^I D_\lambda X^J ) X^{KLM} \Big\}  \, .
\end{align}
The final term ${\rm STr} \{ \bar{\epsilon} \Gamma^{IJKLM} \Gamma^{\mu \nu \lambda} \psi D_\mu ( D_\nu X^I D_\lambda X^J ) X^{KLM} \}$ is zero because 
\begin{align}
{\rm STr} \{ \bar{\epsilon} \Gamma^{IJKLM} \Gamma^{\mu \nu \lambda} \psi D_\mu ( D_\nu X^I D_\lambda X^J ) X^{KLM} \} &= 2 {\rm STr} \{ \bar{\epsilon} \Gamma^{IJKLM} \Gamma^{\mu \nu \lambda} \psi ( \tilde{F}_{\mu \nu} X^I ) X^{KLM} D_\lambda X^J \} 
\end{align}
and the right hand side vanishes after application of \eref{Useful_Id_4}. We can therefore add any multiple of this term as we please.
This allows us to write the variation as
\begin{align}
\nonumber
{\rm STr} \Big\{ &+ i \left( - \sfrac{1}{4} f_1 \right) \bar{\epsilon} \Gamma^{IJKLM} \Gamma^{\mu \nu \lambda} [ X^I , X^J , \psi ] D_\mu X^K D_\nu X^L D_\lambda X^M \\
\nonumber
&+ i \left( - \sfrac{3}{2} f_4 \right) \bar{\epsilon} \Gamma^{IJKLM} \Gamma^{\mu \nu \lambda} \psi [ X^I , X^J , D_\mu X^K ] D_\nu X^L D_\lambda X^M \\
\nonumber
&+ i \left( + \sfrac{1}{2} f_4 \right) \bar{\epsilon} \Gamma^{IJKLM} \Gamma^{\mu \nu \lambda} D_\mu \psi D_\nu X^I D_\lambda X^J X^{KLM} \\
&+ i \left( + \sfrac{3}{4} f_1 \right) \bar{\epsilon} \Gamma^{IJKLM} \Gamma^{\mu \nu \lambda} \psi [ X^I , X^J , D_\mu X^K ] D_\nu X^L D_\lambda X^M \Big\} \\
\nonumber \\
\nonumber
={\rm STr} \Big\{ &+ i \left( - \sfrac{1}{4} f_1 \right) \bar{\epsilon} \Gamma^{IJKLM} \Gamma^{\mu \nu \lambda} [ X^I , X^J , \psi ] D_\mu X^K D_\nu X^L D_\lambda X^M \\
\nonumber
&+ i \left( - \sfrac{1}{2} f_4 \right) \bar{\epsilon} \Gamma^{IJKLM} \Gamma^{\mu \nu \lambda} \psi [ X^I , X^J , D_\mu X^K D_\nu X^L D_\lambda X^M ] \\
\nonumber
&+ i \left( + \sfrac{1}{2} f_4 \right) \bar{\epsilon} \Gamma^{IJKLM} \Gamma^{\mu \nu \lambda} D_\mu \psi D_\nu X^I D_\lambda X^J X^{KLM} \\
&+ i \left( + \sfrac{1}{4} f_1 \right) \bar{\epsilon} \Gamma^{IJKLM} \Gamma^{\mu \nu \lambda} \psi D_\nu X^I D_\lambda X^J D_\mu ( X^{KLM} ) \Big\} \\
\nonumber \\
\nonumber
={\rm STr} \Big\{ &+ i \left( - \sfrac{1}{4} f_1 \right) \bar{\epsilon} \Gamma^{IJKLM} \Gamma^{\mu \nu \lambda} [ X^I , X^J , \psi ] D_\mu X^K D_\nu X^L D_\lambda X^M \\
\nonumber
&+ i \left( - \sfrac{1}{2} f_4 \right) \bar{\epsilon} \Gamma^{IJKLM} \Gamma^{\mu \nu \lambda} \psi [ X^I , X^J , D_\mu X^K D_\nu X^L D_\lambda X^M ] \\
\nonumber
&+ i \left( + \sfrac{1}{2} f_4 \right) \bar{\epsilon} \Gamma^{IJKLM} \Gamma^{\mu \nu \lambda} D_\mu \psi D_\nu X^I D_\lambda X^J X^{KLM} \\
&+ i \left( + \sfrac{1}{4} f_1 \right) \bar{\epsilon} \Gamma^{IJKLM} \Gamma^{\mu \nu \lambda} \psi D_\mu ( D_\nu X^I D_\lambda X^J X^{KLM} ) \Big\} \, .
\end{align}
If $f_4 = \sfrac{1}{2} f_1$ the first two lines vanish through the gauge invariance condition \eqref{Gauge_Inv} and the last two lines are a total derivative. In the analysis of $\tilde{\delta} \mathcal{L}_4$ in chapter \ref{c_Higher_Derivative_BLG} we found $f_1 = + \sfrac{1}{6} \hat{\bold{d}}$ hence invariance demands that $f_4 = + \sfrac{1}{12} \hat{\bold{d}}$.
%
\subsubsection{$\Gamma^{(3)} \Gamma^{\mu \nu \lambda}$ terms}
The $\Gamma^{(3)} \Gamma^{\mu \nu \lambda}$ terms are
\begin{align}
\nonumber
{\rm STr} \Big\{ &+ i \left( - \sfrac{1}{2} f_5 \right) \bar{\epsilon} \Gamma^{IJK} \Gamma^{\mu \nu \lambda} D_\mu \psi D_\nu X^M D_\lambda X^I X^{JKM} \\
\nonumber
&+ i \left( + \sfrac{1}{2} f_5 \right) \bar{\epsilon} \Gamma^{IJK} \Gamma^{\mu \nu \lambda} \psi D_\mu ( D_\nu X^M D_\lambda X^I X^{JKM} ) \\
\nonumber
&+ i \left( + \sfrac{1}{2} g_{12} \right) \bar{\epsilon} \Gamma^{IJK} \Gamma^{\mu \nu \lambda} \psi D_\mu X^M ( \tilde{F}_{\nu \lambda} X^M ) X^{IJK} \\
\nonumber
&+ i \left( + \sfrac{1}{2} g_{13} \right) \bar{\epsilon} \Gamma^{IJK} \Gamma^{\mu \nu \lambda} \psi D_\mu X^I ( \tilde{F}_{\nu \lambda} X^M ) X^{JKM} \\[6pt]
\nonumber
&+ i \left( + \sfrac{3}{2} f_1 + \hat{\bold{j}} \right) \bar{\epsilon} \Gamma^{IJK} \Gamma^{\mu \nu \lambda} [ X^I , X^M , \psi ] D_\mu X^M D_\nu X^J D_\lambda X^K \\
\nonumber
&+ i \left( - \sfrac{3}{2} f_1 - g_7 \right) \bar{\epsilon} \Gamma^{IJK} \Gamma^{\mu \nu \lambda} \psi [ X^I , X^M , D_\mu X^M ] D_\nu X^J D_\lambda X^K \\
&+ i \left( - \sfrac{3}{2} f_1 + \sfrac{1}{2} \hat{\bold{j}} \right) \bar{\epsilon} \Gamma^{IJK} \Gamma^{\mu \nu \lambda} \psi D_\mu X^M [ X^I , X^M , D_\nu X^J D_\lambda X^K ] \Big\} \, .
\end{align}
The identity \eqref{Useful_Id_4} can be used to show
\begin{equation}
+ \sfrac{i}{2} g_{12} {\rm STr} \{ \bar{\epsilon} \Gamma^{IJK} \Gamma^{\mu \nu \lambda} \psi D_\mu X^M ( \tilde{F}_{\nu \lambda} X^M ) X^{IJK} \} = + \sfrac{3i}{2} g_{12} {\rm STr} \{ \bar{\epsilon} \Gamma^{IJK} \Gamma^{\mu \nu \lambda} \psi D_\mu X^M ( \tilde{F}_{\nu \lambda} X^I ) X^{JKM} \} \, .
\end{equation}
Using this result and expanding out the $f_5$ terms leads to
\begin{align}
\nonumber
{\rm STr} \Big\{ &+ i \left( - \sfrac{1}{2} f_5 \right) \bar{\epsilon} \Gamma^{IJK} \Gamma^{\mu \nu \lambda} D_\mu \psi D_\nu X^M D_\lambda X^I X^{JKM} \\
\nonumber
&+ i \left( + \sfrac{1}{2} f_5 + \sfrac{1}{2} g_{13} \right) \bar{\epsilon} \Gamma^{IJK} \Gamma^{\mu \nu \lambda} \psi D_\mu ( D_\nu X^M ) D_\lambda X^I X^{JKM} \\
\nonumber
&+ i \left( + \sfrac{1}{2} f_5 - \sfrac{3}{2} g_{12} \right) \bar{\epsilon} \Gamma^{IJK} \Gamma^{\mu \nu \lambda} \psi D_\nu X^M D_\mu ( D_\lambda X^I ) X^{JKM} \\
\nonumber
&+ i \left( + \sfrac{1}{2} f_5 \right) \bar{\epsilon} \Gamma^{IJK} \Gamma^{\mu \nu \lambda} \psi D_\nu X^M D_\lambda X^I D_\mu ( X^{JKM} ) \\[6pt]
\nonumber
&+ i \left( + \sfrac{3}{2} f_1 + \hat{\bold{j}} \right) \bar{\epsilon} \Gamma^{IJK} \Gamma^{\mu \nu \lambda} [ X^I , X^M , \psi ] D_\mu X^M D_\nu X^J D_\lambda X^K \\
\nonumber
&+ i \left( - \sfrac{3}{2} f_1 - g_7 \right) \bar{\epsilon} \Gamma^{IJK} \Gamma^{\mu \nu \lambda} \psi [ X^I , X^M , D_\mu X^M ] D_\nu X^J D_\lambda X^K \\
&+ i \left( - \sfrac{3}{2} f_1 + \sfrac{1}{2} \hat{\bold{j}} \right) \bar{\epsilon} \Gamma^{IJK} \Gamma^{\mu \nu \lambda} \psi D_\mu X^M [ X^I , X^M , D_\nu X^J D_\lambda X^K ] \Big\} \, .
\end{align}
The final three terms are zero by the gauge invariance condition if 
\begin{equation}
\sfrac{3}{2} f_1 + \hat{\bold{j}} = - \sfrac{3}{2} f_1 - g_7 = - \sfrac{3}{2} f_1 + \sfrac{1}{2} \hat{\bold{j}} \, .
\end{equation}
Solving these simultaneous equations yields $f_1= - \sfrac{1}{6} \hat{\bold{j}}$ and $f_1= + \sfrac{1}{3} g_7$. The first four lines vanish only if $f_5 =0$ which forces $g_{12} = 0$ and $g_{13}=0$. We know from the $\tilde{\delta} \mathcal{L}_4$ terms that $f_1 = + \sfrac{1}{6} \hat{\bold{d}}$ so we find $\hat{\bold{j}}=-\hat{\bold{d}}$ and $g_7 = + \sfrac{1}{2} \hat{\bold{d}}$.
%
%
%
\subsubsection{$\Gamma^{(1)} \Gamma^{\mu \nu \lambda}$ terms}
The $\Gamma^{(1)} \Gamma^{\mu \nu \lambda}$ terms are
\begin{align}
\nonumber
{\rm STr} \Big\{ &+ i \left( + \sfrac{1}{2} f_6 + 3 \bold{c} \right) \bar{\epsilon} \Gamma^I \Gamma^{\mu \nu \lambda} D_\mu \psi D_\nu X^J D_\lambda X^K X^{IJK} \\
\nonumber
&+ i \left( - \sfrac{1}{2} f_6 + \sfrac{1}{2} g_{14} \right) \bar{\epsilon} \Gamma^I \Gamma^{\mu \nu \lambda} \psi D_\mu ( D_\nu X^J D_\lambda X^K ) X^{IJK} \\
\nonumber
&+ i \left( - \sfrac{1}{2} f_6 \right) \bar{\epsilon} \Gamma^I \Gamma^{\mu \nu \lambda} \psi D_\nu X^J D_\lambda X^K [ D_\mu X^I , X^J , X^K ] \\
\nonumber
&+ i \left( - \sfrac{1}{2} f_6 - \sfrac{1}{2} \hat{\bold{j}} \right) \bar{\epsilon} \Gamma^I \Gamma^{\mu \nu \lambda} \psi D_\nu X^J D_\lambda X^K [ X^I , D_\mu X^J , X^K ] \\
\nonumber
&+ i \left( - \sfrac{1}{2} f_6 - \sfrac{1}{2} \hat{\bold{j}} \right) \bar{\epsilon} \Gamma^I \Gamma^{\mu \nu \lambda} \psi D_\nu X^J D_\lambda X^K [ X^I , X^J , D_\mu X^K ] \\[6pt]
\nonumber
&+ i \left( + \sfrac{3}{2} f_1 + 3 \bold{c} + \hat{\bold{i}} + \hat{\bold{j}} \right) \bar{\epsilon} \Gamma^I \Gamma^{\mu \nu \lambda} [X^J,X^K,\psi] D_\mu X^I D_\nu X^J D_\lambda X^K \\
\nonumber
&+ i \left( - \sfrac{3}{2} f_1 - \hat{\bold{i}} \right) \bar{\epsilon} \Gamma^I \Gamma^{\mu \nu \lambda} \psi D_\nu X^J D_\lambda X^K [ X^J , X^K , D_\mu X^I ] \\
&+ i \left( - \sfrac{3}{2} f_1 - \sfrac{1}{2} g_6 + \sfrac{1}{2} \hat{\bold{j}} \right) \bar{\epsilon} \Gamma^I \Gamma^{\mu \nu \lambda} \psi D_\mu X^I  [ X^J , X^K , D_\nu X^J D_\lambda X^K ] \Big\} \, .
\end{align}
To proceed we see that we have two terms which are identical and so we can redistribute their coefficients in the following way
\begin{align}
\nonumber
{\rm STr} \Big\{ &+ i \left( - \sfrac{1}{2} f_6 \right) \bar{\epsilon} \Gamma^I \Gamma^{\mu \nu \lambda} \psi D_\nu X^J D_\lambda X^K [ D_\mu X^I , X^J , X^K ] \\
&+ i \left( - \sfrac{3}{2} f_1 - \hat{\bold{i}} \right) \bar{\epsilon} \Gamma^I \Gamma^{\mu \nu \lambda} \psi D_\nu X^J D_\lambda X^K [ X^J , X^K , D_\mu X^I ] \Big\} \\[10pt]
\nonumber
={\rm STr} \Big\{ &+ i \left( - \sfrac{1}{2} f_6 - \sfrac{1}{2} \hat{\bold{j}} \right) \bar{\epsilon} \Gamma^I \Gamma^{\mu \nu \lambda} \psi D_\nu X^J D_\lambda X^K [ D_\mu X^I , X^J , X^K ] \\
&+ i \left( - \sfrac{3}{2} f_1 - \hat{\bold{i}} + \sfrac{1}{2} \hat{\bold{j}} \right) \bar{\epsilon} \Gamma^I \Gamma^{\mu \nu \lambda} \psi D_\nu X^J D_\lambda X^K [ X^J , X^K , D_\mu X^I ] \Big\} \, .
\end{align}
This allows us to write the $\Gamma^I \Gamma^{\mu \nu \lambda}$ terms as
\begin{align}
\nonumber
{\rm STr} \Big\{&+ i \left( + \sfrac{1}{2} f_6 + 3 \bold{c} \right) \bar{\epsilon} \Gamma^I \Gamma^{\mu \nu \lambda} D_\mu \psi D_\nu X^J D_\lambda X^K X^{IJK} \\
\nonumber
&+ i \left( - \sfrac{1}{2} f_6 + \sfrac{1}{2} g_{14} \right) \bar{\epsilon} \Gamma^I \Gamma^{\mu \nu \lambda} \psi D_\mu ( D_\nu X^J D_\lambda X^K ) X^{IJK} \\
\nonumber
&+ i \left( - \sfrac{1}{2} f_6 - \sfrac{1}{2} \hat{\bold{j}} \right) \bar{\epsilon} \Gamma^I \Gamma^{\mu \nu \lambda} \psi D_\nu X^J D_\lambda X^K D_\mu ( X^{IJK} ) \\[6pt]
\nonumber
&+ i \left( + \sfrac{3}{2} f_1 + 3 \bold{c} + \hat{\bold{i}} + \hat{\bold{j}} \right) \bar{\epsilon} \Gamma^I \Gamma^{\mu \nu \lambda} [X^J,X^K,\psi] D_\mu X^I D_\nu X^J D_\lambda X^K \\
\nonumber
&+ i \left( - \sfrac{3}{2} f_1 - \hat{\bold{i}} + \sfrac{1}{2} \hat{\bold{j}} \right) \bar{\epsilon} \Gamma^I \Gamma^{\mu \nu \lambda} \psi [ X^J , X^K , D_\mu X^I ] D_\nu X^J D_\lambda X^K \\
&+ i \left( - \sfrac{3}{2} f_1 - \sfrac{1}{2} g_6 + \sfrac{1}{2} \hat{\bold{j}} \right) \bar{\epsilon} \Gamma^I \Gamma^{\mu \nu \lambda} \psi D_\mu X^I  [ X^J , X^K , D_\nu X^J D_\lambda X^K ] \Big\} \, .
\end{align}
The first three terms form a total derivative if
\begin{equation}
+ \sfrac{1}{2} f_6 + 3 \bold{c} = - \sfrac{1}{2} f_6 + \sfrac{1}{2} g_{14}  = - \sfrac{1}{2} f_6 - \sfrac{1}{2} \hat{\bold{j}} \, ,
\end{equation}
which are solved by $f_6 = - 3 \bold{c} - \sfrac{1}{2} \hat{\bold{j}}$ and $g_{14} = - \hat{\bold{j}}$. The final three terms are zero by the gauge invariance condition if
\begin{equation}
+ \sfrac{3}{2} f_1 + 3 \bold{c} + \hat{\bold{i}} + \hat{\bold{j}} = - \sfrac{3}{2} f_1 - \hat{\bold{i}} + \sfrac{1}{2} \hat{\bold{j}} = - \sfrac{3}{2} f_1 - \sfrac{1}{2} g_6 + \sfrac{1}{2} \hat{\bold{j}} \, .
\end{equation}
This gives $g_6 = + 2 \hat{\bold{i}}$ and $f_1 = - \bold{c} - \sfrac{2}{3} \hat{\bold{i}} - \sfrac{1}{6} \hat{\bold{j}}$.
Using the value $g_6 = - 2 \hat{\bold{d}}$ from the four derivative section and the value $f_1= \sfrac{1}{6} \hat{\bold{d}}$ from both the four derivative section and the $\Gamma^3 \Gamma^{\mu \nu \lambda}$ section above, we find $\hat{\bold{i}} = - \hat{\bold{d}}$ and $\bold{c} = + \sfrac{2}{3} \hat{\bold{d}}$. This then forces the values $g_{14} = + \hat{\bold{d}}$ and $f_6 = - \sfrac{3}{2} \hat{\bold{d}}$.
%
%
%
\subsubsection{$\Gamma^{(5)} \Gamma^\mu$ terms}
The $\Gamma^{(5)} \Gamma^{\mu}$ terms are
\begin{align}
%
\nonumber
{\rm STr} \Big\{ &+ i \left( + f_4 + s_5 - \sfrac{1}{6} \hat{\bold{d}} - \hat{\bold{f}} \right) \bar{\epsilon} \Gamma^{IJKLM} \Gamma^\mu D^\nu \psi D_\mu X^I D_\nu X^J X^{KLM} \\
\nonumber
&+ i \left( - f_4 + s_5 - \hat{\bold{f}} \right) \bar{\epsilon} \Gamma^{IJKLM} \Gamma^\mu \psi D^\nu ( D_\mu X^I ) D_\nu X^J X^{KLM} \\
\nonumber
&+ i \left( - f_4 \right) \bar{\epsilon} \Gamma^{IJKLM} \Gamma^\mu \psi D_\mu X^I D^\nu ( D_\nu X^J ) X^{KLM} \\
&+ i \left( - f_4 + s_5 - \sfrac{1}{6} \hat{\bold{d}} \right) \bar{\epsilon} \Gamma^{IJKLM} \Gamma^\mu \psi D_\mu X^I D_\nu X^J D^\nu ( X^{KLM} ) \Big\} \, .
\end{align}
This forms a total derivative if
\begin{equation}
+ f_4 + s_5 - \sfrac{1}{6} \hat{\bold{d}} - \hat{\bold{f}} = - f_4 + s_5 - \hat{\bold{f}} = - f_4 = - f_4 + s_5 - \sfrac{1}{6} \hat{\bold{d}} \, .
\end{equation}
The solution to these equations is $f_4 = + \sfrac{1}{12} \hat{\bold{d}}$, $s_5 = + \sfrac{1}{6} \hat{\bold{d}}$ and $\hat{\bold{f}} = + \sfrac{1}{6} \hat{\bold{d}}$.
%
%
%
\subsubsection{$\Gamma^{(3)} \Gamma^\mu$ terms}
The $\Gamma^{(3)} \Gamma^{\mu}$ terms are
\begin{align}
\nonumber
{\rm STr} \Big\{ &+ i \left( - s_8 - \sfrac{1}{2} \hat{\bold{d}} + 3 \hat{\bold{f}} \right) \bar{\epsilon} \Gamma^{IJK} \Gamma^\mu D^\nu \psi D_\mu X^L D_\nu X^I X^{JKL} \\
\nonumber
&+ i \left( - 2 s_8 - \hat{\bold{d}} - \hat{\bold{j}} \right) \bar{\epsilon} \Gamma^{IJK} \Gamma^\mu \psi D_\mu X^L D_\nu X^I [ D^\nu X^J , X^K , X^L ] \\
\nonumber
&+ i \left( - s_7 + \sfrac{1}{2} \hat{\bold{d}} + \hat{\bold{g}} \right) \bar{\epsilon} \Gamma^{IJK} \Gamma^\mu D^\nu \psi D_\mu X^I D_\nu X^L X^{JKL} \\
\nonumber
&+ i \left( - 2 s_7 + \hat{\bold{d}} + \hat{\bold{j}} \right) \bar{\epsilon} \Gamma^{IJK} \Gamma^\mu \psi D_\mu X^I D_\nu X^L [ D^\nu X^J , X^K , X^L ] \\
\nonumber
&+ i \left( + \sfrac{1}{2} f_3 + \sfrac{1}{2} \hat{\bold{d}} \right) \bar{\epsilon} \Gamma^{IJK} \Gamma^\mu \psi D_\mu X^I D_\nu X^L [ X^J , X^K , D^\nu X^L ] \\
\nonumber
&+ i \left( + \hat{\bold{h}} + \hat{\bold{j}} \right) \bar{\epsilon} \Gamma^{IJK} \Gamma^\mu [X^J,X^L,\psi] D^\nu X^I D_\nu X^L D_\mu X^K \\
\nonumber
&+ i \left( - s_7 \right) \bar{\epsilon} \Gamma^{IJK} \Gamma^\mu \psi D_\mu X^I D_\nu X^L [ X^J , X^K , D^\nu X^L ] \\
\nonumber
&+ i \left( - s_8 \right) \bar{\epsilon} \Gamma^{IJK} \Gamma^\mu \psi D_\mu X^L D_\nu X^I [ X^J , X^K , D^\nu X^L ] \\
\nonumber
&+ i \left( - g_8 \right) \bar{\epsilon} \Gamma^{IJK} \Gamma^\mu \psi D_\nu X^K D_\mu X^I [ X^J , X^L , D^\nu X^L ] \\
\nonumber
&+ i \left( + \sfrac{1}{2} f_8 \right) \bar{\epsilon} \Gamma^{IJK} \Gamma^\mu D_\mu \psi D_\nu X^L D^\nu X^I X^{JKL} \\
\nonumber
&+ i \left( - \sfrac{1}{2} f_8 \right) \bar{\epsilon} \Gamma^{IJK} \Gamma^\mu \psi D_\mu ( D_\nu X^L ) D^\nu X^I X^{JKL} \\
\nonumber
&+ i \left( - s_8 + 3 \hat{\bold{f}} \right) \bar{\epsilon} \Gamma^{IJK} \Gamma^\mu \psi D_\nu ( D_\mu X^L ) D^\nu X^I X^{JKL} \\
\nonumber
&+ i \left( - \sfrac{1}{2} f_8 \right) \bar{\epsilon} \Gamma^{IJK} \Gamma^\mu \psi D^\nu X^L D_\mu ( D_\nu X^I ) X^{JKL} \\
\nonumber
&+ i \left( - s_7 - \hat{\bold{g}} \right) \bar{\epsilon} \Gamma^{IJK} \Gamma^\mu \psi D^\nu X^L D_\nu ( D_\mu X^I ) X^{JKL} \\
\nonumber
&+ i \left( - f_8 + \hat{\bold{h}} \right) \bar{\epsilon} \Gamma^{IJK} \Gamma^\mu \psi D_\nu X^L D^\nu X^I [ D_\mu X^J , X^K , X^L ] \\
\nonumber
&+ i \left( + \sfrac{1}{2} f_2 - \sfrac{1}{2} f_8 \right) \bar{\epsilon} \Gamma^{IJK} \Gamma^\mu \psi D_\nu X^L D^\nu X^I [ X^J , X^K , D_\mu X^L ] \\
\nonumber
&+ i \left( - \sfrac{1}{4} f_2 \right) \bar{\epsilon} \Gamma^{IJK} \Gamma^\mu [ X^I , X^J , \psi ] D_\mu X^L D_\nu X^L D^\nu X^K \\
\nonumber
&+ i \left( - \sfrac{1}{4} f_2 \right) \bar{\epsilon} \Gamma^{IJK} \Gamma^\mu \psi [ X^I , X^J , D_\mu X^L ] D_\nu X^L D^\nu X^K \\
\nonumber
&+ i \left( + \sfrac{1}{4} f_2 - \sfrac{1}{2} \hat{\bold{d}} \right) \bar{\epsilon} \Gamma^{IJK} \Gamma^\mu \psi D_\mu X^L [ X^I , X^J , D^\nu X^L ] D_\nu X^K \\
\nonumber
&+ i \left( + \sfrac{1}{4} f_2 + \sfrac{1}{2} \hat{\bold{e}} \right) \bar{\epsilon} \Gamma^{IJK} \Gamma^\mu \psi D_\mu X^L D_\nu X^L [ X^I , X^J , D^\nu X^K ] \\
\nonumber
&+ i \left( - 3 s_6 \right) \bar{\epsilon} \Gamma^{IJK} \Gamma^\mu \psi D_\mu X^L D_\nu X^L [ X^I , X^J , D^\nu X^K ] \\
\nonumber
&+ i \left( - \sfrac{1}{4} f_3 \right) \bar{\epsilon} \Gamma^{IJK} \Gamma^\mu [ X^I , X^J , \psi ] D_\mu X^K D_\nu X^L D^\nu X^L \\
\nonumber
&+ i \left( + \sfrac{1}{2} f_7 \right) \bar{\epsilon} \Gamma^{IJK} \Gamma^\mu D_\mu \psi D_\nu X^L D^\nu X^L X^{IJK} \\
\nonumber
&+ i \left( - f_7 \right) \bar{\epsilon} \Gamma^{IJK} \Gamma^\mu \psi D_\mu ( D_\nu X^L ) D^\nu X^L X^{IJK} \\
\nonumber
&+ i \left( - s_6 - \hat{\bold{f}} \right) \bar{\epsilon} \Gamma^{IJK} \Gamma^\mu \psi D_\nu ( D_\mu X^L ) D^\nu X^L X^{IJK} \\
\nonumber
&+ i \left( + \sfrac{1}{4} f_3 - \sfrac{3}{2} f_7 \right) \bar{\epsilon} \Gamma^{IJK} \Gamma^\mu \psi D_\nu X^L D^\nu X^L [ X^I , X^J , D_\mu X^K ] \\
\nonumber
&+ i \left( - s_6 - \hat{\bold{f}} - \sfrac{1}{6} \hat{\bold{e}} \right) \bar{\epsilon} \Gamma^{IJK} \Gamma^\mu D^\nu \psi D_\mu X^L D_\nu X^L X^{IJK} \\
\nonumber
&+ i \left( + g_9 \right) \bar{\epsilon} \Gamma^{IJK} \Gamma^\mu \psi D^\nu X^L ( \tilde{F}_{\mu \nu} X^L ) X^{IJK} \\
&+ i \left( + g_{10} \right) \bar{\epsilon} \Gamma^{IJK} \Gamma^\mu \psi D^\nu X^I ( \tilde{F}_{\mu \nu} X^L ) X^{JKL} \Big\} \, . 
\end{align}
We know $+ \sfrac{1}{4} f_2 - \sfrac{1}{2} \hat{\bold{d}} = + \sfrac{1}{4} f_2 + \sfrac{1}{2} \hat{\bold{e}} = - \sfrac{1}{4} f_2$ , $+ \sfrac{1}{2} \hat{\bold{d}} + \sfrac{1}{2} f_3 = - \sfrac{1}{2} f_3$, $\hat{\bold{d}} = - \hat{\bold{j}}$, $\hat{\bold{d}} = - \hat{\bold{e}}$ and $\hat{\bold{f}} = - \sfrac{1}{6} \hat{\bold{e}}$. In addition we have previously found $f_5$ and $g_8$ to be zero. Using this coefficient data and swapping the order of the covariant derivatives in some terms at the expense of introducing a field strength we get
\begin{align}
\nonumber
{\rm STr} \Big\{ &+ i \left( - s_6 \right) \bar{\epsilon} \Gamma^{IJK} \Gamma^\mu D^\nu ( \psi D_\mu X^L D_\nu X^L X^{IJK} ) \\
\nonumber
&+ i \left( + s_6 \right) \bar{\epsilon} \Gamma^{IJK} \Gamma^\mu \psi D_\mu X^L X^{IJK} D^\nu ( D_\nu X^L ) \\[6pt]
\nonumber
&+ i \left( - s_7 \right) \bar{\epsilon} \Gamma^{IJK} \Gamma^\mu \psi D_\mu X^I D_\nu X^L [ X^J , X^K , D^\nu X^L ] \\[6pt]
%
%
\nonumber
&+ i \left( - s_8 \right) \bar{\epsilon} \Gamma^{IJK} \Gamma^\mu D^\nu ( \psi D_\mu X^L D_\nu X^I X^{JKL} ) \\
\nonumber
&+ i \left( + s_8 \right) \bar{\epsilon} \Gamma^{IJK} \Gamma^\mu \psi D_\mu X^L X^{JKL} D^\nu ( D_\nu X^I ) \\[6pt]
\nonumber
&+ i \left( - s_7 + \sfrac{1}{2} \hat{\bold{d}} + \hat{\bold{g}} \right) \bar{\epsilon} \Gamma^{IJK} \Gamma^\mu D^\nu \psi D_\mu X^I D_\nu X^L X^{JKL} \\[6pt]
\nonumber
&+ i \left( + \hat{\bold{d}} - \hat{\bold{h}} \right) \bar{\epsilon} \Gamma^{IJK} \Gamma^\mu [X^J,X^L,\psi] D_\mu X^I D^\nu X^K D_\nu X^L \\
\nonumber
&+ i \left( + 2 s_7 \right) \bar{\epsilon} \Gamma^{IJK} \Gamma^\mu \psi D_\mu X^I [ X^J , X^L , D^\nu X^K ] D_\nu X^L \\[6pt]
\nonumber
&+ i \left( + \sfrac{1}{2} \hat{\bold{d}} - \sfrac{1}{2} \hat{\bold{h}} \right) \bar{\epsilon} \Gamma^{IJK} \Gamma^\mu \psi D_\nu X^L D^\nu X^I [ X^J , X^K , D_\mu X^L ] \\[6pt]
\nonumber
&+ i \left( + \sfrac{1}{2} f_8 \right) \bar{\epsilon} \Gamma^{IJK} \Gamma^\mu D_\mu \psi D_\nu X^L D^\nu X^I X^{JKL} \\
\nonumber
&+ i \left( - \sfrac{1}{2} f_8 + \sfrac{1}{2} \hat{\bold{d}} \right) \bar{\epsilon} \Gamma^{IJK} \Gamma^\mu \psi D_\mu ( D_\nu X^L ) D^\nu X^I X^{JKL} \\
%
%
\nonumber
&+ i \left( - \sfrac{1}{2} f_8 - s_7 - \hat{\bold{g}} \right) \bar{\epsilon} \Gamma^{IJK} \Gamma^\mu \psi D^\nu X^L D_\mu ( D_\nu X^I ) X^{JKL} \\
%
%
\nonumber
&+ i \left( - \sfrac{1}{2} f_8 + \sfrac{1}{2} \hat{\bold{h}} \right) \bar{\epsilon} \Gamma^{IJK} \Gamma^\mu \psi D_\nu X^L D^\nu X^I D_\mu ( X^{JKL} ) \\[6pt]
%
%
\nonumber
&+ i \left( + \sfrac{1}{2} f_7 \right) \bar{\epsilon} \Gamma^{IJK} \Gamma^\mu D_\mu \psi D_\nu X^L D^\nu X^L X^{IJK} \\
\nonumber
&+ i \left( - \sfrac{1}{2} f_7 - \sfrac{1}{12} \hat{\bold{d}} \right) \bar{\epsilon} \Gamma^{IJK} \Gamma^\mu \psi D_\mu ( D_\nu X^L D^\nu X^L ) X^{IJK} \\
\nonumber
&+ i \left( - \sfrac{1}{2} f_7 - \sfrac{1}{12} \hat{\bold{d}} \right) \bar{\epsilon} \Gamma^{IJK} \Gamma^\mu \psi D_\nu X^L D^\nu X^L D_\mu ( X^{IJK} ) \\[6pt]
%
%
\nonumber
&+ i \left( + s_7 + \hat{\bold{g}} \right) \bar{\epsilon} \Gamma^{IJK} \Gamma^\mu \psi D^\nu X^L ( \tilde{F}_{\mu \nu} X^I ) X^{JKL} \\
\nonumber
&+ i \left( + g_{10} - \sfrac{1}{2} \hat{\bold{d}} \right) \bar{\epsilon} \Gamma^{IJK} \Gamma^\mu \psi ( \tilde{F}_{\mu \nu} X^L ) D^\nu X^I X^{JKL} \\
&+ i \left( + g_9 + \sfrac{1}{6} \hat{\bold{d}} \right) \bar{\epsilon} \Gamma^{IJK} \Gamma^\mu \psi ( \tilde{F}_{\mu \nu} X^L ) D^\nu X^L X^{IJK} \Big\} \, .
%
%
%
\end{align}
Many of these terms cannot be formed into total derivatives or be removed through the gauge invariance condition and so we must set their coefficients to zero. This tells us 
\begin{equation}
s_6 =0 \, , \qquad s_7 =0 \, , \qquad s_8 =0 \, , \qquad \hat{\bold{g}} = - \sfrac{1}{2} \hat{\bold{d}} \, , \qquad \hat{\bold{h}} = + \hat{\bold{d}} \, .
\end{equation}
Further, we get two total derivatives if
\begin{equation}
+ \sfrac{1}{2} f_8  = - \sfrac{1}{2} f_8 + \sfrac{1}{2} \hat{\bold{d}} = - \sfrac{1}{2} f_8 - s_7 - \hat{\bold{g}} = - \sfrac{1}{2} f_8 + \sfrac{1}{2} \hat{\bold{h}}
\end{equation}
and
\begin{equation}
+ \sfrac{1}{2} f_7 = - \sfrac{1}{2} f_7 - \sfrac{1}{12} \hat{\bold{d}} \, .
\end{equation}
Solving these simultaneous equations we find $f_7 = - \sfrac{1}{12} \hat{\bold{d}}$ and $f_8 = + \sfrac{1}{2} \hat{\bold{d}}$. The remaining terms are
\begin{align}
\nonumber
{\rm STr} \Big\{ &+ i \left( + \hat{\bold{g}} \right) \bar{\epsilon} \Gamma^{IJK} \Gamma^\mu \psi D^\nu X^L ( \tilde{F}_{\mu \nu} X^I ) X^{JKL} \\
\nonumber
&+ i \left( + g_{10} - \sfrac{1}{2} \hat{\bold{d}} \right) \bar{\epsilon} \Gamma^{IJK} \Gamma^\mu \psi ( \tilde{F}_{\mu \nu} X^L ) D^\nu X^I X^{JKL} \\
&+ i \left( + g_9 + \sfrac{1}{6} \hat{\bold{d}} \right) \bar{\epsilon} \Gamma^{IJK} \Gamma^\mu \psi ( \tilde{F}_{\mu \nu} X^L ) D^\nu X^L X^{IJK} \Big\} \\[10pt]
\nonumber
= {\rm STr} \Big\{ &+ i \left( + 3 g_9 + \hat{\bold{g}} + \sfrac{1}{2} \hat{\bold{d}} \right) \bar{\epsilon} \Gamma^{IJK} \Gamma^\mu \psi D^\nu X^L ( \tilde{F}_{\mu \nu} X^I ) X^{JKL} \\
&+ i \left( + g_{10} - \sfrac{1}{2} \hat{\bold{d}} \right) \bar{\epsilon} \Gamma^{IJK} \Gamma^\mu \psi ( \tilde{F}_{\mu \nu} X^L ) D^\nu X^I X^{JKL} \Big\} \, ,
\end{align}
where we have employed \eref{Useful_Id_4}. Each coefficient above must be zero and because $\hat{\bold{g}} + \sfrac{1}{2} \hat{\bold{d}}=0$ we see that $g_9=0$ and $g_{10} = + \sfrac{1}{2} \hat{\bold{d}}$.
%
%
%
\subsubsection{$\Gamma^{(1)} \Gamma^\mu$ terms}
The $\Gamma^{(1)} \Gamma^{\mu}$ terms are
\begin{align}
\nonumber
{\rm STr} \Big\{ &+ i \left( + f_6 + s_9 + \hat{\bold{d}} - 2 \hat{\bold{g}} \right) \bar{\epsilon} \Gamma^I \Gamma^\mu D^\nu \psi D_\mu X^J D_\nu X^K X^{IJK} \\
\nonumber
&+ i \left( - f_6 + s_9 + 2 \hat{\bold{g}} \right) \bar{\epsilon} \Gamma^I \Gamma^\mu \psi D^\nu ( D_\mu X^J ) D_\nu X^K X^{IJK} \\
\nonumber
&+ i \left( - f_6 \right) \bar{\epsilon} \Gamma^I \Gamma^\mu \psi D_\mu X^J D^\nu ( D_\nu X^K ) X^{IJK} \\
\nonumber
&+ i \left( - f_6 + s_9 + \hat{\bold{d}} \right) \bar{\epsilon} \Gamma^I \Gamma^\mu \psi D_\mu X^J D_\nu X^K D^\nu ( X^{IJK} ) \\[6pt]
\nonumber
&+ i \left( - \hat{\bold{h}} - 2 \hat{\bold{i}} + \hat{\bold{j}} \right) \bar{\epsilon} \Gamma^I \Gamma^\mu [X^J,X^{K},\psi] D^\nu X^I D_\mu X^J D_\nu X^K \\
\nonumber
&+ i \left( + 2 \hat{\bold{i}} \right) \bar{\epsilon} \Gamma^I \Gamma^{\mu} \psi [X^J,X^K, D^\nu X^I ] D_\mu X^J D_\nu X^K \\
\nonumber
&+ i \left( + g_5 + \hat{\bold{h}} \right) \bar{\epsilon} \Gamma^I \Gamma^\mu \psi D^\nu X^I [X^J,X^K, D_\mu X^J ] D_\nu X^K \\
\nonumber
&+ i \left( - g_3 + \hat{\bold{j}} \right) \bar{\epsilon} \Gamma^I \Gamma^{\mu} \psi D^\nu X^I D_\mu X^J [X^J,X^K, D_\nu X^K ] \\[6pt]
\nonumber
&+ i \left( + g_4 + \hat{\bold{j}} \right) \bar{\epsilon} \Gamma^I \Gamma^{\mu} \psi [X^J,X^K, D^\nu X^J ] D_\mu X^I D_\nu X^K \\[6pt]
\nonumber
&+ i \left( + \sfrac{1}{2} f_2 - 4 \bold{a} + \hat{\bold{h}} - \hat{\bold{j}} \right) \bar{\epsilon} \Gamma^I \Gamma^\mu [ X^I , X^J , \psi ] D_\mu X^K D_\nu X^K D^\nu X^J \\
\nonumber
&+ i \left( - \sfrac{1}{2} f_2 - \hat{\bold{h}} \right) \bar{\epsilon} \Gamma^I \Gamma^\mu \psi [X^I,X^J, D_\mu X^K ] D^\nu X^K D_\nu X^J \\
\nonumber
&+ i \left( - \sfrac{1}{2} f_2 - \hat{\bold{j}} \right) \bar{\epsilon} \Gamma^I \Gamma^{\mu} \psi D_\mu X^K [X^I,X^J, D^\nu X^K ] D_\nu X^J \\
\nonumber
&+ i \left( - \sfrac{1}{2} f_2 - g_2 \right) \bar{\epsilon} \Gamma^I \Gamma^\mu \psi D_\mu X^K D_\nu X^K [ X^I , X^J , D^\nu X^J ] \\[6pt]
\nonumber
&+ i \left( + \sfrac{1}{2} f_3 - 4 \bold{b} + \hat{\bold{j}} \right) \bar{\epsilon} \Gamma^I \Gamma^\mu [ X^I , X^J , \psi ] D_\mu X^J D_\nu X^K D^\nu X^K \\
\nonumber
&+ i \left( - \sfrac{1}{2} f_3 - g_1 \right) \bar{\epsilon} \Gamma^I \Gamma^\mu \psi [ X^I ,  X^J , D_\mu X^J ] D_\nu X^K D^\nu X^K \\
\nonumber
&+ i \left( - \sfrac{1}{2} f_3 + \sfrac{1}{2} \hat{\bold{j}} \right) \bar{\epsilon} \Gamma^I \Gamma^{\mu} \psi D_\mu X^J [X^I,X^J, D^\nu X^K D_\nu X^K ] \\
&+ i \left( - g_{11} \right) \bar{\epsilon} \Gamma^I \Gamma^\mu \psi D^\nu X^J X^{IJK} ( \tilde{F}_{\mu \nu} X^K ) \Big\} \, .
\end{align}
To these terms we add zero in the following form
\begin{align}
0={\rm STr} \Big\{ &\left( - 2 \hat{\bold{d}} + 2 \hat{\bold{d}} \right) \bar{\epsilon} \Gamma^I \Gamma^\mu \psi D_\mu X^J D_\nu X^K D^\nu ( X^{IJK} ) \Big\} \\[10pt]
\nonumber
={\rm STr} \Big\{ &\left( - 2 \hat{\bold{d}} \right) \bar{\epsilon} \Gamma^I \Gamma^\mu \psi D_\mu X^J D_\nu X^K D^\nu ( X^{IJK} ) \\
\nonumber
& \left( + 2 \hat{\bold{d}} \right) \bar{\epsilon} \Gamma^I \Gamma^\mu \psi D_\mu X^J D_\nu X^K [ D^\nu X^I , X^J , X^K ] \\
\nonumber
& \left( + 2 \hat{\bold{d}} \right) \bar{\epsilon} \Gamma^I \Gamma^\mu \psi D_\mu X^J D_\nu X^K [ X^I , D^\nu X^J , X^K ] \\
& \left( + 2 \hat{\bold{d}} \right) \bar{\epsilon} \Gamma^I \Gamma^\mu \psi D_\mu X^J D_\nu X^K [ X^I , X^J , D^\nu X^K ] \Big\} \, .
\end{align}
We then get after reordering some blocks of terms
\begin{align}
\nonumber
{\rm STr} \Big\{ &+ i \left( + f_6 + s_9 + \hat{\bold{d}} - 2 \hat{\bold{g}} \right) \bar{\epsilon} \Gamma^I \Gamma^\mu D^\nu \psi D_\mu X^J D_\nu X^K X^{IJK} \\
\nonumber
&+ i \left( - f_6 + s_9 + 2 \hat{\bold{g}} \right) \bar{\epsilon} \Gamma^I \Gamma^\mu \psi D^\nu ( D_\mu X^J ) D_\nu X^K X^{IJK} \\
\nonumber
&+ i \left( - f_6 \right) \bar{\epsilon} \Gamma^I \Gamma^\mu \psi D_\mu X^J D^\nu ( D_\nu X^K ) X^{IJK} \\
\nonumber
&+ i \left( - f_6 + s_9 - \hat{\bold{d}} \right) \bar{\epsilon} \Gamma^I \Gamma^\mu \psi D_\mu X^J D_\nu X^K D^\nu ( X^{IJK} ) \\[6pt]
\nonumber
&+ i \left( + \sfrac{1}{2} f_2 - 4 \bold{a} + \hat{\bold{h}} - \hat{\bold{j}} \right) \bar{\epsilon} \Gamma^I \Gamma^\mu [ X^I , X^J , \psi ] D_\mu X^K D_\nu X^K D^\nu X^J \\
\nonumber
&+ i \left( - \sfrac{1}{2} f_2 - \hat{\bold{h}} \right) \bar{\epsilon} \Gamma^I \Gamma^\mu \psi [X^I,X^J, D_\mu X^K ] D^\nu X^K D_\nu X^J \\
\nonumber
&+ i \left( - \sfrac{1}{2} f_2 - 2 \hat{\bold{d}} - \hat{\bold{j}} \right) \bar{\epsilon} \Gamma^I \Gamma^{\mu} \psi D_\mu X^K [X^I,X^J, D^\nu X^K ] D_\nu X^J \\
\nonumber
&+ i \left( - \sfrac{1}{2} f_2 - g_2 \right) \bar{\epsilon} \Gamma^I \Gamma^\mu \psi D_\mu X^K D_\nu X^K [ X^I , X^J , D^\nu X^J ] \\[6pt]
\nonumber
&+ i \left( - \hat{\bold{h}} - 2 \hat{\bold{i}} + \hat{\bold{j}} \right) \bar{\epsilon} \Gamma^I \Gamma^\mu [X^J,X^{K},\psi] D^\nu X^I D_\mu X^J D_\nu X^K \\
\nonumber
&+ i \left( + 2 \hat{\bold{d}} + 2 \hat{\bold{i}} \right) \bar{\epsilon} \Gamma^I \Gamma^{\mu} \psi [X^J,X^K, D^\nu X^I ] D_\mu X^J D_\nu X^K \\
\nonumber
&+ i \left( + g_5 + \hat{\bold{h}} \right) \bar{\epsilon} \Gamma^I \Gamma^\mu \psi D^\nu X^I [X^J,X^K, D_\mu X^J ] D_\nu X^K \\
\nonumber
&+ i \left( - g_3 + \hat{\bold{j}} \right) \bar{\epsilon} \Gamma^I \Gamma^{\mu} \psi D^\nu X^I D_\mu X^J [X^J,X^K, D_\nu X^K ] \\[6pt]
\nonumber
&+ i \left( + g_4 + \hat{\bold{j}} \right) \bar{\epsilon} \Gamma^I \Gamma^{\mu} \psi [X^J,X^K, D^\nu X^J ] D_\mu X^I D_\nu X^K \\[6pt]
\nonumber
&+ i \left( + \sfrac{1}{2} f_3 - 4 \bold{b} + \hat{\bold{j}} \right) \bar{\epsilon} \Gamma^I \Gamma^\mu [ X^I , X^J , \psi ] D_\mu X^J D_\nu X^K D^\nu X^K \\
\nonumber
&+ i \left( - \sfrac{1}{2} f_3 - g_1 \right) \bar{\epsilon} \Gamma^I \Gamma^\mu \psi [ X^I ,  X^J , D_\mu X^J ] D_\nu X^K D^\nu X^K \\
\nonumber
&+ i \left( - \sfrac{1}{2} f_3 + \hat{\bold{d}} + \sfrac{1}{2} \hat{\bold{j}} \right) \bar{\epsilon} \Gamma^I \Gamma^{\mu} \psi D_\mu X^J [X^I,X^J, D^\nu X^K D_\nu X^K ] \\[6pt]
&+i \left( - g_{11} \right) \bar{\epsilon} \Gamma^I \Gamma^\mu \psi D^\nu X^J X^{IJK} ( \tilde{F}_{\mu \nu} X^K ) \Big\} \, .
\end{align}
These terms group into total derivatives or are zero by the gauge invariance condition \eqref{Gauge_Inv} with the exception of the final term which must be zero, hence $g_{11} = 0$. The first block of terms gives us
\begin{equation}
f_6 + s_9 + \hat{\bold{d}} - 2 \hat{\bold{g}} = - f_6 + s_9 + 2 \hat{\bold{g}} = - f_6 = - f_6 + s_9 - \hat{\bold{d}} \, .
\end{equation}
This is solved by $f_6 = - \sfrac{3}{2} \hat{\bold{d}}$, $s_9 = + \hat{\bold{d}}$ and $\hat{\bold{g}}= - \sfrac{1}{2} \hat{\bold{d}}$.
The second block gives
\begin{equation}
+ \sfrac{1}{2} f_2 - 4 \bold{a} + \hat{\bold{h}} - \hat{\bold{j}} = - \sfrac{1}{2} f_2 - \hat{\bold{h}} = - \sfrac{1}{2} f_2 - 2 \hat{\bold{d}} - \hat{\bold{j}} = - \sfrac{1}{2} f_2 - g_2 \, .
\end{equation}
We have previously found $\bold{a} = + \hat{\bold{d}}$, $f_2= + \hat{\bold{d}}$ and $g_2 = + \hat{\bold{d}}$ from $\tilde{\delta} \mathcal{L}_4$. Using these values enables us to identify $\hat{\bold{j}} = - \hat{\bold{d}}$ and $\hat{\bold{h}} = + \hat{\bold{d}}$, which agree with the analysis of earlier sections above.
The third block gives us the conditions
\begin{equation}
- \hat{\bold{h}} - 2 \hat{\bold{i}} + \hat{\bold{j}} = + 2 \hat{\bold{d}} + 2 \hat{\bold{i}} = + g_5 + \hat{\bold{h}} = - g_3 +\hat{\bold{j}} \, .
\end{equation}
Pleasingly these equations are satisfied with our previously discovered values for the coefficients and our new value, $\hat{\bold{h}} = + \hat{\bold{d}}$. The fourth and fifth blocks of terms also confirm values we have found earlier in our analysis.

\subsection{Two Derivative Terms - $\tilde{\delta} \mathcal{L}_2$}

The terms from \eqref{Varied_L_BLG} and \eqref{Varied_L_Higher} which contain a total of two derivatives are
\begin{align}
\nonumber
{\rm STr} \Big\{ &- i s_{10} \, \bar{\epsilon} \Gamma^I D^\mu ( \psi X^{JKL} X^{JKL} ) D_\mu X^I - i s_{11} \, \bar{\epsilon} \Gamma^I D^\mu ( \psi X^{JKL} X^{IJK} ) D_\mu X^I \\
\nonumber
&+ i g_9 \bar{\epsilon} \Gamma^{KLM} \Gamma^{\mu \nu} \psi [ X^I , X^J , D_\mu X^J ] D_\nu X^I X^{KLM} \\
\nonumber
&+ i g_{10} \bar{\epsilon} \Gamma^{JLM} \Gamma^{\mu \nu} \psi [ X^I , X^N , D_\mu X^N ] D_\nu X^J X^{ILM} \\
\nonumber
&+ i g_{11} \bar{\epsilon} \Gamma^M \Gamma^{\mu \nu} \psi [ X^I , X^L , D_\mu X^L ] D_\nu X^J X^{IJM} + i g_{12} \bar{\epsilon} \Gamma^{KLM} \psi [ X^I , X^N , D^\mu X^N ] D_\mu X^I X^{KLM} \\
\nonumber
&+ i g_{13} \bar{\epsilon} \Gamma^{JLM} \psi [ X^I , X^N , D^\mu X^N ] D_\mu X^J X^{ILM} + i g_{14} \bar{\epsilon} \Gamma^M \psi [ X^I , X^L , D^\mu X^L ] D_\mu X^J X^{IJM} \\
\nonumber
&- \sfrac{i}{2} s_1 \bar{\epsilon} \Gamma^{ILM} \Gamma^{\mu \nu} [ X^J , X^K , \psi D_\mu X^L D_\nu X^M ] X^{IJK} - \sfrac{i}{2} s_2 \bar{\epsilon} \Gamma^I \Gamma^{\mu \nu} [ X^J , X^K , \psi D_\mu X^L D_\nu X^I ] X^{JKL} \\
\nonumber
&- \sfrac{i}{2} s_3 \bar{\epsilon} \Gamma^I [ X^J , X^K , \psi D^\mu X^I D_\mu X^L ] X^{JKL} - \sfrac{i}{2} s_4 \bar{\epsilon} \Gamma^I [ X^J , X^K , \psi D^\mu X^L D_\mu X^L ] X^{IJK} \\
\nonumber
&- \sfrac{i}{2} f_9 \bar{\epsilon} \Gamma^I D^\mu \psi D_\mu X^I X^{JKL} X^{JKL} + \sfrac{i}{2} f_9 \bar{\epsilon} \Gamma^I \psi D^\mu ( D_\mu X^I X^{JKL} X^{JKL} ) \\
\nonumber
&+ \sfrac{i}{2} f_9 \bar{\epsilon} \Gamma^I \Gamma^{\mu \nu} D_\mu \psi D_\nu X^I X^{JKL} X^{JKL} - \sfrac{i}{2} f_9 \bar{\epsilon} \Gamma^I \Gamma^{\mu \nu} \psi D_\mu ( D_\nu X^I X^{JKL} X^{JKL} ) \\
\nonumber
&- \sfrac{i}{2} f_{10} \bar{\epsilon} \Gamma^I D^\mu \psi D_\mu X^J X^{JKL} X^{IKL} + \sfrac{i}{2} f_{10} \bar{\epsilon} \Gamma^I \psi D^\mu ( D_\mu X^J X^{JKL} X^{IKL} ) \\
\nonumber
&+ \sfrac{i}{2} f_{10} \bar{\epsilon} \Gamma^I \Gamma^{\mu \nu} D_\mu \psi D_\nu X^J X^{JKL} X^{IKL} - \sfrac{i}{2} f_{10} \bar{\epsilon} \Gamma^I \Gamma^{\mu \nu} \psi D_\mu ( D_\nu X^J X^{JKL} X^{IKL} ) \\
\nonumber
&+ \sfrac{i}{4} f_4 \, \bar{\epsilon} \Gamma^{KLMNO} \Gamma^{IJ} \Gamma^{\mu \nu} [ X^I , X^J , \psi ] D_\mu X^K D_\nu X^L X^{MNO} \\
\nonumber
&- \sfrac{i}{4} f_4 \bar{\epsilon} \Gamma^{KLMNO} \Gamma^{IJ} \Gamma^{\mu \nu} \psi [ X^I , X^J , D_\mu X^K D_\nu X^L X^{MNO} ] \\
\nonumber
&- \sfrac{i}{4} f_5 \, \bar{\epsilon} \Gamma^{KLM} \Gamma^{IJ} \Gamma^{\mu \nu} [ X^I , X^J , \psi ] D_\mu X^N D_\nu X^K X^{LMN} \\
\nonumber
&+ \sfrac{i}{4} f_5 \bar{\epsilon} \Gamma^{KLM} \Gamma^{IJ} \Gamma^{\mu \nu} \psi [ X^I , X^J , D_\mu X^N D_\nu X^K X^{LMN} ] \\
\nonumber
&+ \sfrac{i}{4} f_6 \, \bar{\epsilon} \Gamma^K \Gamma^{IJ} \Gamma^{\mu \nu} [ X^I , X^J , \psi ] D_\mu X^L D_\nu X^M X^{KLM} \\
\nonumber
&- \sfrac{i}{4} f_6 \bar{\epsilon} \Gamma^K \Gamma^{IJ} \Gamma^{\mu \nu} \psi [ X^I , X^J , D_\mu X^L D_\nu X^M X^{KLM} ] \\
\nonumber
&+ \sfrac{i}{4} f_7 \, \bar{\epsilon} \Gamma^{KLM} \Gamma^{IJ} [ X^I , X^J , \psi ] D^\mu X^N D_\mu X^N X^{KLM} \\
\nonumber
&- \sfrac{i}{4} f_7 \bar{\epsilon} \Gamma^{KLM} \Gamma^{IJ} \psi [ X^I , X^J , D^\mu X^N D_\mu X^N X^{KLM} ] \\
\nonumber
&+ \sfrac{i}{4} f_8 \, \bar{\epsilon} \Gamma^{KLM} \Gamma^{IJ} [ X^I , X^J , \psi ] D^\mu X^N D_\mu X^K X^{LMN} \\
\nonumber
&- \sfrac{i}{4} f_8 \bar{\epsilon} \Gamma^{KLM} \Gamma^{IJ} \psi [ X^I , X^J , D^\mu X^N D_\mu X^K X^{LMN} ] \\
\nonumber
&- \sfrac{i}{2} g_{15} \bar{\epsilon} \Gamma^I \Gamma^{\mu \nu} \psi X^{JKL} X^{JKL} ( \tilde{F}_{\mu \nu} X^I ) \\
\nonumber
&- 3 i \bold{c} \, \varepsilon^{\mu \nu \lambda} \, \bar{\epsilon} \Gamma^I \Gamma_\mu [ X^I , X^J , \psi ] X^{JKL} D_\nu X^K D_\lambda X^L \\
\nonumber
&+ 4 i \bold{d} \, \bar{\epsilon} \Gamma^I [ X^J , X^K , \psi ] X^{IJL} D^\mu X^K D_\mu X^L + 2 i \bold{d} \, \bar{\epsilon} \Gamma^I [ X^J , X^K , \psi ] X^{JKL} D^\mu X^I D_\mu X^L \\
\nonumber
&+ 2 i \bold{d} \, \bar{\epsilon} \Gamma^I D^\mu \psi X^{IJK} X^{JKL} D_\mu X^L + 6 i \bold{e} \, \bar{\epsilon} \Gamma^I [ X^J , X^K \psi ] X^{IJK} D^\mu X^L D_\mu X^L \\
\nonumber
&+ 2 i \bold{e} \, \bar{\epsilon} \Gamma^I D^\mu \psi X^{JKL} X^{JKL} D_\mu X^I \\
\nonumber
&- \sfrac{i}{6} \hat{\bold{f}} \, \bar{\epsilon} \Gamma^{MNO} \Gamma^{IJKL} D^\mu \psi X^{IJK} X^{MNO} D_\mu X^L - \sfrac{i}{6} \hat{\bold{f}} \, \bar{\epsilon} \Gamma^{MNO} \Gamma^{IJKL} \psi D^\mu ( X^{MNO} ) X^{IJK} D_\mu X^L \\
\nonumber
&- \sfrac{i}{6} \hat{\bold{g}} \, \bar{\epsilon} \Gamma^{LMN} \Gamma^{IJ} D^\mu \psi X^{IJK} X^{LMN} D_\mu X^K + \sfrac{i}{6} \hat{\bold{g}} \, \bar{\epsilon} \Gamma^{LMN} \Gamma^{IJ} \psi D^\mu ( X^{LMN} ) X^{IJK} D_\mu X^K \\
\nonumber
&- \sfrac{i}{6} \hat{\bold{h}} \, \bar{\epsilon} \Gamma^{LMN} \Gamma^{IJ} [X^J,X^K,\psi] X^{LMN} D^\mu X^I D_\mu X^K \\
\nonumber
&+ \sfrac{i}{6} \hat{\bold{h}} \, \bar{\epsilon} \Gamma^{LMN} \Gamma^{IJ} \psi [X^J ,X^K, X^{LMN} ] D^\mu X^I D_\mu X^K \\
\nonumber
&+ i \hat{\bold{k}} \, \bar{\epsilon} \Gamma^M \Gamma^{IJ} [X^K,X^{L},\psi] D^\mu X^M D_\mu X^I X^{JKL} + i \hat{\bold{k}} \, \bar{\epsilon} \Gamma^M \Gamma^{IJ} \psi [X^K,X^L, D^\mu X^M ] D_\mu X^I X^{JKL} \\
\nonumber
&+ i \hat{\bold{l}} \, \bar{\epsilon} \Gamma^I [X^J,X^K,\psi] D^\mu X^I D_\mu X^L X^{JKL} - i \hat{\bold{l}} \, \bar{\epsilon} \Gamma^I \psi [X^J,X^K, D^\mu X^I ] D_\mu X^L X^{JKL} \\
\nonumber
&+ i \hat{\bold{n}} \, \bar{\epsilon} \Gamma^M \Gamma^{IJ} [X^K,X^L,\psi] X^{IJK} D^\mu X^M D_\mu X^L + i \hat{\bold{n}} \, \bar{\epsilon} \Gamma^M \Gamma^{IJ} \psi [X^K,X^L, D^\mu X^M ] X^{IJK} D_\mu X^L \\
\nonumber
&- \sfrac{i}{6} \hat{\bold{i}} \, \bar{\epsilon} \Gamma^{KLM} \Gamma^{\mu \nu} [X^I,X^J,\psi] X^{KLM} D_\mu X^I D_\nu X^J \\
\nonumber
&+ \sfrac{i}{6} \hat{\bold{i}} \, \bar{\epsilon} \Gamma^{KLM} \Gamma^{\mu \nu} \psi [X^I,X^J, X^{KLM} ] D_\mu X^I D_\nu X^J \\
\nonumber
&- \sfrac{i}{6} \hat{\bold{j}} \, \bar{\epsilon} \Gamma^{LMN} \Gamma^{IJ} \Gamma^{\mu \nu} [X^J,X^K,\psi] X^{LMN} D_\mu X^I D_\nu X^K \\
\nonumber
&- \sfrac{i}{6} \hat{\bold{j}} \, \bar{\epsilon} \Gamma^{LMN} \Gamma^{IJ} \Gamma^{\mu \nu} \psi [X^J,X^K, X^{LMN} ] D_\mu X^I D_\nu X^K \\
\nonumber
&- i \hat{\bold{k}} \, \bar{\epsilon} \Gamma^M \Gamma^{IJ} \Gamma^{\mu \nu} [X^K,X^{L},\psi] D_\nu X^M D_\mu X^I X^{JKL} \\
\nonumber
&- i \hat{\bold{k}} \, \bar{\epsilon} \Gamma^M \Gamma^{IJ} \Gamma^{\mu \nu} \psi [X^K,X^L, D_\nu X^M ] D_\mu X^I X^{JKL} \\
\nonumber
&- i \hat{\bold{l}} \, \bar{\epsilon} \Gamma^I \Gamma^{\mu \nu} [X^J,X^K,\psi] D_\nu X^I D_\mu X^L X^{JKL} + i \hat{\bold{l}} \, \bar{\epsilon} \Gamma^I \Gamma^{\mu \nu} \psi [X^J,X^K, D_\nu X^I ] D_\mu X^L X^{JKL} \\
\nonumber
&- i \hat{\bold{n}} \, \bar{\epsilon} \Gamma^M \Gamma^{IJ} \Gamma^{\mu \nu} [X^K,X^L,\psi] X^{IJK} D_\nu X^M D_\mu X^L \\
&- i \hat{\bold{n}} \, \bar{\epsilon} \Gamma^M \Gamma^{IJ} \Gamma^{\mu \nu} \psi [X^K,X^L, D_\nu X^M ] X^{IJK} D_\mu X^L \Big\} \, .
\end{align}
Once expanded, the $\Gamma$-matrices have the following structure:
\begin{align}
\Gamma^{(7)} \Gamma^{\mu \nu} \, , \quad \Gamma^{(7)} \, , \quad \Gamma^{(5)} \Gamma^{\mu \nu} \, , \quad \Gamma^{(5)} \, , \quad \Gamma^{(3)} \Gamma^{\mu \nu} \, , \quad \Gamma^{(3)} \, , \quad \Gamma^{(1)} \Gamma^{\mu \nu} \, , \quad \Gamma^{(1)} \, .
\end{align}
Once again, we will deal with these in turn.

\subsubsection{$\Gamma^{(7)} \Gamma^{\mu \nu}$ terms}
The $\Gamma^{(7)} \Gamma^{\mu \nu}$ terms are
\begin{align}
\nonumber
{\rm STr} \Big\{ &+ i \left( + \sfrac{1}{4} f_4 \right) \bar{\epsilon} \Gamma^{IJKLMNO} \Gamma^{\mu \nu} [ X^I , X^J , \psi ] D_\mu X^K D_\nu X^L X^{MNO} \\
&+ i \left( - \sfrac{1}{4} f_4 \right) \bar{\epsilon} \Gamma^{IJKLMNO} \Gamma^{\mu \nu} \psi [ X^I , X^J , D_\mu X^K D_\nu X^L X^{MNO} ] \Big\} \, .
\end{align}
Using Eqs.\,\eqref{Useful_Id_5} and \eqref{Useful_Id_4} as well as the fundamental identity we can show that each of these terms is identically zero.

\subsubsection{$\Gamma^{(7)} $ terms}
Similarly, we find the $\Gamma^{(7)} $ terms are also zero
\begin{align}
\nonumber
0= {\rm STr} \Big\{ &+ i \left( - \sfrac{1}{6} \hat{\bold{f}} \right) \bar{\epsilon} \Gamma^{IJKLMNO} D^\mu \psi X^{IJK} X^{MNO} D_\mu X^L \\
&+ i \left( - \sfrac{1}{2} \hat{\bold{f}} \right) \bar{\epsilon} \Gamma^{IJKLMNO} \psi [ D^\mu X^M , X^N , X^O ] X^{IJK} D_\mu X^L \Big\} \, .
\end{align}
%

\subsubsection{$\Gamma^{(5)} \Gamma^{\mu \nu}$ terms}
The $\Gamma^{(5)} \Gamma^{\mu \nu}$ terms are (we have used $f_5=0$ from the $\Gamma^{(3)} \Gamma^{\mu\nu\lambda}$ section of $\tilde{\mathcal{L}}_3$)
\begin{align}
\nonumber
{\rm STr} \Big\{ &+ i \left( - \sfrac{1}{6} \hat{\bold{j}} - f_4 \right) \, \bar{\epsilon} \Gamma^{IJKLM} \Gamma^{\mu \nu} [X^I,X^N,\psi] X^{JKL} D_\mu X^M D_\nu X^N \\
\nonumber
&+ i \left( - \sfrac{3}{2} f_4 \right) \bar{\epsilon} \Gamma^{IJKLM} \Gamma^{\mu \nu} [ X^I , X^N , \psi ] X^{JKN} D_\mu X^L D_\nu X^M \\
\nonumber
&+ i \left( - \sfrac{1}{6} \hat{\bold{j}} + f_4 \right) \, \bar{\epsilon} \Gamma^{IJKLM} \Gamma^{\mu \nu} \psi [X^I,X^N, X^{JKL} ] D_\mu X^M D_\nu X^N \\
\nonumber
&+ i \left( + f_4 \right) \bar{\epsilon} \Gamma^{IJKLM} \Gamma^{\mu \nu} \psi X^{JKL} [ X^I , X^N , D_\mu X^M ] D_\nu X^N \\
\nonumber
&+ i \left( + f_4 \right) \bar{\epsilon} \Gamma^{IJKLM} \Gamma^{\mu \nu} \psi X^{JKL} D_\mu X^M [ X^I , X^N , D_\nu X^N ] \\
&+ i \left( + \sfrac{3}{2} f_4 \right) \bar{\epsilon} \Gamma^{IJKLM} \Gamma^{\mu \nu} \psi [ X^I , X^N , X^{JKN} D_\mu X^L D_\nu X^M ] \Big\} \, .
\end{align}
Using the identity \eqref{Useful_Id2} we can show 
\begin{align}
{\rm STr} \{ \bar{\epsilon} \Gamma^{IJKLM} \Gamma^{\mu \nu} [ X^I , X^N , \psi ] X^{JKN} D_\mu X^M D_\nu X^N \} =& 0 \, , \\[6pt]
{\rm STr} \{ \bar{\epsilon} \Gamma^{IJKLM} \Gamma^{\mu \nu} [ X^I , X^N , D_\mu X^L D_\nu X^M ] X^{JKN} \} =& 0 \, , \\[6pt]
{\rm STr} \{ \bar{\epsilon} \Gamma^{IJKLM} \Gamma^{\mu \nu} [X^I,X^N,\psi] X^{JKL} D_\mu X^M D_\nu X^N \} =& 0 \, , \\[6pt]
{\rm STr} \{ \bar{\epsilon} \Gamma^{IJKLM} \Gamma^{\mu \nu} \psi X^{JKL} [ X^I , X^N , D_\mu X^M ] D_\nu X^N \} =& 0 \, , \\[6pt]
{\rm STr} \{ \bar{\epsilon} \Gamma^{IJKLM} \Gamma^{\mu \nu} \psi X^{JKL} [ X^I , X^N , D_\nu X^N ] D_\mu X^M \} =& 0 \, .
\end{align}
It then follows by the gauge invariance condition that
\begin{align}
{\rm STr} \{ \bar{\epsilon} \Gamma^{IJKLM} \Gamma^{\mu \nu} \psi [ X^I , X^N , X^{JKN} ] D_\mu X^L D_\nu X^M \} =& 0 \, , \\[6pt]
{\rm STr} \{ \bar{\epsilon} \Gamma^{IJKLM} \Gamma^{\mu \nu} \psi [ X^I , X^N , X^{JKL} ] D_\mu X^M D_\nu X^N \} =& 0 \, .
\end{align}
Consequently all the $\Gamma^{(5)} \Gamma^{\mu \nu}$ terms are zero.

\subsubsection{$\Gamma^{(5)}$ terms}
The $\Gamma^{(5)}$ terms are 
\begin{align}
\nonumber
{\rm STr} \Big\{ &+ i \left( - \sfrac{3}{2} \hat{\bold{f}} \right) \bar{\epsilon} \Gamma^{IJKLM} D^\mu \psi X^{IJN} X^{LMN} D_\mu X^K \\
\nonumber
&+ i \left( - 3 \hat{\bold{f}} \right) \bar{\epsilon} \Gamma^{IJKLM} \psi [ D^\mu X^I , X^J , X^N ] X^{LMN} D_\mu X^K \\
\nonumber
&+ i \left( - \sfrac{1}{4} f_8 + \sfrac{1}{2} \hat{\bold{g}} \right) \bar{\epsilon} \Gamma^{IJKLM} \psi D^\mu X^N [ X^I , X^J , D_\mu X^K ] X^{LMN} \\
%
%
%
\nonumber
&+ i \left( - \sfrac{1}{4} f_8 - \sfrac{3}{2} \hat{\bold{f}} \right) \bar{\epsilon} \Gamma^{IJKLM} \psi [ X^I , X^J , D^\mu X^N ] X^{LMN} D_\mu X^K \\
\nonumber
&+ i \left( - \sfrac{1}{6} \hat{\bold{h}} \right) \bar{\epsilon} \Gamma^{IJKLM} [X^L,X^N,\psi] X^{IJM} D^\mu X^K D_\mu X^N \\
\nonumber
&+ i \left( + \sfrac{1}{6} \hat{\bold{h}} \right) \bar{\epsilon} \Gamma^{IJKLM} \psi [X^L ,X^N, X^{IJM} ] D^\mu X^K D_\mu X^N \\
\nonumber
&+ i \left( + \sfrac{1}{4} f_8 \right) \bar{\epsilon} \Gamma^{IJKLM} [ X^I , X^J , \psi ] D^\mu X^N D_\mu X^K X^{LMN} \\
&+ i \left( - \sfrac{1}{4} f_8 \right) \bar{\epsilon} \Gamma^{IJKLM} \psi D^\mu X^N D_\mu X^K [ X^I , X^J , X^{LMN} ] \Big\} \, .
\end{align}
Once again the identities \eqref{Useful_Id_5} and \eqref{Useful_Id_4} simplify matters because 
\begin{align}
- \sfrac{3i}{2} \hat{\bold{f}} \, &{\rm STr} \{ \bar{\epsilon} \Gamma^{IJKLM} D^\mu \psi X^{IJN} X^{LMN} D_\mu X^K \} = 0 \, , \\[10pt]
- 3 i \hat{\bold{f}} \, &{\rm STr} \{ \bar{\epsilon} \Gamma^{IJKLM} \psi [ D^\mu X^I , X^J , X^N ] X^{LMN} D_\mu X^K \} = 0 \, , \\[10pt]
+ i \left( - \sfrac{1}{4} f_8 + \sfrac{1}{2} \hat{\bold{g}} \right) &{\rm STr} \{ \bar{\epsilon} \Gamma^{IJKLM} \psi D^\mu X^N [ X^I , X^J , D_\mu X^K ] X^{LMN} \} = 0 \, , \\[10pt]
+ i \left( - \sfrac{1}{4} f_8 - \sfrac{3}{2} \hat{\bold{f}} \right) &{\rm STr} \{ \bar{\epsilon} \Gamma^{IJKLM} \psi [ X^I , X^J , D^\mu X^N ] X^{LMN} D_\mu X^K \} = 0 \, , \\[10pt]
- \sfrac{i}{6} \hat{\bold{h}} \, &{\rm STr} \{ \bar{\epsilon} \Gamma^{IJKLM} [X^L,X^N,\psi] X^{IJM} D^\mu X^K D_\mu X^N \} = 0 \, , \\[10pt]
+ \sfrac{i}{4} f_8 \, &{\rm STr} \{ \bar{\epsilon} \Gamma^{IJKLM} [ X^I , X^J , \psi ] D^\mu X^N D_\mu X^K X^{LMN} \} = 0 \, .
\end{align}
In particular we see that any multiple of ${\rm STr} \{ \bar{\epsilon} \Gamma^{IJKLM} [ X^I , X^J , \psi D^\mu X^N D_\mu X^K ] X^{LMN} \}$ is zero.
In addition the fundamental identity tells us
\begin{align}
\nonumber
+ \sfrac{i}{6} \hat{\bold{h}} \, {\rm STr} \{ \bar{\epsilon} \Gamma^{IJKLM} \psi [X^L ,X^N, X^{IJM} ] D^\mu X^K D_\mu X^N \} \\
= - \sfrac{i}{2} \hat{\bold{h}} \, {\rm STr} \{ \bar{\epsilon} \Gamma^{IJKLM} \psi &[X^I ,X^J, X^{LMN} ] D^\mu X^K D_\mu X^N \} \, .
\end{align}
Hence, we are left with
\begin{align}
\nonumber
{\rm STr} \Big\{ &+ i \left( - \sfrac{1}{4} f_8 - \sfrac{1}{2} \hat{\bold{h}} \right) \bar{\epsilon} \Gamma^{IJKLM} \psi D^\mu X^N D_\mu X^K [ X^I , X^J , X^{LMN} ] \Big\} \\
\nonumber
= {\rm STr} \Big\{ &+ i \left( - \sfrac{1}{4} f_8 - \sfrac{1}{2} \hat{\bold{h}} \right) \bar{\epsilon} \Gamma^{IJKLM} [ X^I , X^J , \psi D^\mu X^N D_\mu X^K ] X^{LMN}\\
&+ i \left( - \sfrac{1}{4} f_8 - \sfrac{1}{2} \hat{\bold{h}} \right) \bar{\epsilon} \Gamma^{IJKLM} \psi D^\mu X^N D_\mu X^K [ X^I , X^J , X^{LMN} ] \Big\} \, ,
\end{align}
which vanishes because of gauge invariance.

\subsubsection{$\Gamma^{(3)} \Gamma^{\mu \nu}$ terms}
The $\Gamma^{(3)} \Gamma^{\mu \nu}$ terms are 
\begin{align}
\nonumber
{\rm STr} \Big\{ &+ i \left( - \sfrac{3}{2} f_4 - \sfrac{1}{2} s_1 + \hat{\bold{k}} \right) \bar{\epsilon} \Gamma^{IJK} \Gamma^{\mu \nu} [ X^L , X^M , \psi ] D_\mu X^J D_\nu X^K X^{ILM} \\
\nonumber
&+ i \left( + \sfrac{3}{2} f_4 - \sfrac{1}{2} s_1 + \sfrac{1}{2} \hat{\bold{k}} \right) \bar{\epsilon} \Gamma^{IJK} \Gamma^{\mu \nu} \psi [ X^L , X^M , D_\mu X^J D_\nu X^K ] X^{ILM} \\
\nonumber
&+ i \left( + \sfrac{3}{2} f_4 \right) \bar{\epsilon} \Gamma^{IJK} \Gamma^{\mu \nu} \psi D_\mu X^J D_\nu X^K [ X^L , X^M , X^{ILM} ] \\[6pt]
\nonumber
&+ i \left( + 3 f_4 + \sfrac{1}{2} \hat{\bold{j}} + \hat{\bold{n}} \right) \bar{\epsilon} \Gamma^{IJK} \Gamma^{\mu \nu} [X^L,X^M,\psi] D_\mu X^K D_\nu X^M X^{IJL} \\
\nonumber
&+ i \left( - 3 f_4 + \hat{\bold{n}} \right) \bar{\epsilon} \Gamma^{IJK} \Gamma^{\mu \nu} \psi [X^L,X^M, D_\mu X^K D_\nu X^M ] X^{IJL} \\
\nonumber
&+ i \left( - 3 f_4 + \sfrac{1}{2} \hat{\bold{j}} \right) \bar{\epsilon} \Gamma^{IJK} \Gamma^{\mu \nu} \psi D_\mu X^K D_\nu X^M [ X^L , X^M , X^{IJL} ] \\[6pt]
\nonumber
&+ i \left( - \sfrac{1}{2} f_4 + \sfrac{1}{12} f_6 - \sfrac{1}{6} \hat{\bold{i}} - \sfrac{1}{6} \hat{\bold{j}} \right) \bar{\epsilon} \Gamma^{IJK} \Gamma^{\mu \nu} [X^L,X^M,\psi] D_\mu X^L D_\nu X^M X^{IJK} \\
\nonumber
&+ i \left( + \sfrac{1}{2} f_4 - \sfrac{1}{12} f_6 - \sfrac{1}{2} g_9 \right) \bar{\epsilon} \Gamma^{IJK} \Gamma^{\mu \nu} \psi [ X^L , X^M , D_\mu X^L D_\nu X^M ] X^{IJK} \\
\nonumber
&+ i \left( + \sfrac{1}{2} f_4 - \sfrac{1}{12} f_6 + \sfrac{1}{6} \hat{\bold{i}} - \sfrac{1}{6} \hat{\bold{j}} \right) \bar{\epsilon} \Gamma^{IJK} \Gamma^{\mu \nu} \psi D_\mu X^L D_\nu X^M [X^L,X^M, X^{IJK} ] \\[6pt]
%
%
%
%
%
&+ i \left( - g_{10} - \hat{\bold{n}} \right) \bar{\epsilon} \Gamma^{IJK} \Gamma^{\mu \nu} \psi [ X^L , X^M , D_\nu X^M ] D_\mu X^K X^{IJL} \Big\} \, , \label{L_2_trans_3_wvol_2}
\end{align}
where we have used
\begin{align}
\nonumber
+ \sfrac{i}{4} f_6 \, {\rm STr} \{ \bar{\epsilon} \Gamma^{IJK} \Gamma^{\mu \nu} &[ X^I , X^J , \psi ] D_\mu X^L D_\nu X^M X^{KLM} \} \\
=&+ \sfrac{i}{12} f_6 \, {\rm STr} \{ \bar{\epsilon} \Gamma^{IJK} \Gamma^{\mu \nu} [ X^L , X^M , \psi ] D_\mu X^L D_\nu X^M X^{IJK} \} \, , \\[10pt]
\nonumber
- \sfrac{i}{4} f_6 \, {\rm STr} \{ \bar{\epsilon} \Gamma^{IJK} \Gamma^{\mu \nu} \psi &[ X^I , X^J , D_\mu X^L D_\nu X^M ] X^{KLM} \} \\
=&- \sfrac{i}{12} \, f_6 {\rm STr} \{ \bar{\epsilon} \Gamma^{IJK} \Gamma^{\mu \nu} \psi [ X^L , X^M , D_\mu X^L D_\nu X^M ] X^{IJK} \} \, , \\[10pt]
\nonumber
- \sfrac{i}{4} f_6 \, {\rm STr} \{ \bar{\epsilon} \Gamma^{IJK} \Gamma^{\mu \nu} \psi D_\mu X^L &D_\nu X^M [ X^I , X^J , X^{KLM} ] \} \\
=& - \sfrac{i}{12} f_6 \, {\rm STr} \{  \bar{\epsilon} \Gamma^{IJK} \Gamma^{\mu \nu} \psi D_\mu X^L D_\nu X^M [ X^L , X^M , X^{IJK} ] \} \, , \\[10pt]
\nonumber
- \sfrac{i}{2} \hat{\bold{j}} \, {\rm STr} \{ \bar{\epsilon} \Gamma^{IJK} \Gamma^{\mu \nu} \psi D_\mu X^M &D_\nu X^L [X^I,X^L, X^{JKM} ] \} \\
=& - \sfrac{i}{6} \hat{\bold{j}} \, {\rm STr} \{ \bar{\epsilon} \Gamma^{IJK} \Gamma^{\mu \nu} \psi D_\mu X^L D_\nu X^M [X^L,X^M, X^{IJK} ] \} \, .
\end{align}
The first block of terms in \eqref{L_2_trans_3_wvol_2} is zero through gauge invariance if
\begin{align}
- \sfrac{3}{2} f_4 - \sfrac{1}{2} s_1 + \hat{\bold{k}} = + \sfrac{3}{2} f_4 - \sfrac{1}{2} s_1 + \sfrac{1}{2} \hat{\bold{k}} = + \sfrac{3}{2} f_4 \, ,
\end{align}
which is solved by $\hat{\bold{k}} = s_1 = + \sfrac{1}{2} \hat{\bold{d}}$ and $f_4 = + \sfrac{1}{6} s_1$. The second block is zero if
\begin{align}
+ 3 f_4 + \sfrac{1}{2} \hat{\bold{j}} + \hat{\bold{n}} = - 3 f_4 + \hat{\bold{n}} = - 3 f_4 + \sfrac{1}{2} \hat{\bold{j}} \, ,
\end{align}
and the third block is zero for
\begin{align}
- \sfrac{1}{2} f_4 + \sfrac{1}{12} f_6 - \sfrac{1}{6} \hat{\bold{i}} - \sfrac{1}{6} \hat{\bold{j}} = + \sfrac{1}{2} f_4 - \sfrac{1}{12} f_6 - \sfrac{1}{2} g_9 = + \sfrac{1}{2} f_4 - \sfrac{1}{12} f_6 + \sfrac{1}{6} \hat{\bold{i}} - \sfrac{1}{6} \hat{\bold{j}} \, .
\end{align}
The second block constraints are solved when $f_4 = - \sfrac{1}{12} \hat{\bold{j}}$ and $\hat{\bold{n}} = - 6 f_4 = - \sfrac{1}{2} \hat{\bold{d}}$. To solve the third block constraints we use the previously found value $f_6 = - \sfrac{3}{2} \hat{\bold{d}}$ from $\tilde{\delta} \mathcal{L}_3$ together with $f_4 = - \sfrac{1}{12} \hat{\bold{j}}$. We then find $g_9 = 0$ together with $f_4 = - \sfrac{1}{18} f_6$ and $\hat{\bold{i}} = \hat{\bold{j}}$.
The final term in \eqref{L_2_trans_3_wvol_2} is zero if $g_{10} =- \hat{\bold{n}}= + \sfrac{1}{2} \hat{\bold{d}}$.

\subsubsection{$\Gamma^{(3)}$ terms}
The $\Gamma^{(3)}$ terms are 
\begin{align}
\nonumber
{\rm STr} \Big\{ &+ i \left( + 3 \hat{\bold{f}} + \hat{\bold{g}} \right) \bar{\epsilon} \Gamma^{IJK} D^\mu \psi X^{IJL} X^{KLM} D_\mu X^M \\[6pt]
\nonumber
&+ i \left( + f_8 - \sfrac{1}{2} \hat{\bold{h}} - \hat{\bold{n}} \right) \bar{\epsilon} \Gamma^{IJK} [ X^I , X^L , \psi ] D^\mu X^K D_\mu X^M X^{JLM} \\
\nonumber
&+ i \left( - f_8 - 2 \hat{\bold{g}} - 2 \hat{\bold{n}} - 6 \hat{\bold{f}} \right) \bar{\epsilon} \Gamma^{IJK} \psi [ X^I , X^L , D^\mu X^K ] D_\mu X^M X^{JLM} \\
\nonumber
&+ i \left( - f_8 + 6 \hat{\bold{f}} \right) \bar{\epsilon} \Gamma^{IJK} \psi D^\mu X^K [ X^I , X^L , D_\mu X^M ] X^{JLM} \\
\nonumber
&+ i \left( - f_8 + \hat{\bold{h}} \right) \bar{\epsilon} \Gamma^{IJK} \psi D^\mu X^K D_\mu X^M [ X^I , X^L , X^{JLM} ] \\[6pt]
\nonumber
&+ i \left( - \sfrac{1}{2} f_8 - \sfrac{1}{2} \hat{\bold{h}} \right) \bar{\epsilon} \Gamma^{IJK} [ X^I , X^L , \psi ] D^\mu X^L D_\mu X^M X^{JKM} \\
\nonumber
&+ i \left( + \sfrac{1}{2} f_8 - \hat{\bold{h}} \right) \bar{\epsilon} \Gamma^{IJK} \psi [ X^I , X^L , D^\mu X^L ] D_\mu X^M X^{JKM} \\
\nonumber
&+ i \left( + \sfrac{1}{2} f_8 - \hat{\bold{h}} \right) \bar{\epsilon} \Gamma^{IJK} \psi D^\mu X^L [ X^I , X^L , D_\mu X^M ] X^{JKM} \\
\nonumber
&+ i \left( - \sfrac{1}{2} f_8 - \sfrac{1}{2} \hat{\bold{h}} \right)  \bar{\epsilon} \Gamma^{IJK} \psi D^\mu X^L D_\mu X^M [ X^I , X^L , X^{JKM} ] \\[6pt]
%
%
\nonumber
&+ i \left( - 3 g_{12} - 3 \hat{\bold{f}} - \hat{\bold{g}} \right) \bar{\epsilon} \Gamma^{IJK} \psi [ X^I , X^J , D^\mu X^L ] D_\mu X^M X^{KLM} \\
\nonumber
&+ i \left( + 3 \hat{\bold{f}} - \hat{\bold{k}} \right) \bar{\epsilon} \Gamma^{IJK} \psi [ X^L , X^M , D^\mu X^I ] D_\mu X^K X^{JLM} \\
\nonumber
&+ i \left( + g_{13} \right) \bar{\epsilon} \Gamma^{IJK} \psi [ X^L , X^M , D^\mu X^M ] D_\mu X^K X^{IJL} \\[6pt]
\nonumber
&+ i \left( - \sfrac{3}{2} f_7 \right) \bar{\epsilon} \Gamma^{IJK} [ X^I , X^L , \psi ] D^\mu X^M D_\mu X^M X^{JKL} \\
\nonumber
&+ i \left( + \sfrac{3}{2} f_7 + \sfrac{3}{2} \hat{\bold{f}} \right) \bar{\epsilon} \Gamma^{IJK} \psi [ X^I , X^L , D^\mu X^M D_\mu X^M ] X^{JKL} \\
&+ i \left( - \sfrac{3}{2} f_7 \right) \bar{\epsilon} \Gamma^{IJK} \psi D^\mu X^M D_\mu X^M [ X^I , X^L , X^{JKL} ] \Big\} \, , \label{L_2_trans_3_wvol_0}
\end{align}
where have used the following
\begin{align}
\nonumber
+ i \left( + \sfrac{1}{2} \hat{\bold{h}} + \hat{\bold{n}} \right) {\rm STr} \{ \bar{\epsilon} \Gamma^{IJK} &[X^L , X^M ,\psi] D^\mu X^K D_\mu X^M X^{IJL} \} \\
=& + i \left( - \sfrac{1}{2} \hat{\bold{h}} - \hat{\bold{n}} \right) {\rm STr} \{ \bar{\epsilon} \Gamma^{IJK} [X^I , X^L ,\psi] D^\mu X^K D_\mu X^M X^{JLM} \} \, , 
\end{align}
\begin{align}
\nonumber
- i g_{12} \, {\rm STr} \{ \bar{\epsilon} \Gamma^{IJK} \psi [ X^L , X^M , &D^\mu X^L ] D_\mu X^M X^{IJK} \\
=& - 3 i g_{12} \, {\rm STr} \{ \bar{\epsilon} \Gamma^{IJK} \psi [ X^I , X^J , D^\mu X^L ] D_\mu X^M X^{KLM} \} \, , \\[10pt]
%
%
\nonumber
- \sfrac{i}{2} \hat{\bold{h}} \, {\rm STr} \{ \bar{\epsilon} \Gamma^{IJK} \psi D^\mu X^K D_\mu X^M &[X^L ,X^M, X^{IJL} ] \} \\
=& + i \hat{\bold{h}} \, {\rm STr} \{ \bar{\epsilon} \Gamma^{IJK} \psi D^\mu X^K D_\mu X^M [X^I ,X^L, X^{JLM} ] \} \, , \\[10pt]
%
%
\nonumber
+ i \hat{\bold{n}} \, {\rm STr} \{ \bar{\epsilon} \Gamma^{IJK} \psi [X^L,X^M, D^\mu X^K ] &X^{IJL} D_\mu X^M \\
=& - 2 i \hat{\bold{n}} \, {\rm STr} \{  \bar{\epsilon} \Gamma^{IJK} \psi [X^I,X^L, D^\mu X^K ] D_\mu X^M X^{JLM} \} \, , \\[10pt]
%
%
\nonumber
+ 3 i \hat{\bold{f}} \, {\rm STr} \{ \bar{\epsilon} \Gamma^{IJK} \psi [ X^L , X^M , D^\mu X^I ] &D_\mu X^M X^{JKL} \\
=& - 6 i \hat{\bold{f}} \, {\rm STr} \{  \bar{\epsilon} \Gamma^{IJK} \psi [ X^I , X^L , D^\mu X^K ] D_\mu X^M X^{JLM} \} \, , \\[10pt]
\nonumber
+ i {\rm STr} \{ \bar{\epsilon} \Gamma^{IJK} \psi [ X^I , X^L , D^\mu X^L ] &D_\mu X^M X^{JKM} \} \\
=& - i {\rm STr} \{ \bar{\epsilon} \Gamma^{IJK} \psi D^\mu X^L [ X^I , X^L , D_\mu X^M ] X^{JKM} \} 
\end{align}
and
\begin{align}
+ \sfrac{3i}{2} f_7 \, {\rm STr} \{ \bar{\epsilon} \Gamma^{IJK} \psi D^\mu X^M D_\mu X^M [ X^I , X^L , X^{JKL} ] \} =& 0 \, , \\[10pt] 
%
%
%
+ i \left( + \sfrac{1}{2} f_8 + \sfrac{1}{2} \hat{\bold{h}} \right)  \bar{\epsilon} \Gamma^{IJK} \psi D^\mu X^L D_\mu X^M [ X^I , X^L , X^{JKM} ] =& 0 \, .
\end{align}
We know the values of all the coefficients appearing in \eqref{L_2_trans_3_wvol_0} from our work in previous sections. With these values we see that the $\Gamma^{(3)}$ terms are zero by the gauge invariance condition.

\subsubsection{$\Gamma^{(1)} \Gamma^{\mu \nu}$ terms}
The $\Gamma^{(1)} \Gamma^{\mu \nu}$ terms are
\begin{align}
\nonumber
{\rm STr} \Big\{ &+ i \left( - \sfrac{1}{2} f_6 - 3 \bold{c} \right) \bar{\epsilon} \Gamma^I \Gamma^{\mu \nu} [ X^I , X^J , \psi ] D_\mu X^K D_\nu X^L X^{JKL} \\
\nonumber
&+ i \left( + \sfrac{1}{2} f_6 - 2 f_{10} - \hat{\bold{n}} \right) \bar{\epsilon} \Gamma^I \Gamma^{\mu \nu} \psi [ X^I , X^J , D_\mu X^K D_\nu X^L ] X^{JKL} \\
\nonumber
&+ i \left( + \sfrac{1}{2} f_6 + \sfrac{1}{2} \hat{\bold{j}} \right) \bar{\epsilon} \Gamma^I \Gamma^{\mu \nu} \psi D_\mu X^K D_\nu X^L [ X^I , X^J , X^{JKL} ] \\[6pt]
%
%
\nonumber
&+ i \left( + \hat{\bold{j}} - 2 \hat{\bold{n}} \right) \bar{\epsilon} \Gamma^I \Gamma^{\mu \nu} [X^J,X^K,\psi] D_\mu X^K D_\nu X^L X^{IJL} \\
\nonumber
&+ i \left( + 2 f_{10} + g_{11} + 2 \hat{\bold{n}} \right) \bar{\epsilon} \Gamma^I \Gamma^{\mu \nu} \psi [ X^J , X^K , D_\mu X^K ] D_\nu X^L X^{IJL} \\[6pt]
%
%
%
%
\nonumber
&+ i \left( + \hat{\bold{k}} + \hat{\bold{l}} \right) \bar{\epsilon} \Gamma^I \Gamma^{\mu \nu} [X^K,X^L,\psi] D_\mu X^I D_\nu X^J X^{JKL} \\
\nonumber
&+ i \left( - f_{10} - \hat{\bold{l}} \right) \bar{\epsilon} \Gamma^I \Gamma^{\mu \nu} \psi [ X^K , X^L , D_\mu X^I ] D_\nu X^J X^{JKL} \\
\nonumber
&+ i \left( + 6 f_9 + \hat{\bold{k}} \right) \bar{\epsilon} \Gamma^I \Gamma^{\mu \nu} \psi D_\mu X^I [ X^K , X^L , D_\nu X^J ] X^{JKL} \\[6pt]
%
%
%
%
%
\nonumber
&+ i \left( - f_{10} + \hat{\bold{k}} \right) \bar{\epsilon} \Gamma^I \Gamma^{\mu \nu} \psi [ X^K, X^L , D_\mu X^J ] D_\nu X^J X^{IKL} \\[6pt]
\nonumber
%
%
%
%
\nonumber
&+ i \left( + \sfrac{1}{2} f_9 \right) \bar{\epsilon} \Gamma^I \Gamma^{\mu \nu} D_\mu \psi D_\nu X^I X^{JKL} X^{JKL} \\
\nonumber
&+ i \left( + \sfrac{1}{2} f_9 \right) \bar{\epsilon} \Gamma^I \Gamma^{\mu \nu} \psi D_\mu ( D_\nu X^I ) X^{JKL} X^{JKL} \\
\nonumber
&+ i \left( + \sfrac{1}{2} f_9 \right) \bar{\epsilon} \Gamma^I \Gamma^{\mu \nu} \psi D_\nu X^I D_\mu ( X^{JKL} X^{JKL} ) \\[6pt]
\nonumber
&+ i \left( + \sfrac{1}{2} f_{10} \right) \bar{\epsilon} \Gamma^I \Gamma^{\mu \nu} D_\mu \psi D_\nu X^J X^{JKL} X^{IKL} \\
\nonumber
&+ i \left( + \sfrac{1}{2} f_{10} \right) \bar{\epsilon} \Gamma^I \Gamma^{\mu \nu} \psi D_\mu ( D_\nu X^J ) X^{JKL} X^{IKL} \\
\nonumber
&+ i \left( + \sfrac{1}{2} f_{10} \right) \bar{\epsilon} \Gamma^I \Gamma^{\mu \nu} \psi D_\nu X^J D_\mu ( X^{JKL} X^{IKL} ) \\[6pt]
&+ i \left( - \sfrac{1}{2} f_9 - \sfrac{1}{6} f_{10} - \sfrac{1}{2} g_{15} \right) \bar{\epsilon} \Gamma^I \Gamma^{\mu \nu} \psi ( \tilde{F}_{\mu \nu} X^I ) X^{JKL} X^{JKL} \Big\} \, , \label{L_2_trans_1_wvol_2}
\end{align}
where we have used 
\begin{align}
\nonumber
+ i \hat{\bold{j}} \, {\rm STr} \{ \bar{\epsilon} \Gamma^I \Gamma^{\mu \nu} \psi D_\mu X^L D_\nu X^M &[X^M,X^J, X^{JLI} ] \} \\
=&+ \sfrac{i}{2} \hat{\bold{j}} \, {\rm STr} \{ \bar{\epsilon} \Gamma^I \Gamma^{\mu \nu} \psi D_\mu X^K D_\nu X^L [ X^I , X^J , X^{JKL} ] \} \, ,
\end{align}
\begin{align}
\nonumber
+ i \left( - 2 f_{10} - 2 \hat{\bold{n}} \right) {\rm STr} \{ \bar{\epsilon} \Gamma^I \Gamma^{\mu \nu} &\psi [ X^J , X^K , D_\mu X^L ] D_\nu X^J X^{IKL} \} \\
\nonumber
=&+ i \left( - 2 f_{10} - 2 \hat{\bold{n}} \right) {\rm STr} \{ \bar{\epsilon} \Gamma^I \Gamma^{\mu \nu} \psi [ X^K , X^L , D_\mu X^L ] D_\nu X^J X^{IJK} \} \\
&+ i \left( - f_{10} - \hat{\bold{n}} \right) {\rm STr} \{ \bar{\epsilon} \Gamma^I \Gamma^{\mu \nu} \psi [ X^I , X^J , D_\mu X^K D_\nu X^L ] X^{JKL} \} \, ,
\end{align}
and
\begin{align}
{\rm STr} \{ \bar{\epsilon} \Gamma^I \Gamma^{\mu \nu} \psi ( \tilde{F}_{\mu \nu} X^J ) X^{JKL} X^{IKL} \} = \sfrac{1}{3} {\rm STr} \{ \bar{\epsilon} \Gamma^I \Gamma^{\mu \nu} \psi ( \tilde{F}_{\mu \nu} X^I ) X^{JKL} X^{JKL} \} \, .
\end{align}
The terms in \eqref{L_2_trans_1_wvol_2} are invariant (up to total derivatives) if
\begin{equation}
- \sfrac{1}{2} f_6 - 3 \bold{c} = + \sfrac{1}{2} f_6 - 2 f_{10} - \hat{\bold{n}} = + \sfrac{1}{2} f_6 + \sfrac{1}{2} \hat{\bold{j}} \, ,
\end{equation}
\begin{equation}
0 = + \hat{\bold{j}} - 2 \hat{\bold{n}} = + 2 f_{10} + g_{11} + 2 \hat{\bold{n}} \, ,
\end{equation}
\begin{equation}
0 = + \hat{\bold{k}} + \hat{\bold{l}} = - f_{10} - \hat{\bold{l}} = + 6 f_9 + \hat{\bold{k}} \, , 
\end{equation}
\begin{equation}
0= - f_{10} + \hat{\bold{k}} \, ,
\end{equation}
\begin{equation}
0 = - \sfrac{1}{2} f_9 - \sfrac{1}{6} f_{10} - \sfrac{1}{2} g_{15}  \, .
\end{equation}
We have previously found values for $f_6$, $\bold{c}$, $\hat{\bold{j}}$, $\hat{\bold{k}}$ and $\hat{\bold{n}}$. Using these values here gives 
\begin{align}
\hat{\bold{l}} =& - \hat{\bold{k}}  = - \sfrac{1}{2} \hat{\bold{d}} \, , \\[10pt]
f_{10} =& + \hat{\bold{k}}  = + \sfrac{1}{2} \hat{\bold{d}} \, , \\[10pt]
f_9 =& - \sfrac{1}{6} f_{10} = - \sfrac{1}{12} \hat{\bold{d}} \, , \\[10pt]
g_{11} =& \ 0 \, , \\[10pt]
g_{15} =& - f_9 - \sfrac{1}{3} f_{10} = - \sfrac{1}{12} \hat{\bold{d}} \, .
\end{align}
%

\subsubsection{$\Gamma^{(1)}$ terms}
The $\Gamma^{(1)}$ terms are 
\begin{align}
\nonumber
{\rm STr} \Big\{ &+ i \left( - \sfrac{3}{2} f_7 - \sfrac{1}{2} s_4 + 6 \bold{e} + \hat{\bold{k}} \right) \bar{\epsilon} \Gamma^I [ X^J , X^K , \psi ] D^\mu X^L D_\mu X^L X^{IJK} \\
\nonumber
&+ i \left( + \sfrac{3}{2} f_7 - \sfrac{1}{2} s_4 - \sfrac{3}{2} \hat{\bold{f}} + \sfrac{1}{2} \hat{\bold{g}} + \sfrac{1}{2} \hat{\bold{k}} \right) \bar{\epsilon} \Gamma^I \psi [ X^J , X^K , D^\mu X^L D_\mu X^L ] X^{IJK} \\
\nonumber
&+ i \left( + \sfrac{3}{2} f_7 \right) \bar{\epsilon} \Gamma^I \psi D^\mu X^L D_\mu X^L [ X^J , X^K , X^{IJK} ] \\[6pt]
\nonumber
&+ i \left( - \sfrac{1}{2} f_8 - \sfrac{1}{2} s_3 + 2 \bold{d} - \hat{\bold{k}} + \hat{\bold{l}} \right) \bar{\epsilon} \Gamma^I [ X^J , X^K , \psi ] D^\mu X^I D_\mu X^L X^{JKL} \\
\nonumber
&+ i \left( + \sfrac{1}{2} f_8 - \sfrac{1}{2} s_3 - \hat{\bold{l}} \right) \bar{\epsilon} \Gamma^I \psi [ X^J , X^K , D^\mu X^I ] D_\mu X^L X^{JKL} \\
\nonumber
&+ i \left( + \sfrac{1}{2} f_8 - \sfrac{1}{2} s_3 - \hat{\bold{l}} \right) \bar{\epsilon} \Gamma^I \psi D^\mu X^I [ X^J , X^K , D_\mu X^L ] X^{JKL} \\
\nonumber
&+ i \left( + \sfrac{1}{2} f_8 \right) \bar{\epsilon} \Gamma^I \psi D^\mu X^I D_\mu X^L [ X^J , X^K , X^{JKL} ] \\[6pt]
\nonumber
&+ i \left( - f_8 + 4 \bold{d} - \hat{\bold{h}} + 2 \hat{\bold{n}} \right) \bar{\epsilon} \Gamma^I [ X^J , X^K , \psi ] D^\mu X^K D_\mu X^L X^{IJL} \\
\nonumber
&+ i \left( + f_8 + g_{14} \right) \bar{\epsilon} \Gamma^I \psi [ X^J , X^K , D^\mu X^K ] D_\mu X^L X^{IJL} \\
\nonumber
&+ i \left( + f_8 + 6 \hat{\bold{f}} - 2 \hat{\bold{g}} + 2 \hat{\bold{n}} \right) \bar{\epsilon} \Gamma^I \psi D_\mu X^K [X^J , X^K , D^\mu X^L ] X^{IJL} \\
\nonumber
&+ i \left( + f_8 + \hat{\bold{h}} \right) \bar{\epsilon} \Gamma^I \psi D^\mu X^K D_\mu X^L [X^J ,X^K, X^{IJL} ] \\[6pt]
\nonumber
&+ i \left( - \sfrac{1}{2} f_{10} - s_{11} + 2 \bold{d} \right) \bar{\epsilon} \Gamma^I D^\mu \psi X^{IJK} X^{JKL} D_\mu X^L \\
\nonumber
&+ i \left( + \sfrac{1}{2} f_{10} - s_{11} - \hat{\bold{g}} \right) \bar{\epsilon} \Gamma^I \psi D^\mu ( X^{JKL} ) X^{IJK} D_\mu X^L \\
\nonumber
&+ i \left( + \sfrac{1}{2} f_{10} - s_{11} - \hat{\bold{g}} \right) \bar{\epsilon} \Gamma^I \psi X^{JKL} D^\mu ( X^{IJK} ) D_\mu X^L \\
\nonumber
&+ i \left( + \sfrac{1}{2} f_{10} \right) \bar{\epsilon} \Gamma^I \psi X^{JKL} X^{IJK} D^\mu ( D_\mu X^L ) \\[6pt]
\nonumber
&+ i \left( - \sfrac{1}{2} f_9 - s_{10} + 2 \bold{e} + \hat{\bold{f}} \right) \bar{\epsilon} \Gamma^I D^\mu \psi X^{JKL} X^{JKL} D_\mu X^I \\
\nonumber
&+ i \left( + \sfrac{1}{2} f_9 - s_{10} + \sfrac{1}{2} \hat{\bold{f}} - \sfrac{1}{6} \hat{\bold{k}} + \sfrac{1}{6} \hat{\bold{l}} \right) \bar{\epsilon} \Gamma^I \psi D^\mu ( X^{JKL} X^{JKL} ) D_\mu X^I \\
&+ i \left( + \sfrac{1}{2} f_9 \right) \bar{\epsilon} \Gamma^I \psi X^{JKL} X^{JKL} D^\mu ( D_\mu X^I ) \Big\} \, .
\end{align}
Therefore for invariance we require
\begin{equation}
- \sfrac{3}{2} f_7 - \sfrac{1}{2} s_4 + 6 \bold{e} + \hat{\bold{k}}  =  + \sfrac{3}{2} f_7 - \sfrac{1}{2} s_4 - \sfrac{3}{2} \hat{\bold{f}} + \sfrac{1}{2} \hat{\bold{g}} + \sfrac{1}{2} \hat{\bold{k}} =  + \sfrac{3}{2} f_7 \, ,
\end{equation}
\begin{equation}
- \sfrac{1}{2} f_8 - \sfrac{1}{2} s_3 + 2 \bold{d} - \hat{\bold{k}} + \hat{\bold{l}}  =  + \sfrac{1}{2} f_8 - \sfrac{1}{2} s_3 - \hat{\bold{l}}  =  + \sfrac{1}{2} f_8 \, ,
\end{equation}
\begin{equation}
- f_8 + 4 \bold{d} - \hat{\bold{h}} + 2 \hat{\bold{n}}  =  + f_8 + g_{14}  =  + f_8 + 6 \hat{\bold{f}} - 2 \hat{\bold{g}}  + 2 \hat{\bold{n}} =  + f_8 + \hat{\bold{h}} \, ,
\end{equation}
\begin{equation}
- \sfrac{1}{2} f_{10} - s_{11} + 2 \bold{d}  =  + \sfrac{1}{2} f_{10} - s_{11} - \hat{\bold{g}}  =  + \sfrac{1}{2} f_{10} \, ,
\end{equation}
\begin{equation}
- \sfrac{1}{2} f_9 - s_{10} + 2 \bold{e} + \hat{\bold{f}}  =  + \sfrac{1}{2} f_9 - s_{10} + \sfrac{1}{2} \hat{\bold{f}} - \sfrac{1}{6} \hat{\bold{k}} + \sfrac{1}{6} \hat{\bold{l}}  =  + \sfrac{1}{2} f_9 \, .
\end{equation}
Using the values we have previously found for $f_7, f_8, s_3 , s_4, g_{14}, \hat{\bold{f}} , \hat{\bold{g}} , \hat{\bold{h}} , \hat{\bold{k}} , \hat{\bold{l}}$ and $\hat{\bold{n}}$ we can solve the equations above to find
\begin{equation}
\bold{d} = + \hat{\bold{d}} \, , \quad \bold{e} = - \sfrac{1}{6} \hat{\bold{d}} \, , \quad s_{10} = - \sfrac{1}{12} \hat{\bold{d}} \, , \quad s_{11} = + \sfrac{1}{2} \hat{\bold{d}} \, , \quad f_9 = - \sfrac{1}{12} \hat{\bold{d}} \, , \quad f_{10} = + \sfrac{1}{2} \hat{\bold{d}} \, .
\end{equation}

\subsection{One Derivative Terms - $\tilde{\delta} \mathcal{L}_1$}
The terms from \eqref{Varied_L_BLG} and \eqref{Varied_L_Higher} which contain a single derivative are
\begin{align}
\nonumber
{\rm STr} \Big\{ &- i g_{15} \bar{\epsilon} \Gamma^{I} \Gamma_{\mu} \psi X^{JKL} X^{JKL} [ X^I , X^M , D^\mu X^M ] \\
\nonumber
&- \sfrac{i}{2} s_5 \bar{\epsilon} \Gamma^{IJKLM} \Gamma^{\mu} [ \psi D_{\mu} X^{J} X^{KLM} , X^O , X^P ] X^{IOP} \\
\nonumber
&- \sfrac{i}{2} s_6 \bar{\epsilon} \Gamma^{KLM} \Gamma^{\mu} [ \psi D_{\mu} X^{I} X^{KLM} , X^O , X^P ] X^{IOP} \\
\nonumber
&- \sfrac{i}{2} s_7 \bar{\epsilon} \Gamma^{JLM} \Gamma^{\mu} [ \psi D_{\mu} X^{J} X^{ILM} , X^O , X^P ] X^{IOP} \\
\nonumber
&- \sfrac{i}{2} s_8 \bar{\epsilon} \Gamma^{ILM} \Gamma^{\mu} [ \psi D_{\mu} X^{J} X^{JLM} , X^O , X^P ] X^{IOP} \\
\nonumber
&- \sfrac{i}{2} s_9 \bar{\epsilon} \Gamma^{M} \Gamma^{\mu} [ \psi D_{\mu} X^{J} X^{IJM} , X^O , X^P ] X^{IOP} \\
\nonumber
&+ \sfrac{i}{2} f_{11} \bar{\epsilon} \Gamma^{NOP} \Gamma^\mu D_\mu \psi X^{JKL} X^{JKL} X^{NOP} \\
\nonumber
&- \sfrac{i}{2} f_{11} \bar{\epsilon} \Gamma^{NOP} \Gamma^\mu \psi D_\mu ( X^{JKL} X^{JKL} X^{NOP} ) \\
\nonumber
&- \sfrac{i}{4} f_9 \bar{\epsilon} \Gamma^J \Gamma^{OP} \Gamma^\mu [ X^O , X^P , \psi ] D_\mu X^J X^{KLM} X^{KLM} \\
\nonumber
&+ \sfrac{i}{4} f_9 \bar{\epsilon} \Gamma^J \Gamma^{OP} \Gamma^\mu \psi [ X^O , X^P , D_\mu X^J X^{KLM} X^{KLM} ] \\
\nonumber
&- \sfrac{i}{4} f_{10} \bar{\epsilon} \Gamma^M \Gamma^{OP} \Gamma^\mu [ X^O , X^P , \psi ] D_\mu X^J X^{JKL} X^{KLM} \\
\nonumber
&+ \sfrac{i}{4} f_{10} \bar{\epsilon} \Gamma^M \Gamma^{OP} \Gamma^\mu \psi [ X^O , X^P , D_\mu X^J X^{JKL} X^{KLM} ] \\ 
\nonumber
&+ 2i \bold{d} \, \bar{\epsilon} \Gamma^N \Gamma_\mu [ X^K , X^N , \psi ] D^\mu X^L  X^{IJK} X^{IJL} \\
\nonumber 
&+ 2i \bold{e} \, \bar{\epsilon} \Gamma^N \Gamma_\mu [ X^L , X^N , \psi ] D^\mu X^L X^{IJK} X^{IJK} \\
\nonumber
&- \sfrac{i}{6} \hat{\bold{k}} \, \bar{\epsilon} \Gamma^{NOP} \Gamma^{IJ} \Gamma_\mu [X^K,X^{L},\psi] D^\mu X^{I} X^{NOP} X^{JKL} \\
\nonumber
&- \sfrac{i}{6} \hat{\bold{k}} \, \bar{\epsilon} \Gamma^{NOP} \Gamma^{IJ} \Gamma_\mu \psi [X^K,X^{L}, X^{NOP} ] {D}^\mu X^{I} X^{JKL} \\
\nonumber
&- \sfrac{i}{6} \hat{\bold{l}} \, \bar{\epsilon} \Gamma^{NOP} \Gamma_\mu [X^I,X^{J},\psi] {D}^\mu X^{K} X^{NOP} X^{IJK} \\
\nonumber
&+ \sfrac{i}{6} \hat{\bold{l}} \, \bar{\epsilon} \Gamma^{NOP} \Gamma_\mu \psi [X^I,X^{J}, X^{NOP} ] {D}^\mu X^{K}X^{IJK} \\
\nonumber
&- \sfrac{i}{6} \hat{\bold{n}} \, \bar{\epsilon} \Gamma^{NOP} \Gamma^{IJ} \Gamma_\mu [X^K,X^{L},\psi] {D}^\mu X^L X^{NOP} X^{IJK}\\
\nonumber
&- \sfrac{i}{6} \hat{\bold{n}} \, \bar{\epsilon} \Gamma^{NOP} \Gamma^{IJ} \Gamma_\mu \psi [X^K,X^{L}, X^{NOP} ] {D}^\mu X^L X^{IJK}\\
\nonumber
&+ i \hat{\bold{p}} \, \bar{\epsilon} \Gamma^N \Gamma^{IJ} \Gamma_\mu [X^K,X^{L},\psi] D^\mu X^N X^{IJM}X^{KLM} \\
&- i \hat{\bold{p}} \, \bar{\epsilon} \Gamma^N \Gamma^{IJ} \Gamma_\mu \psi [X^K,X^{L}, D^\mu X^N ] X^{IJM}X^{KLM} \Big\} \, .
\end{align}
Once expanded, the $\Gamma$-matrices have the following structure:
\begin{align}
\Gamma^{(5)} \Gamma^{\mu} \, , \qquad \Gamma^{(3)} \Gamma^{\mu} \, , \qquad \Gamma^{(1)} \Gamma^{\mu} \, .
\end{align}
As before, we will deal with these in turn.
%
\subsubsection{$\Gamma^{(5)} \Gamma^\mu$ terms}
The terms with five transverse $\Gamma$-matrix indices are
\begin{align}
\nonumber
{\rm STr} \Big\{ &+ i \left( - \sfrac{1}{2} s_5 + \sfrac{1}{6} \hat{\bold{k}} \right) \bar{\epsilon} \Gamma^{IJKLM} \Gamma^{\mu} [X^N,X^O,\psi] D_{\mu} X^{I} X^{JKL} X^{MNO} \\
\nonumber
&+ i \left( - \sfrac{1}{2} s_5 \right) \bar{\epsilon} \Gamma^{IJKLM} \Gamma^{\mu} \psi [ X^N , X^O , D_{\mu} X^{I} ] X^{JKL} X^{MNO} \\
\nonumber
&+ i \left( - \sfrac{1}{2} s_5 + \sfrac{1}{6} \hat{\bold{k}} \right) \bar{\epsilon} \Gamma^{IJKLM} \Gamma^{\mu} \psi [X^N,X^O, X^{JKL} ] X^{MNO} D_{\mu} X^{I} \\
%
%
\nonumber
&+ i \left( + \sfrac{1}{6} \hat{\bold{n}} \right) \bar{\epsilon} \Gamma^{IJKLM} \Gamma_\mu [X^N,X^O,\psi] {D}^\mu X^N X^{IJK} X^{LMO} \\
&+ i \left( + \sfrac{1}{6} \hat{\bold{n}} \right) \bar{\epsilon} \Gamma^{IJKLM} \Gamma_\mu \psi [X^N,X^O, X^{IJK} ] {D}^\mu X^N X^{LMO} \Big\} \, .
\end{align}
We note that terms which contain $\Gamma^{IJKLM} X^{JKL} X^{MNO}$ or $\Gamma^{IJKLM} X^{IJK} X^{LMO}$ are zero because of \eref{Useful_Id_4} and so we are left with
\begin{align}
\nonumber
{\rm STr} \Big\{ &+ i \left( - \sfrac{1}{2} s_5 + \sfrac{1}{6} \hat{\bold{k}} \right) \bar{\epsilon} \Gamma^{IJKLM} \Gamma^{\mu} \psi [ X^N , X^O , X^{JKL} ] D_{\mu} X^{I} X^{MNO} \\
&+ i \left( + \sfrac{1}{6} \hat{\bold{n}} \right) \bar{\epsilon} \Gamma^{IJKLM} \Gamma^\mu \psi [X^N,X^O, X^{IJK} ] {D}_\mu X^N X^{LMO} \Big\} \, .
\end{align}
Taking $s_5 = + \sfrac{1}{3} \hat{\bold{k}} = + \sfrac{1}{6} \hat{\bold{d}}$ removes the first term. The second term is identically zero as we will now show.
As terms with $\Gamma^{IJKLM} X^{IJK} X^{LMO}$ are zero we have by gauge invariance
\begin{align}
\nonumber
0 =&+ {\rm STr} \Big\{ \bar{\epsilon} \Gamma^{IJKLM} \Gamma_\mu [X^N,X^O, \psi ] {D}^\mu X^N X^{IJK} X^{LMO} \Big\} \\
\nonumber
&+ {\rm STr} \Big\{ \bar{\epsilon} \Gamma^{IJKLM} \Gamma_\mu \psi [X^N,X^O, {D}^\mu X^N ] X^{IJK} X^{LMO} \Big\} \\
\nonumber
&+ {\rm STr} \Big\{ \bar{\epsilon} \Gamma^{IJKLM} \Gamma_\mu \psi {D}^\mu X^N [X^N,X^O, X^{IJK} ] X^{LMO} \Big\} \\
&+ {\rm STr} \Big\{ \bar{\epsilon} \Gamma^{IJKLM} \Gamma_\mu \psi {D}^\mu X^N X^{IJK} [X^N,X^O, X^{LMO} ] \Big\} \\[10pt]
\nonumber
=&+ {\rm STr} \Big\{ \bar{\epsilon} \Gamma^{IJKLM} \Gamma_\mu \psi {D}^\mu X^N [X^N,X^O, X^{IJK} ] X^{LMO} \Big\} \\
&+ {\rm STr} \Big\{ \bar{\epsilon} \Gamma^{IJKLM} \Gamma_\mu \psi {D}^\mu X^N X^{IJK} [X^N,X^O, X^{LMO} ] \Big\} \, . \label{TotalDeriv}
\end{align}
However, using the fundamental identity we can show that
\begin{align}
\nonumber
+ {\rm STr} \Big\{ \bar{\epsilon} \Gamma^{IJKLM} \Gamma_\mu \psi {D}^\mu X^N &X^{IJK} [X^N,X^O, X^{LMO} ] \Big\} \\
&= - 2 {\rm STr} \Big\{ \bar{\epsilon} \Gamma^{IJKLM} \Gamma_\mu \psi {D}^\mu X^N X^{IJK} [X^L,X^O, X^{MNO} ] \Big\} \, .
\end{align}
For the other term we use the fundamental identity and then \eref{Useful_Id2} to get
\begin{align}
\nonumber
+ {\rm STr} \Big\{ \bar{\epsilon} \Gamma^{IJKLM} \Gamma_\mu \psi {D}^\mu X^N [X^N,&X^O, X^{IJK} ] X^{LMO} \Big\} \\
=& + 3 {\rm STr} \Big\{ \bar{\epsilon} \Gamma^{IJKLM} \Gamma_\mu \psi {D}^\mu X^N [X^I,X^J, X^{KNO} ] X^{LMO} \Big\} \\[10pt]
=& + {\rm STr} \Big\{ \bar{\epsilon} \Gamma^{IJKLM} \Gamma_\mu \psi {D}^\mu X^N [X^L,X^O, X^{MNO} ] X^{IJK} \Big\} \, .
\end{align}
Hence we see that \eqref{TotalDeriv} is
\begin{align}
\nonumber
0=&+ {\rm STr} \Big\{ \bar{\epsilon} \Gamma^{IJKLM} \Gamma_\mu \psi {D}^\mu X^N [X^L,X^O, X^{MNO} ] X^{IJK} \Big\} \\
&- 2 {\rm STr} \Big\{ \bar{\epsilon} \Gamma^{IJKLM} \Gamma_\mu \psi {D}^\mu X^N X^{IJK} [X^L,X^O, X^{MNO} ] \Big\} \\[10pt]
=&- {\rm STr} \Big\{ \bar{\epsilon} \Gamma^{IJKLM} \Gamma_\mu \psi {D}^\mu X^N [X^L,X^O, X^{MNO} ] X^{IJK} \Big\}  \\[10pt]
=&- {\rm STr} \Big\{ \bar{\epsilon} \Gamma^{IJKLM} \Gamma_\mu \psi {D}^\mu X^N [X^N,X^O, X^{IJK} ] X^{LMO} \Big\} \, .
\end{align}
Therefore
\begin{align}
{\rm STr} \Big\{ + \sfrac{i}{6} \hat{\bold{n}} \, \bar{\epsilon} \Gamma^{IJKLM} \Gamma^\mu \psi [X^N,X^O, X^{IJK} ] {D}_\mu X^N X^{LMO} \Big\} = 0 \, .
\end{align}

\subsubsection{$\Gamma^{(3)} \Gamma^\mu$ terms}

The terms with three transverse $\Gamma$-matrix indices are
\begin{align}
\nonumber
{\rm STr} \Big\{ &+ i \left( + \sfrac{1}{2} f_{11} \right) \bar{\epsilon} \Gamma^{IJK} \Gamma^\mu D_\mu \psi X^{LMN} X^{LMN} X^{IJK} \\
\nonumber
&+ i \left( - 3 f_{11} \right) \bar{\epsilon} \Gamma^{IJK} \Gamma^\mu \psi X^{IJK} X^{LMN} [ X^M , X^N , D_\mu X^L ] \\
\nonumber
&+ i \left( + \sfrac{1}{2} f_{11} \right) \bar{\epsilon} \Gamma^{IJK} \Gamma^\mu \psi D_\mu ( X^{IJK} ) X^{LMN} X^{LMN} \\[6pt]
\nonumber
&+ i \left( - \sfrac{1}{4} f_9 \right) \bar{\epsilon} \Gamma^{IJK} \Gamma^\mu [ X^I , X^J , \psi ] D_\mu X^K X^{LMN} X^{LMN} \\
\nonumber
&+ i \left( + \sfrac{1}{4} f_9 - 3 f_{11} \right) \bar{\epsilon} \Gamma^{IJK} \Gamma^\mu \psi [ X^I , X^J ,D_\mu X^K ]  X^{LMN} X^{LMN} \\
\nonumber
&+ i \left( + \sfrac{1}{4} f_9 \right) \bar{\epsilon} \Gamma^{IJK} \Gamma^\mu \psi D_\mu X^K [ X^I , X^J , X^{LMN} X^{LMN} ] \\[6pt]
\nonumber
&+ i \left( - \sfrac{1}{4} f_{10} \right) \bar{\epsilon} \Gamma^{IJK} \Gamma^\mu [ X^I , X^J , \psi ] D_\mu X^L X^{LMN} X^{KMN} \\
\nonumber
&+ i \left( + \sfrac{1}{4} f_{10} \right) \bar{\epsilon} \Gamma^{IJK} \Gamma^\mu \psi [ X^I , X^J , D_\mu X^L ] X^{LMN} X^{KMN} \\
\nonumber
&+ i \left( + \sfrac{1}{4} f_{10} \right) \bar{\epsilon} \Gamma^{IJK} \Gamma^\mu \psi D_\mu X^L [ X^I , X^J , X^{LMN} ] X^{KMN} \\
\nonumber
&+ i \left( + \sfrac{1}{4} f_{10} \right) \bar{\epsilon} \Gamma^{IJK} \Gamma^\mu \psi D_\mu X^L X^{LMN} [ X^I , X^J , X^{KMN} ] \\[6pt]
\nonumber
&+ i \left( + \sfrac{1}{2} \hat{\bold{k}} + \hat{\bold{p}} \right) \bar{\epsilon} \Gamma^{IJK} \Gamma^\mu [X^L,X^M,\psi] D_\mu X^K X^{IJN} X^{LMN} \\
\nonumber
&+ i \left( - \hat{\bold{p}} \right) \bar{\epsilon} \Gamma^{IJK} \Gamma^\mu \psi [X^L,X^M, D_\mu X^K ] X^{IJN} X^{LMN} \\
\nonumber
&+ i \left( + \sfrac{1}{2} \hat{\bold{k}} \right)  \bar{\epsilon} \Gamma^{IJK} \Gamma^\mu \psi D^\mu X^K [X^L,X^M, X^{IJN} ] X^{LMN} \\[6pt]
\nonumber
&+ i \left( - \sfrac{1}{2} \hat{\bold{k}} \right) \bar{\epsilon} \Gamma^{IJK} \Gamma^\mu [X^M,X^N,\psi] D_\mu X^L X^{IJL} X^{KMN} \\
\nonumber
&+ i \left( - \sfrac{1}{6} \hat{\bold{l}} \right) \bar{\epsilon} \Gamma^{IJK} \Gamma^\mu [X^M,X^N,\psi] D_\mu X^L X^{IJK} X^{LMN} \\
%
%
\nonumber
&+ i \left( - \sfrac{1}{2} \hat{\bold{k}} \right) \bar{\epsilon} \Gamma^{IJK} \Gamma^\mu \psi D_\mu X^L [X^M,X^N, X^{IJL} ] X^{KMN} \\
%
%
\nonumber
&+ i \left( + \sfrac{1}{6} \hat{\bold{l}} \right)  \bar{\epsilon} \Gamma^{IJK} \Gamma^\mu \psi D_\mu X^L [X^M,X^N, X^{IJK} ] X^{LMN} \\
&+ i \left( + \hat{\bold{n}} \right)  \bar{\epsilon} \Gamma^{IJK} \Gamma^\mu \psi D_\mu X^L [X^L,X^M, X^{IJN} ] X^{KMN} \Big\} \, ,
\end{align}
where we have discarded terms with the coefficients $s_6$, $s_7$ and $s_8$ which are zero. Using \eref{Useful_Id_5} and \eref{Useful_Id_4} we can rewrite several of the terms to find
\begin{align}
\nonumber
{\rm STr} \Big\{ &+ i \left( + \sfrac{1}{2} f_{11} \right) \bar{\epsilon} \Gamma^{IJK} \Gamma^\mu D_\mu \psi X^{IJK} X^{LMN} X^{LMN} \\
\nonumber
&+ i \left( + \sfrac{1}{2} f_{11} \right) \bar{\epsilon} \Gamma^{IJK} \Gamma^\mu \psi D_\mu ( X^{IJK} ) X^{LMN} X^{LMN} \\
\nonumber
&+ i \left( + \sfrac{1}{2} f_{11} \right) \bar{\epsilon} \Gamma^{IJK} \Gamma^\mu \psi X^{IJK} D_\mu ( X^{LMN} X^{LMN} ) \\[10pt]
\nonumber
&+ i \left( - \sfrac{1}{4} f_9 + \sfrac{1}{6} \hat{\bold{k}} + \sfrac{1}{3} \hat{\bold{p}} \right) \bar{\epsilon} \Gamma^{IJK} \Gamma^\mu [ X^I , X^J , \psi ] D_\mu X^K X^{LMN} X^{LMN} \\
\nonumber
&+ i \left( + \sfrac{1}{4} f_9 - 3 f_{11} - \sfrac{1}{3} \hat{\bold{p}} \right) \bar{\epsilon} \Gamma^{IJK} \Gamma^\mu \psi [ X^I , X^J ,D_\mu X^K ]  X^{LMN} X^{LMN} \\
\nonumber
&+ i \left( + \sfrac{1}{4} f_9 + \sfrac{1}{12} \hat{\bold{k}} \right) \bar{\epsilon} \Gamma^{IJK} \Gamma^\mu \psi D_\mu X^K [ X^I , X^J , X^{LMN} X^{LMN} ] \\[10pt]
\nonumber
&+ i \left( - \sfrac{1}{4} f_{10} - \sfrac{1}{2} \hat{\bold{k}} - \sfrac{1}{2} \hat{\bold{l}} \right) \bar{\epsilon} \Gamma^{IJK} \Gamma^\mu [ X^I , X^J , \psi ] D_\mu X^L X^{LMN} X^{KMN} \\
\nonumber
&+ i \left( + \sfrac{1}{4} f_{10} - 18 f_{11} \right) \bar{\epsilon} \Gamma^{IJK} \Gamma^\mu \psi [ X^I , X^J , D_\mu X^L ] X^{LMN} X^{KMN} \\
\nonumber
&+ i \left( + \sfrac{1}{4} f_{10} \right) \bar{\epsilon} \Gamma^{IJK} \Gamma^\mu \psi D_\mu X^L [ X^I , X^J , X^{LMN} ] X^{KMN} \\
\nonumber
&+ i \left( + \sfrac{1}{4} f_{10} + \sfrac{1}{2} \hat{\bold{l}} \right) \bar{\epsilon} \Gamma^{IJK} \Gamma^\mu \psi D_\mu X^L X^{LMN} [ X^I , X^J , X^{KMN} ] \\[10pt]
%
%
%
%
%
%
%
\nonumber
&+ i \left( - \sfrac{1}{2} \hat{\bold{k}} \right) \bar{\epsilon} \Gamma^{IJK} \Gamma^\mu \psi D_\mu X^L [X^M,X^N, X^{IJL} ] X^{KMN} \\
%
%
%
&+ i \left( + \hat{\bold{n}} \right)  \bar{\epsilon} \Gamma^{IJK} \Gamma^\mu \psi D_\mu X^L [X^L,X^M, X^{IJN} ] X^{KMN} \Big\} \, . \label{L_1_trans_3_wvol_1}
\end{align}
The fundamental identity can be used to show
\begin{align}
\nonumber
\bar{\epsilon} \Gamma^{IJK} \Gamma^\mu \psi D_\mu X^L [ X^I , X^J , X^{LMN} ] X^{KMN} =&+ \bar{\epsilon} \Gamma^{IJK} \Gamma^\mu \psi D_\mu X^L [X^M,X^N, X^{IJL} ] X^{KMN} \\
&+ 2 \bar{\epsilon} \Gamma^{IJK} \Gamma^\mu \psi D_\mu X^L [X^L,X^M, X^{IJN} ] X^{KMN} \, .
\end{align}
With the knowledge that $\hat{\bold{n}} = - \hat{\bold{k}}$, the final two terms in \eqref{L_1_trans_3_wvol_1} combine with one of the $f_{10}$ terms to leave
\begin{align}
\nonumber
{\rm STr} \Big\{ &+ i \left( + \sfrac{1}{2} f_{11} \right) \bar{\epsilon} \Gamma^{IJK} \Gamma^\mu D_\mu \psi X^{IJK} X^{LMN} X^{LMN} \\
\nonumber
&+ i \left( + \sfrac{1}{2} f_{11} \right) \bar{\epsilon} \Gamma^{IJK} \Gamma^\mu \psi D_\mu ( X^{IJK} ) X^{LMN} X^{LMN} \\
\nonumber
&+ i \left( + \sfrac{1}{2} f_{11} \right) \bar{\epsilon} \Gamma^{IJK} \Gamma^\mu \psi X^{IJK} D_\mu ( X^{LMN} X^{LMN} ) \\[10pt]
\nonumber
&+ i \left( - \sfrac{1}{4} f_9 + \sfrac{1}{6} \hat{\bold{k}} + \sfrac{1}{3} \hat{\bold{p}} \right) \bar{\epsilon} \Gamma^{IJK} \Gamma^\mu [ X^I , X^J , \psi ] D_\mu X^K X^{LMN} X^{LMN} \\
\nonumber
&+ i \left( + \sfrac{1}{4} f_9 - 3 f_{11} - \sfrac{1}{3} \hat{\bold{p}} \right) \bar{\epsilon} \Gamma^{IJK} \Gamma^\mu \psi [ X^I , X^J ,D_\mu X^K ]  X^{LMN} X^{LMN} \\
\nonumber
&+ i \left( + \sfrac{1}{4} f_9 + \sfrac{1}{12} \hat{\bold{k}} \right) \bar{\epsilon} \Gamma^{IJK} \Gamma^\mu \psi D_\mu X^K [ X^I , X^J , X^{LMN} X^{LMN} ] \\[10pt]
\nonumber
&+ i \left( - \sfrac{1}{4} f_{10} - \sfrac{1}{2} \hat{\bold{k}} - \sfrac{1}{2} \hat{\bold{l}} \right) \bar{\epsilon} \Gamma^{IJK} \Gamma^\mu [ X^I , X^J , \psi ] D_\mu X^L X^{LMN} X^{KMN} \\
\nonumber
&+ i \left( + \sfrac{1}{4} f_{10} - 18 f_{11} \right) \bar{\epsilon} \Gamma^{IJK} \Gamma^\mu \psi [ X^I , X^J , D_\mu X^L ] X^{LMN} X^{KMN} \\
\nonumber
&+ i \left( + \sfrac{1}{4} f_{10} - \sfrac{1}{2} \hat{\bold{k}} \right) \bar{\epsilon} \Gamma^{IJK} \Gamma^\mu \psi D_\mu X^L [ X^I , X^J , X^{LMN} ] X^{KMN} \\
&+ i \left( + \sfrac{1}{4} f_{10} + \sfrac{1}{2} \hat{\bold{l}} \right) \bar{\epsilon} \Gamma^{IJK} \Gamma^\mu \psi D_\mu X^L X^{LMN} [ X^I , X^J , X^{KMN} ] \Big\} \, .
\end{align}
The $f_{11}$ terms form a total derivative. The remaining blocks of terms vanish when
\begin{equation}
- \sfrac{1}{4} f_9 + \sfrac{1}{6} \hat{\bold{k}} + \sfrac{1}{3} \hat{\bold{p}} = + \sfrac{1}{4} f_9 - 3 f_{11} - \sfrac{1}{3} \hat{\bold{p}} = + \sfrac{1}{4} f_9 + \sfrac{1}{12} \hat{\bold{k}} \, ,
\end{equation}
\begin{equation}
- \sfrac{1}{4} f_{10} - \sfrac{1}{2} \hat{\bold{k}} - \sfrac{1}{2} \hat{\bold{l}} = + \sfrac{1}{4} f_{10} - 18 f_{11} = + \sfrac{1}{4} f_{10} - \sfrac{1}{2} \hat{\bold{k}} = + \sfrac{1}{4} f_{10} + \sfrac{1}{2} \hat{\bold{l}} \, .
\end{equation}
From these equations we find $f_{11} = + \sfrac{1}{36} \hat{\bold{k}} = + \sfrac{1}{72} \hat{\bold{d}}$ and $\hat{\bold{p}} = - \sfrac{1}{2} \hat{\bold{k}} = - \sfrac{1}{4} \hat{\bold{d}}$.

\subsubsection{$\Gamma^{(1)} \Gamma^\mu$ terms}

The terms with a single transverse $\Gamma$-matrix index are
\begin{align}
\nonumber
{\rm STr} \Big\{ &+ i \left( + \sfrac{1}{2} f_9 - 2 \bold{e} \right) \bar{\epsilon} \Gamma^I \Gamma^\mu [ X^I , X^J , \psi ] D_\mu X^J X^{KLM} X^{KLM} \\
\nonumber
&+ i \left( - \sfrac{1}{2} f_9 - g_{15} \right) \bar{\epsilon} \Gamma^I \Gamma_\mu \psi [ X^I , X^J , D^\mu X^J ] X^{KLM} X^{KLM} \\
\nonumber
&+ i \left( - \sfrac{1}{2} f_9 \right) \bar{\epsilon} \Gamma^I \Gamma^\mu \psi D_\mu X^J [ X^I , X^J , X^{KLM} X^{KLM} ] \\[6pt]
\nonumber
&+ i \left( - \sfrac{1}{2} f_{10} + 2 \bold{d} \right) \bar{\epsilon} \Gamma^I \Gamma^\mu [ X^I , X^L , \psi ] D_\mu X^J X^{JKM} X^{KLM} \\
\nonumber
&+ i \left( + \sfrac{1}{2} f_{10} \right) \bar{\epsilon} \Gamma^I \Gamma^\mu \psi [ X^I , X^L , D_\mu X^J ] X^{JKM} X^{KLM} \\
\nonumber
&+ i \left( + \sfrac{1}{2} f_{10} \right) \bar{\epsilon} \Gamma^I \Gamma^\mu \psi D_\mu X^J [ X^I , X^L , X^{JKM} ] X^{KLM} \\
\nonumber
&+ i \left( + \sfrac{1}{2} f_{10} \right) \bar{\epsilon} \Gamma^I \Gamma^\mu \psi D_\mu X^J X^{JKM} [ X^I , X^L , X^{KLM} ] \\[6pt]
\nonumber
&+ i \left( + \sfrac{1}{2} s_9 + \hat{\bold{k}} - 2 \hat{\bold{p}} \right) \bar{\epsilon} \Gamma^I \Gamma^\mu [ X^K , X^L , \psi ] D_\mu X^J X^{IJM} X^{KLM} \\
\nonumber
&+ i \left( + \sfrac{1}{2} s_9 + 2 \hat{\bold{p}} \right) \bar{\epsilon} \Gamma^I \Gamma^\mu \psi [ X^K , X^L , D_\mu X^J ] X^{IJM} X^{KLM} \\
\nonumber
&+ i \left( + \sfrac{1}{2} s_9 + \hat{\bold{k}} \right) \bar{\epsilon} \Gamma^I \Gamma^\mu \psi D_\mu X^J [ X^K , X^L , X^{IJM} ] X^{KLM} \\[6pt]
\nonumber
&+ i \left( - \hat{\bold{n}} \right) \bar{\epsilon} \Gamma^I \Gamma^\mu [ X^J , X^K , \psi ] D_\mu X^J X^{ILM} X^{KLM} \\
&+ i \left( - \hat{\bold{n}} \right) \bar{\epsilon} \Gamma^I \Gamma^\mu \psi D_\mu X^J [ X^J , X^K , X^{ILM} ] X^{KLM} \Big\} \, . \label{L_1_trans1_wvol1}
\end{align}
The $[X,X,\psi]$ terms are not independent. They are related by
\begin{align}
+ \bar{\epsilon} \Gamma^I \Gamma^\mu [ X^I , X^J , \psi ] D_\mu X^J X^{KLM} X^{KLM} =& - 3 \bar{\epsilon} \Gamma^I \Gamma^\mu [ X^I , X^L , \psi ] D_\mu X^J X^{JKM} X^{KLM} \\[10pt]
=& + 3 \Gamma^I \Gamma^\mu [ X^K , X^L , \psi ] D_\mu X^J X^{IJM} X^{KLM} \\[10pt]
=& - 3 \bar{\epsilon} \Gamma^I \Gamma^\mu [ X^J , X^K , \psi ] D_\mu X^J X^{ILM} X^{KLM} \, .
\end{align}
We can similarly show
\begin{align}
+ \bar{\epsilon} \Gamma^I \Gamma^\mu \psi D_\mu X^J [ X^I , X^J , X^{KLM} ] X^{KLM} =& + 3 \bar{\epsilon} \Gamma^I \Gamma^\mu \psi D_\mu X^J [ X^K , X^L , X^{IJM} ] X^{KLM} \\[10pt]
=& -3 \bar{\epsilon} \Gamma^I \Gamma^\mu \psi D_\mu X^J [ X^J , X^K , X^{ILM} ] X^{KLM} \, .
\end{align}
Using these results we can write the terms in \eref{L_1_trans1_wvol1} as
\begin{align}
\nonumber
{\rm STr} \Big\{ &+ i \left( + \sfrac{1}{2} f_9 + \sfrac{1}{3} f_{10} + \sfrac{1}{6} s_9 - \sfrac{2}{3} \bold{d} - 2 \bold{e} + \sfrac{1}{3} \hat{\bold{k}} + \sfrac{1}{3} \hat{\bold{n}} - \sfrac{2}{3} \hat{\bold{p}} \right) \bar{\epsilon} \Gamma^I \Gamma^\mu [ X^I , X^J , \psi ] D_\mu X^J X^{KLM} X^{KLM} \\
\nonumber
&+ i \left( - \sfrac{1}{2} f_9 - g_{15} \right) \bar{\epsilon} \Gamma^I \Gamma_\mu \psi [ X^I , X^J , D^\mu X^J ] X^{KLM} X^{KLM} \\
\nonumber
&+ i \left( - \sfrac{1}{2} f_9 + \sfrac{1}{12} s_9 + \sfrac{1}{6} \hat{\bold{k}} + \sfrac{1}{6} \hat{\bold{n}} \right) \bar{\epsilon} \Gamma^I \Gamma^\mu \psi D_\mu X^J [ X^I , X^J , X^{KLM} X^{KLM} ] \\[6pt]
\nonumber
&+ i \left( + \sfrac{1}{2} f_{10} \right) \bar{\epsilon} \Gamma^I \Gamma^\mu [ X^I , X^L , \psi ] D_\mu X^J X^{JKM} X^{KLM} \\
\nonumber
&+ i \left( + \sfrac{1}{2} f_{10} \right) \bar{\epsilon} \Gamma^I \Gamma^\mu \psi [ X^I , X^L , D_\mu X^J ] X^{JKM} X^{KLM} \\
\nonumber
&+ i \left( + \sfrac{1}{2} f_{10} \right) \bar{\epsilon} \Gamma^I \Gamma^\mu \psi D_\mu X^J [ X^I , X^L , X^{JKM} ] X^{KLM} \\
\nonumber
&+ i \left( + \sfrac{1}{2} f_{10} \right) \bar{\epsilon} \Gamma^I \Gamma^\mu \psi D_\mu X^J X^{JKM} [ X^I , X^L , X^{KLM} ] \\[6pt]
%
&+ i \left( + \sfrac{1}{2} s_9 + 2 \hat{\bold{p}} \right) \bar{\epsilon} \Gamma^I \Gamma^\mu \psi [ X^K , X^L , D_\mu X^J ] X^{IJM} X^{KLM} \Big\} \, .
\end{align}
We have previously found values for all the coefficients appearing above. Using this data we see the final term disappears because its coefficient is zero and the remaining two blocks of terms vanish through the gauge invariance condition.

\subsection{Zero Derivative Terms - $\tilde{\delta} \mathcal{L}_0$}
The terms from \eqref{Varied_L_BLG} and \eqref{Varied_L_Higher} which contain no covariant derivatives are
\begin{align}
\nonumber
{\rm STr} \Big\{ &- \sfrac{i}{2} s_{10} \bar{\epsilon} \Gamma^{I} [ X^Q , X^R , \psi X^{JKL} X^{JKL} ] X^{IQR} \\
\nonumber
&- \sfrac{i}{2} s_{11} \bar{\epsilon} \Gamma^{L} [ X^Q , X^R , \psi X^{JKL} X^{JKI} ] X^{IQR} \\
\nonumber
&+ \sfrac{i}{4} f_{11} \bar{\epsilon} \Gamma^{NOP} \Gamma^{QR} [ X^Q , X^R , \psi ] X^{JKL} X^{JKL} X^{NOP} \\
\nonumber
&- \sfrac{i}{4} f_{11} \bar{\epsilon} \Gamma^{NOP} \Gamma^{QR} \psi [ X^Q , X^R , X^{JKL} X^{JKL} X^{NOP} ] \\	
\nonumber
&+ 12 i \bold{f} \, \bar{\epsilon} \Gamma^I [ \psi , X^J ,X^K ] X^{IJK} X^{LMN}X^{LMN} \\
\nonumber
&- \sfrac{i}{6} \hat{\bold{p}} \, \bar{\epsilon} \Gamma^{PQR} \Gamma^{IJ} [X^K,X^{L},\psi] X^{PQR} X^{IJM}X^{KLM} \\
&+ \sfrac{i}{6} \hat{\bold{p}} \, \bar{\epsilon} \Gamma^{PQR} \Gamma^{IJ} \psi [X^K,X^{L}, X^{PQR} ] X^{IJM}X^{KLM} \Big\} \, . 
\end{align}
We deal separately with the $\Gamma^{(5)}$, $\Gamma^{(3)}$ and $\Gamma^{(1)}$ terms which result from expanding out the $\Gamma$-matrices below. 
%
%
\subsubsection{$\Gamma^{(5)}$ terms}
The terms with five $\Gamma$-matrix indices are
\begin{align}
\nonumber
{\rm STr} \Big\{ &+ i \left( + \sfrac{1}{4} f_{11} \right) \bar{\epsilon} \Gamma^{IJKLM} [ X^I , X^J , \psi ] X^{KLM} X^{NOP} X^{NOP} \\
\nonumber
&+ i \left( - \sfrac{1}{4} f_{11} \right) \bar{\epsilon} \Gamma^{IJKLM} \psi [ X^I , X^J , X^{KLM} ] X^{NOP} X^{NOP} \\
\nonumber
&+ i \left( - \sfrac{1}{2} f_{11} \right) \bar{\epsilon} \Gamma^{IJKLM} \psi X^{KLM} [ X^I , X^J , X^{NOP} ] X^{NOP} \\[6pt]
\nonumber
&+ i \left( - \sfrac{1}{6} \hat{\bold{p}} \right) \bar{\epsilon} \Gamma^{IJKLM} [ X^N , X^O , \psi ] X^{IJK} X^{LMP} X^{NOP} \\
&+ i \left( + \sfrac{1}{6} \hat{\bold{p}} \right) \bar{\epsilon} \Gamma^{IJKLM} \psi [ X^N , X^O , X^{IJK} ] X^{LMP} X^{NOP} \Big\} \, .
\end{align}
Using the fundamental identity and antisymmetry in the $\Gamma$-matrix indices we can show that
\begin{align}
{\rm STr} \{ \bar{\epsilon} \Gamma^{IJKLM} \psi [ X^I , X^J , X^{KLM} ] X^{NOP} X^{NOP} \} = 0 \, .
\end{align}
Applying \eref{Useful_Id2} with the indices $I_1 I_2 = IJ$ and  $J_1 J_2 J_3 = KLM$ we find
\begin{align}
{\rm STr} \{ \bar{\epsilon} \Gamma^{IJKLM} [ X^I , X^J , \psi ] X^{KLM} X^{NOP} X^{NOP} \} = 0 \, .
\end{align}
When we combine these two expressions with the gauge invariance condition in \eref{Gauge_Inv} we conclude
\begin{align}
{\rm STr} \{  \bar{\epsilon} \Gamma^{IJKLM} \psi X^{KLM} [ X^I , X^J , X^{NOP} ] X^{NOP} \} = 0 \, ,
\end{align}
and hence all the $f_{11}$ terms are zero. 

Focusing now on the $\hat{\bold{p}}$ terms. We find 
\begin{align}
{\rm STr} \{ \bar{\epsilon} \Gamma^{IJKLM} [ X^N , X^O , \psi ] X^{IJK} X^{LMP} X^{NOP} \} =0 \, ,
\end{align}
after applying \eref{Useful_Id2}.
We can also show
\begin{align}
{\rm STr} \{ \bar{\epsilon} \Gamma^{IJKLM} \psi [ X^N , X^O , X^{LMP} ] X^{IJK} X^{NOP} \} =& \sfrac{1}{3} {\rm STr} \{ \bar{\epsilon} \Gamma^{IJKLM} \psi [ X^I , X^J , X^{NOP} ] X^{KLM} X^{NOP} \} \, , \\
=& 0 \, , \\[10pt]
{\rm STr} \{ \bar{\epsilon} \Gamma^{IJKLM} \psi X^{IJK} X^{LMP} [ X^N , X^O , X^{NOP} ] \} =& 0 \, .
\end{align}
Therefore the $\Gamma^{(5)}$ terms can be written as
\begin{align}
\nonumber
{\rm STr} \Big\{ &+ i \left( + \sfrac{1}{6} \hat{\bold{p}} \right) \bar{\epsilon} \Gamma^{IJKLM} [ X^N , X^O , \psi ] X^{IJK} X^{LMP} X^{NOP} \\
\nonumber
&+ i \left( + \sfrac{1}{6} \hat{\bold{p}} \right) \bar{\epsilon} \Gamma^{IJKLM} \psi [ X^N , X^O , X^{IJK} ] X^{LMP} X^{NOP} \\
\nonumber
&+ i \left( + \sfrac{1}{6} \hat{\bold{p}} \right) \bar{\epsilon} \Gamma^{IJKLM} \psi X^{IJK} [ X^N , X^O , X^{LMP} ] X^{NOP} \\
&+ i \left( + \sfrac{1}{6} \hat{\bold{p}} \right) \bar{\epsilon} \Gamma^{IJKLM} \psi X^{IJK} X^{LMP} [ X^N , X^O , X^{NOP} ] \Big\} \, ,
\end{align}
and are zero as a consequence of gauge invariance.

\subsubsection{$\Gamma^{(3)}$ terms}
The terms with three $\Gamma$-matrix indices are
\begin{align}
\nonumber
{\rm STr} \Big\{ &+ i \left( - \sfrac{3}{2} f_{11} \right) \bar{\epsilon} \Gamma^{IJK} [ \psi , X^I , X^L ] X^{JKL} X^{MNO} X^{MNO} \\
\nonumber
&+ i \left( + \sfrac{3}{2} f_{11} \right) \bar{\epsilon} \Gamma^{IJK} \psi [ X^I , X^L , X^{JKL} ] X^{MNO} X^{MNO} \\
\nonumber
&+ i \left( + 3 f_{11} \right) \bar{\epsilon} \Gamma^{IJK} \psi X^{JKL} X^{MNO} [ X^I , X^L , X^{MNO} ] \\
\nonumber
&+ i \left( + \hat{\bold{p}} \right) \bar{\epsilon} \Gamma^{IJK} [ \psi , X^L , X^M ] X^{IJN} X^{KNO} X^{LMO} \\
&+ i \left( - \hat{\bold{p}} \right) \bar{\epsilon} \Gamma^{IJK} \psi [ X^L , X^M , X^{IJN} ] X^{KNO} X^{LMO} \Big\} \, .
\end{align}
We can then show   
\begin{align}
{\rm STr} \{ \bar{\epsilon} \Gamma^{IJK} [ X^I , X^L , \psi ] X^{JKL} X^{MNO} X^{MNO} ] \} =& 0 \, , \\[10pt]
{\rm STr} \{ \bar{\epsilon} \Gamma^{IJK} \psi [ X^I , X^L , X^{JKL} ] X^{MNO} X^{MNO} \} =& 0 \, , \\[10pt]
{\rm STr} \{ \bar{\epsilon} \Gamma^{IJK} \psi X^{JKL} [ X^I , X^L , X^{MNO} ] X^{MNO} \} =& 0 \, ,
\end{align}
so the $f_{11}$ terms are all removed as was the case in the $\Gamma^{(5)}$ subsection above. We also find   
\begin{align}
{\rm STr} \{ \bar{\epsilon} \Gamma^{IJK} [ X^L , X^M , \psi ] X^{IJN} X^{KNO} X^{LMO} \} =&0 \, , \\[10pt]
{\rm STr} \{ \bar{\epsilon} \Gamma^{IJK} \psi X^{IJN} X^{KNO} [ X^L , X^M , X^{LMO} ] \} =&0 \, , \\[10pt]
%
{\rm STr} \{ \bar{\epsilon} \Gamma^{IJK} \psi X^{IJN} [ X^L , X^M , X^{KNO} ] X^{LNO} \} =& - \sfrac{1}{3} {\rm STr} \{ \bar{\epsilon} \Gamma^{IJK} \psi X^{JKL} [ X^I , X^L , X^{MNO} ] X^{MNO} \} \\
=& 0 \, .
\end{align}
Therefore the $\Gamma^{(3)}$ terms can be written as
\begin{align}
\nonumber
{\rm STr} \Big\{ &+ i \left( - \hat{\bold{p}} \right) \bar{\epsilon} \Gamma^{IJK} [ \psi , X^L , X^M ] X^{IJN} X^{KNO} X^{LMO} \\
\nonumber
&+ i \left( - \hat{\bold{p}} \right) \bar{\epsilon} \Gamma^{IJK} \psi [ X^L , X^M , X^{IJN} ] X^{KNO} X^{LMO} \\
\nonumber
&+ i \left( - \hat{\bold{p}} \right) \bar{\epsilon} \Gamma^{IJK} \psi X^{IJN} [ X^L , X^M , X^{KNO} ] X^{LMO} \\
&+ i \left( - \hat{\bold{p}} \right) \bar{\epsilon} \Gamma^{IJK} \psi X^{IJN} X^{KNO} [ X^L , X^M , X^{LMO} ] \Big\} \, ,
\end{align}
and vanish because of the gauge invariance condition.
%
\subsubsection{$\Gamma^{(1)}$ terms}
The terms with a single $\Gamma$-matrix index are
\begin{align}
\nonumber
{\rm STr} \Big\{ &+i \left( - \sfrac{3}{2} f_{11} - \sfrac{1}{2} s_{10} + 12 \bold{f} \right) \bar{\epsilon} \Gamma^{I} [ X^J , X^K , \psi ] X^{LMN} X^{LMN} X^{IJK} \\
\nonumber
&+ i \left( + \sfrac{3}{2} f_{11} - \sfrac{1}{2} s_{10} \right) \bar{\epsilon} \Gamma^{I} \psi [ X^J , X^K , X^{LMN} X^{LMN} ] X^{IJK} \\
\nonumber
&+i \left( + \sfrac{3}{2} f_{11} \right) \bar{\epsilon} \Gamma^I \psi [ X^J , X^K , X^{IJK} ] X^{LMN} X^{LMN} \\
\nonumber
&+i \left( - \sfrac{1}{2} s_{11} + \hat{\bold{p}} \right) \bar{\epsilon} \Gamma^{I} [ X^J , X^K , \psi ] X^{JKN} X^{LMN} X^{ILM} \\
\nonumber
&+i \left( - \sfrac{1}{2} s_{11} - \hat{\bold{p}} \right) \bar{\epsilon} \Gamma^{I} \psi [ X^J , X^K , X^{ILM} ] X^{LMN} X^{JKN} \\
&+ i \left( - \sfrac{1}{2} s_{11} \right) \bar{\epsilon} \Gamma^{I} \psi [ X^J , X^K , X^{LMN} ]  X^{ILM} X^{JKN} \Big\} \, .
\end{align}
We then apply \eref{Useful_Id} to the two terms which contain $X^{ILM} X^{JKN}$ to find 
\begin{align}
\nonumber
\left( - \sfrac{1}{2} s_{11} + \hat{\bold{p}} \right) {\rm STr} \{ \bar{\epsilon} \Gamma^I &[ \psi , X^J , X^K ] X^{ILM} X^{JKN} X^{LMN} \} \\
&= \left( - \sfrac{1}{6} s_{11} + \sfrac{1}{3} \hat{\bold{p}} \right) {\rm STr} \{ \bar{\epsilon} \Gamma^I [ \psi , X^J , X^K ] X^{IJK} X^{LMN} X^{LMN} \} \, ,
\end{align}
and
\begin{align}
- \sfrac{1}{2} s_{11} \, {\rm STr} \{ \bar{\epsilon} \Gamma^I \psi [ X^J , X^K , X^{LMN} ] X^{ILM} X^{JKN} \} = - \sfrac{1}{6} s_{11} \, {\rm STr} \{ \bar{\epsilon} \Gamma^I \psi [ X^J , X^K , X^{LMN} ] X^{IJK} X^{LMN} \} \, .
\end{align}
Hence, we are left with
\begin{align}
\nonumber
{\rm STr} \Big\{ &+i \left( - \sfrac{3}{2} f_{11} - \sfrac{1}{2} s_{10} - \sfrac{1}{6} s_{11} + 12 \bold{f} + \sfrac{1}{3} \hat{\bold{p}} \right) \bar{\epsilon} \Gamma^{I} [ X^J , X^K , \psi ] X^{LMN} X^{LMN} X^{IJK} \\
\nonumber
&+ i \left( + \sfrac{3}{2} f_{11} - \sfrac{1}{2} s_{10} - \sfrac{1}{12} s_{11} \right) \bar{\epsilon} \Gamma^{I} \psi [ X^J , X^K , X^{LMN} X^{LMN} ] X^{IJK} \\
\nonumber
&+i \left( + \sfrac{3}{2} f_{11} \right) \bar{\epsilon} \Gamma^I \psi X^{LMN} X^{LMN} [ X^J , X^K , X^{IJK} ] \\[6pt]
&+i \left( - \sfrac{1}{2} s_{11} - \hat{\bold{p}} \right) \bar{\epsilon} \Gamma^{I} \psi [ X^J , X^K , X^{ILM} ] X^{LMN} X^{JKN} \Big\} \, .
\end{align}
The last line vanishes because $s_{11} = - 2 \hat{\bold{p}} = + \sfrac{1}{2} \hat{\bold{d}}$.
The remaining terms are then zero by the gauge invariance condition if
\begin{equation}
- \sfrac{3}{2} f_{11} - \sfrac{1}{2} s_{10} - \sfrac{1}{6} s_{11} + 12 \bold{f} + \sfrac{1}{3} \hat{\bold{p}} = + \sfrac{3}{2} f_{11} - \sfrac{1}{2} s_{10} - \sfrac{1}{12} s_{11} = + \sfrac{3}{2} f_{11} \, .
\end{equation}
Using previously found values this condition identifies $\bold{f} = + \sfrac{1}{72} \hat{\bold{d}}$.

\newpage
\section{Higher Derivative BLG Closure Calculations}\label{a_HDer_closure}


In this appendix we show in detail how the supersymmetries detailed in chapter \ref{c_Higher_Derivative_BLG} close on the scalar and gauge fields. The steps in the calculations are repeated in each of the subsections and are as follows; we first separate out certain terms according to their number of covariant derivatives and then we insert the relevant supersymmetry transformations. Next, we use the relation $\{ \Gamma^\mu , \Gamma^I \} = 0$ to group all worldvolume $\Gamma$-matrices together and then expand them out using the Clifford algebra relation. Following this, we perform the $( 1 \leftrightarrow 2 )$ antisymmetrisation in the supersymmetry parameters making heavy use of \eref{Commutator_Relation}. The transverse $\Gamma$-matrix algebra is performed next and our calculations have been aided by using the symbolic computer package Cadabra \cite{Peeters:2007wn,Peeters:2006kp}. Finally, we simplify the remaining expressions wherever possible using the identities in Eqs.\,\eqref{Useful_Id_5} and \eqref{Useful_Id_4}.

\subsection{Closure on the Scalar Fields}\label{s_Closure_scalar}

\subsubsection{$3 \ DX$ terms}

Firstly we consider terms which involve three covariant derivatives. 
\begin{align}
\nonumber
T_{M2} \, \big( \delta_1 \delta_2' + \delta'_1 \delta_2 ) X^I_a &- (1 \leftrightarrow 2) \big)_{3DX} \\
\nonumber
=&+ i s_1 ( \bar{\epsilon}_2 \Gamma^{IJK} \Gamma^{\mu \nu} \Gamma^\lambda \Gamma^L \epsilon_1) D_{\lambda} X^{L}_b  D_{\mu} X^{J}_c D_{\nu} X^{K}_d \ d^{bcd}{}_{a} \\
\nonumber
&+ i s_2 ( \bar{\epsilon}_2 \Gamma^{J} \Gamma^{\mu \nu} \Gamma^\lambda \Gamma^L \epsilon_1) D_{\lambda} X^{L}_b D_{\mu} X^{I}_c D_{\nu} X^{J}_d \ d^{bcd}{}_{a} \\
\nonumber
&+ i s_3 ( \bar{\epsilon}_2 \Gamma^{J} \Gamma^\lambda \Gamma^L \epsilon_1) D_{\lambda} X^{L}_b D_{\mu} X^{I}_c D^{\mu} X^{J}_d \ d^{bcd}{}_{a} \\
\nonumber
&+ i s_4 ( \bar{\epsilon}_2 \Gamma^{I} \Gamma^\lambda \Gamma^L \epsilon_1) D_{\lambda} X^{L}_b D_{\mu} X^{J}_c D^{\mu} X^{J}_d \ d^{bcd}{}_{a} \\
\nonumber
&+ i f_1 ( \bar{\epsilon}_2 \Gamma^I \Gamma^{JKL} \Gamma^{\mu \nu \lambda} \epsilon_1 ) D_{\mu} X^{J}_b D_{\nu} X^{K}_c D_{\lambda} X^{L}_d \ d^{bcd}{}_{a} \\
\nonumber
&+ i f_2  ( \bar{\epsilon}_2 \Gamma^I \Gamma^{K} \Gamma^{\mu} \epsilon_1 ) D_{\mu} X^{J}_b D_{\nu} X^{J}_c D^{\nu} X^{K}_d \ d^{bcd}{}_{a} \\
\nonumber
&+ i f_3  ( \bar{\epsilon}_2 \Gamma^I  \Gamma^{K} \Gamma^{\mu} \epsilon_1 ) D_{\mu} X^{K}_b D_{\nu} X^{J}_c D^{\nu} X^{J}_d \ d^{bcd}{}_{a} \\
&- ( 1 \leftrightarrow 2) \\
\nonumber \\
\nonumber
=& +i ( 6 f_1 - 2 s_1 - 2 s_2 ) ( \bar{\epsilon}_2 \Gamma^{JK} \Gamma^{\mu \nu \lambda} \epsilon_1) D_{\mu} X^{I}_b D_{\nu} X^{J}_c D_{\lambda} X^{K}_d \ d^{bcd}{}_{a} \\
\nonumber	
&+ i ( 2 f_2 + 2 s_2 - 2 s_3 ) ( \bar{\epsilon}_2 \Gamma^\mu \epsilon_1) D_{\mu} X^{J}_b D_{\nu} X^{I}_c D^{\nu} X^{J}_d \ d^{bcd}{}_{a} \\
&+ i ( 2 f_3 -  2 s_2 - 2 s_4) ( \bar{\epsilon}_2 \Gamma^{\mu} \epsilon_1) D_{\mu} X^{I}_b D_{\nu} X^{J}_c D^{\nu} X^{J}_d \ d^{bcd}{}_{a} \, .
\end{align}

\subsubsection{$2 \ DX$ terms}

The two covariant derivative terms are
\begin{align}
\nonumber
T_{M2} \, \big( \delta_1 \delta'_2 + \delta_1' \delta_2 ) X^I_a &- ( 1 \leftrightarrow 2) \big)_{2DX} \\
\nonumber
=& -\sfrac{i}{6} s_1 ( \bar{\epsilon}_2 \Gamma^{IJK} \Gamma^{\mu \nu} \Gamma^{LMN} \epsilon_1 ) X^{LMN}_b D_{\mu} X^{J}_c D_{\nu} X^{K}_d \ d^{bcd}{}_{a} \\
\nonumber
&- \sfrac{i}{6} s_2 ( \bar{\epsilon}_2 \Gamma^{J} \Gamma^{\mu \nu} \Gamma^{LMN} \epsilon_1 ) X^{LMN}_b D_{\mu} X^{I}_c D_{\nu} X^{J}_d \ d^{bcd}{}_{a} \\
\nonumber
&- \sfrac{i}{6} s_3 ( \bar{\epsilon}_2 \Gamma^{J} \Gamma^{LMN} \epsilon_1 ) X^{LMN}_b D_{\mu} X^{I}_c D^{\mu} X^{J}_d \ d^{bcd}{}_{a} \\
\nonumber
&- \sfrac{i}{6} s_4 ( \bar{\epsilon}_2 \Gamma^{I} \Gamma^{LMN} \epsilon_1 ) X^{LMN}_b D_{\mu} X^{J}_c D^{\mu} X^{J}_d \ d^{bcd}{}_{a} \\
\nonumber
&+i s_5 ( \bar{\epsilon}_2 \Gamma^{IJKLM} \Gamma^{\mu} \Gamma^{\nu} \Gamma^N \epsilon_1 ) D_\nu X^N_b D_{\mu} X^{J}_c X^{KLM}_d \ d^{bcd}{}_{a} \\
\nonumber
&+ i s_6 ( \bar{\epsilon}_2 \Gamma^{KLM} \Gamma^{\mu} \Gamma^{\nu} \Gamma^N \epsilon_1 ) D_\nu X^N_b D_{\mu} X^{I}_c X^{KLM}_d \ d^{bcd}{}_{a} \\
\nonumber
&+ i s_7 ( \bar{\epsilon}_2 \Gamma^{JLM} \Gamma^{\mu} \Gamma^{\nu} \Gamma^N \epsilon_1 ) D_\nu X^N_b D_{\mu} X^{J}_c X^{ILM}_d \ d^{bcd}{}_{a} \\
\nonumber
&+ i s_8 ( \bar{\epsilon}_2 \Gamma^{ILM} \Gamma^{\mu} \Gamma^{\nu} \Gamma^N \epsilon_1 ) D_\nu X^N_b D_{\mu} X^{J}_c X^{JLM}_d \ d^{bcd}{}_{a} \\
\nonumber
&+ i s_9 ( \bar{\epsilon}_2 \Gamma^{M} \Gamma^{\mu} \Gamma^{\nu} \Gamma^N \epsilon_1 ) D_\nu X^N_b D_{\mu} X^{J}_c X^{IJM}_d \ d^{bcd}{}_{a} \\
\nonumber
&+i f_4 ( \epsilon_2 \Gamma^I \Gamma^{JKLMN} \Gamma^{\mu \nu} \epsilon_1 )  D_\mu X^J_b D_\nu X^K_c X^{LMN}_d \ d^{bcd}{}_a \\
\nonumber
&+ i f_5 ( \epsilon_2 \Gamma^I \Gamma^{KLM} \Gamma^{\mu \nu} \epsilon_1 ) D_\mu X^J_b D_\nu X^K_c X^{JLM}_d \ d^{bcd}{}_a \\
\nonumber
&+ i f_6 ( \epsilon_2 \Gamma^I \Gamma^M \Gamma^{\mu \nu} \epsilon_1 ) D_\mu X^J_b D_\nu X^K_c X^{JKM}_d \ d^{bcd}{}_a \\
\nonumber
&+ i f_7 ( \epsilon_2 \Gamma^I \Gamma^{KLM} \epsilon_1 ) D_\mu X^J_b D^\mu X^J_c X^{KLM}_d \ d^{bcd}{}_a \\
\nonumber
&+ i f_8 ( \epsilon_2 \Gamma^I \Gamma^{KLM} \epsilon_1 ) D_\mu X^J_b D^\mu X^K_c X^{JLM}_d \ d^{bcd}{}_a \\
&- ( 1 \leftrightarrow 2 ) \\
\nonumber \\
\nonumber
=&+ i ( 6 f_4 - s_1 + 2 s_7 ) ( \bar{\epsilon}_2 \Gamma^{JKLM} \Gamma^{\mu \nu} \epsilon_1 ) D_{\mu} X^{J}_b D_{\nu} X^{K}_c X^{ILM}_d \ d^{bcd}{}_{a} \\
\nonumber
&+ i \left( 4 f_4 - 2 s_5 - 2 s_6 - \sfrac{1}{3} s_2 \right) ( \bar{\epsilon}_2 \Gamma^{JKLM} \Gamma^{\mu \nu} \epsilon_1 ) D_{\mu} X^{I}_b D_\nu X^J_c X^{KLM}_d \ d^{bcd}{}_{a} \\		
\nonumber
&+ i ( 2 f_5 + 2 s_1 + 2 s_8 - 6 s_5 ) ( \bar{\epsilon}_2 \Gamma^{IKLM} \Gamma^{\mu \nu} \epsilon_1 ) D_{\mu} X^{J}_b D_{\nu} X^{K}_c X^{JLM}_d \ d^{bcd}{}_{a} \\		
\nonumber
&+ i ( 2 f_6 + 2 s_1 + 2 s_9 ) ( \bar{\epsilon}_2 \Gamma^{\mu \nu} \epsilon_1 ) D_{\mu} X^{J}_b D_{\nu} X^{K}_c X^{IJK}_d \ d^{bcd}{}_{a} \\
\nonumber
&+ i ( 6 f_7 - s_4 + 2 s_7 ) ( \bar{\epsilon}_2 \Gamma^{JK} \epsilon_1 ) D_{\mu} X^{L}_b D^{\mu} X^{L}_c X^{IJK}_d \ d^{bcd}{}_{a} \\			
\nonumber
&+ i ( 2 f_8 - s_3 + 2 s_8 + 6 s_6 ) ( \bar{\epsilon}_2 \Gamma^{KL} \epsilon_1 ) D_{\mu} X^{I}_b D^{\mu} X^{J}_c X^{JKL}_d \ d^{bcd}{}_{a} \\	
&+ i ( - 4 f_8 + 2 s_9 + 4 s_7 ) ( \bar{\epsilon}_2 \Gamma^{KL} \epsilon_1 ) D^\mu X^J_b D_{\mu} X^{L}_c X^{IJK}_d \ d^{bcd}{}_{a} \, .
\end{align}

\subsubsection{$1 \ DX$ terms}

The single covariant derivative terms are
\begin{align}
\nonumber
T_{M2} \, \big( \delta_1 \delta'_2 + \delta'_1 \delta_2 ) X^I_a &- ( 1 \leftrightarrow 2) \big)_{1DX} \\
\nonumber
=&- \sfrac{i}{6} s_5 ( \bar{\epsilon}_2 \Gamma^{IJKLM} \Gamma^{\mu} \Gamma^{PQR} \epsilon_1 ) X^{PQR}_b D_{\mu} X^{J}_c X^{KLM}_d \ d^{bcd}{}_{a} \\
\nonumber
&- \sfrac{i}{6} s_6 ( \bar{\epsilon}_2 \Gamma^{KLM} \Gamma^{\mu} \Gamma^{PQR} \epsilon_1 ) X^{PQR}_b D_{\mu} X^{I}_c X^{KLM}_d \ d^{bcd}{}_{a} \\
\nonumber
&- \sfrac{i}{6} s_7 ( \bar{\epsilon}_2 \Gamma^{JLM} \Gamma^{\mu} \Gamma^{PQR} \epsilon_1 ) X^{PQR}_b D_{\mu} X^{J}_c X^{ILM}_d \ d^{bcd}{}_{a} \\
\nonumber
&- \sfrac{i}{6} s_8 ( \bar{\epsilon}_2 \Gamma^{ILM} \Gamma^{\mu} \Gamma^{PQR} \epsilon_1 ) X^{PQR}_b D_{\mu} X^{J}_c X^{JLM}_d \ d^{bcd}{}_{a} \\
\nonumber
&- \sfrac{i}{6} s_9 ( \bar{\epsilon}_2 \Gamma^{M} \Gamma^{\mu} \Gamma^{PQR} \epsilon_1 ) X^{PQR}_b D_{\mu} X^{J}_c X^{IJM}_d \ d^{bcd}{}_{a} \\
\nonumber
&+ i s_{10} ( \bar{\epsilon}_2 \Gamma^{I} \Gamma^\mu \Gamma^P \epsilon_1 ) D_\mu X^P_b X^{JKL}_c X^{JKL}_d \ d^{bcd}{}_{a} \\
\nonumber
&+ i s_{11} ( \bar{\epsilon}_2 \Gamma^{L} \Gamma^\mu \Gamma^P \epsilon_1 ) D_\mu X^P_b X^{JKL}_c X^{JKI}_d \ d^{bcd}{}_{a} \\
\nonumber
&+i f_{11} ( \bar{\epsilon}_2 \Gamma^I \Gamma^{LMN} \Gamma^\mu \epsilon_1 ) D_\mu X^J_b X^{JKL}_c X^{KMN}_d \, d^{bcd}{}_a \\
\nonumber
&+i f_9 ( \bar{\epsilon}_2 \Gamma^I \Gamma^J \Gamma^\mu \epsilon_1 ) D_\mu X^J_b X^{KLM}_c X^{KLM}_d \, d^{bcd}{}_a \\
&- ( 1 \leftrightarrow 2 ) \\
\nonumber \\
\nonumber
=&+ i ( 3 s_6 ) ( \bar{\epsilon}_2 \Gamma^{KLMN} \Gamma^{\mu} \epsilon_1 )  D_\mu X^I_b X^{JKL}_c X^{JMN}_d \, d^{bcd}{}_a \\
\nonumber
&+ i ( 6 s_5 - 2 s_7 ) ( \bar{\epsilon}_2 \Gamma^{JMNO} \Gamma^{\mu} \epsilon_1 )  D_\mu X^J_b X^{IKM}_c X^{KNO}_d \, d^{bcd}{}_a \\
\nonumber
&+ i \left( 2 s_5 + \sfrac{1}{3} s_9 \right) ( \bar{\epsilon}_2 \Gamma^{KMNO} \Gamma^{\mu} \epsilon_1 ) D_\mu X^J_b X^{IJK}_c X^{MNO}_d \, d^{bcd}{}_a \\
\nonumber
&+ i ( s_7 + s_8 ) ( \bar{\epsilon}_2 \Gamma^{KLMN} \Gamma^{\mu} \epsilon_1 ) D_\mu X^J_b X^{IKL}_c X^{JMN}_d \, d^{bcd}{}_a \\
\nonumber
&+ i ( - 6 s_5 - 2 s_8 ) ( \bar{\epsilon}_2 \Gamma^{ILMN} \Gamma^\mu \epsilon_1 ) D_\mu X^J_b X^{JKL}_c X^{KMN}_d \, d^{bcd}{}_a \\
\nonumber
&+ i ( 2 f_9 - 2 s_6 - 2 s_{10} ) ( \bar{\epsilon}_2 \Gamma^\mu \epsilon_1 ) D_\mu X^I_b X^{JKL}_c X^{JKL}_d \, d^{bcd}{}_a \\
&+ i ( 2 f_{10} - 2 s_7 - 2 s_8 - 2 s_{11} ) ( \bar{\epsilon}_2 \Gamma^\mu \epsilon_1 ) D_\mu X^J_b X^{IKL}_c X^{JKL}_d \, d^{bcd}{}_a \, .
\end{align}
The presence of two 3-brackets in each of these terms allows us use of \eref{Useful_Id}. As an example of how this helps us we take the first of the preceding terms with the choice $I_1 I_2 I_3 = KLJ$, \ $J_1 J_2 J_3 = MNJ$ in \eref{Useful_Id},
\begin{align}
\nonumber
i ( &3 s_6 ) ( \bar{\epsilon}_2 \Gamma^{KLMN} \Gamma^{\mu} \epsilon_1 ) D_\mu X^I_b X^{JKL}_c X^{JMN}_d \, d^{bcd}{}_a \\[10pt]
&=+ i ( 3 s_6 ) ( \bar{\epsilon}_2 \Gamma^{KLMN} \Gamma^{\mu} \epsilon_1 ) D_\mu X^I_b ( X^{MLJ}_c X^{KNJ}_d + X^{KMJ}_c X^{LNJ}_d + X^{KLM}_c X^{JNJ}_d ) \, d^{bcd}{}_a \\[10pt]
&=- 2 i ( 3 s_6 ) ( \bar{\epsilon}_2 \Gamma^{KLMN} \Gamma^{\mu} \epsilon_1 ) D_\mu X^I_b X^{JKL}_c X^{JMN}_d \, d^{bcd}{}_a \, ,
\end{align}
where in the last line we used antisymmetry in $K \leftrightarrow L$ and also $L \leftrightarrow M$. Thus, 
\begin{equation}
+ i ( 3 s_6 ) ( \bar{\epsilon}_2 \Gamma^{KLMN} \Gamma^{\mu} \epsilon_1 ) D_\mu X^I_b X^{JKL}_c X^{JMN}_d \, d^{bcd}{}_a = 0 \, .
\end{equation}
In a similar fashion we can show that the rest of the terms with four transverse $\Gamma$-matrix indices vanish. Hence, we are left with
\begin{align}
\nonumber
T_{M2} \, \big( \delta_1 \delta'_2 + \delta'_1 \delta_2 ) X^I_a &- ( 1 \leftrightarrow 2) \big)_{1DX} \\
\nonumber=&+ i ( 2 f_9 - 2 s_6 - 2 s_{10} ) ( \bar{\epsilon}_2 \Gamma^\mu \epsilon_1 ) D_\mu X^I_b X^{JKL}_c X^{JKL}_d \, d^{bcd}{}_a \\
&+ i ( 2 f_{10} - 2 s_7 - 2 s_8 - 2 s_{11} ) ( \bar{\epsilon}_2 \Gamma^\mu \epsilon_1 ) D_\mu X^J_b X^{IKL}_c X^{JKL}_d \, d^{bcd}{}_a \, .
\end{align}

\subsubsection{$0 \ DX$ terms}

The terms which feature no covariant derivatives are
\begin{align}
\nonumber
T_{M2} \, \big( \delta_1 \delta'_2 + \delta'_1 \delta_2 ) X^I_a -  (1 \leftrightarrow 2) \big)_{0DX} =&- \sfrac{i}{6} s_{10} ( \bar{\epsilon}_2 \Gamma^{I} \Gamma^{PQR} \epsilon_1 ) X^{PQR}_b X^{JKL}_c X^{JKL}_d \ d^{bcd}{}_{a} \\
\nonumber
&- \sfrac{i}{6} s_{11} ( \bar{\epsilon}_2 \Gamma^{L} \Gamma^{PQR} \epsilon_1 ) X^{PQR}_b X^{JKL}_c X^{JKI}_d \ d^{bcd}{}_{a} \\
\nonumber
&+ i f_{11} ( \bar{\epsilon}_2 \Gamma^I \Gamma^{PQR} \epsilon_1 ) X^{JKL}_b X^{JKL}_c X^{PQR}_d \, d^{bcd}{}_a \\
&- (1 \leftrightarrow 2 ) \\
\nonumber \\
\nonumber
=&+ i \left( 6 f_{11} - s_{10} \right) ( \epsilon_2 \Gamma^{JK} \epsilon_1 ) X^{IJK}_b X^{LMN}_c X^{LMN}_d \, d^{bcd}{}_a \\
&- i s_{11} ( \bar{\epsilon}_2 \Gamma^{MN} \epsilon_1 ) X^{LMN}_b X^{JKL}_c X^{IJK}_d \ d^{bcd}{}_{a} \, .
\end{align}
After applying \eref{Useful_Id_5} twice to the $s_{11}$ term and relabelling the transverse Lorentz indices, we get
\begin{align}
%
- i s_{11} ( \bar{\epsilon}_2 \Gamma^{MN} \epsilon_1 ) X^{LMN}_b X^{JKL}_c X^{IJK}_d \ d^{bcd}{}_{a} =& - \sfrac{i}{3} s_{11} ( \bar{\epsilon}_2 \Gamma^{JK} \epsilon_1 ) X^{LMN}_c X^{IJK}_b X^{LMN}_d \ d^{bcd}{}_{a} \, .
\end{align}
Thus,
\begin{align}
%
T_{M2} \, \big( \delta_1 \delta'_2 + \delta'_1 \delta_2 ) X^I_a &- (1 \leftrightarrow 2) \big)_{0DX} = + i \left( 6 f_{11} - s_{10} - \sfrac{1}{3} s_{11} \right) ( \epsilon_2 \Gamma^{JK} \epsilon_1 ) X^{IJK}_b X^{LMN}_c X^{LMN}_d \, d^{bcd}{}_a \, .
\end{align}

\subsection{Closure on the Gauge Field}\label{s_Closure_gauge}

\subsubsection{$3 \ DX$ terms}

The terms with three covariant derivatives are
\begin{align}
\nonumber
T_{M2} \big( ( \delta_1 \delta'_2 + \delta'_1 \delta_2 ) &\tilde{A}_{\mu}{}^b{}_a - ( 1 \leftrightarrow 2 ) \big)_{3DX} \\
\nonumber
=&+ i g_1 ( \bar{\epsilon}_2 \Gamma_\mu \Gamma^I \Gamma_\rho \Gamma^L \epsilon_1 ) D^\rho X^L_e D_\nu X^J_f D^\nu X^J_g X^I_c \, d^{efg}{}_d f^{cdb}{}_a \\
\nonumber
&+ i g_2 ( \bar{\epsilon}_2 \Gamma^\nu \Gamma^I \Gamma^\rho \Gamma^L \epsilon_1 ) D_\rho X^L_e D_\mu X^J_f D_\nu X^J_g X^I_c \, d^{efg}{}_d f^{cdb}{}_a \\
\nonumber
&+ i g_3 ( \bar{\epsilon}_2 \Gamma^\nu \Gamma^J \Gamma^\rho \Gamma^L \epsilon_1 ) D_\rho X^L_e D_\mu X^J_f D_\nu X^I_g X^I_c \, d^{efg}{}_d f^{cdb}{}_a \\
\nonumber
&+ i g_4 ( \bar{\epsilon}_2 \Gamma^\nu \Gamma^J \Gamma^\rho \Gamma^L \epsilon_1 ) D_\rho X^L_e D_\mu X^I_f D^\nu X^J_g X^I_c \, d^{efg}{}_d f^{cdb}{}_a \\
\nonumber
&+ i g_5 ( \bar{\epsilon}_2 \Gamma_\mu \Gamma^J \Gamma_\rho \Gamma^L \epsilon_1 ) D^\rho X^L_e D_\nu X^J_f D^\nu X^I_g X^I_c \, d^{efg}{}_d f^{cdb}{}_a \\
\nonumber
&+ i g_6 ( \bar{\epsilon}_2 \Gamma_{\mu \nu \lambda} \Gamma^J \Gamma_\rho \Gamma^L \epsilon_1 ) D^\rho X^L_e D^\nu X^J_f D^\lambda X^I_g X^I_c \, d^{efg}{}_d f^{cdb}{}_a \\
\nonumber
&+ i g_7 ( \bar{\epsilon}_2 \Gamma_{\mu \nu \lambda} \Gamma^{IJK} \Gamma_\rho \Gamma^L \epsilon_1 ) D^\rho X^L_e D^\nu X^J_f D^\lambda X^K_g X^I_c \, d^{efg}{}_d f^{cdb}{}_a \\
\nonumber
&+ i g_8 ( \bar{\epsilon}_2 \Gamma^{\nu} \Gamma^{IJK} \Gamma^\rho \Gamma^L \epsilon_1 ) D_\rho X^L_e D_\mu X^J_f D^\nu X^K_g X^I_c \, d^{efg}{}_d f^{cdb}{}_a \\
\nonumber
 &+i f_1 ( \bar{\epsilon}_2 \Gamma_\mu \Gamma^I \Gamma^{JKL} \Gamma_{\nu \lambda \rho} \epsilon_1 ) D^{\nu} X^{J}_e D^{\lambda} X^{K}_f D^{\rho} X^{L}_g X^I_c \, d^{efg}{}_{d} f^{cdb}{}_a \\
\nonumber
&+ i f_2 ( \bar{\epsilon}_2 \Gamma_\mu \Gamma^I \Gamma^{K} \Gamma_{\nu} \epsilon_1 ) D^{\nu} X^{J}_e D_{\lambda} X^{J}_f D^{\lambda} X^{K}_g X^I_c \, d^{efg}{}_{d} f^{cdb}{}_a \\
\nonumber
&+ i f_3 ( \bar{\epsilon}_2 \Gamma_\mu \Gamma^I \Gamma^{K} \Gamma_{\nu} \epsilon_1 ) D^{\nu} X^{K}_e D_{\lambda} X^{J}_f D^{\lambda} X^{J}_g X^I_c \, d^{efg}{}_{d} f^{cdb}{}_a \\
&- ( 1 \leftrightarrow 2 ) \\
\nonumber \\
\nonumber
=&+ i ( 2 f_2 - 2 g_5 - 2 g_6 ) \varepsilon_{\mu \nu \lambda} ( \bar{\epsilon}_2 \Gamma^{\nu} \epsilon_1 ) X^I_c D^{\rho} X^{I}_e D^{\lambda} X^{J}_f D_{\rho} X^{J}_g \, d^{efg}{}_{d} f^{cdb}{}_a \\
\nonumber
&+ i ( 2 f_3 - 2 g_1 + 2 g_6 ) \varepsilon_{\mu \nu \lambda} ( \bar{\epsilon}_2 \Gamma^{\nu} \epsilon_1 ) X^I_c D^{\lambda} X^{I}_e D_{\rho} X^{J}_f D^{\rho} X^{J}_g \, d^{efg}{}_{d} f^{cdb}{}_a \\
\nonumber
&+ i ( - 2 g_2 + 2 g_3 - 2 g_6 ) \varepsilon_{\nu \lambda \rho} ( \bar{\epsilon}_2 \Gamma^\nu \epsilon_1 ) X^I_c D^\lambda X^I_e D_\mu X^J_f D^\rho X^J_g \, d^{efg}{}_d f^{cdb}{}_a \\
\nonumber
&+ i ( - 2 g_3 + 2 g_5 - 2 g_8 ) ( \bar{\epsilon}_2 \Gamma^{JK} \epsilon_1 ) D_\mu X^J_e D_\nu X^K_f D^\nu X^I_g X^I_c \, d^{efg}{}_d f^{cdb}{}_a \\			
\nonumber
&+ i ( - 6 f_1 + 2 g_7 + 2 g_8 ) \varepsilon^{\nu \lambda \rho} ( \bar{\epsilon}_2 \Gamma^{IJKL} \Gamma_\rho \epsilon_1 ) X^I_c D_{\mu} X^J_e D_{\nu} X^K_f D_{\lambda} X^L_g \, d^{efg}{}_{d} f^{cdb}{}_a \\	
\nonumber
&+ i ( 2 f_3 - 2 g_1 - 2 g_8 ) ( \bar{\epsilon}_2 \Gamma^{IJ} \epsilon_1 ) X^I_c D_{\mu} X^{J}_e D_{\nu} X^{K}_f D^{\nu} X^{K}_g \, d^{efg}{}_{d} f^{cdb}{}_a \\			
&+ i ( 2 f_2 - 2 g_2 + 2 g_8 ) ( \bar{\epsilon}_2 \Gamma^{IK} \epsilon_1 ) X^I_c D_{\mu} X^{J}_e D_{\nu} X^{J}_f D^{\nu} X^{K}_g \, d^{efg}{}_{d} f^{cdb}{}_a \, .
\end{align}

\subsubsection{$2 \ DX$ terms}

The terms with two covariant derivatives are
\begin{align}
\nonumber
T_{M2} \big( ( \delta_1 \delta'_2 + \delta'_1 &\delta_2 ) \tilde{A}_{\mu}{}^b{}_a - ( 1 \leftrightarrow 2 ) \big)_{2DX} \\
\nonumber
=& - \sfrac{i}{6} g_1 ( \bar{\epsilon}_2 \Gamma_\mu \Gamma^I \Gamma^{PQR} \epsilon_1 ) X^{PQR}_e D_\nu X^J_f D^\nu X^J_g X^I_c \, d^{efg}{}_d f^{cdb}{}_a \\
\nonumber
&- \sfrac{i}{6} g_2 ( \bar{\epsilon}_2 \Gamma^\nu \Gamma^I \Gamma^{PQR} \epsilon_1 ) X^{PQR}_e D_\mu X^J_f D_\nu X^J_g X^I_c \, d^{efg}{}_d f^{cdb}{}_a \\
\nonumber
&- \sfrac{i}{6} g_3 ( \bar{\epsilon}_2 \Gamma^\nu \Gamma^J \Gamma^{PQR} \epsilon_1 ) X^{PQR}_e D_\mu X^J_f D_\nu X^I_g X^I_c \, d^{efg}{}_d f^{cdb}{}_a \\
\nonumber
&- \sfrac{i}{6} g_4 ( \bar{\epsilon}_2 \Gamma^\nu \Gamma^J \Gamma^{PQR} \epsilon_1 ) X^{PQR}_e D_\mu X^I_f D^\nu X^J_g X^I_c \, d^{efg}{}_d f^{cdb}{}_a \\
\nonumber
&- \sfrac{i}{6} g_5 ( \bar{\epsilon}_2 \Gamma_\mu \Gamma^J \Gamma^{PQR} \epsilon_1 ) X^{PQR}_e D_\nu X^J_f D^\nu X^I_g X^I_c \, d^{efg}{}_d f^{cdb}{}_a \\
\nonumber
&- \sfrac{i}{6} g_6 ( \bar{\epsilon}_2 \Gamma_{\mu \nu \lambda} \Gamma^J \Gamma^{PQR} \epsilon_1 ) X^{PQR}_e D^\nu X^J_f D^\lambda X^I_g X^I_c \, d^{efg}{}_d f^{cdb}{}_a \\
\nonumber
&- \sfrac{i}{6} g_7 ( \bar{\epsilon}_2 \Gamma_{\mu \nu \lambda} \Gamma^{IJK} \Gamma^{PQR} \epsilon_1 ) X^{PQR}_e D^\nu X^J_f D^\lambda X^K_g X^I_c \, d^{efg}{}_d f^{cdb}{}_a \\
\nonumber
&- \sfrac{i}{6} g_8 ( \bar{\epsilon}_2 \Gamma^{\nu} \Gamma^{IJK} \Gamma^{PQR} \epsilon_1 ) X^{PQR}_e D_\mu X^J_f D^\nu X^K_g X^I_c \, d^{efg}{}_d f^{cdb}{}_a \\
\nonumber
&+i g_9 ( \bar{\epsilon}_2 \Gamma_{\mu \nu} \Gamma^{KLM} \Gamma_{\lambda} \Gamma^P \epsilon_1 ) D^\lambda X^P_e D^\nu X^I_f X^{KLM}_g X^I_c \, d^{efg}{}_d f^{cdb}{}_a \\
\nonumber
&+i g_{10} ( \bar{\epsilon}_2 \Gamma_{\mu \nu} \Gamma^{JLM} \Gamma_{\lambda} \Gamma^P \epsilon_1 ) D^\lambda X^P_e D^\nu X^J_f X^{ILM}_g X^I_c \, d^{efg}{}_d f^{cdb}{}_a \\
\nonumber
&+i g_{11} ( \bar{\epsilon}_2 \Gamma_{\mu \nu} \Gamma^{M} \Gamma_{\lambda} \Gamma^P \epsilon_1 ) D^\lambda X^P_e D^\nu X^J_f X^{IJM}_g X^I_c \, d^{efg}{}_d f^{cdb}{}_a \\
\nonumber
&+i g_{12} ( \bar{\epsilon}_2 \Gamma^{KLM} \Gamma_{\lambda} \Gamma^P \epsilon_1 ) D^\lambda X^P_e D_\mu X^I_f X^{KLM}_g X^I_c \, d^{efg}{}_d f^{cdb}{}_a \\	
\nonumber
&+i g_{13} ( \bar{\epsilon}_2 \Gamma^{JLM} \Gamma_{\lambda} \Gamma^P \epsilon_1 ) D^\lambda X^P_e D_\mu X^J_f X^{ILM}_g X^I_c \, d^{efg}{}_d f^{cdb}{}_a \\	
\nonumber
&+i g_{14} ( \bar{\epsilon}_2 \Gamma^{M} \Gamma_{\lambda} \Gamma^P \epsilon_1 ) D^\lambda X^P_e D_\mu X^J_f X^{IJM}_g X^I_c \, d^{efg}{}_d f^{cdb}{}_a \\
\nonumber
 &+i f_4 ( \bar{\epsilon}_2 \Gamma_\mu \Gamma^I \Gamma^{JKLMN} \Gamma_{\nu \lambda} \epsilon_1 )  D^\nu X^J_e D^\lambda X^K_f X^{LMN}_g X^I_c \, d^{efg}{}_d f^{cdb}{}_a \\
\nonumber
&+ i f_5 ( \bar{\epsilon}_2 \Gamma_\mu \Gamma^I  \Gamma^{KLM} \Gamma_{\nu \lambda} \epsilon_1 ) D^\nu X^J_e D^\lambda X^K_f X^{JLM}_g X^I_c \, d^{efg}{}_d f^{cdb}{}_a \\
\nonumber
&+ i f_6 ( \bar{\epsilon}_2 \Gamma_\mu \Gamma^I  \Gamma^M \Gamma_{\nu \lambda} \epsilon_1 ) D^\nu X^J_e D^\lambda X^K_f X^{JKM}_g X^I_c \, d^{efg}{}_d f^{cdb}{}_a \\
\nonumber
&+ i f_7 ( \bar{\epsilon}_2 \Gamma_\mu \Gamma^I  \Gamma^{KLM} \epsilon_1 ) D_\nu X^J_e D^\nu X^J_f X^{KLM}_g X^I_c \, d^{efg}{}_d f^{cdb}{}_a \\
\nonumber
&+ i f_8 ( \bar{\epsilon}_2 \Gamma_\mu \Gamma^I  \Gamma^{KLM} \epsilon_1 ) D_\nu X^J_e D^\nu X^K_f X^{JLM}_g X^I_c \, d^{efg}{}_d f^{cdb}{}_a \\
& - ( 1 \leftrightarrow 2 ) \\
\nonumber \\
\nonumber
=&+ i ( 4 f_6 + 2 g_8 - 2 g_{11} - 2 g_{14} ) ( \bar{\epsilon}_2 \Gamma^{\nu} \epsilon_1 ) D_\mu X^J_e D_\nu X^K_f X^{IJK}_g X^I_c \, d^{efg}{}_d f^{cdb}{}_a \\[6pt]
\nonumber
&+ i ( 2 f_5 - g_6 + 6 g_9 - 2 g_{12} ) \varepsilon_{\mu \nu \lambda} ( \bar{\epsilon}_2 \Gamma^{KL} \epsilon_1 ) D^\nu X^I_e D^\lambda X^J_f X^{JKL}_g X^I_c \, d^{efg}{}_d f^{cdb}{}_a \\
\nonumber
&+ i ( - 2 f_6 - 2 g_7 - 4 g_{12} ) \varepsilon_{\mu \nu \lambda} ( \bar{\epsilon}_2 \Gamma^{IJ} \epsilon_1 ) D^\nu X^K_e D^\lambda X^L_f X^{JKL}_g X^I_c \, d^{efg}{}_d f^{cdb}{}_a \\[6pt]
\nonumber
&+i ( + 4 f_5 + 4 g_7 - 4 g_{10} + 2 g_{11} ) \varepsilon_{\mu \nu \lambda} ( \bar{\epsilon}_2 \Gamma^{KL} \epsilon_1 ) D^\nu X^J_e D^\lambda X^L_f X^{IJK}_g X^I_c \, d^{efg}{}_d f^{cdb}{}_a \\[6pt]
\nonumber
&+ i ( 2 f_8 ) ( \bar{\epsilon}_2 \Gamma_\mu \Gamma^{IJKL} \epsilon_1 ) D^\nu X^J_e D_\nu X^M_f X^{KLM}_g X^I_c \, d^{efg}{}_d f^{cdb}{}_a \\
\nonumber
&+ i ( 2 g_9 - \sfrac{1}{3} g_5 ) ( \bar{\epsilon}_2 \Gamma_{\mu} \Gamma^{JKLM} \epsilon_1 ) D^\nu X^I_e D_\nu X^J_f X^{KLM}_g X^I_c \, d^{efg}{}_d f^{cdb}{}_a \\[6pt]	
\nonumber
&+ i ( - 2 f_5 - g_8 ) ( \bar{\epsilon}_2 \Gamma^{\nu} \Gamma^{IJLM} \epsilon_1 ) D_\nu X^K_e D_\mu X^J_f X^{KLM}_g X^I_c \, d^{efg}{}_d f^{cdb}{}_a \\
\nonumber
&+ i ( -4 f_4 ) ( \bar{\epsilon}_2 \Gamma_{\nu} \Gamma^{JKLM} \epsilon_1 ) D^\nu X^I_e D_\mu X^J_f X^{KLM}_g X^I_c \, d^{efg}{}_d f^{cdb}{}_a \\
\nonumber
&+ i ( - \sfrac{1}{3} g_3 - 2 g_9 ) ( \bar{\epsilon}_2 \Gamma^{\nu} \Gamma^{JKLM} \epsilon_1 ) D_\nu X^I_e D_\mu X^J_f X^{KLM}_g X^I_c \, d^{efg}{}_d f^{cdb}{}_a \\[6pt]
\nonumber
&+ i ( 2 f_5 + g_8 ) ( \bar{\epsilon}_2 \Gamma^{\nu} \Gamma^{IJKL} \epsilon_1 ) D^\nu X^J_e D_\mu X^M_f X^{KLM}_g X^I_c \, d^{efg}{}_d f^{cdb}{}_a \\
\nonumber
&+ i ( 4 f_4 ) ( \bar{\epsilon}_2 \Gamma_{\nu} \Gamma^{JKLM} \epsilon_1 ) D^\nu X^J_e D_\mu X^I_f X^{KLM}_g X^I_c \, d^{efg}{}_d f^{cdb}{}_a \\
\nonumber
&+ i ( - \sfrac{1}{3} g_4 + 2 g_{12} ) ( \bar{\epsilon}_2 \Gamma^\nu \Gamma^{JKLM} \epsilon_1 ) D_\nu X^J_e D_\mu X^I_f X^{KLM}_g X^I_c \, d^{efg}{}_d f^{cdb}{}_a \\[6pt]	
\nonumber
&+ i ( 12 f_4 - g_8 - 2 g_{10} - 2 g_{13} ) ( \bar{\epsilon}_2 \Gamma^{\nu} \Gamma^{JKLM} \epsilon_1 ) D_\mu X^J_e D_\nu X^K_f X^{ILM}_g X^I_c \, d^{efg}{}_d f^{cdb}{}_a \\[6pt]
\nonumber
&+ i ( 2 f_7 - \sfrac{1}{3} g_1 ) ( \bar{\epsilon}_2 \Gamma_\mu \Gamma^{IJKL} \epsilon_1 ) D_\nu X^M_e D^\nu X^M_f X^{JKL}_g X^I_c \, d^{efg}{}_d f^{cdb}{}_a \\
\nonumber
&+ i ( - \sfrac{1}{3} g_2 ) ( \bar{\epsilon}_2 \Gamma^{\nu} \Gamma^{IJKL} \epsilon_1 ) D_\nu X^M_e D_\mu X^M_f X^{JKL}_g X^I_c \, d^{efg}{}_d f^{cdb}{}_a \\
&+ i ( 2 f_4 - \sfrac{1}{3} g_7 ) ( \bar{\epsilon}_2 \Gamma_{\mu \nu \lambda} \Gamma^{IJKLMN}\epsilon_1 ) D^\nu X^J_e D^\lambda X^K_f X^{LMN}_g X^I_c \, d^{efg}{}_d f^{cdb}{}_a .
\end{align}
Applying \eref{Useful_Id_4} to these terms we find that the last three lines are identically zero and several of the other terms combine. Ultimately we are left with
\begin{align}
\nonumber
T_{M2} \big( ( \delta_1 \delta'_2 &+ \delta'_1 \delta_2 ) \tilde{A}_{\mu}{}^b{}_a - ( 1 \leftrightarrow 2 ) \big)_{2DX} \\
	\nonumber
	=&+ i ( 4 f_6 + 2 g_8 - 2 g_{11} - 2 g_{14} ) ( \bar{\epsilon}_2 \Gamma^{\nu} \epsilon_1 ) X^I_c D_\mu X^J_e D_\nu X^K_f X^{IJK}_g \, d^{efg}{}_d f^{cdb}{}_a \\	
	\nonumber
	&+ i ( 2 f_5 + 2 f_6 - g_6 + 2 g_7 + 6 g_9 ) \varepsilon_{\mu \nu \lambda} ( \bar{\epsilon}_2 \Gamma^{KL} \epsilon_1 ) X^I_c D^\nu X^I_e D^\lambda X^J_f X^{JKL}_g \, d^{efg}{}_d f^{cdb}{}_a \\
	\nonumber
	&+ i ( 4 f_5 + 4 g_7 - 4 g_{10} + 2 g_{11} ) \varepsilon_{\mu \nu \lambda} ( \bar{\epsilon}_2 \Gamma^{KL} \epsilon_1 ) X^I_c D^\nu X^J_e D^\lambda X^L_f X^{IJK}_g \, d^{efg}{}_d f^{cdb}{}_a \\
	\nonumber
	&+ i ( 2 f_8 + g_5 - 6 g_9 ) ( \bar{\epsilon}_2 \Gamma_\mu \Gamma^{IJKL} \epsilon_1 ) X^I_c D^\nu X^J_e D_\nu X^M_f X^{KLM}_g \, d^{efg}{}_d f^{cdb}{}_a \\	
	\nonumber
	&+ i ( 12 f_4 - 2 f_5 + g_3 - g_8 + 6 g_9 ) ( \bar{\epsilon}_2 \Gamma^{\nu} \Gamma^{IJLM} \epsilon_1 ) X^I_c D_\mu X^J_e D_\nu X^K_f X^{KLM}_g \, d^{efg}{}_d f^{cdb}{}_a \\
	\nonumber
	&+ i ( - 12 f_4 + 2 f_5 + g_4 + g_8 - 6 g_{12} ) ( \bar{\epsilon}_2 \Gamma^{\nu} \Gamma^{IJKL} \epsilon_1 ) X^I_c D_\nu X^J_e D_\mu X^M_f X^{KLM}_g \, d^{efg}{}_d f^{cdb}{}_a \\
	&+ i ( 12 f_4 - g_8 - 2 g_{10} - 2 g_{13} ) ( \bar{\epsilon}_2 \Gamma^{\nu} \Gamma^{JKLM} \epsilon_1 ) X^I_c D_\mu X^J_e D_\nu X^K_f X^{ILM}_g \, d^{efg}{}_d f^{cdb}{}_a \, .
\end{align}

\subsubsection{$1 \ DX$ terms}

The terms with a single covariant derivative are
\begin{align}
\nonumber
T_{M2} \big( ( \delta_1 \delta'_2 + \delta'_1 \delta_2 ) &\tilde{A}_{\mu}{}^b{}_a - ( 1 \leftrightarrow 2 ) \big)_{1DX} \\
\nonumber
=&- \sfrac{i}{6} g_9 ( \bar{\epsilon}_2 \Gamma_{\mu \nu} \Gamma^{KLM} \Gamma^{PQR} \epsilon_1 ) X^{PQR}_e D^\nu X^I_f X^{KLM}_g X^I_c \, d^{efg}{}_d f^{cdb}{}_a \\
\nonumber
&- \sfrac{i}{6} g_{10} ( \bar{\epsilon}_2 \Gamma_{\mu \nu} \Gamma^{JLM} \Gamma^{PQR} \epsilon_1 ) X^{PQR}_e D^\nu X^J_f X^{ILM}_g X^I_c \, d^{efg}{}_d f^{cdb}{}_a \\
\nonumber
&-\sfrac{i}{6} g_{11} ( \bar{\epsilon}_2 \Gamma_{\mu \nu} \Gamma^{M} \Gamma^{PQR} \epsilon_1 ) X^{PQR}_e D^\nu X^J_f X^{IJM}_g X^I_c \, d^{efg}{}_d f^{cdb}{}_a \\
\nonumber
&-\sfrac{i}{6} g_{12} ( \bar{\epsilon}_2 \Gamma^{KLM} \Gamma^{PQR} \epsilon_1 ) X^{PQR}_e D_\mu X^I_f X^{KLM}_g X^I_c \, d^{efg}{}_d f^{cdb}{}_a \\	
\nonumber
&-\sfrac{i}{6} g_{13} ( \bar{\epsilon}_2 \Gamma^{JLM} \Gamma^{PQR} \epsilon_1 ) X^{PQR}_e D_\mu X^J_f X^{ILM}_g X^I_c \, d^{efg}{}_d f^{cdb}{}_a \\	
\nonumber
&-\sfrac{i}{6} g_{14} ( \bar{\epsilon}_2 \Gamma^{M} \Gamma^{PQR} \epsilon_1 ) X^{PQR}_e D_\mu X^J_f X^{IJM}_g X^I_c \, d^{efg}{}_d f^{cdb}{}_a \\
\nonumber
&+i g_{15} ( \bar{\epsilon}_2 \Gamma_{\mu} \Gamma^{I} \Gamma_\nu \Gamma^P \epsilon_1 ) D^\nu X^P_e X^{JKL}_f X^{JKL}_g X^I_c \, d^{efg}{}_d f^{cdb}{}_a \\
\nonumber
&+ i f_{9} ( \bar{\epsilon}_2 \Gamma_{\mu} \Gamma^I \Gamma^J \Gamma_\nu \epsilon_1 ) D^\nu X^J_e X^{KLM}_f X^{KLM}_g X^I_c \, d^{efg}{}_d f^{cdb}{}_a \\
\nonumber
&+ i f_{10} ( \bar{\epsilon}_2 \Gamma_{\mu} \Gamma^I \Gamma^M \Gamma_\nu \epsilon_1 ) D^\nu X^J_e X^{JKL}_f X^{KLM}_g X^I_c \, d^{efg}{}_d f^{cdb}{}_a \\
&- ( 1 \leftrightarrow 2 ) \\
\nonumber \\
\nonumber
=&+i ( 2 f_{9} + 2 g_9 - 2 g_{15} ) ( \bar{\epsilon}_2 \Gamma^{\nu} \epsilon_1 ) \varepsilon_{\mu \nu \lambda} D^\lambda X^I_e X^{JKL}_f X^{JKL}_g X^I_c \, d^{efg}{}_d f^{cdb}{}_a \\	
\nonumber
&+ i ( 2 f_{10} + 2 g_{10} ) ( \bar{\epsilon}_2 \Gamma^{\nu} \epsilon_1 ) \varepsilon_{\mu \nu \lambda} D^\lambda X^J_e X^{JKL}_f X^{IKL}_g X^I_c \, d^{efg}{}_d f^{cdb}{}_a \\			
\nonumber
&+ i ( 2 f_{9} - 2 g_{15} ) ( \bar{\epsilon}_2 \Gamma^{IJ} \epsilon_1 ) D_\mu X^J_e X^{KLM}_f X^{KLM}_g X^I_c \, d^{efg}{}_d f^{cdb}{}_a \\	
\nonumber
&+ i ( - 2 g_{13} ) ( \bar{\epsilon}_2 \Gamma^{LM} \epsilon_1 ) D_\mu X^M_e X^{IJK}_f X^{JKL}_g  X^I_c \, d^{efg}{}_d f^{cdb}{}_a \\
\nonumber
&+ i ( - g_{14} ) ( \bar{\epsilon}_2 \Gamma^{LM} \epsilon_1 ) D_\mu X^J_e X^{IJK}_f X^{KLM}_g X^I_c \, d^{efg}{}_d f^{cdb}{}_a \\	
\nonumber
&+ i ( 2 f_{10} ) ( \bar{\epsilon}_2 \Gamma^{IJ} \epsilon_1 ) D_\mu X^M_e X^{JKL}_f X^{KLM}_g X^I_c \, d^{efg}{}_d f^{cdb}{}_a \\
\nonumber
&+ i ( 4 g_{13} ) ( \bar{\epsilon}_2 \Gamma^{KL} \epsilon_1 ) D_\mu X^M_e X^{IJK}_f X^{JLM}_g X^I_c \, d^{efg}{}_d f^{cdb}{}_a \\
\nonumber
&+ i ( - 3 g_9  ) ( \bar{\epsilon}_2 \Gamma^{JKLM} \Gamma_{\mu \nu} \epsilon_1 ) D^\nu X^I_e X^{JKN}_f X^{LMN}_g X^I_c \, d^{efg}{}_d f^{cdb}{}_a \\	
\nonumber
&+ i ( - 2 g_{10} ) ( \bar{\epsilon}_2 \Gamma^{JKLM} \Gamma_{\mu \nu} \epsilon_1 ) D^\nu X^J_e X^{IMN}_f X^{KLN}_g X^I_c \, d^{efg}{}_d f^{cdb}{}_a \\
\nonumber
&+ i ( - \sfrac{1}{3} g_{11} ) ( \bar{\epsilon}_2 \Gamma^{KLMN} \Gamma_{\mu \nu} \epsilon_1 ) D^\nu X^J_e X^{IJK}_e X^{LMN}_g X^I_c \, d^{efg}{}_d f^{cdb}{}_a \\	
\nonumber
&+ i ( - g_{10} ) ( \bar{\epsilon}_2 \Gamma^{JKLM} \Gamma_{\mu \nu} \epsilon_1 ) D^\nu X^N_e X^{IJK}_f X^{LMN}_g X^I_c \, d^{efg}{}_d f^{cdb}{}_a \\		
&+ i ( - \sfrac{1}{3} g_{13} ) ( \bar{\epsilon}_2  \Gamma^{JKLMNO} \epsilon_1 ) D_\mu X^O_e X^{IJK}_f X^{LMN}_g X^I_c \, d^{efg}{}_d f^{cdb}{}_a \, .	
\end{align}
The presence of two 3-brackets means we can use both \eref{Useful_Id_5} and \eref{Useful_Id_4} here. Using \eref{Useful_Id_5} we can show that the terms with four or more transverse $\Gamma$-matrix indices are all identically zero. After using \eref{Useful_Id_4} on the remaining terms we are left with
\begin{align}
\nonumber
T_{M2} \, (\delta_1 \delta_2' \tilde{A}_\mu{}^b{}_a &+ \delta'_1 \delta_2 \tilde{A}_\mu{}^b{}_a )_{1DX} - (1 \leftrightarrow 2) \\
	\nonumber
	=&+ i ( 2 f_{9} + \sfrac{2}{3} f_{10} + 2 g_9 + \sfrac{2}{3} g_{10} - 2 g_{15} ) ( \bar{\epsilon}_2 \Gamma^{\nu} \epsilon_1 ) \varepsilon_{\mu \nu \lambda} D^\lambda X^I_e X^{JKL}_f X^{JKL}_g X^I_c \, d^{efg}{}_d f^{cdb}{}_a \\	
	\nonumber
	&+ i ( 2 f_{9} - \sfrac{2}{3} g_{13} - 2 g_{15} ) ( \bar{\epsilon}_2 \Gamma^{IJ} \epsilon_1 ) D_\mu X^J_e X^{KLM}_f X^{KLM}_g X^I_c \, d^{efg}{}_d f^{cdb}{}_a \\
	&+ i ( 2 f_{10} + 2 g_{13} - g_{14} ) ( \bar{\epsilon}_2 \Gamma^{LM} \epsilon_1 ) D_\mu X^J_e X^{IJK}_f X^{KLM}_g X^I_c \, d^{efg}{}_d f^{cdb}{}_a \, .
\end{align}

\subsubsection{$0 \ DX$ terms}

The terms with feature no covariant derivatives are
\begin{align}
\nonumber
T_{M2} \big( ( \delta_1 \delta'_2 + \delta'_1 \delta_2 ) &\tilde{A}_{\mu}{}^b{}_a - ( 1 \leftrightarrow 2 ) \big)_{0DX} \\
\nonumber
=& -\sfrac{i}{6} g_{15} ( \bar{\epsilon}_2 \Gamma_{\mu} \Gamma^{I} \Gamma^{PQR} \epsilon_1 ) X^{PQR}_e X^{JKL}_f X^{JKL}_g X^I_c \, d^{efg}{}_d f^{cdb}{}_a \\
\nonumber
&+ i f_{11} ( \bar{\epsilon}_2 \Gamma_\mu \Gamma^I  \Gamma^{NOP} \epsilon_1 ) X^{JKL}_e X^{JKL}_f X^{NOP}_g X^I_c \, d^{efg}{}_d f^{cdb}{}_a \\
&- ( 1 \leftrightarrow 2 ) \\
\nonumber \\
=&+ i ( 2 f_{11} -\sfrac{1}{3} g_{15} ) ( \bar{\epsilon}_2 \Gamma_{\mu} \Gamma^{IJKL} \epsilon_1 ) X^{JKL}_e X^{MNO}_f X^{MNO}_g X^I_c \, d^{efg}{}_d f^{cdb}{}_a \, .		
\end{align}
Using \eref{Useful_Id_4} with $J_1 J_2 J_3 = JKL$ and then exploiting the antisymmetry in $IJKL$ shows that the above term is zero.

\end{appendices}

\clearpage
\addcontentsline{toc}{section}{References}

%
%


\bibliographystyle{utphys}
\bibliography{PhDThesis}


\end{spacing}

\end{document}